# PONTIFICIA UNIVERSIDAD CATÓLICA DEL PERÚ
## ESCUELA DE POSGRADO

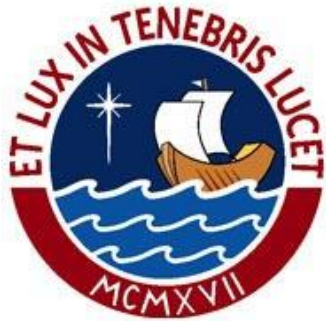

## SOLUCIÓN EXACTA PARA UN MODELO SIMPLIFICADO DE UN SISTEMA CUÁNTICO ABIERTO

Tesis para optar el grado académico de magíster en física

Presentado por:

Lic. Sotelo Bazan, Eduardo Franco

Asesor:

Dr. Castillo Egoavil, Hernan Alfredo

Jurado:

Dr. De Zela Martinez, Francisco Antonio

Dr. Massoni Kamimoto, Eduardo Ruben

Lima, Mayo 2021

DEDICATORIA:

A mis padres.

# AGRADECIMIENTOS

# RESUMEN

En este trabajo se desarrolló un modelo simplificado de un oscilador inicialmente excitado como un sistema cuántico interactuando con un gran número de osciladores como un reservorio. Todos estos osciladores están en su estado fundamental y sin acoplamientos entre sí, en el límite de acoplamiento débil entre el sistema y el reservorio. Este sistema podría ser un oscilador excitado en una micro cavidad que interactúa con el vacío del campo electromagnético a temperatura cero. El principal objetivo de este trabajo es obtener la solución exacta para la matriz de densidad del sistema en estas condiciones.

El planteamiento general consiste en calcular la evolución de todos los osciladores como una única entidad aislada mediante el operador $e^{-i\hat{H}t}$, donde $\hat{H}$ es el hamiltoniano total. Partiendo de un estado inicial total factorizable entre el sistema y el reservorio, la evolución es unitaria y se toma la traza parcial en los grados de libertad del entorno para obtener la matriz de densidad del sistema en cualquier instante del tiempo; este procedimiento requiere diagonalizar[1] $\hat{H}$.

Se desarrollan técnicas generales que pueden ser extendidas a versiones más elaboradas del modelo, se inicia con la descomposición del espacio de Hilbert total $\mathcal{H} = \mathcal{H}_0 \otimes \mathcal{H}_1 \otimes \cdots \mathcal{H}_N$, que es el producto tensorial de los subespacios de Hilbert de cada oscilador $\mathcal{H}_i$, en subespacios $\mathcal{H}(\Sigma)$ llamados *subespacio de número de excitación definido*, que corresponde al conjunto de todos los estados $|\vec{n}\rangle \in \mathcal{H}$ que tienen el mismo número de excitación colectiva $\Sigma$; cumpliéndose: $\mathcal{H} = \mathcal{H}(0) \oplus \mathcal{H}(1) \oplus \mathcal{H}(2) \cdots \oplus \mathcal{H}(N+1)$, donde $N$ es el número de osciladores del entorno. Se introducen diagramas compuestos de nodos y flechas para representar la acción del hamiltoniano en cada subespacio $\mathcal{H}(\Sigma)$. Se plantea una notación para trabajar en estos subespacios y calcular la sumatoria asociada a la traza parcial.

Los resultados son evaluados para un reservorio de $N = 1000$ osciladores, valores particulares de la fuerza de acoplamiento y orden óhmico de la densidad espectral, contrastados con la correspondiente solución markoviana, descrita en la sección [2.3.1].

---

[1] La diagonalización del Hamiltoniano es realizada en el Anexo B.

# ABSTRACT

A simplified model of an initially excited oscillator as a quantum system interacting with a large number of oscillators acting as a reservoir has been developed in this work. All these oscillators are in their ground state uncoupled each other and at the limit of the weak coupling between the system and the reservoir. This system could be an oscillator excited in a microcavity that interacts with the vacuum's electromagnetic field at zero temperature. This work's primary goal is to obtain the system's density matrix's exact solution in these conditions.

The general approach calculates all oscillators' evolution as a single isolated entity using the operator $e^{-i\hat{H}t}$, when $\hat{H}$ is the total hamiltonian. Starting from a total initial state that can be factored between the system and the reservoir, the evolution is unitary, and the partial trace is taken in all the degrees of freedom of the environment to obtain the density matrix of the system at any instant of time; this procedure requires diagonalizing[2] $\hat{H}$.

General techniques are developed and would be extended to more elaborate versions of the model, starting with the decomposition of the total Hilbert space $\mathcal{H} = \mathcal{H}_0 \otimes \mathcal{H}_1 \otimes \cdots \mathcal{H}_N$, which is the tensor product of the Hilbert subspaces of each oscillator $\mathcal{H}_i$, into subspaces $\mathcal{H}(\Sigma)$ called a *subspace of defined excitation number*, which corresponds to the set of all states $|\vec{n}\rangle \in \mathcal{H}$ that have the same collective excitation number $\Sigma$; being fulfilling: $\mathcal{H} = \mathcal{H}(0) \oplus \mathcal{H}(1) \oplus \mathcal{H}(2) \cdots \oplus \mathcal{H}(N+1)$, $N$ is the environment's oscillators number. Diagrams composed of nodes and arrows are introduced to represent the Hamiltonian's action in each subspace $\mathcal{H}(\Sigma)$. A notation is also proposed to work on these subspaces and calculate the sum associated with the partial trace.

The results are evaluated for a reservoir of $N = 1000$ oscillators, particular values of the coupling force, and ohmic order of the spectral density, contrasted with the corresponding Markovian solution described in section [2.3.1].

---

[2] The diagonalization of the hamiltonian is performed in Annex B.

# Índice

# LISTA DE FIGURAS

# LISTA DE TABLAS

# CAPÍTULO I

# INTRODUCCIÓN

**1.1 Perspectiva crítica sobre la preservación de la unitaridad en la irreversibilidad de los sistemas abiertos**

El problema de la irreversibilidad cuántica es tratado en la mecánica cuántica de sistemas abiertos partiendo que un microsistema aislado evoluciona unitariamente, pero tras su interacción con el entorno (como un reservorio fuera del equilibrio, un baño termal, o un aparato de medición), es todo el sistema aislado *microsistema-entorno* el que evoluciona unitariamente, mientras que el microsistema abierto evoluciona irreversiblemente, por ejemplo, de acuerdo a la ecuación maestra de Lindblad [2], [3], [5], [6], ver sección [2.3.1].

Sin embargo, cuando el entorno puede contener sistemas clásicos tales como aparatos de medición, es posible asumir una perspectiva crítica sobre el paradigma establecido de que la unitaridad es preservada también en este caso. La obtención de la ecuación maestra de Lindblad no está basada únicamente en considerar la evolución unitaria de todo el sistema general microsistema-entorno y luego tomar la traza parcial en los grados de libertad del entorno para obtener una ecuación maestra que nos describa la evolución del microsistema abierto; adicionalmente se hacen consideraciones o aproximaciones, en el caso de la ecuación maestra de Lindblad, son la aproximación de *Born, Markov y Rotating wave* [3], [5], [6], que no son derivables[3] de la unitaridad cuántica, y en consecuencia sus incorporaciones vician el razonamiento de la irreversibilidad de la dinámica en el microsistema abierto debido únicamente a la unitaridad del sistema global *microsistema-entorno*; estas aproximaciones pierden información, contrario a la preservación de la unitaridad[4].

Finalmente, a diferencia de la conservación de la energía o de la cantidad de movimiento, la conservación de la información[5] [9], [10], el determinismo o la preservación de la unitaridad o reversibilidad no han sido demostrados experimentalmente para sistemas de muchas partículas o grados de libertad, debido a la imposibilidad práctica de realizar mediciones sobre las predicciones exactas para este tipo de sistemas globales, que incluyen el entorno con muchos grados de libertad; en cambio, se acepta o asume que la unitaridad se sigue cumpliendo en el sistema total; así, es posible que realmente la unitaridad no sea preservada en la realidad física, aunque si en la descripción teórica, así por ejemplo, serían las aproximaciones markovianas las que "corregirían" la preservación teórica de la unitaridad, para la consistencia experimental.

---

[3] En el sentido que son introducidas para la obtención de la dinámica markoviana.
[4] Se pierde la información de las correlaciones entre el sistema y el reservorio, ver sección [2.3.1].
[5] Los teoremas de *no*-cloning y *no-deleting* proveen la conservación de la información [9]; el teorema de *no-hidding* aborda el problema de la pérdida de información [10].

### 1.1.1 Controversia sobre el postulado del colapso

Las implicancias de la discusión de la sección precedente sobre la preservación, o no, de la unitaridad en la escala macroscópica, donde se encuentran los sistemas clásicos[6] que emergen al aumentar los grados de libertad en el entorno, como parte de la transición al mundo clásico, como una sola unidad aislada, tiene implicancia sobre los fundamentos de la mecánica cuántica: acerca de si el postulado del colapso lo es realmente o, al contrario, si fuese un teorema, derivable del resto de postulados [11]. Si en el acto de medición, el sistema total aislado *microsistema-aparato de medición-entorno* evoluciona unitariamente, cualquier estado posterior en el tiempo estaría determinado a partir de las condiciones iniciales generales, el resultado de una medición sería predecible si se conoce la información completa al inicio, entonces el colapso del estado cuántico sería derivado de la evolución unitaria total, y dejaría de ser un postulado para ser un teorema; esta situación se puede expresar matemáticamente como sigue: sean $|\psi\rangle$ y $|A\rangle$ los estados de un microsistema y el aparato de medición al tiempo $t = 0$ inicialmente separables, al tiempo $t_0 > 0$ se realiza el acto de medición e inmediatamente después el sistema queda en el autoestado $|a_n\rangle$ de un observable, a la vez que el aparato de medición evoluciona a $|A_n\rangle$; sea $\hat{H}$ el hamiltoniano del sistema global *microsistema-aparato de medición* como una sola entidad aislada, entonces debe cumplirse que:

$$\lim_{t \to t_0^+} e^{-i\hat{H}t} |\psi\rangle|A\rangle = |a_n\rangle|A_n\rangle \qquad (1.1)$$

---

[6] Todos los sistemas clásicos están compuestos de muchos microsistemas regidos por las leyes de la mecánica cuántica; la evolución de un sistema global que incluye sistemas clásicos es también un problema cuántico de muchos microsistemas, y la preservación, o no, de la unitaridad en él es una discusión legítima.

Es importante precisar que el postulado del colapso va a asociado con el postulado de la regla de Born, de manera que la preservación de la unitaridad global requiere demostrar matemáticamente no solo la ecuación (1.1), sino también que la probabilidad $\wp_n$ que el microsistema evolucione parcialmente de $|\psi\rangle$ a $|a_n\rangle$ para una configuración aleatoria del microestado de $|A\rangle$ para un mismo macroestado[7], es $\wp_n = |\langle a_n|\psi\rangle|^2$. Por el contrario, si el colapso del estado cuántico y la regla de Born no pueden ser derivados de los postulados restantes, sino que deben ser introducidos como otros postulados, entonces son fenómenos físicos incompatibles con la evolución unitaria global[8].

**1.2 El uso de un modelo simplificado para evaluar la unitaridad**

La discusión sobre los fundamentos de la mecánica cuántica está relacionada con el estudio de la mecánica cuántica misma como objeto de estudio; esto es, estudiar cómo esta teoría, con sus principios, postulados, y herramientas matemáticas, funciona para explicar el mundo microscópico y su extensión o transición al comportamiento clásico, a decir, qué resultados predice o conduce la teoría cuando aborda un sistema con muchas partículas; para este fin se puede emplear un modelo simplificado: un objeto teórico mínimo y suficiente para evaluar el efecto de preservar la unitaridad en el sistema global en la evolución del sistema bajo estudio; para esto se requiere que el modelo sea resoluble exactamente, sin aproximaciones para compararlo con la correspondiente solución markoviana para el mismo problema, y evidenciar el efecto de dichas aproximaciones.

Este modelo simplificado consiste en muchos osciladores cuánticos acoplados, uno de ellos se toma como el sistema bajo estudio y el resto forman su entorno, iniciando con una distribución de energía: el sistema excitado, y el entorno en su estado fundamental. Con el paso del tiempo los osciladores arriban a un estado de equilibrio global (debería ser el equilibrio térmico en consistencia con la teoría clásica), y se evalúa si la evolución del sistema se aproxima a la dinámica markoviana para el límite de acoples débiles.

---

[7] El macroestado del aparato de medida registra el resultado de la medición, pero siempre se tiene una incertidumbre de su microestado. Al momento de la observación la incertidumbre del microestado haría de fuente de ruido para la evolución del microsistema, provocando el indeterminismo (desde la perspectiva del observador) en el resultado de su medida, pero la cual es determinista si se conoce el microestado.

[8] La cuestión sobre si el colapso del estado cuántico es o no un postulado no es trivial, ni un asunto filosófico fuera de la teoría física, su consistencia elemental, y su matemática; así, por ejemplo, la discusión sobre si el 5to postulado de Euclides era o no un postulado, conllevó al desarrollo de las geometrías no euclidianas.

# CAPÍTULO II
# REVISIÓN DEL MARCO TEÓRICO

## 2.1 Operadores y superoperadores

### 2.1.1 Operadores en mecánica cuántica

En la mecánica cuántica solo se emplean operadores lineales, los cuales al actuar sobre un ket $|\psi\rangle \in \mathcal{H}$ lo transforman en otro ket del mismo espacio de Hilbert $\mathcal{H}$; la linealidad de estos operadores se muestra a continuación (ecuación izquierda):

$$\hat{A}_j\left(\sum_i \beta_i |\psi_i\rangle\right) = \sum_i \beta_i \hat{A}_j |\psi_i\rangle \quad \left(\sum_j \alpha_j \hat{A}_j\right)|\psi_i\rangle = \sum_j \alpha_j \hat{A}_j |\psi_i\rangle \quad \begin{array}{l} \alpha_j, \beta_i \in \mathbb{C} \\ |\psi_i\rangle, \hat{A}_j|\psi_i\rangle \in \mathcal{H} \end{array} \quad (2.1)$$

Donde $\alpha_i$ y $\beta_i$ son escalares complejos, la ecuación del centro de (2.1) corresponde a la definición de la suma de operadores, los cuales podrían ser no lineales; dentro del conjunto de los operadores lineales se encuentran los operadores unitarios, normales, hermíticos, proyectores, positivos e identidad, como se muestran en la Figura 2.1:

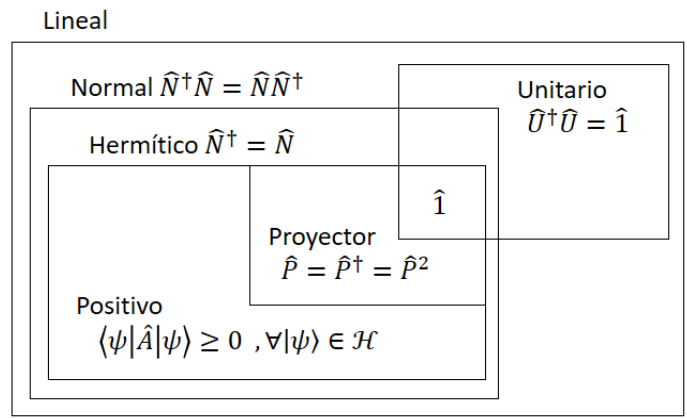

**Figura 2.1** Intersección de conjuntos de los tipos de operadores. Fuente: *J. Audretsch, Entangled Systems: New Directions in Quantum Physics* [4].

La definición de estos operadores se muestra en la Tabla 2.1:

| Tipo de operador $\hat{A}$ | Definición | $\hat{A} = \sum_k \hat{P}_k \lambda_k$ |
|---|---|---|
| Normal | $\hat{A} = \sum_i |\psi_i\rangle\langle\psi_i|\lambda_i \quad \lambda_i \in \mathbb{C}$ | $\lambda_k \in \mathbb{C}$ |
| Unitario | $\hat{A}^\dagger \hat{A} = \hat{1}$ | $\lambda_k = e^{i\alpha_k}$ |
| Hermítico | $\hat{A} = \sum_i \hat{P}_i \lambda_i \quad \lambda_i \in \mathbb{R}$ | $\lambda_k \in \mathbb{R}$ |
| Positivo | $\langle\varphi|\hat{A}|\varphi\rangle \geq 0 \quad \forall |\varphi\rangle \in \mathcal{H}$ | $\lambda_k \geq 0$ |
| Proyector | $\hat{P}_i = |\psi_i\rangle\langle\psi_i|$ | $\lambda_k \in \{0\,;\,1\}$ |
| Identidad | $\hat{A} = \sum_i \hat{P}_i$ | $\lambda_k = 1$ |

**Tabla 2.1** Comparación de los distintos tipos de operadores, en la segunda columna se tiene su definición general, en la tercera columna se tiene la definición de sus autovalores [4].

De acuerdo a la Figura 2.1, los operadores normales son todos aquellos que se pueden diagonalizar; es decir, que tienen autovalores no nulos, si estos autovalores son reales entonces se tienen los operadores hermíticos, y si estos autovalores son positivos entonces el operador es positivo, si estos autovalores solo puede tomar los valores { 0 , 1 } entonces el operador es un proyector, y si todos sus autovalores son la unidad, entonces este operador es la identidad, por otro lado, si el operador es normal, pero sus autovalores no son reales, sino de la forma $e^{i\alpha}$ entonces viene a ser un operador unitario [4].

Es posible construir funciones de operadores $f(\hat{A})$ que a su vez son otros operadores $\hat{f}$, al expresar la función en una serie de potencias, cuyos autovalores son también la función evaluada en los correspondientes autovalores y autovectores, como se muestra:

$$f(x) = \sum_{n=0}^{\infty} c_n x^n \qquad c_i \in \mathbb{C} \qquad \hat{B} = f(\hat{A}) = \sum_{n=0}^{\infty} c_n \hat{A}^n \qquad (2.2)$$

$$\hat{A}|\psi_i\rangle = a_i|\psi_i\rangle \qquad\qquad \hat{B}|\psi_i\rangle = f(a_i)|\psi_i\rangle$$

Es posible concebir el espacio vectorial complejo de Liouville $\mathbb{L}$ cuyos elementos $|A)$, $|B)$, $|C)$, $\cdots \in \mathbb{L}$ son los operadores $\hat{A}$, $\hat{B}$, $\hat{C}$ que actúan en $\mathcal{H}$; se define el producto escalar como la traza de un operador con la adjunta del otro [4]:

$$(A|B) = Tr[\hat{A}^\dagger \hat{B}] \qquad\qquad (A_i|A_j) = \delta_{ij} \qquad (2.3)$$

A la derecha de (2.3) se tiene que los operadores $\hat{A}_i$ forman una base ortonormal en $\mathbb{L}$.

### 2.1.2 Operador de densidad

Los kets $|\psi\rangle$, elementos del espacio de Hilbert, permiten representar la superposición cuántica; esto es, un vector complejo normado (cuyas componentes corresponden a cada autoestado de un observable) que contiene toda la información del sistema físico, y que por lo tanto no se desconoce o ignora alguna información sobre él, y no se necesita asociar alguna distribución de probabilidad para modelar nuestra ignorancia del sistema, esto corresponde a un estado puro $|\psi\rangle\langle\psi|$; sin embargo, los kets no permiten representar una ignorancia sobre el estado del sistema, a decir, asociar una distribución de probabilidad sobre qué ket o estado cuántico podría tener el sistema, a esto se le llama una mezcla (estadística o probabilística) de estados; así, se puede construir un operador probabilístico que representa la ignorancia o incertidumbre sobre la información o el estado del sistema: sea un sistema bajo estudio del cual se ignora su estado cuántico exacto, solo se sabe que podría estar en el estado $|\psi_i\rangle$ con probabilidad $p_i$, entonces el estado mezcla viene a ser representado por el operador probabilístico siguiente [4]:

$$\hat{\rho} = \sum_i p_i|\psi_i\rangle\langle\psi_i| \qquad\qquad \langle\hat{A}\rangle = Tr[\hat{\rho}\hat{A}] \qquad (2.4)$$

$\hat{\rho}$ es un estado puro solo si se puede escribir en la forma $\hat{\rho} = |\varphi\rangle\langle\varphi|$ que corresponde a un operador proyección, lo cual conduce a la idempotencia $\hat{\rho}^2 = \hat{\rho}$, siendo $\hat{\rho}^2 < \hat{\rho}$ para el caso mixto; $\hat{\rho}$ es hermítico, positivo y de traza igual a la unidad [4]; el lado derecho de (2.4) establece el valor medio de un observable $\hat{A}$ como la traza de $\hat{\rho}\hat{A}$, y para $\hat{A} = \hat{1}$ se obtiene que $\hat{\rho}$ tiene traza igual a la unidad, en este caso el valor medio de $\hat{1}$ concuerda con la exigencia de que $\hat{\rho}$ tiene traza igual a la unidad.

Desde que el operador de densidad recoge la ignorancia sobre el estado del sistema, se emplea la entropía de Von Neumann para medir dicha ignorancia, siendo nula para un estado puro (pues, por definición no se ignora algo sobre el sistema) y aumentando conforme se acentúa la mezcla [6]:

$$S(\hat{\rho}) = -Tr[\,\hat{\rho}\ln\hat{\rho}\,] \qquad S(\widehat{U}\hat{\rho}\widehat{U}^\dagger) = S(\hat{\rho}) \qquad (2.5)$$

La entropía $S(\hat{\rho})$ es invariante ante transformaciones unitarias de $\hat{\rho}$, de manera que no importa la base elegida de $\hat{\rho}$; la entropía de Von Neumann es la versión cuántica de la entropía de Shannon en nats $H = -\sum_k p_k \ln p_k$ la cual está definida solo para una distribución de probabilidad clásica $\{\,p_k\,\}$, de hecho, ambas coinciden cuando los estados puros de (2.4) son ortogonales $\langle\psi_i|\psi_j\rangle = \delta_{ij} \rightarrow S(\hat{\rho}) = H(p_i)$.

### 2.1.3 Superoperadores en mecánica cuántica

Los superoperadores son operadores lineales en el espacio de Liouville $\mathbb{L}$ que actúan en los elementos $|A)$, $\cdots$, transformándolos dentro del espacio $\mathbb{L}$, o equivalentemente, son operadores que transforman un operador en otro que actúan en el mismo espacio $\mathcal{H}$; sea un superoperador $L$, en general debe ser lineal [4]:

$$L: \hat{A}_i \rightarrow \hat{B}_i \qquad \hat{A}_i, \hat{B}_i \in \mathbb{L} \qquad L\cdot\left(\sum_i \alpha_i \hat{A}_i\right) = \sum_i \alpha_i\, L\cdot\hat{A}_i \qquad \alpha_i \in \mathbb{C} \qquad (2.6)$$

Es posible encontrar una forma general para la expresión del superoperador en términos de suma y multiplicación de operadores $\widehat{D}_i$, $\hat{I}_i$ y escalares complejos $\beta_i$:

$$L\cdot\hat{A} = \sum_i \beta_i\, \hat{I}_i \cdot \hat{A} \cdot \widehat{D}_i = \hat{B} \qquad \beta_i \in \mathbb{C} \qquad \widehat{D}_i, \hat{I}_i, \hat{A}, \hat{B} \in \mathbb{L} \qquad (2.7)$$

Los operadores $\hat{I}_i$ y $\widehat{D}_i$ son operadores que actúan por la "izquierda" y "derecha" de $\hat{A}$ respectivamente, esta es la forma más general, pero se pueden elegir los operadores $\hat{I}_i = \hat{1}$ o $\hat{D}_i = \hat{1}$ para quedarse con términos en $L \cdot \hat{A}$ donde solo actúan sobre $\hat{A}$ por la derecha o por la izquierda respectivamente; por ejemplo, el conmutador del hamiltoniano con cualquier operador es un superoperador: $L = [\hat{H},] \rightarrow L \cdot \hat{A} = \hat{H} \cdot \hat{A} - \hat{A} \cdot \hat{H}$.

Al igual que los operadores, la potencia de un superoperador es un nuevo superoperador:

$$L^n = L \cdot L \cdots L \ (n \text{ veces}) \qquad L^n: \hat{A} \rightarrow \hat{X} \qquad \hat{A}, \hat{X} \in \mathbb{L} \qquad (2.8)$$

Por ejemplo, las primeras potencias de $L = [\hat{H},]$ en $\hat{A}$ viene a ser:

$$L^0 \cdot \hat{A} = \hat{A} \qquad\qquad L^1 \cdot \hat{A} = [\hat{H}, \hat{A}]$$
$$L^2 \cdot \hat{A} = \left[\hat{H}, [\hat{H}, \hat{A}]\right] \qquad\qquad L^3 \cdot \hat{A} = \left[\hat{H}, \left[\hat{H}, [\hat{H}, \hat{A}]\right]\right] \qquad (2.9)$$

En el caso de $L^0$ es evidente que es el superoperador identidad.

Desde que se puede definir potencias del superoperador, es posible definir una función de un superoperador como un desarrollo en serie de potencias:

$$f(L) = \sum_{n=0}^{\infty} c_i L^n \qquad f(L): \hat{A} \rightarrow \hat{P} \qquad c_i \in \mathbb{C} \qquad \hat{A}, \hat{P} \in \mathbb{L} \qquad (2.10)$$

Por ejemplo, la exponencial de $L = [\hat{H},]$ viene a ser la suma como se muestra:

$$e^{sL} = e^{s[\hat{H},]} = \hat{1} + s[\hat{H},] + \frac{s^2}{2}\left[\hat{H}, [\hat{H},]\right] + \frac{s^3}{3!}\left[\hat{H}, \left[\hat{H}, [\hat{H},]\right]\right] + \cdots \qquad (2.11)$$

Al igual que los operadores cumplen una ecuación de autovalores, los superoperadores también cumplen una ecuación de autovalores, y, en consecuencia, una función de ese superoperador también:

$$L \cdot \hat{A} = l\, \hat{A} \qquad\qquad f(L) \cdot \hat{A} = f(l) \cdot \hat{A} \qquad\qquad l \in \mathbb{C} \qquad (2.12)$$

En general, las propiedades de los operadores como la adjunta, la hermiticidad, positividad y unitaridad pueden ser extendidas también a los superoperadores [4].

## 2.2 Evolución temporal mediante superoperadores

### 2.2.1 Superoperadores en la evolución unitaria, grupos

Los superoperadores pueden representar la evolución unitaria del operador de densidad, generalizando la evolución unitaria del ket de estado a las mezclas estadísticas; para el ket de estado, la evolución temporal está dado por la ecuación de Schrödinger, y puede ser representado por la acción del operador de evolución $\hat{U}(t)$:

$$\hat{U}(t) \cdot |\psi(t_0)\rangle = |\psi(t_0 + t)\rangle \qquad i\hbar \frac{d}{dt}|\psi(t)\rangle = \hat{H}(t)|\psi(t)\rangle \qquad (2.13)$$

(2.13) permite obtener la forma diferencial del operador de evolución $\hat{U}(t)$, en general para cualquier hamiltoniano dependiente del tiempo $\hat{H}(t)$:

$$\hat{U}(t + dt) = \left(\hat{1} - \frac{i}{\hbar}\hat{H}(t)dt\right) \cdot \hat{U}(t) \qquad (2.14)$$

Los operadores de evolución forman un grupo $(U, \cdot)$ con $U = \{\hat{U}(t), t \in \mathbb{R}\}$ al cumplir las propiedades generales de grupo: asociatividad $\hat{U}(t_1) \cdot \hat{U}(t_2) \in U$, existencia del elemento neutro $\hat{U}(0) = \hat{1}$ e inverso $\hat{U}^{-1}(t_1) = \hat{U}(-t_1)$; estos operadores $\hat{U}(t)$ pueden ser expresado en términos de su generador (cuando el Hamiltoniano no depende del tiempo), $\hat{U}(t) = \exp[-(i/\hbar)\hat{H}t]$, donde $-(i/\hbar)\hat{H}$ es el generador del grupo $(U, \cdot)$, y el tiempo $t \in \mathbb{R}$ es el parámetro de la "rotación" en el espacio de Hilbert, que corresponde a la evolución temporal.

La generalización de (2.13) a estados mezclas se hace mediante el uso del operador de densidad $\hat{\rho}(t)$, y viene a ser la ecuación de Von Neumann:

$$\hat{U}(t) \cdot \hat{\rho}(t_0) \cdot \hat{U}^{\dagger}(t) = \hat{\rho}(t_0 + t) \qquad \frac{d}{dt}\hat{\rho}(t) = -\frac{i}{\hbar}[\hat{H}, \hat{\rho}(t)] \qquad (2.15)$$

(2.15) se puede representar con los superoperadores $\Lambda_t^U$ y $\mathcal{L}_U$:

$$\Lambda_t^U \cdot \hat{\rho}(t_0) = \hat{U}(t) \cdot \hat{\rho}(t_0) \cdot \hat{U}^{\dagger}(t) \qquad \mathcal{L}_U = -\frac{i}{\hbar}[\hat{H}, \ ] \qquad (2.16)$$

$$\Lambda_t^U \cdot \hat{\rho}(t_0) = \hat{\rho}(t_0 + t) \qquad \qquad \frac{d}{dt}\hat{\rho}(t) = \mathcal{L}_U \cdot \hat{\rho}(t)$$

(2.16) permite generalizar (2.14) para estados descritos por operadores de densidad[9]:

$$\Lambda_{t+\delta t}^U = (1 + \mathcal{L}_U \delta t) \cdot \Lambda_t^U \;\rightarrow\; \mathcal{L}_U = \frac{\Lambda_{\delta t}^U - 1}{\delta t} \qquad (2.17)$$

Donde $1 = \Lambda_0^U$ es el superoperador identidad. Es posible establecer un grupo ($\Lambda^U$, $\cdot$) donde $\Lambda^U = \{\Lambda_t^U, t \in \mathbb{R}\}$ es el conjunto de los superoperadores de evolución:

$$\Lambda_{t_1}^U \cdot \Lambda_{t_2}^U \cdot \Lambda_{t_3}^U = \left(\Lambda_{t_1}^U \cdot \Lambda_{t_2}^U\right) \cdot \Lambda_{t_3}^U = \Lambda_{t_1}^U \cdot \left(\Lambda_{t_2}^U \cdot \Lambda_{t_3}^U\right) = \Lambda_{t_1+t_2+t_3}^U \qquad \forall \Lambda_{t_i}^U \in \Lambda^U$$

$$1 \cdot \Lambda_t^U = \Lambda_t^U \cdot 1 = \Lambda_t^U \qquad 1 = \Lambda_0^U \in \Lambda^U \qquad \forall \Lambda_t^U \in \Lambda^U \qquad (2.18)$$

$$(\Lambda_t^U)^{-1} \cdot \Lambda_t^U = \Lambda_t^U \cdot (\Lambda_t^U)^{-1} = 1 \qquad (\Lambda_t^U)^{-1} = \Lambda_{-t}^U \in \Lambda^U \qquad \forall \Lambda_t^U \in \Lambda^U$$

Si el hamiltoniano no depende del tiempo, en (2.16) $\mathcal{L}_U$ tampoco depende del tiempo y (2.17) puede resolverse como la exponencial $\Lambda_t^U = e^{\mathcal{L}_U t}$, donde $\mathcal{L}_U$ es el generador del grupo ($\Lambda^U$, $\cdot$), con el tiempo $t \in \mathbb{R}$ como parámetro de la rotación.

Es importante resaltar que debido a que la evolución unitaria es reversible, los elementos $\widehat{U}(t)$ o $\Lambda_t^U$ siempre tienen inversa, y es requerido el elemento neutro o identidad, y por eso siempre forman un grupo.

La entropía de Von Neumann $S(\hat{\rho})$ se preserva frente a las transformaciones unitarias $\hat{\rho} \rightarrow \widehat{U}\hat{\rho}\widehat{U}^\dagger$, lo que conduce a la conservación de la entropía tras la acción del superoperador $\Lambda_t^U$:

$$S(\Lambda_t^U \cdot \hat{\rho}) = S(\hat{\rho}) \qquad \qquad \forall t \in \mathbb{R} \qquad (2.19)$$

Así, la conservación de la entropía en (2.19) significa la conservación de la información (o la ignorancia) de $\hat{\rho}$ bajo la acción de $\Lambda_t^U$, lo cual es también equivalente a la reversibilidad[10] de $\Lambda_t^U$.

---

[9] Para obtener la ecuación de la derecha de (2.17) se hace $t \rightarrow 0$, quedando solo un diferencial $\delta t$ no nulo.
[10] $\Lambda_t^U$ es reversible porque conserva la información, $\Lambda_t^U \cdot \hat{\rho}$ contiene la misma información que $\hat{\rho}$ (solo que transformado), lo que permite obtener/conocer nuevamente $\hat{\rho}$ a partir de $\Lambda_t^U \cdot \hat{\rho}$, de lo contrario (si se

### 2.2.2 Superoperadores en la evolución irreversible, semigrupos

Cuando un microsistema interactúa difusivamente con su entorno, como en la termalización, su evolución temporal es irreversible y no unitaria[11] [2], [3], [6], y ya no puede ser descrita mediante los superoperadores $\Lambda_t^U$, por el contrario la dinámica del sistema es irreversible, lo que conlleva a una descripción de la evolución temporal solo "hacia adelante" en el tiempo, pero no "hacia atrás"[12]; es decir, se deben usar nuevos superoperadores $\Lambda_t$ que ya no pueden formar un grupo porque su inversa temporal (hacer negativo el parámetro tiempo) no puede ser considerada como parte de la descripción física del sistema bajo estudio[13], así, como los superoperadores solo se pueden componer $\Lambda_{t_1} \cdot \Lambda_{t_2} \cdots$ para dar el superoperador $\Lambda_{t_1+t_2+\cdots}$, para valores positivos del parámetro tiempo $t_k$, el nuevo conjunto de superoperadores $\Lambda = \{\Lambda_t, t > 0\}$ forman un semigrupo $(\Lambda, \cdot)$ [6].

A la condición de que los $\Lambda_t$ forman un semigrupo, se añade la positividad (que el nuevo operador de densidad $\hat{\rho}(t) = \Lambda_t \cdot \hat{\rho}(0)$ también es positivo) y la preservación de la traza $Tr[\hat{\rho}(t)] = Tr[\hat{\rho}(0)]$.

La relación entre $\Lambda_t^U$ y $\mathcal{L}_U$ de (2.16) se puede generalizar a esta dinámica irreversible, expresando $\mathcal{L}$ como la derivada temporal de $\Lambda_t$, como se puede ver en la referencia [1]:

$$\mathcal{L}(t) \cdot \Lambda_t = \lim_{\delta t \to 0^+} \frac{\Lambda_{t+\delta t} - \Lambda_t}{\delta t} \qquad \forall t > 0 \qquad (2.20)$$

$\mathcal{L}$ viene a ser el generador del semigrupo, si se hace actuar (2.20) en el operador probabilístico $\hat{\rho}(0)$ entonces se genera un ecuación diferencial sobre la dinámica irreversible del sistema:

---

perdiera información) no se podría recuperar el estado inicial; aquí es posible establecer las equivalencias entre conservación de información, reversibilidad y determinismo.

[11] Aunque, en principio, la evolución de todo el sistema general aislado microsistema-entorno sigue evolucionando unitariamente, y es solo el microsistema que de forma parcial es irreversible y no-unitario; la asunción de que el sistema global, que incluye sistemas clásicos-macroscópicos como el aparato de medición, evoluciona unitariamente viene a ser en realidad un supuesto, que es discutido en la introducción, sección [1.1], de este trabajo.

[12] Una teoría de procesos irreversibles solo puede describir el estado futuro a partir del estado pasado, pero no lo contrario, por ejemplo, tras la información del estado de equilibrio térmico de un sistema, no se puede describir el estado inicial fuera del equilibrio del que inició.

[13] Estos procesos difusivos pierden información del estado inicial del sistema, de manera que solo tiene sentido emplear superoperadores que avanzan en el tiempo ($t > 0$), pero no que retroceden en el tiempo ($t < 0$).

$$\frac{d}{dt}\hat{\rho}(t) = \mathcal{L}(t) \cdot \hat{\rho}(t) \qquad \forall t > 0 \qquad (2.21)$$

La solución general de (2.21) es como se muestra, usando el *time-ordering* $\mathcal{T}$ [6]:

$$\Lambda_t = \mathcal{T}\left\{\exp\int_0^t ds\, \mathcal{L}(s)\right\} \qquad \mathcal{T}\left\{\prod_{i=1}^n \mathcal{L}(s_i)\right\} = \prod_{j=1}^n \mathcal{L}(\tilde{s}_j) \qquad \forall t > 0 \qquad (2.22)$$

$$\{s_i\} \to \{\tilde{s}_1 > \tilde{s}_2 > \cdots \tilde{s}_n\}$$

Donde $\mathcal{T}\{\cdots\}$ ordena los $\mathcal{L}(s_i)$ como el producto de superoperadores de mayor a menor argumento $s_i$ (el conjunto $\{s_i\}$ es transformado en otro conjunto $\{\tilde{s}_1 > \tilde{s}_2 > \cdots \tilde{s}_n\}$ con los mismos elementos, pero con los índices ordenados apropiadamente para el funcionamiento del ordenamiento temporal); si $\mathcal{L}$ no depende del tiempo, se puede integrar[14] $\mathcal{L}(t)$ o simplificar (2.22), dando la relación entre $\Lambda_t$ y $\mathcal{L}$:

$$\Lambda_t = e^{t\mathcal{L}} \qquad \dot{\mathcal{L}} = 0 \qquad \forall t > 0 \qquad (2.23)$$

De esta manera la evolución temporal es $\hat{\rho}(t) = e^{t\mathcal{L}} \cdot \hat{\rho}(0)$ ; en general, $\mathcal{L}$ puede ser expresado como la suma de la parte puramente unitaria $\mathcal{L}_U = -\frac{i}{\hbar}[\hat{H},\ ]$ de (2.16) y puramente difusiva $\mathcal{L}_D$:

$$\mathcal{L} = -\frac{i}{\hbar}[\hat{H},\ ] + \mathcal{L}_D \qquad \forall t > 0 \qquad (2.24)$$

La dinámica general irreversible aumenta la entropía de Von Newman (2.5), pues se pierde la coherencia del estado puro inicial:

$$S(\Lambda_t \cdot \hat{\rho}) \geq S(\hat{\rho}) \qquad \forall t > 0 \qquad (2.25)$$

Donde la condición de igualdad se cumple para $\mathcal{L}_D = 0$ en (2.24), o cuando $\hat{\rho}$ arriba a un estado mezcla de equilibrio $\hat{\rho}_\infty$, entonces es invariante a la acción de $\Lambda_t$ y pueden cumplir una ecuación de autovalores (2.12) con autovalor uno:

---

[14] Se puede verificar que el desarrollo en serie de potencias del exponencial $\Lambda_t = e^{t\mathcal{L}}$ satisface (2.21) solo cuando el superoperador $\mathcal{L}$ no depende del tiempo.

$$\Lambda_t \cdot \hat{\rho}_\infty = \hat{\rho}_\infty \qquad\qquad \forall t > 0 \qquad (2.26)$$

De acuerdo a (2.24), si $\mathcal{L}$ no depende del tiempo, entonces $e^{t\mathcal{L}} \cdot \hat{\rho}_\infty = \hat{\rho}_\infty$, y se tiene:

$$\mathcal{L} \cdot \hat{\rho}_\infty = 0 \qquad\qquad \dot{\mathcal{L}} = 0 \qquad (2.27)$$

(2.27) es la condición para encontrar estados mezcla de equilibrios o convergencia temporal de (2.21) cuando $\mathcal{L}$ no depende del tiempo.

### 2.2.3 Dinámica de semigrupos completamente positivos y que preservan la traza

El superoperador $\Lambda_t$ debe formar un semigrupo para representar la dinámica irreversible del sistema, pero además debe preservar la traza y la positividad del operador de densidad sobre el que actúa; sin embargo, es necesario establecer una condición más fuerte que preservar la positividad, $\Lambda_t$ debe hacer un mapeo completamente positivo; esto es, el superoperador $\Lambda_t$ extendido al producto tensorial $\Lambda_t \otimes 1_n$ donde $1_n$ actúa en el espacio de Hilbert de un entorno arbitrario con $n$ grados de libertad, realiza a su vez un mapeo positivo para cualquier $n$; esta condición "más fuerte" de positividad es requerido por que es posible la existencia de mapeos positivos que no son completamente positivos [1]; así, si el superoperador $\Lambda_t$ es extensible a mantener $\hat{\rho}_{s+\varepsilon} \in \mathcal{H}_s \otimes \mathcal{H}_\varepsilon$ siempre positivo, con $\mathcal{H}_s$ y $\mathcal{H}_\varepsilon$ los espacios de Hilbert del sistema y cualquier entorno, se dice que es completamente positivo (CP), y como también debe preservar la traza de $\hat{\rho}_s$ se dice que $\Lambda_t$ realiza un mapeo CPTP (*Completely Positive and Trace Preserving*) [6], [8].

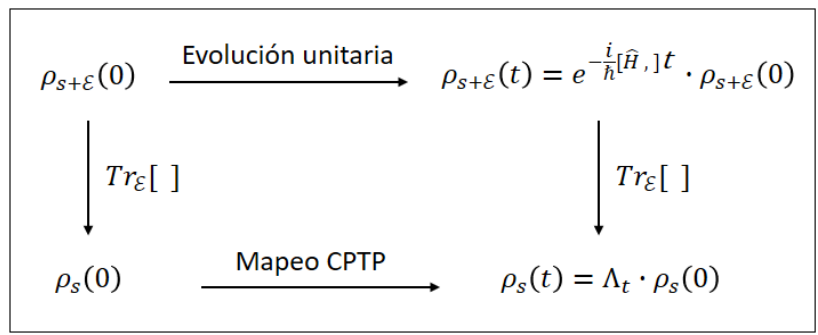

**Figura 2.2.** El superoperador $\Lambda_t$ hace evolucionar el estado $\rho_s$ del sistema abierto preservando su traza, a la vez que garantiza la positividad del estado total $\rho_{s+\varepsilon}$.

La definición de *completamente positivo* se da en el contexto de los mapeos lineales; sea el superoperador $\phi_t$ : $\phi_t \cdot \rho = \rho_t$, entonces $\phi_t$ es positivo si tanto $\rho$ como $\rho_t$ son positivos; la interacción "ruidosa" del sistema con el entorno es descrita con un mapeo estocástico $\phi: (\mathcal{H}_s) \to \mathfrak{B}'(\mathcal{H}_s)$ que transforma las matrices de $\rho$ en las de $\rho_t$, $\mathfrak{B}$ y $\mathfrak{B}'$ son los $C^*-álgebras$, que corresponden al álgebra de matrices complejas (espacio de Banach) [1], [8]; al considerar el entorno se tiene una extensión de $\mathfrak{B}(\mathcal{H}_s)$ a $\mathfrak{B}(\mathcal{H}_s) \otimes \mathfrak{B}(\mathcal{H}_\mathcal{E})_n$, donde $\mathfrak{B}(\mathcal{H}_\mathcal{E})_n$ indica que el espacio de Hilbert del entorno es de dimensión $n$ (o que las matrices de $\hat{\rho}_\mathcal{E}$ son de orden $n \times n$), entonces el mapeo viene a ser:

$$\phi \otimes 1_{\mathcal{E},n} : \mathfrak{B}(\mathcal{H}_s) \otimes \mathfrak{B}(\mathcal{H}_\mathcal{E})_n \to \mathfrak{B}'(\mathcal{H}_s) \otimes \mathfrak{B}(\mathcal{H}_\mathcal{E})_n \tag{2.28}$$

Se tiene el mapeo identidad $1_{\mathcal{E},n}$ : $\mathfrak{B}(\mathcal{H}_\mathcal{E})_n \to \mathfrak{B}(\mathcal{H}_\mathcal{E})_n$ que deja invariante cualquier matriz de ese espacio; si para un $n$ dado en (2.28) el superoperador $\phi \otimes 1_{\mathcal{E},n}$ hace un mapeo positivo, entonces $\phi$ es $n-$positivo; $\phi$ es completamente positivo si $\phi \otimes 1_{\mathcal{E},n}$ es positivo para todos los $n$; esto es, para cualquier número de grados de libertad del entorno.

Kraus [7] demostró que existe un conjunto de operadores $\hat{A}_k$ que componen al superoperador $\phi_t$ para un $t$ dado, completamente positivo, de acuerdo a (2.7):

$$\phi_t \cdot \rho = \sum_k \hat{A}_k^\dagger \rho \hat{A}_k \qquad \sum_k \hat{A}_k \hat{A}_k^\dagger \leq 1 \tag{2.29}$$

Este es llamado el teorema de Kraus. Estos operadores $\hat{A}_k$ son conocidos como *operadores de Kraus*, y tienen un uso general en cualquier *operación cuántica*.

En general en cualquier operación cuántica como la descrita en (2.29), la traza cumple: $0 \leq Tr[\phi_t \cdot \rho] \leq Tr[\rho]$ para cualquier $\rho$, la condición de igualdad se corresponde con la condición de igualdad en (2.29); es decir, para que la dinámica de semigrupo, caracterizado por $\Lambda_t$, además de ser completamente positivo, también conserve la traza se tiene que establecer en términos de los operadores de Kraus dependientes del tiempo $\hat{A}_k(t)$ como un parámetro positivo (para la evolución temporal de la dinámica irreversible) como se muestra [7]:

$$\Lambda_t \cdot \rho = \sum_k \hat{A}_k^\dagger(t) \rho \hat{A}_k(t) \qquad \sum_k \hat{A}_k(t) \hat{A}_k^\dagger(t) = 1 \qquad \forall t > 0 \tag{2.30}$$

Las ecuaciones (2.21) y (2.30) nos hablan de las propiedades matemáticas que debe tener $\Lambda_t$ para que físicamente describa la evolución temporal de un sistema cuántico abierto (en interacción) con el entorno, de manera que justifica usar $\Lambda_t$ para representar la evolución temporal de la traza parcial en todos los grados de libertad del entorno de $\hat{\rho}_{S+\mathcal{E}} \in \mathcal{H}_S \otimes \mathcal{H}_\mathcal{E}$.

## 2.3 Ecuaciones maestras para sistemas cuánticos abiertos

### 2.3.1 Aproximación markoviana, ecuación maestra de Lindblad

La ecuación maestra markoviana es una herramienta para extraer la dinámica de un subsistema de un sistema mucho más grande, el sistema cuántico de interés, bajo estudio, etiquetado como $S$, tiene el espacio de Hilbert $\mathcal{H}_S$, mientras que el entorno es un reservorio (si está en equilibrio térmico es un baño termal), etiquetado como $R$, es considerado un sistema cuántico infinito, tiene el espacio de Hilbert $\mathcal{H}_R$; el sistema total $S + R$ es considerado como un gran sistema cuántico aislado, de manera que su operador de densidad evoluciona unitariamente de acuerdo a (2.16):

$$\frac{d}{dt}\hat{\rho}_{S+R} = \mathcal{L}_U \cdot \hat{\rho}_{S+R} \qquad \mathcal{L}_U = -\frac{i}{\hbar}[\hat{H},] \qquad (2.31)$$
$$\hat{\rho}_{S+R}, \hat{H} \in \mathcal{H}_S \otimes \mathcal{H}_R$$

$\hat{H}$ es el hamiltoniano total de $S + R$, asumimos de la forma dada por (2.32), ver [5]:

$$\hat{H} = \hat{H}_S \otimes \hat{1}_R + \hat{1}_S \otimes \hat{H}_R + \lambda \hat{H}_{SR} \qquad \hat{H}_S \in \mathcal{H}_S \qquad \hat{H}_{SR} \in \mathcal{H}_S \otimes \mathcal{H}_R$$
$$\hat{H}_{SR} = \sum_j \hat{V}_j^S \otimes \hat{V}_j^R \qquad \hat{H}_R \in \mathcal{H}_R \qquad \lambda \ll 1 \qquad (2.32)$$

$\hat{H}_S$ y $\hat{H}_R$ son los hamiltonianos libres del sistema y reservorio, $\hat{V}_j^S$ y $\hat{V}_j^R$ son auto adjuntos ($\hat{V}_j^{S\dagger} = \hat{V}_j^S$), $\lambda$ es un parámetro adimensional muy pequeño, así la interacción entre el sistema y el reservorio es perturbativa; a partir de (2.32) se construyen los superoperadores:

$$H_S \cdot = [\hat{H}_S, ] \qquad H_R \cdot = [\hat{H}_R, ] \qquad H_{SR} \cdot = [\hat{H}_{SR}, ]$$

$$H \cdot = H_S \cdot \otimes \mathcal{I}^R + \mathcal{I}^S \otimes H_R \cdot + \lambda H_{SR} \cdot = H^0 \cdot + \lambda H_{SR} \cdot \qquad (2.33)$$

Donde $\mathcal{I}^S$, $\mathcal{I}^R$ son los mapeos identidad actuando en los espacios $\mathfrak{B}(\mathcal{H}_S)$ y $\mathfrak{B}(\mathcal{H}_R)$ respectivamente; es conveniente introducir los mapeos [5]:

$$\mathcal{I} = \mathcal{I}^S \otimes \mathcal{I}^R \qquad \mathcal{A}: \mathfrak{B}(\mathcal{H}_S) \to \mathfrak{B}(\mathcal{H}_S) \otimes \mathfrak{B}(\mathcal{H}_{R_0}) \qquad \wp = \mathcal{A}\, Tr_R[\ ] \qquad (2.34)$$

Donde $\mathcal{A}$ es el *amplificador* que lleva del espacio asociado al sistema, al espacio asociado al sistema y el reservorio con un estado dado por el operador de densidad $\hat{R}_0$, así por ejemplo, $\mathcal{A} \cdot \hat{\rho}_S = \hat{\rho}_S \otimes \hat{R}_0$, también para un mapeo cualquiera del sistema $A \in \mathfrak{B}(\mathcal{H}_S)$ se puede expresar: $Tr_S[(\Lambda_t \cdot \hat{\rho}_S)A] = Tr_{S+R}[(\Lambda_t^U \cdot \hat{\rho}_S \otimes \hat{R}_0)(A \otimes \mathcal{I}^R)]$ donde $\Lambda_t$ y $\Lambda_t^U$ actúan en los espacios del sistema y el sistema-reservorio respectivamente; también se puede expresar $\Lambda_t \cdot \hat{\rho}_S = Tr_R[\Lambda_t^U \cdot \mathcal{A} \cdot \hat{\rho}_S]$, o también: $\mathcal{A} \cdot (\Lambda_t \cdot \hat{\rho}_S) = \wp \Lambda_t^U \cdot \mathcal{A} \cdot \hat{\rho}_S$.

Es evidente el rol de $\wp$ sobre un estado arbitrario no factorizable $\hat{\rho} \in \mathcal{H}_S \otimes \mathcal{H}_R$, este es transformado como se muestra $\wp: \hat{\rho} \to Tr_R[\hat{\rho}] \otimes \hat{R}_0$; esto no solo pierde las correlaciones entre el entorno y el sistema $\mathcal{X}_{S,R} = \hat{\rho} - Tr_R[\hat{\rho}] \otimes Tr_S[\hat{\rho}]$, sino que además fija el estado del entorno haciéndolo inmutable: $Tr_S[\hat{\rho}] \to \hat{R}_0$; así, la diferencia entre $\mathcal{I}$ y $\wp$ viene a ser [5]:

$$(\mathcal{I} - \wp) \cdot \hat{\rho} = \hat{\rho} - Tr_R[\hat{\rho}] \otimes \hat{R}_0 \qquad (2.35)$$

El superoperador $(\mathcal{I} - \wp) \cdot$ tiene una relevancia física importante, es parte fundamental de la aproximación markoviana, y consiste en suponer que el reservorio permanece inalterado, en su estado inicial, y siempre separable con el sistema.

El estado del sistema viene a ser la traza parcial en los grados de libertad del entorno de $\hat{\rho}_{S+R}$, así la ecuación master para el sistema se obtiene de tomar la traza parcial en (2.31):

$$\hat{\rho}_S = Tr_R[\hat{\rho}_{S+R}] \qquad \qquad \frac{d}{dt}\hat{\rho}_S = -\frac{i}{\hbar} Tr_R[H \cdot \hat{\rho}_{S+R}] \qquad (2.36)$$

Se considera que al inicio, $t_0 = 0$, el sistema y el reservorio son independientes y separables, de manera que el estado inicial total viene a ser [5]:

$$\hat{\rho}_{S+R}(0) = \hat{\rho}_S(0) \otimes \hat{R}_0 \qquad (2.37)$$

Bajo esta condición, la segunda ecuación de (2.36) viene a ser [5]:

$$\begin{aligned} \frac{d}{dt}\hat{\rho}_S(t) &= \mathcal{L}_{U,ef} \cdot \hat{\rho}_S(t) + \lambda^2 \int_0^t d\tau\, \mathcal{K}(\tau) \cdot \hat{\rho}_S(t-\tau) \\ \mathcal{L}_{U,ef} &= -\frac{i}{\hbar}\big[\hat{H}_{S,ef}, \big] \qquad \hat{H}_{S,ef} = \hat{H}_S + \lambda \sum_j \omega^R\big(\hat{V}_j^R\big)\hat{V}_j^S \\ \mathcal{K}(\tau) \cdot &= -Tr_R\big[H_{SR} \cdot (\mathcal{I}-\wp) \cdot \Lambda_\tau^{U'} \cdot (\mathcal{I}-\wp) \cdot H_{SR} \cdot \mathcal{A}\big] \\ \Lambda_\tau^{U'} \cdot &= \exp -\frac{i}{\hbar}H'\tau \qquad H' = (\mathcal{I}-\wp) \cdot H \cdot (\mathcal{I}-\wp) \cdot \end{aligned} \qquad (2.38)$$

Donde $\omega^R\big(\hat{V}_j^R\big)$ es la función de correlación del reservorio que depende de los operadores $\hat{V}_j^R$ del reservorio, $\hat{H}_{S,ef}$ es una modificación del hamiltoniano libre del sistema $\hat{H}_S$ debido al acoplamiento con el reservorio y se desvanece cuando se anulan las correlaciones $\omega^R\big(\hat{V}_j^R\big) = 0$, la diferencia es del orden de la perturbación $\lambda$; $\mathcal{K}(\tau)$ es el *kernel* o núcleo de la integral, el cual se puede expandir en una serie de potencias en la constante de acoplamiento $\lambda$, que se encuentra en $\Lambda_\tau^{U'}$, como se muestra [5]:

$$\begin{aligned} \mathcal{K}(\tau) &= e^{-iH_S\tau}\left\{\mathcal{K}_0(\tau) + \sum_{n=1}^{\infty}(-i\lambda)^n \mathcal{K}_n(\tau)\int_t \prod_{i=1}^n dt_i\, \mathcal{K}_n(\tau|t_1,\cdots t_n)\right\} \\ t:\; 0 &\leq t_n \leq \cdots \leq t_1 \leq \tau \qquad \mathcal{K}_0(\tau) = -Tr_R[H_{SR}(\tau) \cdot (\mathcal{I}-\wp) \cdot H_{SR} \cdot \mathcal{A}] \\ \mathcal{K}_n(\tau|t_1,\cdots t_n) &= -Tr_R\left[H_{SR}(\tau) \cdot (\mathcal{I}-\wp) \cdot \prod_{i=1}^n H_{SR}(t_i) \cdot (\mathcal{I}-\wp) \cdot \cdot H_{SR} \cdot \mathcal{A}\right] \end{aligned} \qquad (2.39)$$

Donde $H_{SR}(\tau) = \big[\hat{H}_{SR}(\tau), \big]$, con $\hat{H}_{SR}(t) = e^{i\hat{H}_0 t}\hat{H}_{SR}e^{-i\hat{H}_0 t}$, es $H_{SR}$ en la imagen de interacción; los núcleos $\mathcal{K}_n(\tau|t_1,\cdots t_n)$ dependen de las funciones de correlación temporal múltiple $\omega^R\big(\hat{V}_{j_1}^R(t_{i_1}) \cdots \hat{V}_{j_k}^R(t_{i_k})\big)$ de los operadores del reservorio, donde $\hat{V}_j^R(t)$ es $\hat{V}_j^R$ en la imagen de interacción: $\hat{V}_j^R(t) = e^{i\hat{H}_R t}\, \hat{V}_j^R\, e^{-i\hat{H}_R t}$ ; es decir, depende de los estados anteriores en el tiempo $t_i \leq \tau \leq t$, desde que la aproximación de Born en (2.38) mantiene a segundo orden en $\lambda$, supone acotar a orden cero en (2.39) para mantener a segundo orden en (2.38), así se tiene $\mathcal{K}(\tau) \to e^{-iH_s\tau}\mathcal{K}_0(\tau)$; así, en (2.38) y (2.24), el término disipativo toma la forma $\mathcal{L}_D \cdot \hat{\rho}_S(t) = \lambda^2 \int_0^t d\tau\, e^{-iH_s\tau}\mathcal{K}_0(\tau) \cdot \hat{\rho}_S(t-\tau)$.

Se considera entonces una característica singular en la que para tiempos suficientemente largos, el efecto memoria se desvanece en (2.38), esto se justifica si se considera que las correlaciones entre el sistema y el reservorio decaen con un tiempo característico $\tau_R$ que es mucho menor que el tiempo característico de variación del sistema $\tau_S \gg \tau_R$, es muy importante mantener bien separados estas dos escalas de tiempo: $\tau_S/\tau_R \to \infty$, para esto se consideran dos situaciones límites [5]:

Por un lado, se tiene el **límite del acople débil** $\lambda \to 0$, se hace una escala de tiempo donde $\tau = \lambda^2 t$, entonces $\tau_R$ permanece constante y $\tau_S$ tiende a infinito, este límite permite el paso en el kernel: $\mathcal{K}(\tau) \to e^{-iH_s\tau}\mathcal{K}_0(\tau)$; un efecto importante de esta aproximación es que como el reservorio evoluciona cuasi-libre, anula las funciones de correlación $\omega^R\big(\hat{V}_{j_1}^R(t_{i_1}) \cdots \hat{V}_{j_k}^R(t_{i_k})\big)$ superiores a orden 2.

Por otro lado, se tiene el **límite del reservorio singular** $\tau_R \to 0$, este límite requiere de *reservorios singulares*, y permite obtener la dinámica markoviana puesto que en este límite $\mathcal{K}_0(\tau)$ tiende a $\mathcal{K}_0\delta(\tau)$, y se anulan las correcciones de orden superior.

Los detalles de los cálculos siguientes pueden ser consultados en la publicación respectiva, ver la referencia [5], para el límite del reservorio singular, el generador $\mathcal{L}$ de la dinámica reducida de semigrupos es dada por:

$$\frac{d}{dt}\hat{\rho}(t) = \mathcal{L} \cdot \hat{\rho}(t) \qquad \hat{H}_1 = -\frac{1}{\sqrt{N}}\sum_{k=1}^{N^2-1} \mathbb{Im}\big[\gamma_{kN^2}\hat{V}_k^s\big]$$

$$\mathcal{L}\cdot\hat{\rho} = -\frac{i}{\hbar}\big[\hat{H}_s + \hat{H}_1, \hat{\rho}\big] + \frac{1}{2}\sum_{i,j=1}^{N^2-1} \gamma_{ij}\big\{[\hat{V}_i^s\hat{\rho}, \hat{V}_j^s] + [\hat{V}_i^s, \hat{\rho}\hat{V}_j^s]\big\}$$

(2.40)

Donde los coeficientes $\gamma_{ij}$ son reales y simétricos, y construido a partir de términos que componen los operadores del entorno $\hat{V}_k^R$. $\mathcal{L}$ contiene la parte unitaria y difusiva, esta última es un superoperador compuesto solo de los operadores $\hat{V}_i^s$ del sistema, desde que en (2.32) se definen a los operadores del sistema como auto adjuntos $\hat{V}_j^{s\dagger} = \hat{V}_j^s$, la estructura de $\mathcal{L}_D$ en (2.40) no distingue entre $\hat{V}_j^s$ y $\hat{V}_j^{s\dagger}$.

En la referencia [6] se plantea la ecuación de Lindblad en términos de los operadores de saltos del sistema $\hat{A}_j$ en general no auto adjunto, viene dado por el superoperador $\mathcal{L}$:

$$\mathcal{L} \cdot \hat{\rho} = -\frac{i}{\hbar}[\hat{H}', \hat{\rho}] + \sum_i \gamma_i \left( \hat{A}_i \hat{\rho} \hat{A}_i^\dagger - \frac{1}{2}\{\hat{A}_i^\dagger \hat{A}_i, \hat{\rho}\} \right) \quad (2.41)$$

Donde $\{\hat{A}_i^\dagger \hat{A}_i, \hat{\rho}\} = \hat{A}_i^\dagger \hat{A}_i \hat{\rho} + \hat{\rho} \hat{A}_i^\dagger \hat{A}_i$ es el anti conmutador, $\hat{H}' \in \mathcal{H}_S$ es el generador de la evolución de la parte coherente, el cual no necesariamente es igual al hamiltoniano libre del sistema $\hat{H}_s$, y $\gamma_i$ son los coeficientes de relajación asociados a los operadores $\hat{A}_i$ del sistema; por ejemplo, para un átomo de dos niveles excitado, en el estado $|+\rangle$, que se acopla con el campo electromagnético del vacío se tiene $i = 1$, $\hat{A}_1 = \hat{\sigma}_- = |-\rangle\langle+|$, mientras que $\gamma_1$ es el coeficiente $A$ de Einstein, $\hat{H}' = \hbar\omega'\hat{\sigma}_3/2$, además $\hat{\sigma}_+ = \hat{\sigma}_-^\dagger$, $\hat{\sigma}_+\hat{\sigma}_- = |+\rangle\langle+|$, y $\hat{\sigma}_3 = |+\rangle\langle+| - |-\rangle\langle-|$ es la matriz $z$ de Pauli, así la emisión espontánea viene descrita por la ecuación maestra:

$$\frac{d}{dt}\hat{\rho} = -i\frac{\omega'}{2}[\hat{\sigma}_3, \hat{\rho}] + \gamma\left( \hat{\sigma}_-\hat{\rho}\hat{\sigma}_+ - \frac{1}{2}\{\hat{\sigma}_+\hat{\sigma}_-, \hat{\rho}\} \right) \quad (2.42)$$

Como el sistema se encuentra al inicio en el estado excitado, entonces, la solución de (2.42) para $\hat{\rho}(0) = |+\rangle\langle+|$ viene a ser $\hat{\rho}(t) = e^{-\gamma t}|+\rangle\langle+| + (1 - e^{-\gamma t})|-\rangle\langle-|$; este estado converge al estado puro fundamental: $\hat{\rho}(t \to \infty) = |-\rangle\langle-|$, referencia [6].

En *An Open Systems Approach to Quantum Optics*, referencia [3], se muestra una derivación de la ecuación maestra de Lindblad para un hamiltoniano general como (2.32), se considera que el estado total $\hat{\rho}_{S+R}$ sufre la **aproximación de Born**: puede ser aproximadamente factorizado en todo el tiempo, con el reservorio inalterado:

$$\hat{\rho}_{S+R} \to \wp \cdot \hat{\rho}_{S+R} = \hat{\rho}_S \otimes \hat{R}_0 \qquad \hat{\rho}_S = Tr_R[\hat{\rho}_{S+R}] \quad (2.43)$$

Se considera una segunda **aproximación, la de Markov**, según la cual la dinámica actual de $\hat{\rho}_S(t)$ no depende de sus estados pasados $\hat{\rho}_S(t-\tau)$, sino solo del presente, esto es equivalente a reemplazar $\hat{\rho}_S(t-\tau)$ por $\hat{\rho}_S(t)$ en la integral de (2.38).

Para un sistema consistente en un oscilador armónico con un único modo de oscilación, inmerso en una cavidad óptica (entorno), se plantea para el hamiltoniano de (2.32) los términos:

$$\hat{H}_S = \hbar\omega_0 \hat{a}^\dagger \hat{a} \qquad \hat{H}_R = \hbar \sum_{j\neq 0} \omega_j \hat{b}_j^\dagger \hat{b}_j \qquad \lambda \hat{H}_{SR} = \hbar \sum_{j\neq 0}(V_j \hat{a}^\dagger \hat{b}_j + V_j^* \hat{a} \hat{b}_j^\dagger) \qquad (2.44)$$

Donde $\hat{a}$ y $\hat{b}_j$ son los operadores de destrucción para el sistema y los osciladores (por donde se propagan las ondas planas de los fotones) del reservorio, cada uno con frecuencia propia $\omega_j$ que sigue una distribución de densidad de estados $g(\omega)$ tal que $g(\omega)d\omega$ es el número de osciladores con frecuencias entre $\omega$ y $\omega + d\omega$; el sistema se acopla a los osciladores del reservorio mediante los acoples $V_j$, los cuales toman la forma continua $V_j \to V(\omega)$; la ecuación maestra markoviana toma la forma, para el operador de densidad del sistema $\hat{\rho}$, [2], [3]:

$$\dot{\hat{\rho}} = -i\omega_0'[\hat{a}^\dagger \hat{a}, \hat{\rho}] + \kappa(2\hat{a}\hat{\rho}\hat{a}^\dagger - \{\hat{a}^\dagger \hat{a}, \hat{\rho}\})$$
$$+ 2\kappa \bar{n}(\hat{a}\hat{\rho}\hat{a}^\dagger + \hat{a}^\dagger \hat{\rho}\hat{a} - \hat{a}^\dagger \hat{a}\hat{\rho} - \hat{\rho}\hat{a}\hat{a}^\dagger)$$

$$\omega_0' = \omega_0 + P\left[\int_0^\infty \frac{d\omega}{2\pi} \frac{J(\omega)}{\omega_0 - \omega}\right] \qquad \kappa = \frac{J(\omega_0)}{2} \qquad \bar{n} = n(\omega_0, T)$$

(2.45)

Donde $J(\omega)$ es la densidad espectral como una función continua, la cual tiene correspondencia con su definición discreta en término de los acoples $V_j$ [2], [3]:

$$J(\omega) = 2\pi \sum_j |V_j|^2 \delta(\omega - \omega_j) \quad \to \quad J(\omega) = 2\pi g(\omega)|V(\omega)|^2 \qquad (2.46)$$

$P$ es el valor principal de Cauchy de la integral, y proporciona el corrimiento Lamb $\Delta\omega_0 = \omega_0' - \omega_0$ de la frecuencia del sistema, $n(\omega_j, T) = Tr_R[\hat{R}_0 \hat{b}_j^\dagger \hat{b}_j] = Z^{-1} \exp -\hbar\omega_j/k_B T$, $Z = Tr\left[\exp -\frac{\hbar\omega_j}{k_B T}\right]$, es el número medio de fotones para un oscilador con frecuencia $\omega_j$ del reservorio a temperatura $T$; para temperatura cero $\bar{n}$ se anula, ya que $\bar{n}(\omega_j, 0) = 0$, dejando (2.45) simplemente como $\dot{\hat{\rho}} = -i\omega_0'[\hat{a}^\dagger\hat{a}, \hat{\rho}] + \kappa(2\hat{a}\hat{\rho}\hat{a}^\dagger - \{\hat{a}^\dagger\hat{a}, \hat{\rho}\})$; así, para un oscilador en el estado puro inicial (de dos niveles) $|\psi(0)\rangle = \alpha|0\rangle + \beta|1\rangle$, donde $\hat{a}^\dagger\hat{a}|n\rangle = n|n\rangle$, se tiene la solución de (2.45), para un $J(\omega)$ dado a temperatura nula:

$$\hat{\rho}(t) = \begin{pmatrix} 1 - p(t) & q^*(t) \\ q(t) & p(t) \end{pmatrix}$$

$$p(t) = |\beta|^2 e^{-J(\omega_0)t} \qquad q(t) = e^{-\frac{J(\omega_0)}{2}t}\alpha^*\beta e^{-i\omega_0' t}$$

(2.47)

Donde la probabilidad del estado excitado $p(t)$ decae exponencialmente con parámetro $J(\omega_0)$, mientras que el término $q(t)$ decae la mitad de rápido que $p(t)$, pero con su fase oscilando en el tiempo con frecuencia $\omega_0' \neq \omega_0$.

En general para un estado inicial mixto arbitrario, en general mezcla, (2.47) se sigue cumpliendo, solo hay que reemplazar: $|\beta|^2 \to p(0)$ y $\alpha^*\beta \to q(0)$.

### 2.3.2 Ecuación maestra general, dinámica no markoviana

Los fenómenos disipativos en un sistema cuántico abierto en una cavidad han sido bien estudiados cuando el acoplamiento entre el sistema y el reservorio es lo suficientemente débil para considerar una dinámica perturbativa. Si el tiempo de decaimiento de las correlaciones en el entorno es más pequeño que el tiempo característico del sistema, lo suficiente para desvanecer el efecto memoria, entonces se realiza la aproximación markoviana y la ecuación maestra (2.45) es válida. Sin embargo, cuando el acoplamiento es lo suficientemente fuerte o el efecto memoria ya no puede ser despreciado con el paso del tiempo, la aproximación markoviana ya no es aplicable y una nueva ecuación maestra más general, que considere el acoplamiento fuerte o el efecto memoria para tiempos largos, es requerida [2].

Considerando el hamiltoniano total $\hat{H} \in \mathcal{H}_S \otimes \mathcal{H}_R$ que resulta de reemplazar (2.44) en (2.32), para los acoplamientos $V_j$ en general no perturbativos y siempre reales, en cual el sistema y reservorio evolucionan unitariamente según $\hat{\rho}_{S+R}(t) = e^{-i[\hat{H},\ ]t/\hbar} \cdot \hat{\rho}_{S+R}(0)$, el entorno se encuentra inicialmente en equilibrio térmico $\hat{\rho}_R(t_0) = Z^{-1}e^{-\hat{H}_R/k_B T}$ y separable con el sistema: $\hat{\rho}_{S+R}(t_0) = \hat{\rho}_S(t_0) \otimes \hat{\rho}_R(t_0)$, la ecuación master exacta para el sistema es la traza parcial en los grados de libertad del entorno $\dot{\hat{\rho}}_S = -\frac{i}{\hbar} Tr_R[\hat{H}, \hat{\rho}_{S+R}(t)]$, tiene una forma similar a (2.45) pero con los coeficientes $\kappa(t)$, $\tilde{\kappa}(t)$ y $\omega_0'(t)$ dependientes del tiempo, que se obtienen de resolver un sistema de ecuaciones diferenciales dependientes de $J(\omega)$ y $\bar{n}(\omega, T)$, que son presentados en la referencia [2]:

$$\dot{\hat{\rho}} = -i\omega_0'(t)[\hat{a}^\dagger \hat{a}, \hat{\rho}] + \kappa(t)(2\hat{a}\hat{\rho}\hat{a}^\dagger - \{\hat{a}^\dagger \hat{a}, \hat{\rho}\}) \\ + 2\tilde{\kappa}(t)(\hat{a}\hat{\rho}\hat{a}^\dagger + \hat{a}^\dagger\hat{\rho}\hat{a} - \hat{a}^\dagger\hat{a}\hat{\rho} - \hat{\rho}\hat{a}\hat{a}^\dagger) \tag{2.48}$$

De acuerdo con los autores, H. Xiong et al, [2], se recupera la dinámica markoviana para el límite de acople débil $V(\omega) \to 0$, haciendo un desarrollo perturbativo a segundo orden en los acoplamientos $|V(\omega)|^2$, y luego se tomando el límite markoviano de tiempos largos; con estas aproximaciones se puede ignorar el efecto memoria, y transformar (2.48) en (2.45).

### 2.4 División en suma de subespacios para las oscilaciones de Rabi

Esta sección es una revisión del modelo de James-Cummings para emplear la división en subespacios de Hilbert en el que se plantea una ecuación de autovalores, como método base a extenderlo para abordar el modelo simplificado.

Sea el hamiltoniano de James-Cummings tras la aproximación de la onda rotante [12]:

$$\hat{H}_{JC} = \frac{\hbar\omega_0}{2}\hat{\sigma}_3 \otimes \hat{1} + \hbar\omega \hat{1} \otimes \hat{a}^\dagger \hat{a} + \hbar\lambda(\hat{\sigma}_+ \hat{a} + \hat{\sigma}_- \hat{a}^\dagger) \qquad \hat{H}_{JC} \in \mathcal{H}_a \otimes \mathcal{H}_c \tag{2.49}$$

Donde $\hat{\sigma}_3 = |+\rangle\langle+| - |-\rangle\langle-|$, $\hat{\sigma}_+ = |+\rangle\langle-|$, $\hat{\sigma}_- = \hat{\sigma}_+^\dagger$ y la siguiente notación por simplicidad: $\hat{\sigma}_+ \hat{a} = \hat{\sigma}_+ \otimes \hat{a}$; el hamiltoniano muestra la interacción de un átomo, cuyo espacio es $\mathcal{H}_a$, de dos niveles: fundamental $|-\rangle$ y excitado $|+\rangle$, con un único oscilador del campo electromagnético, con espacio $\mathcal{H}_c$, con frecuencia $\omega$ y $|n\rangle$ como autoestado de $\hat{a}^\dagger \hat{a}$ con autovalor $n$ que corresponde al número de fotones o de excitación [12].

Es importante establecer una cantidad conservada de interés, que corresponde a un observable que conmuta con el hamiltoniano (2.49):

$$\hat{N} = \hat{\sigma}_+\hat{\sigma}_- \otimes \hat{1} + \hat{1} \otimes \hat{a}^\dagger \hat{a} \qquad [\hat{N}, \hat{H}_{JC}] = 0 \rightarrow \frac{d}{dt}\hat{N} = 0 \qquad (2.50)$$

$\hat{N}$ es el observable que suma el número de excitación que hay en el átomo y en el oscilador del campo, lo que nos dice (2.50) es que si se tiene un estado inicial $|\psi(0)\rangle \in \mathcal{H} = \mathcal{H}_a \otimes \mathcal{H}_c$ que es autoestado de $\hat{N}$; esto es, $\hat{N}|\psi(0)\rangle = N|\psi(0)\rangle$, entonces el estado en cualquier tiempo posterior $|\psi(t)\rangle = \exp{-\frac{i}{\hbar}\hat{H}_{JC}t}\,|\psi(0)\rangle$ debe ser también autoestado de $\hat{N}$ conservando el autovalor $N$. Así, la evolución temporal de $|\psi(t)\rangle$ está confinada a un subespacio conformado por los autoestados de $\hat{N}$:

$$|\psi(t)\rangle \in \mathcal{H}(N) \subset \mathcal{H} \qquad \mathcal{H}(N) = \{\, |\sigma,n\rangle \in \mathcal{H}\,, \hat{N}|\sigma,n\rangle = N|\sigma,n\rangle \,\} \qquad (2.51)$$

Donde $|\sigma,n\rangle = |\sigma\rangle|n\rangle$ y $\sigma = \pm$, es evidente que $\mathcal{H}(N)$ está conformado por $|-,N\rangle$ y $|+,N-1\rangle$, es decir $\mathcal{H}(N)$ es de dos dimensiones para $N > 0$; el espacio general $\mathcal{H}$ se puede expresar como la suma de todos los sub espacios $\mathcal{H}(N)$:

$$\mathcal{H} = \mathcal{H}_a \otimes \mathcal{H}_c = \bigoplus_{n=0}^{\infty} \mathcal{H}(n) \qquad (2.52)$$

Es posible contar desde $\mathcal{H}(0) = \{\,|-\rangle|0\rangle\,\}$ que es sólo de una dimensión.

El hamiltoniano en los sub espacios $\mathcal{H}(N)$, para $N > 0$, se reduce a $\hat{H}_{jc}^N \in \mathcal{H}(N)$ usando la completitud en este sub espacio $\hat{1}_N \in \mathcal{H}(N)$:

$$\hat{H}_{JC} \rightarrow \hat{H}_{jc}^N = \hat{1}_N \hat{H}_{JC} \hat{1}_N \qquad \hat{1}_N = |-,N\rangle\langle-,N| + |+,N-1\rangle\langle+,N-1| \qquad (2.53)$$

Así (2.49) en (2.53) viene a ser expresado en matrices $2 \times 2$ porque el sub espacio es de dimensión 2, además el hamiltoniano total viene a ser: $\hat{H}_{JC} = \hat{H}_0 + \hat{H}_I$, donde $\hat{H}_0$ y $\hat{H}_I$ son el hamiltoniano libre y el hamiltoniano de interacción respectivamente. Se usa una nueva base para $\mathcal{H}(N)$ donde $|1\rangle = |-,N\rangle$ y $|2\rangle = |+,N-1\rangle$, es conveniente notar que $|1\rangle$ y $|2\rangle$ son autoestados de $\hat{H}_0$, mientras que $\hat{H}_I$ es no diagonal, luego sumándolos:

$$\hat{H}_{jc}^N = \hbar \begin{pmatrix} N\omega - \omega_0/2 & \lambda\sqrt{N} \\ \lambda\sqrt{N} & (N-1)\omega + \omega_0/2 \end{pmatrix} \qquad |1\rangle = \begin{pmatrix} 1 \\ 0 \end{pmatrix} \qquad |2\rangle = \begin{pmatrix} 0 \\ 1 \end{pmatrix} \qquad (2.54)$$

El hamiltoniano $\widehat{H}_{jc}^N$ puede ser diagonalizado para encontrar sus autovectores $|E_\pm\rangle$ (también llamados "estados vestidos", *dressed states* [12]) y autovalores y $\omega_\pm$, las cuales vienen a ser:

$$\omega_\pm = \left(N - \frac{1}{2}\right)\omega \pm \sqrt{\lambda^2 N + \frac{(\omega - \omega_0)^2}{4}} \qquad |E_-\rangle = \begin{pmatrix} \sin\theta/2 \\ \cos\theta/2 \end{pmatrix}$$

$$\tan\theta = \frac{2\lambda}{(\omega - \omega_0)} \qquad |E_+\rangle = \begin{pmatrix} \cos\theta/2 \\ -\sin\theta/2 \end{pmatrix} \qquad (2.55)$$

De forma similar se representa el estado del sistema en esta base como $|\psi(t)\rangle = \alpha(t)|1\rangle + \beta(t)|2\rangle$, con esto se escribe la ecuación de Schrödinger en este sub espacio $i\hbar|\dot{\psi}(t)\rangle = \widehat{H}_{jc}^N|\psi(t)\rangle$, cuya solución general se obtiene encontrando los autovalores $\hbar\omega_-$ y $\hbar\omega_+$ y autoestados $|E_-\rangle$ y $|E_+\rangle$, que son estacionarios en el tiempo, así si el sistema se encuentra en el estado inicial $|\psi(0)\rangle = |2\rangle$, en la base de los autoestados es: $|\psi(0)\rangle = \langle E_-|2\rangle|E_-\rangle + \langle E_+|2\rangle|E_+\rangle$, su evolución temporal viene a ser: $|\psi(t)\rangle = \langle E_-|2\rangle e^{-i\omega_- t}|E_-\rangle + \langle E_+|2\rangle e^{-i\omega_+ t}|E_+\rangle$, y las amplitudes se hallan como:

$$\alpha(t) = \langle 1|\psi(t)\rangle = e^{-i\omega_- t}\langle E_-|2\rangle\langle 1|E_-\rangle + be^{-i\omega_+ t}\langle E_+|2\rangle\langle 1|E_+\rangle$$

$$\beta(t) = \langle 2|\psi(t)\rangle = e^{-i\omega_- t}|\langle 2|E_-\rangle|^2 + e^{-i\omega_+ t}|\langle 2|E_+\rangle|^2 \qquad (2.56)$$

Las amplitudes de (2.56) se pueden generalizar a cualquier estado inicial, desde que son oscilaciones se pueden colocar un desfase temporal para recrear el estado inicial: $\{\alpha,\beta\}(t) \to \{\alpha,\beta\}(t+\theta)$, de manera que $\{\alpha,\beta\}(\theta) = \{\langle 1|\psi(0)\rangle, \langle 2|\psi(0)\rangle\}$.

Usando (2.55) en (2.56) para obtener $p_2(t) = |\beta(t)|^2$:

$$p_2(t) = 1 - \sin^2\theta\,(1 - \cos^2\Omega t) \qquad \Omega = \sqrt{\lambda^2 N + \frac{(\omega-\omega_0)^2}{4}} \qquad (2.57)$$

(2.57) es la forma general para el estado inicial en $|+, N-1\rangle$, con desafinación de frecuencias, $\Omega$ es la frecuencia generalizada de Rabi; en el caso resonante $\omega = \omega_0$, se tiene la frecuencia de Rabi $\Omega = \lambda\sqrt{N}$, y la probabilidad es $p_2(t) = \cos^2\lambda\sqrt{N}t$, $(N > 0)$.

Esta es la dinámica cuando el estado inicial se encuentra en un sub espacio $|\psi(0)\rangle \in \mathcal{H}(N)$, el caso general en que $|\psi(0)\rangle$ es una superposición de estados de diferentes sub espacios, requiere considerar que el Hamiltoniano total se puede expresar como la suma de todos los hamiltonianos en cada sub espacio:

$$\hat{H}_{JC} = \oplus_n \hat{H}_{jc}^n \qquad \begin{aligned} |\psi(t)\rangle &= \oplus_n C_n |\psi_n(t)\rangle \\ |\psi_n(t)\rangle &\in \mathcal{H}(n) \end{aligned} \qquad \begin{aligned} i\hbar \frac{d}{dt}|\psi(t)\rangle &= \hat{H}_{JC}|\psi(t)\rangle \\ i\hbar \frac{d}{dt}|\psi_n(t)\rangle &= \hat{H}_{jc}^n |\psi_n(t)\rangle \end{aligned} \qquad (2.58)$$

En (2.58) $C_n$ no depende del tiempo, pues $\langle \hat{N} \rangle = \langle \psi(t)|\hat{N}|\psi(t)\rangle = \sum_n C_n n$ debe conservarse; $\dot{C}_n = 0$ permite descomponer la ecuación de Schrödinger en varias ecuaciones en paralelo para cada sub espacio, lo cual puede ser resuelto con (2.54), (2.55). Por ejemplo, para el estado inicial del sistema y campo: $c_g|-\rangle + c_e|+\rangle$ y $\sum_{n=0}^{\infty} c_n |n\rangle$, el estado total inicial $|\psi(0)\rangle = \sum_{n=1}^{\infty}\left(c_g c_n |-, n\rangle + c_e c_{n-1}|+, n-1\rangle\right) + c_g c_0 |-, 0\rangle$ es:

$$|\psi(0)\rangle = \sum_{n=1}^{\infty} \tilde{c}_n |\psi_n(0)\rangle + c_g c_0 |-, 0\rangle$$

$$|\psi_n(t)\rangle = \alpha_n(t + \theta_n)|-, n\rangle + \beta_n(t + \theta_n)|+, n-1\rangle \qquad (2.59)$$

donde los ket $|\psi_n(t)\rangle$ resulta de la evolución temporal de $|\psi_n(0)\rangle$, de acuerdo a (2.56); es evidente que: $\alpha_n(\theta_n) = c_g c_n / \tilde{c}_n$ y $\beta_n(\theta_n) = c_e c_{n-1}/\tilde{c}_n$.

# CAPÍTULO III

# MODELO SIMPLIFICADO DE MUCHOS OSCILADORES ACOPLADOS

## 3.1 Marco teórico

Como se señaló en la introducción, el propósito de este modelo simplificado es tener un objeto teórico para estudiar las leyes y propiedades de la mecánica cuántica para sistemas de muchas partículas, y en particular explorar la irreversibilidad cuántica mediante la obtención de soluciones exactas.

### 3.1.1 La base del Hamiltoniano libre

Sea un conjunto de muchos osciladores armónicos cuánticos, cada uno con un solo modo de vibración, distinguibles mediante el índice $i$ como etiqueta: $i = 0, 1, 2, \cdots N$, entonces podemos decir que cada oscilador tiene su propio espacio de Hilbert $\mathcal{H}_i$, en el cual se define su operador de aniquilación $\hat{a}_i$ y su frecuencia angular $\omega_i$, en consecuencia, su hamiltoniano libre $\hat{H}_i$ es derivado y cumple una ecuación de autovalores:

$$\hat{H}_i |n_i\rangle = E_{n_i}^i |n_i\rangle \qquad \begin{aligned} \hat{H}_i &= \omega_i \hat{a}_i^\dagger \hat{a}_i & |n_i\rangle &\in \mathcal{H}_i \\ E_{n_i}^i &= \omega_i n_i & \hat{a}_i |n_i\rangle &\in \mathcal{H}_i \end{aligned} \qquad (3.1)$$

Donde se hace $\hbar \equiv 1$ por simplicidad de la notación; en este espacio se tiene la identidad $\sum_{n_i} |n_i\rangle\langle n_i| = \hat{1}_i$ que establece la completitud en $\mathcal{H}_i$; se define la forma del hamiltoniano $\hat{H}_i = \omega_i \hat{a}_i^\dagger \hat{a}_i$ sin energía del estado fundamental por simplicidad, $E_{n_i}^i$ es la energía del oscilador $i$ con número cuántico de excitación $n_i$; al considerar a todos los osciladores como un conjunto se hace uso del índice vector $\vec{n}$ que tiene información de todos los números cuánticos $n_i$ de cada oscilador:

$$\begin{aligned} |\vec{n}\rangle &= |n_0, n_1, n_2, \cdots n_N\rangle \\ |\vec{n}\rangle &= |n_0\rangle \otimes |n_1\rangle \otimes \ldots |n_N\rangle \end{aligned} \qquad \vec{n} = (n_0, n_1, n_2, \cdots n_N) \qquad \begin{aligned} \mathcal{H} &= \otimes_{i=0}^N \mathcal{H}_i \\ |\vec{n}\rangle &\in \mathcal{H} \end{aligned} \qquad (3.2)$$

Así, la completitud en $\mathcal{H}$ viene a ser $\sum_{\vec{n}}|\vec{n}\rangle\langle\vec{n}| = \hat{1}$; el hamiltoniano libre de todos los osciladores aislados viene a ser:

$$\hat{H}^0 = \sum_{i=0}^{N} \hat{1}_0 \otimes \hat{1}_1 \otimes \cdots \hat{H}_i \otimes \cdots \hat{1}_{N-1} \otimes \hat{1}_N \qquad \hat{H}^0|\vec{n}\rangle = \sum_{i=0}^{N} E_{n_i}^i |\vec{n}\rangle \qquad \begin{array}{l} \hat{H}_i \in \mathcal{H}_i \\ \hat{1}_j \in \mathcal{H}_j \end{array} \qquad (3.3)$$

El estado del sistema total es en general no factorizable:

$$|\psi\rangle = \sum_{\vec{n}} \psi_{\vec{n}} |\vec{n}\rangle \qquad (3.4)$$

Si se dejara evolucionar este sistema de osciladores, simplemente permanecerían inalterados en el tiempo, cada oscilador conservaría su energía independiente del resto; para $t > 0$ se añade el hamiltoniano de interacción que no depende del tiempo, y que puede tomar la siguiente forma (por simplicidad de la notación, debe entenderse que en el producto de los operadores $\hat{a}_{j_1}\hat{a}_{j_2}$, cada operador se ubica en su respectivo espacio de Hilbert $\mathcal{H}_{j_1}$ y $\mathcal{H}_{j_2}$ ordenados según $\mathcal{H} = \bigotimes_{i=0}^{N} \mathcal{H}_i$, y se los completa con el producto tensorial de la identidad $\otimes \hat{1}_k$ para que el hamiltoniano de interacción $\hat{H}^I$ siempre pertenezca correctamente a $\mathcal{H}$, además los índices $j_1, j_2, j_3, \cdots$ denotan índices distintos, a decir: $i, j, k, \cdots$ esta notación permite generalizar cualquier conjunto de índices):

$$\hat{H}^I = \sum_{n=2}^{N+1} \hat{H}_n^I \qquad \hat{H}_n^I = \sum_{j_n \cdots > j_2 > j_1 = 0}^{N} \left( g_{j_1 j_2 \cdots j_n} \hat{a}_{j_1}^{\dagger\, n-1} \hat{a}_{j_2} \cdots \hat{a}_{j_n} + h.c. \right) \qquad (3.5)$$

Donde $\hat{H}_n^I$ es un hamiltoniano de interacción entre $n$ osciladores cualesquiera: $\hat{a}_{j_1}^{\dagger\, n}$ indica que el oscilador $j_1$ se excita $n-1$ veces, mientras los $n-1$ osciladores restantes $j_2, \cdots j_n$ se relajan una vez cada uno, el hermítico conjugado se asocia al proceso inverso, $g_{j_1 j_2 \cdots j_n}$ es el acoplamiento de esa interacción, aplicando iterativamente estos procesos, incluso solo con $\hat{H}_2^I$, en principio es posible conseguir cualquier redistribución de la energía; de hecho, es posible concebir otros términos de interacción, como por ejemplo: $\hat{a}_{j_1}^{\dagger}\hat{a}_{j_2}^{\dagger}\hat{a}_{j_3}\hat{a}_{j_4}$, $\hat{a}_{j_1}^{\dagger\, 2}\hat{a}_{j_2}^{\dagger}\hat{a}_{j_3}\hat{a}_{j_4}\hat{a}_{j_5}$, $\hat{a}_{j_1}^{\dagger}\hat{a}_{j_2}^{\dagger}\hat{a}_{j_3}^{\dagger}\hat{a}_{j_4}\hat{a}_{j_5}\hat{a}_{j_6}$, $\hat{a}_{j_1}^{\dagger\, 3}\hat{a}_{j_2}^{\dagger\, 2}\hat{a}_{j_3}^{\dagger}\hat{a}_{j_4}\hat{a}_{j_5}\hat{a}_{j_6}\hat{a}_{j_7}\hat{a}_{j_8}$, $\cdots$, con sus respectivos acoples y sumados a sus hermíticos conjugados; sin embargo, todos tienen la característica de conservar el número de excitaciones, esta es una característica deseable que se explotará en el desarrollo de este modelo.

Los osciladores interactúan entre sí y dejan de estar aislados, pero el sistema total permanece aislado, con hamiltoniano total $\hat{H} = \hat{H}^0 + \hat{H}^I$, que es independiente del tiempo, y conserva la energía total; la evolución del colectivo viene dada por el operador unitario $\hat{U}(t) = \exp{-i\hat{H}t}$, entonces, considerando (3.4) como el estado inicial al tiempo $t = 0$, el estado en un tiempo posterior $t > 0$ viene a ser:

$$|\psi(t)\rangle = \sum_{\vec{n}} \psi_{\vec{n}}\, e^{-i\hat{H}t}|\vec{n}\rangle \qquad (3.6)$$

El estado general $|\psi(t)\rangle$ es representado en la base de $\{\,|\vec{n}\rangle\,\}$, que son los autoestados del hamiltoniano libre $\hat{H}^0$; la dificultad de desarrollar (3.6) es que al no ser $|\vec{n}\rangle$ autoestado de $\hat{H}$, para muchas partículas $N \gg 1$ se hace muy complicado su cálculo en esta base.

### 3.1.2 La base de los modos normales o vibraciones colectivas

En la mecánica clásica, un sistema de osciladores acoplados puede representar cualquier movimiento interno ya sea en las coordenadas $q_i$ de cada oscilador, o en la base de modos normales, de coordenadas $\tilde{q}_i$ que son combinaciones lineales[15] de $q_i$, que describen vibraciones colectivas, con la misma frecuencia y diferencia de fase, en todo el tiempo; así, cualquier movimiento interno del sistema se puede descomponer y representar en la base de modos normales, que son linealmente independientes; de hecho, si el sistema se encuentra oscilando en solo uno de esos modos, en principio permanecerá así para siempre.

Es posible desarrollar el análogo cuántico de los modos normales de oscilación para un sistema de muchos osciladores acoplados; de la mecánica cuántica se tiene que todo autoestado del hamiltoniano (independiente del tiempo) permanece estacionario en el tiempo, pues es también autoestado del operador de evolución $\hat{U}(t) = \exp{-i\hat{H}t}$, entonces para el espacio de Hilbert $\mathcal{H} = \otimes_{i=0}^{N} \mathcal{H}_i$ de dimensión $N_\mathcal{H}$ se tiene que el hamiltoniano $\hat{H} \in \mathcal{H}$ proporciona $N_\mathcal{H}$ autovectores $|\phi_{\vec{k}}\rangle$, combinaciones lineales de $|\vec{n}\rangle$, con autovalores $\omega_{\vec{k}}$, con los que también se puede formar una base completa de $\mathcal{H}$, entonces los estados $|\phi_{\vec{k}}\rangle$ pueden representar cualquier evolución temporal, donde cada uno de ellos permanece estacionario en el tiempo haciendo que una superposición de kets $|\vec{n}\rangle$ oscilen en el tiempo con una misma frecuencia $\omega_{\vec{k}}$.

---

[15] Esto se debe a que se plantean todas las ecuaciones de movimiento para cada oscilador en términos de sus coordenadas propias $q_i$, $p_i$, estas son ecuaciones diferenciales lineales de segundo orden en $q_i$, las cuales se pueden expresar como una ecuación matricial $\mathbb{M}\vec{x} = \ddot{\vec{x}}$ (no hay derivadas de primer orden $\dot{\vec{x}}$ porque ellas introducen disipaciones y pérdidas de energía, lo que no es el caso), donde $\vec{x}$ tiene información de las coordenadas $q_i$ y $\mathbb{M}$ de los acoples y frecuencias propias, si se busca la solución $\ddot{\vec{x}} = -\omega^2 \vec{x}$ entonces se tiene dos cuestiones importantes: por un lado, todas las coordenadas $q_i$ están oscilando simultáneamente con la misma frecuencia angular y la misma diferencia de fase, estas son las oscilaciones colectivas; y por otro lado, la ecuación matricial queda como una ecuación de autovalores $\mathbb{M}\vec{x} = -\omega^2 \vec{x}$; es decir, las nuevas coordenadas, o base que oscila colectivamente, son los autovectores de $\mathbb{M}$, y por lo tanto combinaciones lineales de nuestra base inicial, las coordenadas $q_i$.

Sea la ecuación de autovalores del Hamiltoniano total:

$$\hat{H}|\phi_{\vec{k}}\rangle = \omega_{\vec{k}}|\phi_{\vec{k}}\rangle \qquad |\phi_{\vec{k}}\rangle = \sum_{\vec{n}} \beta_{\vec{n},\vec{k}} |\vec{n}\rangle \qquad \sum_{\vec{k}} |\phi_{\vec{k}}\rangle\langle\phi_{\vec{k}}| = \hat{1} \qquad (3.7)$$

Las amplitudes $\beta_{\vec{n},\vec{k}}$ cumplen las siguientes relaciones de ortonormalidad:

$$\langle\vec{n}|\phi_{\vec{k}}\rangle = \beta_{\vec{n},\vec{k}} \qquad \sum_{\vec{n}} \beta^*_{\vec{n},\vec{k}} \beta_{\vec{n},\vec{k}'} = \delta_{\vec{k},\vec{k}'} \qquad \sum_{\vec{k}} \beta^*_{\vec{n},\vec{k}} \beta_{\vec{n}',\vec{k}} = \delta_{\vec{n},\vec{n}'} \qquad (3.8)$$

las dos últimas ecuaciones de (3.8) se obtienen de $\langle\phi_{\vec{k}}|\hat{1}_{\vec{n}}|\phi_{\vec{k}'}\rangle$ y $\langle\vec{n}|\hat{1}_{\phi_{\vec{k}}}|\vec{n}'\rangle$, con $\hat{1}_{\vec{n}} = \sum_{\vec{n}}|\vec{n}\rangle\langle\vec{n}|$ y $\hat{1}_{\phi_{\vec{k}}} = \sum_{\vec{k}}|\phi_{\vec{k}}\rangle\langle\phi_{\vec{k}}|$; así, $|\vec{n}\rangle = \sum_k \beta^*_{\vec{n},\vec{k}}|\phi_{\vec{k}}\rangle$, y de la primera ecuación de (3.7) se deduce: $e^{-i\hat{H}t}|\vec{n}\rangle = \sum_k \beta^*_{\vec{n},\vec{k}} e^{-i\hat{H}t}|\phi_{\vec{k}}\rangle = \sum_k \beta^*_{\vec{n},\vec{k}} e^{-i\omega_{\vec{k}}t}|\phi_{\vec{k}}\rangle$, reemplazando en el ket de estado de (3.6):

$$|\psi(t)\rangle = \sum_{\vec{n}} \psi_{\vec{n}} \sum_{\vec{k}} \beta^*_{\vec{n},\vec{k}} e^{-i\omega_{\vec{k}}t} |\phi_{\vec{k}}\rangle \qquad (3.9)$$

(3.9) retorna a la base $\{|\vec{n}\rangle\}$ mediante (3.7):

$$|\psi(t)\rangle = \sum_{\vec{n}} \psi_{\vec{n}}(t)|\vec{n}\rangle \qquad \psi_{\vec{n}}(t) = \sum_{\vec{k}} \left(\sum_{\vec{n}'} \psi_{\vec{n}'} \beta^*_{\vec{n}',\vec{k}}\right) \beta_{\vec{n},\vec{k}} e^{-i\omega_{\vec{k}}t} \qquad (3.10)$$

$$\psi_{\vec{n}}(0) = \psi_{\vec{n}}$$

(3.10) provee la amplitud dependiente del tiempo $\psi_{\vec{n}}(t)$, donde $\psi_{\vec{n}}(0) = \psi_{\vec{n}}$ en consistencia con (3.4) como el estado inicial, esto es posible mediante las amplitudes $\beta_{\vec{n},\vec{k}}$ que son las componentes de los autoestados $|\phi_{\vec{k}}\rangle$ en la base $\{|\vec{n}\rangle\}$.

Es importante notar que toda la información de los acoples e interacciones entre osciladores está resumida en las amplitudes $\beta_{\vec{n},\vec{k}}$ de los autoestados y los autovalores $\omega_{\vec{k}}$.

### 3.1.3 Operador de densidad del oscilador $i$

Las ecuaciones (3.10) proveen un estado entrelazado entre los osciladores del sistema, incluso si el estado inicial era completamente factorizado, sin embargo, al realizar una medición de la energía en cualquier oscilador a un tiempo $t > 0$ se tienen las probabilidades de encontrar al oscilador $i$ en un estado energético $|n_i\rangle$, estas probabilidades vienen dadas por el operador de densidad que resulta de tomar la traza parcial al estado puro $|\psi(t)\rangle\langle\psi(t)|$ en todos los grados de libertad del entorno del oscilador $i$, denotado por el índice vectorial $\vec{\xi}_i$ :

$$\vec{\xi}_i = (n'_0, n'_1, \cdots n'_{j \neq i} \cdots n'_N) \qquad \dim \vec{\xi}_i = N \qquad |\vec{\xi}_i\rangle \in \mathcal{H}$$
$$|\vec{\xi}_i\rangle = |n'_0\rangle \otimes |n'_1\rangle \otimes \cdots |n'_{i-1}\rangle \otimes \hat{1} \otimes |n'_{i+1}\rangle \otimes \cdots |n'_N\rangle \tag{3.11}$$

El braket $\langle\vec{\xi}_i|\vec{n}\rangle$ viene a ser:

$$\langle\vec{\xi}_i|\vec{n}\rangle = |n_i\rangle \prod_{\forall j \neq i} \delta_{n_j, n'_j} \tag{3.12}$$

Entonces, el braket entre $|\psi(t)\rangle$ de (3.10) y $|\vec{\xi}_i\rangle$ de (3.11), usando (3.12) es:

$$\langle\vec{\xi}_i|\psi(t)\rangle = \sum_{n_i} \psi_{n_i, \vec{\xi}_i}(t) |n_i\rangle \qquad \vec{n} \to (n_i, \vec{\xi}_i) \tag{3.13}$$

$\vec{n} \to (n_i, \vec{\xi}_i)$ es la notación que dice: todos los índices $n_j$ de $\vec{n}$, introducido en (3.2), son expresados como el conjunto del índice $n_i$ y todos los índices $n'_k$ de $\vec{\xi}_i$ de (3.11).

Así, el operador de densidad resultante de tomar la traza parcial en todos los grados de libertad del entorno al oscilador $i$, que es descrito por el índice vector $\vec{\xi}_i$ viene a ser:

$$\hat{\rho}_i(t) = Tr_{\vec{\xi}_i} |\psi(t)\rangle\langle\psi(t)| = \sum_{\vec{\xi}_i} \langle\vec{\xi}_i|\psi(t)\rangle\langle\psi(t)|\vec{\xi}_i\rangle = \sum_{\sigma,\sigma'} \sum_{\vec{\xi}_i} \psi_{\sigma,\vec{\xi}_i}(t) \psi^*_{\sigma',\vec{\xi}_i}(t) |\sigma\rangle\langle\sigma'|$$

Donde $|\sigma\rangle = |n_i\rangle$ es el estado de (3.13) y solo se ha cambiado de letra $n_i \to \sigma$ por simplicidad en la escritura; entonces, usando (3.10) en los términos $\psi_{\sigma,\vec{\xi}_i}$ de la expresión anterior se obtiene $\hat{\rho}_i(t)$:

$$\hat{\rho}_i(t) = \sum_{\sigma,\sigma'} \rho^i_{\sigma,\sigma'}(t) |\sigma\rangle\langle\sigma'|$$

$$\rho^i_{\sigma,\sigma'}(t) = \sum_{\vec{k},\vec{k}'} \left( \sum_{\vec{n}',\vec{n}''} \psi_{\vec{n}'} \psi^*_{\vec{n}''} \beta^*_{\vec{n}',\vec{k}} \beta_{\vec{n}'',\vec{k}'} \right) \left( \sum_{\vec{\xi}_i} \beta_{\sigma,\vec{\xi}_i,\vec{k}} \beta^*_{\sigma',\vec{\xi}_i,\vec{k}'} \right) e^{-i(\omega_{\vec{k}} - \omega_{\vec{k}'})t}$$

(3.14)

$\rho^i_{\sigma,\sigma'}(t)$ se puede separar en dos sumas, una que es independiente del tiempo, y otra con dependencia temporal, para esto se introduce el conjunto $D$ de todos los pares $(\vec{k}, \vec{k}')$ que etiquetan autovalores degenerados, que conducen al término $\chi^D_{\sigma,\sigma'}$ que no depende del tiempo, y el cual es considerado en la parte no dependiente del tiempo $\chi_{\sigma,\sigma'}$, y excluido de la parte con dependencia temporal $\Gamma_{\sigma,\sigma'}(t)$:

$$\rho^i_{\sigma,\sigma'}(t) = \chi_{\sigma,\sigma'} + \Gamma_{\sigma,\sigma'}(t) \qquad D = \left\{ (\vec{k}, \vec{k}') / \vec{k} \neq \vec{k}' \wedge \omega_{\vec{k}} = \omega_{\vec{k}'} \right\}$$

$$\chi^D_{\sigma,\sigma'} = \sum_{\vec{k},\vec{k}' \in D} \left( \sum_{\vec{n}',\vec{n}''} \psi_{\vec{n}'} \psi^*_{\vec{n}''} \beta^*_{\vec{n}',\vec{k}} \beta_{\vec{n}'',\vec{k}'} \right) \sum_{\vec{\xi}_i} \beta_{\sigma,\vec{\xi}_i,\vec{k}} \beta^*_{\sigma',\vec{\xi}_i,\vec{k}'}$$

$$\chi_{\sigma,\sigma'} = \sum_{\vec{k}} \left( \sum_{\vec{n}',\vec{n}''} \psi_{\vec{n}'} \psi^*_{\vec{n}''} \beta^*_{\vec{n}',\vec{k}} \beta_{\vec{n}'',\vec{k}} \right) \sum_{\vec{\xi}_i} \beta_{\sigma,\vec{\xi}_i,\vec{k}} \beta^*_{\sigma',\vec{\xi}_i,\vec{k}} + \chi^D_{\sigma,\sigma'}$$

$$\Gamma_{\sigma,\sigma'}(t) = \sum_{\vec{k} \neq \vec{k}'} \left( \sum_{\vec{n}',\vec{n}''} \psi_{\vec{n}'} \psi^*_{\vec{n}''} \beta^*_{\vec{n}',\vec{k}} \beta_{\vec{n}'',\vec{k}'} \right) \sum_{\vec{\xi}_i} \beta_{\sigma,\vec{\xi}_i,\vec{k}} \beta^*_{\sigma',\vec{\xi}_i,\vec{k}'} e^{-i(\omega_{\vec{k}} - \omega_{\vec{k}'})t} - \chi^D_{\sigma,\sigma'}$$

(3.15)

$\chi^D_{\sigma,\sigma'}$ es el término independiente del tiempo debido a la degeneración, y se anula cuando no hay degeneración. Los términos de la diagonal de $\rho^i_{\sigma,\sigma'}(t)$ de (3.15) son:

$$\chi^D_{\sigma,\sigma} = 2\mathbb{Re}\left[ \sum_{\vec{k}' > \vec{k} \in D} \sum_{\vec{n}',\vec{n}''} \psi_{\vec{n}'} \psi^*_{\vec{n}''} \beta^*_{\vec{n}',\vec{k}} \beta_{\vec{n}'',\vec{k}'} \sum_{\vec{\xi}_i} \beta_{\sigma,\vec{\xi}_i,\vec{k}} \beta^*_{\sigma,\vec{\xi}_i,\vec{k}'} \right]$$

(3.16)

$$\chi_{\sigma,\sigma} = \sum_{\vec{k}} \sum_{\vec{n}',\vec{n}''} \psi_{\vec{n}'} \psi^*_{\vec{n}''} \beta^*_{\vec{n}',\vec{k}} \beta_{\vec{n}'',\vec{k}} \sum_{\vec{\xi}_i} \left|\beta_{\sigma,\vec{\xi}_i,\vec{k}}\right|^2 + \chi^D_{\sigma,\sigma}$$

$$\Gamma_{\sigma,\sigma}(t) = 2\mathbb{Re}\left[\sum_{\vec{k}'>\vec{k}} \sum_{\vec{n}',\vec{n}''} \psi_{\vec{n}'} \psi^*_{\vec{n}''} \beta^*_{\vec{n}',\vec{k}} \beta_{\vec{n}'',\vec{k}'} \sum_{\vec{\xi}_i} \beta_{\sigma,\vec{\xi}_i,\vec{k}} \beta^*_{\sigma,\vec{\xi}_i,\vec{k}'} e^{-i(\omega_{\vec{k}}-\omega_{\vec{k}'})t}\right] - \chi^D_{\sigma,\sigma}$$

En resumen de esta sección, el operador de densidad del oscilador $i$ viene dado por $\hat{\rho}_i(t)$ de (3.14), cuyos elementos $\rho^i_{\sigma,\sigma'}(t)$ son descritos en la parte estacionaria y temporal de (3.15), y en particular, los elementos de la diagonal son dados en (3.16).

### 3.1.4 Promedio temporal y arribo a estado de equilibrio

Si se realiza un promedio temporal para tiempos largos en (3.15), se anula el término $\Gamma_{\sigma,\sigma'}(t)$ mientras que sobreviven $\chi^D_{\sigma,\sigma}$ y $\chi_{\sigma,\sigma}$ por no tener dependencia temporal, pues el promedio del factor exponencial compleja se anula:

$$\overline{f(t)} = \lim_{\Delta t \to \infty} \frac{1}{\Delta t} \int_t^{t+\Delta t} f(t')dt' \qquad \overline{e^{i\omega t}} = \begin{cases} 1, \omega = 0 \\ 0, \omega \neq 0 \end{cases} \qquad (3.17)$$

Para $\Delta t \gg \max|\omega_{\vec{k}} - \omega_{\vec{k}'}|^{-1}$ el promedio temporal de $\Gamma_{\sigma,\sigma'}(t)$ se anula:

$$\overline{\Gamma_{\sigma,\sigma'}(t)} = 0 \qquad \Delta t \gg \max|\omega_{\vec{k}} - \omega_{\vec{k}'}|^{-1} \qquad (3.18)$$

De manera que el promedio temporal en esta escala de tiempo del operador probabilístico viene a ser, usando (3.14), (3.15) y (3.18):

$$\hat{\rho}_i^{eq} = \overline{\hat{\rho}_i(t)} = \sum_{\sigma,\sigma'} \chi_{\sigma,\sigma'} |\sigma\rangle\langle\sigma'| \qquad \Delta t \gg \max|\omega_{\vec{k}} - \omega_{\vec{k}'}|^{-1} \qquad (3.19)$$

La justificación del promedio temporal puede ser debido a un cambio de escala de tiempos característicos de la evolución del sistema, del orden de $\tau \sim |\omega_{\vec{k}} - \omega_{\vec{k}'}|^{-1}$, a una escala de tiempo de orden superior $\Delta t$; por ejemplo, el operador de densidad $\hat{\rho}_i(t)$ en la escala microscópica del tiempo $\tau$ puede ser descrito como oscilando cuasi aleatoriamente alrededor del valor medio $\hat{\rho}_i^{eq}$, en esta escala se puede apreciar el cambio suave y continuo de $\hat{\rho}_i(t)$ con el paso del tiempo, mientras que desde la escala clásica ya no se puede observar con suficiente resolución temporal a $\hat{\rho}_i(t)$ y al realizar una medición se tiene una distribución de probabilidad en un intervalo grande de tiempo, respecto a $\tau$, en el que se tiene incertidumbre sobre si se va a medir el estado $\hat{\rho}_i(t_1)$ o $\hat{\rho}_i(t_1 + \Delta t)$ (y desde luego, todos los posibles estados intermedios $\hat{\rho}_i(t)$, $t \in [t_1, t_1 + \Delta t]$ ), de manera que se debe promediar entre ellos en el tiempo.

El operador de densidad $\hat{\rho}_i^{eq} = \overline{\hat{\rho}_i(t)}$ puede ser considerado como el estado de equilibrio al que arriba el sistema tras su evolución temporal para tiempos suficientemente largos, y puede ser interpretado como una evolución irreversible hacia un estado de equilibrio; es posible deducir el estado de equilibrio para el sistema total al realizar el promedio temporal de $|\psi(t)\rangle\langle\psi(t)|$, usando (3.9) y (3.17):

$$\hat{\rho}^{eq} = \overline{|\psi(t)\rangle\langle\psi(t)|} = \sum_{\vec{k}} \rho_{\vec{k},\vec{k}} |\phi_{\vec{k}}\rangle\langle\phi_{\vec{k}}| + \sum_{\vec{k},\vec{k}' \in D} \rho_{\vec{k},\vec{k}'} |\phi_{\vec{k}}\rangle\langle\phi_{\vec{k}'}|$$

$$\rho_{\vec{k},\vec{k}'} = \varphi_{\vec{k}} \varphi_{\vec{k}'}^* \qquad \varphi_{\vec{k}} = \sum_{\vec{n}} \psi_{\vec{n}} \beta_{\vec{n},\vec{k}}^* = \langle\phi_{\vec{k}}|\psi(0)\rangle$$

(3.20)

De acuerdo a (3.20), el promedio temporal de $|\psi(t)\rangle\langle\psi(t)|$ elimina los términos de coherencia (dejando solo los asociados a autoestados degenerados $\vec{k}, \vec{k}' \in D$) del estado inicial, dejando inalteradas las probabilidades de la diagonal; se puede entender a $\hat{\rho}^{eq}$ como el estado de equilibrio cuántico al que todo el sistema de osciladores arriba irreversiblemente.

Estos resultados (3.19) y (3.20) deberían ser consistentes con los estados mezcla correspondientes al equilibrio térmico, dado por $\hat{\rho} = Z^{-1} e^{-\beta \hat{H}}$, $Z = Tr[e^{-\beta \hat{H}}]$, si este fuera el caso, la evolución unitaria, por si sola, explicaría la termalización de un conjunto de osciladores con una distribución arbitraria de energía e información cuántica.

En (3.19) y (3.20) el operador de densidad de equilibrio, promediado en el tiempo, tiene dependencia de las amplitudes $\psi_{\vec{n}}$, que corresponde al estado inicial del sistema total, de manera que $\hat{\rho}_i^{eq}$ y $\hat{\rho}^{eq}$ tienen memoria del estado inicial, mientras que en la termalización el estado de equilibrio térmico pierde toda la información del estado inicial, excepto por la información de las cantidades conservadas, como la energía total.

## 3.2 Modelo de osciladores de 2 niveles: sistema - entorno

En la sección anterior [3.1.4] se evidenció que los estados de un oscilador $i$ y de todo el sistema al arribar al equilibrio $\hat{\rho}_i^{eq}$ y $\hat{\rho}^{eq}$ contienen parte de la información del estado inicial total[16], de las amplitudes $\varphi_{\vec{k}}$ en la base de los modos normales; abordar este efecto memoria inherente es un problema de mucho interés para el modelo simplificado; sin embargo, por simplicidad se formula un caso especial donde todos los osciladores están restringidos a ser sistemas de dos niveles (de manera que pueden encontrarse en el estado fundamental o excitado), se hace la distinción entre el sistema (para el oscilador $i = 0$) y el entorno compuesto de $N$ osciladores (para los demás osciladores, $i = 1:N$); este es uno de los modelos más simples que puede adoptar el modelo simplificado general, y aun así evaluar de forma exacta las ecuaciones (3.15), (3.16) y (3.19) para el sistema.

### 3.2.1 Única frecuencia y sin interacción entre osciladores del entorno

Sean los ket $|n_i\rangle \in \mathcal{H}_i$ de (3.1) de dos niveles:

$$|0\rangle = \begin{pmatrix}1\\0\end{pmatrix} \qquad |1\rangle = \begin{pmatrix}0\\1\end{pmatrix} \qquad n_i \equiv 0, 1 \qquad (3.21)$$

Así, los ket $|\vec{n}\rangle$ vienen a ser una sucesión en binario: $|\vec{n}\rangle = |10110101\cdots 0\rangle$; por exceso de simplicidad del modelo, se puede partir de un hamiltoniano donde todos los osciladores tienen la misma frecuencia del sistema $\omega_0 = \omega$, y el hamiltoniano de interacción solo permite acople entre el sistema y cada oscilador del entorno:

$$\begin{aligned}\hat{H} &= \hat{H}_s + \hat{H}_\varepsilon + \hat{H}_I & \hat{H}_s &= \omega \hat{a}_0^\dagger \hat{a}_0 \\ \hat{H}_0 &= \hat{H}_s + \hat{H}_\varepsilon & \hat{H}_\varepsilon &= \omega \sum_{i=1}^{N} \hat{a}_i^\dagger \hat{a}_i & \hat{H}_I &= \sum_{i=1}^{N}\left(g_i \hat{a}_0 \hat{a}_i^\dagger + hc.\right)\end{aligned} \qquad (3.22)$$

Esta es una situación muy restrictiva, y es inviable para representar un reservorio que contiene osciladores de muchas frecuencias distintas, sin embargo, puede ser útil para entender cómo la energía (y la coherencia) del sistema se redistribuye al entorno.

---

[16] Si bien se pierde la información cuántica asociado a la coherencia, tanto en (3.19) como en (3.20) se tiene dependencia de las amplitudes iniciales $\psi_{\vec{n}}$, puesto que los términos que sobreviven al promedio temporal $\chi_{\sigma,\sigma'}$ no son independientes de las amplitudes del estado inicial.

### 3.2.2 Hamiltoniano sin interacción entre osciladores del entorno

Se considera que todos los osciladores tienen frecuencias distintas, entonces el hamiltoniano total (3.22) se plantea de forma análoga a (2.44) en (2.32):

$$\widehat{H} = \widehat{H}_s + \widehat{H}_\varepsilon + \widehat{H}_I \qquad \widehat{H}_s = \omega_0 \hat{a}_0^\dagger \hat{a}_0$$

$$\widehat{H}_0 = \widehat{H}_s + \widehat{H}_\varepsilon \qquad \widehat{H}_\varepsilon = \sum_{i=1}^{N} \omega_i \hat{a}_i^\dagger \hat{a}_i \qquad \widehat{H}_I = \sum_{i=1}^{N} \left( g_i \hat{a}_0 \hat{a}_i^\dagger + hc. \right) \qquad (3.23)$$

$\widehat{H}_0$ es el hamiltoniano libre: $\widehat{H}_0 = \sum_{i=0}^{N} \omega_i \hat{a}_i^\dagger \hat{a}_i$, la interacción está reducida a que el sistema se relaje a la vez que un oscilador del entorno se excite, y viceversa, pero sin interacción entre los osciladores del entorno, desde que solo pueden tener dos niveles energéticos se descarta interacciones como $\hat{a}_{j_1}^{\dagger\,2} \hat{a}_{j_2} \hat{a}_{j_3}$, en efecto $\widehat{H}_I$ de (3.23) viene a ser $\widehat{H}_2^I$ de (3.5) pero para $i_1 = i$, $i_2 = 0$; esta descripción es apropiada para un oscilador en interacción con $N$ fotones, ya que los fotones no interactúan entre sí, aunque incluso si el modelo no fuera de dos estados para cada oscilador, el hamiltoniano $\widehat{H}_I$ de (3.23) se mantiene en consistencia con $\widehat{H}_I$ de (2.44).

### 3.2.3 Hamiltoniano con interacción entre osciladores del entorno

Se modifica (3.23) para añadir términos de interacción entre los osciladores del entorno, de acuerdo a $\widehat{H}_2^I$ de (3.5):

$$\widehat{H} = \widehat{H}_s + \widehat{H}_\varepsilon + \widehat{H}_I \qquad \widehat{H}_s = \omega_0 \hat{a}_0^\dagger \hat{a}_0$$

$$\widehat{H}_0 = \widehat{H}_s + \widehat{H}_\varepsilon \qquad \widehat{H}_\varepsilon = \sum_{i=1}^{N} \omega_i \hat{a}_i^\dagger \hat{a}_i \qquad \widehat{H}_I = \sum_{j>i=0}^{N} \left( g_{ij} \hat{a}_i \hat{a}_j^\dagger + hc. \right) \qquad (3.24)$$

Es posible escribir (3.24) de forma compacta, al expresar $\widehat{H} = \widehat{H}_0 + \widehat{H}_I$ en una sola sumatoria, y generalizar los acoples $g_{ij}$:

$$\widehat{H} = \sum_{i,j=0}^{N} g_{ij} \hat{a}_j^\dagger \hat{a}_i \qquad g_{ji} = g_{ij}^* \qquad g_{ii} = \omega_i \qquad (3.25)$$

pues la condición $g_{ji} = g_{ij}^*$ permite que el hermítico conjugado de $\widehat{H}_I$ en (3.24) corresponde a la sumatoria de $\widehat{H}_I$ para $i > j = 0:N$, y $g_{ii} = \omega_i$ permite que $\widehat{H}_0$ se integre a ambas sumatorias, donde $\forall i \neq j$ se tiene $[\hat{a}_i, \hat{a}_j] = 0$ porque los operadores pertenecen a espacios distintos.

Este hamiltoniano corresponde a un modelo donde los demás osciladores del entorno también interactúan entre sí, de manera que es más efectivo la redistribución de la energía e información del sistema al entorno.

### 3.2.4 Sistema fijado a $i = 0$

En esta sección se considera el modelo sistema – entorno, con el sistema fijado en $i = 0$; esto es, que el sistema es el oscilador etiquetado con $i = 0$, todos los demás osciladores etiquetados por los índices $i = 1, 2, 3 \cdots N$ conforman el entorno; entonces se escribe $\rho_{\sigma,\sigma'}^{i=0} = \rho_{\sigma,\sigma'}$ y para el índice vector del entorno $\vec{\xi}_0 = \vec{\xi}$, así, $i = 0$ en (3.15):

$$
\left.\begin{aligned}
\rho_{\sigma,\sigma'}(t) &= \chi_{\sigma,\sigma'} + \Gamma_{\sigma,\sigma'}(t) \qquad\qquad D = \left\{ (\vec{k}, \vec{k}') \,/\, \omega_{\vec{k}} = \omega_{\vec{k}'} \right\} \\
\chi_{\sigma,\sigma'}^D &= \sum_{\vec{k}, \vec{k}' \in D} \left( \sum_{\vec{n}', \vec{n}''} \psi_{\vec{n}'} \psi_{\vec{n}''}^* \beta_{\vec{n}', \vec{k}}^* \beta_{\vec{n}'', \vec{k}'} \right) \sum_{\vec{\xi}} \beta_{\sigma, \vec{\xi}, \vec{k}} \beta_{\sigma', \vec{\xi}, \vec{k}'}^* \\
\chi_{\sigma,\sigma'} &= \sum_{\vec{k}} \left( \sum_{\vec{n}', \vec{n}''} \psi_{\vec{n}'} \psi_{\vec{n}''}^* \beta_{\vec{n}', \vec{k}}^* \beta_{\vec{n}'', \vec{k}} \right) \sum_{\vec{\xi}} \beta_{\sigma, \vec{\xi}, \vec{k}} \beta_{\sigma', \vec{\xi}, \vec{k}}^* + \chi_{\sigma,\sigma'}^D \\
\Gamma_{\sigma,\sigma'}(t) &= \sum_{\vec{k} \neq \vec{k}'} \left( \sum_{\vec{n}', \vec{n}''} \psi_{\vec{n}'} \psi_{\vec{n}''}^* \beta_{\vec{n}', \vec{k}}^* \beta_{\vec{n}'', \vec{k}'} \right) \sum_{\vec{\xi}} \beta_{\sigma, \vec{\xi}, \vec{k}} \beta_{\sigma', \vec{\xi}, \vec{k}'}^* e^{-i(\omega_{\vec{k}} - \omega_{\vec{k}'})t} - \chi_{\sigma,\sigma'}^D
\end{aligned}\right\} \quad (3.26)
$$

De acuerdo a (3.19), $\chi_{\sigma,\sigma'}$ es el término que sobrevive al promediar el operador de densidad del sistema, mientras que $\Gamma_{\sigma,\sigma'}(t)$ es la suma de términos oscilantes en el tiempo, los cuales introducen desviaciones, que puede ser análogas a un ruido estocástico, y que hacen variar el estado del sistema $\rho_{\sigma,\sigma'}(t)$ alrededor de $\chi_{\sigma,\sigma'}$.

# CAPÍTULO IV

# DIAGRAMAS Y ALGORITMOS PARA RESOLVER EL MODELO

## 4.1 Esquema general

Las ecuaciones (3.15) contienen toda la información del operador de densidad del sistema a un tiempo $t > 0$, como la traza parcial en todos los grados del entorno. Para avanzar en estos cálculos se necesita conocer las amplitudes $\beta_{\vec{n},\vec{k}}$ de los autoestados y los autovalores $\omega_{\vec{k}}$. El espacio de Hilbert total $\mathcal{H}$ tiene dimensión $2^{N+1}$, el índice $\vec{k}$ toma $2^{N+1}$ valores distintos, etiquetando los autoestados y autovalores; así, para $N$ suficientemente grande, se requiere descomponer el problema general en "sub problemas" más fáciles de abordar. Al extender el método empleado en la sección [2.4] dividiendo el espacio total en sub espacios en los que se tiene hamiltonianos reducidos para esa dinámica, se plantean allí las ecuaciones de autovalores para obtener los autovectores y autovalores que forman una base para esos sub espacios. Este proceso se resuelve de forma numérica pues los sub espacios siguen teniendo una dimensión grande, al menos del orden de $N$ (para sub espacios no triviales, los cuales solo tienen una dimensión); así, las valores numéricos de $\beta_{\vec{n},\vec{k}}$ y $\omega_{\vec{k}}$ son reemplazados en (3.15) para evaluar la evolución temporal del operador de densidad del sistema de forma numérica, esto no supone un problema de cálculo numérico relevante, ya que las ecuaciones (3.15) ya dan la expresión analítica para cualquier instante de tiempo, y solo recae la atención en la precisión del cálculo numérico de autovectores, autovalores, y las sumas y productos requeridos en (3.15).

También se necesita las condiciones iniciales para introducir en (3.15), en la Figura 4.1 se considera la elección de estas condiciones iniciales expresadas en los sub espacios que, al evaluarlas en el modelo simplificado, reducen las ecuaciones (3.15) a nuevas ecuaciones que reducen enormemente la complejidad del problema, y en las que se reemplazan los valores numéricos de $\beta_{\vec{n},\vec{k}}$ y $\omega_{\vec{k}}$ en estos sub espacios, al final se evalúa y grafica la evolución temporal, así como los valores medios que resulta de promediar temporalmente.

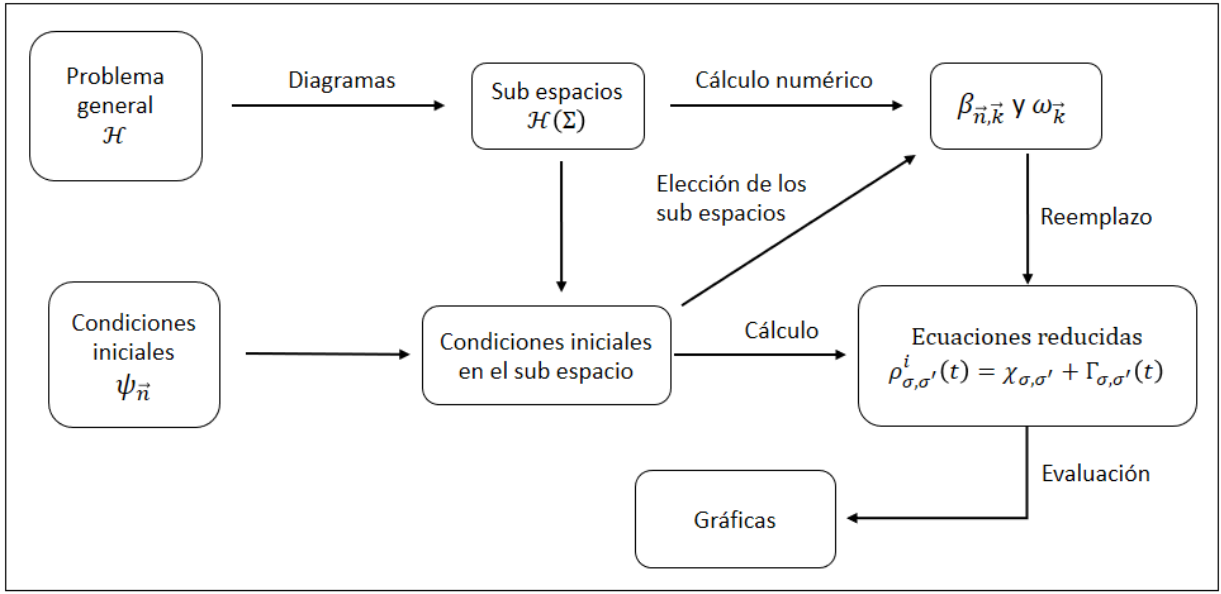

**Figura 4.1** Diagrama sobre las etapas generales para evaluar y graficar el comportamiento de las ecuaciones (3.15) para unas condiciones iniciales dadas.

## 4.2 Descomposición del espacio de Hilbert en suma de subespacios de número de excitación definido

Esta sección es una extensión del método empleado en la sección [2.4] para cuando el entorno tiene muchos osciladores.

De forma análoga a (2.50), se establece el número de excitación $\hat{\Sigma}$:

$$\hat{\Sigma} = \sum_{i=0}^{N} \hat{a}_i^\dagger \hat{a}_i \qquad \hat{\Sigma}|\vec{n}\rangle = \Sigma|\vec{n}\rangle \qquad [\hat{H}, \hat{\Sigma}] = 0 \rightarrow \frac{d}{dt}\hat{\Sigma} = 0 \qquad (4.1)$$

$$\Sigma = \Sigma_{\vec{n}} = \sum_{i=0}^{N} n_i \qquad \Sigma = 0{:}N+1$$

Donde el autovalor $\Sigma$ solo toma valores enteros, desde $0$ hasta $N+1$; como $\hat{\Sigma}$ conmuta con el hamiltoniano es una constante en la evolución temporal, así la evolución temporal de un autoestado de $\hat{H}_0$ siempre se va a encontrar en una superposición de otros autoestados de $\hat{H}_0$ que tengan el mismo $\Sigma_{\vec{n}}$ ; es decir, solo pueden evolucionar entre los autoestados de $\hat{\Sigma}$ que tiene el mismo autovalor degenerado $\Sigma_{\vec{n}}$ :

$$e^{-i\hat{H}t}|\vec{n}\rangle = \sum_{\vec{n}'\in\mathcal{N}(\Sigma)} \alpha_{\vec{n}'}(t)|\vec{n}'\rangle \qquad \mathcal{N}(\Sigma) = \{\vec{n} \;/\; \hat{\Sigma}|\vec{n}\rangle = \Sigma|\vec{n}\rangle\} \qquad (4.2)$$

Donde $\mathcal{N}(\Sigma)$ es el conjunto de todos los $\vec{n}$ tal que $\hat{\Sigma}|\vec{n}\rangle = \Sigma|\vec{n}\rangle$; si se promedia $\hat{\Sigma}$ en el estado $e^{-i\hat{H}t}|\vec{n}\rangle$ de (4.2), se obtiene $\langle\hat{\Sigma}\rangle = \Sigma$, el cual permanece constante en todo el tiempo, de forma consistente con la conmutación de (4.1).

Puesto que la evolución temporal de cada ket $|\vec{n}\rangle$ está confinada a un subespacio formado por todas las combinaciones lineales de los ket $|\vec{n}'\rangle$, $\vec{n}' \in \mathcal{N}(\Sigma)$, se define ese sub espacio de Hilbert como $\mathcal{H}(\Sigma)$, llamado *Subespacio de número de excitación definido*, tal que la suma directa de todos los subespacios da el espacio total de Hilbert de los $N+1$ osciladores:

$$\mathcal{H}(\Sigma) = \{|\vec{n}\rangle \in \mathcal{H} \;/\; \vec{n} \in \mathcal{N}(\Sigma)\}$$
$$\# = \#_\Sigma = \dim \mathcal{H}(\Sigma) = C_\Sigma^{N+1} \qquad \mathcal{H} = \bigoplus_{\Sigma=0}^{N+1} \mathcal{H}(\Sigma) = \bigotimes_{i=0}^{N} \mathcal{H}_i \qquad (4.3)$$

Donde $\# = \#_\Sigma$ es la dimensión de $\mathcal{H}(\Sigma)$, el número de elementos de $\mathcal{N}(\Sigma)$, o el grado de degeneración asociado al autovalor $\Sigma$, y corresponde con el número de combinaciones de $\Sigma$ unos en $N+1$ osciladores; $\mathcal{H}(0)$ y $\mathcal{H}(N+1)$ son subespacios triviales que contienen un solo elemento: $|\vec{0}\rangle = |000\cdots0\rangle$ y $|\vec{1}\rangle = |111\cdots1\rangle$ respectivamente.

En cada $\mathcal{H}(\Sigma)$ se puede definir la identidad $\hat{1}_\Sigma \in \mathcal{H}(\Sigma)$, que es el proyector del subespacio, con los que se puede reducir el hamiltoniano total, proyectando $\hat{H}$ en $\mathcal{H}(\Sigma)$, análogo a (2.53):

$$\hat{1}_\Sigma = \sum_{\vec{n}\in\mathcal{N}(\Sigma)} |\vec{n}\rangle\langle\vec{n}| \qquad \hat{H}^\Sigma = \hat{1}_\Sigma \hat{H} \hat{1}_\Sigma \;\; \in \mathcal{H}(\Sigma) \qquad (4.4)$$

Así, $\hat{1}_\Sigma \hat{H}_0 \hat{1}_\Sigma$ da términos en la diagonal, mientras que $\hat{1}_\Sigma \hat{H}_I \hat{1}_\Sigma$ da términos fuera de la diagonal, en la base de $|\vec{n}\rangle$; $\hat{H}^\Sigma = \hat{1}_\Sigma \hat{H} \hat{1}_\Sigma$ es el hamiltoniano total reducido al subespacio $\mathcal{H}(\Sigma)$, los autovectores $|\phi_{\vec{k}}\rangle$ de $\hat{H}^\Sigma$ también pertenecen al mismo subespacio:

$$\widehat{H}^{\Sigma}|\phi_{\vec{k}}\rangle = \omega_{\vec{k}}|\phi_{\vec{k}}\rangle \qquad\qquad |\phi_{\vec{k}}\rangle \in \mathcal{H}(\Sigma) \qquad (4.5)$$

Así, tanto $|\vec{n}\rangle$ como $|\phi_{\vec{k}}\rangle$ componen la base de un mismo subespacio $\mathcal{H}(\Sigma)$, esto significa que existen $\# = C_{\Sigma}^{N+1}$ autoestados o modos de oscilaciones, de vibraciones colectivas con frecuencias $\omega_{\vec{k}}$.

(4.5) provee una información muy importante sobre las amplitudes $\beta_{\vec{n},\vec{k}} = \langle \vec{n}|\phi_{\vec{k}}\rangle$:

$$\begin{aligned} |\phi_{\vec{k}}\rangle \in \mathcal{H}(\Sigma) \\ |\vec{n}\rangle \in \mathcal{H}(\Sigma') \end{aligned} \rightarrow \beta_{\vec{n},\vec{k}} = \delta_{\Sigma,\Sigma'} \beta_{\vec{n},\vec{k}} \qquad |\phi_{\vec{k}}\rangle = \sum_{\vec{n}\in \mathcal{N}(\Sigma)} \beta_{\vec{n},\vec{k}} |\vec{n}\rangle \qquad (4.6)$$

Si $|\phi_{\vec{k}}\rangle$ y $|\vec{n}\rangle$ pertenecen a subespacios distintos, su braket se anula; es decir, el ket $|\phi_{\vec{k}}\rangle$ tiene amplitudes nulas $\beta_{\vec{n},\vec{k}} = 0$ en las componentes $|\vec{n}\rangle \notin \mathcal{H}(\Sigma)$.

### 4.3 Diagramas de acción del hamiltoniano reducido

En esta sección se presentan diagramas que representan de forma esquemática la acción del hamiltoniano, para posteriormente generalizar un algoritmo que genere las componentes matriciales del hamiltoniano reducido $\widehat{H}^{\Sigma}$.

Estos diagramas contienen toda la información del Hamiltoniano reducido.

#### 4.3.1 Hamiltoniano de interacción de (3.22)

El hamiltoniano de interacción de (3.22) viene a ser:

$$\widehat{H}_I = \sum_{i=1}^{N} g_i a a_i^{\dagger} + hc. \qquad (4.7)$$

Donde el término $g_i a a_i^{\dagger}$ representa la relajación del sistema y la excitación del oscilador $i$, cuantificado con el acople $g_i$, su conjugada hermitiana representa el proceso inverso, con acople $g_i^*$, como se muestra:

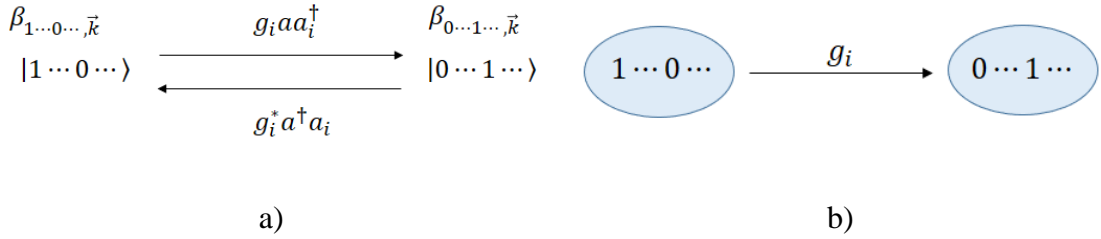

**Figura 4.2** Diagrama esquemático del término $g_i a a_i^\dagger$ y su adjunta.

La Figura 4.2 dice que una flecha entrante a un nodo se puede convertir a una flecha saliente, pero con el conjugado complejo de su acople.

Ambos diagramas de a) y b) de la Figura 4.2 contienen la misma información, pues una se puede construir de la otra inequívocamente, en a) se tiene más detallado y explícito los elementos que participan, mientras que en b) está resumido; los nodos corresponden a los índices vectores $\vec{n}$ que caracterizan los ket $|\vec{n}\rangle$ y las amplitudes $\beta_{\vec{n},\vec{k}}$ (el índice $\vec{k}$ es el mismo para todos los nodos, y se mantiene incógnito), y la flecha representa el sentido de la acción del término $g_i a a_i^\dagger$ del hamiltoniano, es evidente que el proceso inverso corresponde a una flecha en reversa con el acople conjugado $g_i^*$, lo que es $g_i^* a^\dagger a_i$.

Es importante notar que el término de interacción $g_i a a_i^\dagger$ no provoca ninguna flecha saliente en el nodo derecho de b) debido a que $g_i a a_i^\dagger |0 \cdots 1 \cdots\rangle = 0$ (el 1 se encuentra en la posición $i$), en efecto, las flechas salientes de los nodos representan las acciones no destructivas del hamiltoniano, mientras que las flechas entrantes representan que esos estados son creados por el hamiltoniano; por simplicidad, en los diagramas sólo se representan las flechas correspondientes a $\sum_{i=1}^{N} g_i a a_i^\dagger$, queda implícito que los conjugados hermíticos son todas las flechas invertidas con los acoples conjugados.

A continuación, se presenta los diagramas completos para los primeros valores de $N$.

Para $\boldsymbol{N = 2}$, se tiene en total 3 osciladores, entonces el ket es $|\vec{n}\rangle = |n_0, n_1, n_2\rangle$, y el número de excitación puede tomar solo 4 valores, desde 0 hasta $N + 1 = 3$; es decir, $\Sigma \equiv 0, 1, 2, 3$; los subespacios triviales $\mathcal{H}(0)$ y $\mathcal{H}(3)$ tienen los estados $|000\rangle$ y $|111\rangle$ respectivamente; para $\Sigma = 1$ se tienen los tres vectores degenerados $|001\rangle, |010\rangle, |100\rangle$ que pertenecen a $\mathcal{H}(1)$, mientras que para $\Sigma = 2$, se tiene la base $|110\rangle, |101\rangle, |011\rangle$ que pertenecen a $\mathcal{H}(2)$, como se muestra:

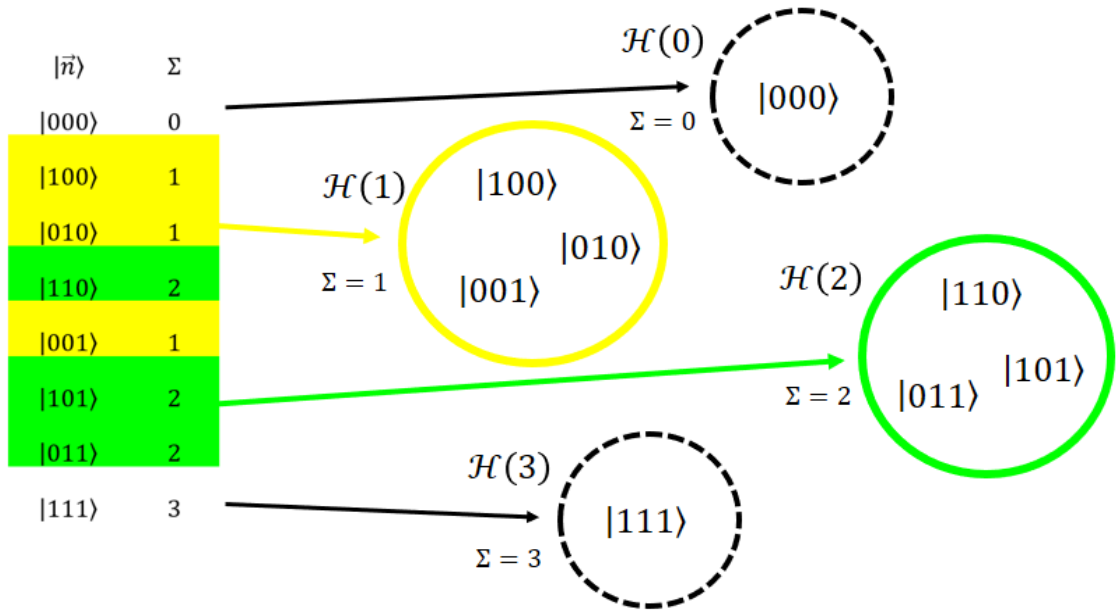

**Figura 4.3** Agrupación de los ket $|\vec{n}\rangle$ en 4 subespacios $\mathcal{H}(\Sigma)$.

Así, para $N = 2$ en (4.7) se tiene el hamiltoniano de interacción siguiente:

$$\widehat{H}_I = g_1 a a_1^\dagger + g_2 a a_2^\dagger + hc.$$

Los diagramas de la acción de $\widehat{H}_I$ en los subespacios $\mathcal{H}(1)$ y $\mathcal{H}(2)$ son:

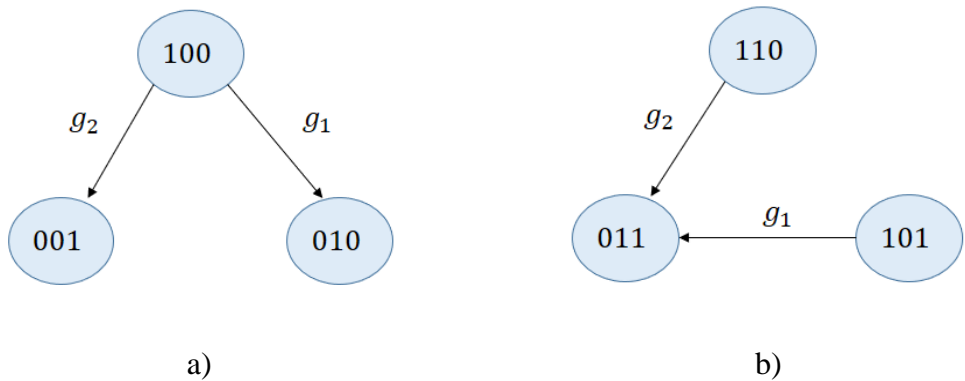

**Figura 4.4** Gráficas correspondiente a $g_i a a_i^\dagger$, que vienen a ser las transformaciones $1 \to 0$ del primer dígito (sistema), y $0 \to 1$ del segundo o tercer dígito (entorno).

Es importante notar que no existen flechas entre los otros dos nodos porque no hay términos de interacción $g_{12} a_1 a_2^\dagger$ en el hamiltoniano (4.7) que permita acoplarlos.

La construcción general de estos diagramas es como sigue: se colocan los nodos de forma circular, ordenados ascendentemente[17] de forma horaria, luego se representan las flechas correspondientes a la parte $\sum_{i=0}^{N} g_i a a_i^\dagger$ del hamiltoniano, la elección de esta parte en vez de su hermitiana es debido a que el operador de destrucción está en el sistema y el operador de creación está en el oscilador $i > 0$ del entorno (lo cual se invierte con la adjunta), esta es una elección arbitraria pero se lo mantiene así para fijar un orden; entonces las flechas simplemente representa la acción[18] de $\sum_{i=0}^{N} g_i a a_i^\dagger$ en cada uno de los nodos; otra forma de representar las flechas es eligiendo los nodos de partida y llegada de las flechas, y graficando solo si se cumple una condición compatible con la acción de $\sum_{i=0}^{N} g_i a a_i^\dagger$, para esto se plantea el siguiente algoritmo, aprovechando el orden establecido:

**Algoritmo 1**

Paso 1.- Se elige el nodo de partida, se empieza por el primer nodo, se copia su índice vector como $\vec{n}_0$

Paso 2.- Se elige el nodo de llegada, se empieza por el nodo que sigue al nodo de partida, se copia su índice vector como $\vec{n}_1$

Paso 3.- Se realiza la resta: $\Delta \vec{n} = \vec{n}_1 - \vec{n}_0$

Paso 4.- Se evalúa la condición: si $\Delta \vec{n} = (-1, 0, 0, \cdots 0, 1, 0, \cdots 0)$, donde $-1$ está en la posición 0, mientras que 1 está en la posición $i$ de $\Delta \vec{n}$ (todas las demás componentes son ceros). Si la condición es verdadera (si existe un $i$ que cumpla la condición), entonces se establece una flecha del nodo de partida al nodo de llegada, y se etiqueta el acople $g_i$; si esta condición es falsa no se hace nada.

Paso 5.- Se elige el nodo de llegada como el siguiente del nodo de llegada anterior, se copia su índice vector como $\vec{n}_1$ y se regresa al paso 3; si el nodo de llegada anterior es el último nodo, se pasa al paso 6.

Paso 6.- Se elige el nodo de partida como el siguiente del nodo de partida anterior, se copia su índice vector como $\vec{n}_0$ y se regresa al paso 2; si el nodo de partida anterior es el penúltimo nodo, se acaba el algoritmo.

---

[17] Como los nodos contienen secuencia de unos y ceros, se los convierte a número binario leído de izquierda a derecha; así, por ejemplo: $100 \to 1$ , $010 \to 2$ , $110 \to 3$ , $001 \to 4$ , $\cdots$, este orden coincide con el de la Tabla de la Figura 4.3.

[18] De acuerdo a la Figura 4.2

Con este algoritmo se pueden construir los diagramas para cualquier valor de $N$ y $\Sigma$, con $\# = C_\Sigma^{N+1}$ nodos $\vec{n} \in \mathcal{N}(\Sigma)$; a continuación algunos ejemplos en la Figura 4.5:

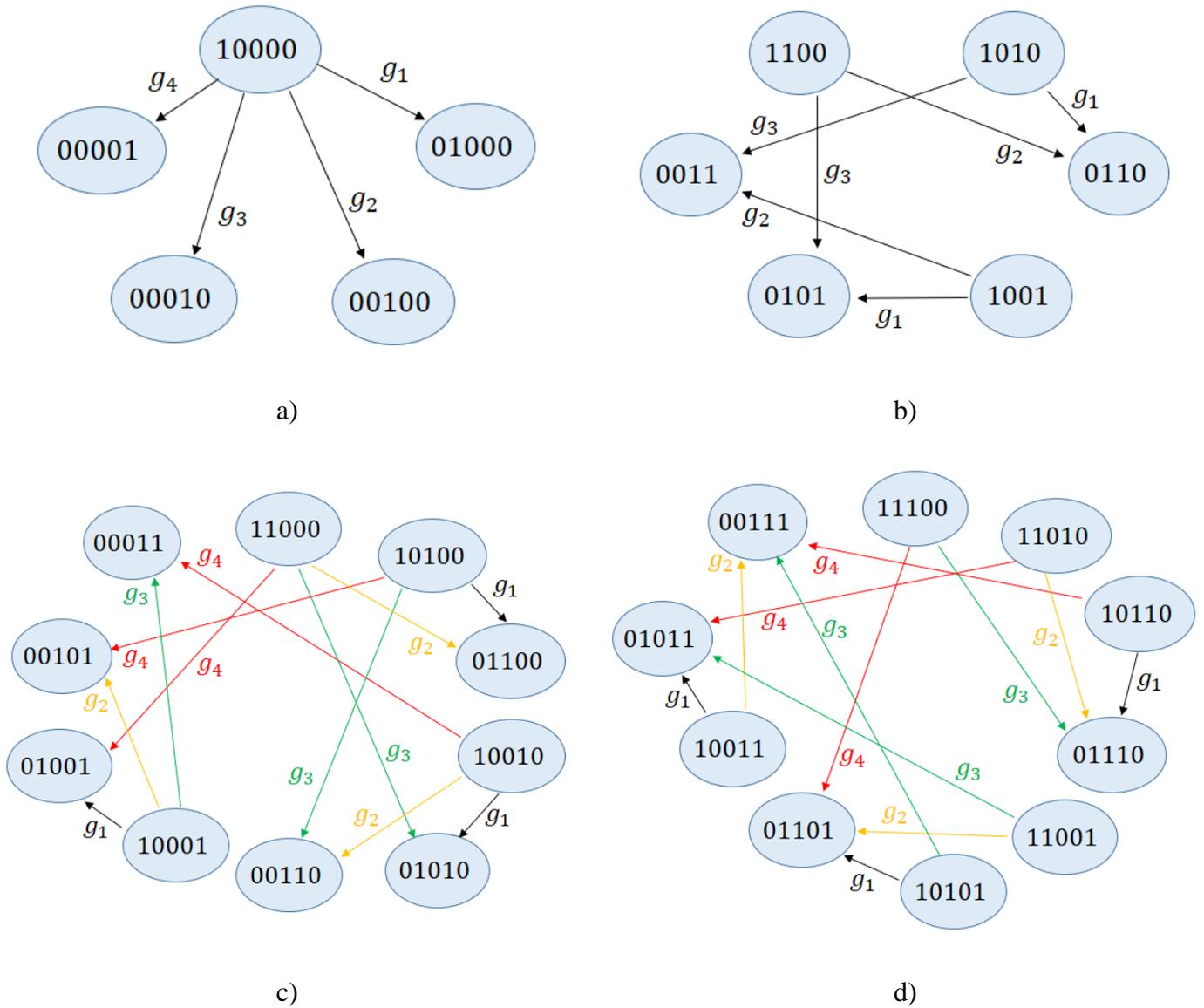

**Figura 4.5** Diagramas del Hamiltoniano (4.7) para distintas configuraciones de $N$ y $\Sigma$.

En a) se tiene $N = 4$ y $\Sigma = 1$, entonces $\# = C_1^5 = 5$.

En b) se tiene $N = 3$ y $\Sigma = 2$, entonces $\# = C_2^4 = 6$.

En c) se tiene $N = 4$ y $\Sigma = 2$, entonces $\# = C_2^5 = 10$.

En d) se tiene $N = 4$ y $\Sigma = 3$, entonces $\# = C_3^5 = 10$.

Con estos diagramas se puede construir el hamiltoniano reducido $\widehat{H}^{\Sigma}$, para el diagrama a) de la Figura 4.4 se tiene el subespacio $\mathcal{H}(1)$, la acción del hamiltoniano $\widehat{H}_I$ en cada uno de los 3 ket $|\vec{n}\rangle$, asociados a los nodos, viene a ser:

$$\widehat{H}_I|100\rangle = g_1|010\rangle + g_2|001\rangle \quad \widehat{H}_I|010\rangle = g_1^*|100\rangle \quad \widehat{H}_I|001\rangle = g_2^*|100\rangle \quad (4.8)$$

(4.8) dice que el diagrama a) de la Figura 4.4 se lee como sigue: el hamiltoniano actuando en el nodo $\vec{n}$ se expresa como $\widehat{H}_I|\vec{n}\rangle$, entonces las líneas que conectan este nodo $\vec{n}$ con otro nodo $\vec{n}'$ se expresa como sigue: si la línea sale del nodo $\vec{n}$ y entra al nodo $\vec{n}'$ se coloca $g_i|\vec{n}'\rangle$, si la línea sale del nodo $\vec{n}'$ y entra al nodo $\vec{n}$ se coloca el acople conjugado $g_i^*|\vec{n}'\rangle$, si el nodo $\vec{n}$ se conectan con varios otros nodos $\vec{n}_k'$, se suman todos ellos $g_{i_1}|\vec{n}_1'\rangle + g_{i_2}|\vec{n}_2'\rangle + g_{i_3}|\vec{n}_3'\rangle + \cdots$, esta suma debe estar igualada a $\widehat{H}_I|\vec{n}\rangle$, en consistencia con (4.8) y el diagrama a) de la Figura 4.4.

En general, la ecuación de autovalores en el espacio $\mathcal{H}(\Sigma)$ es de la forma $\widehat{H}|\phi_{\vec{k}}\rangle = (\widehat{H}_0 + \widehat{H}_I)|\phi_{\vec{k}}\rangle = \omega_{\vec{k}}|\phi_{\vec{k}}\rangle$, de acuerdo a (4.1) y (3.22) $\widehat{H}_0 = \omega \sum_{i=0}^{N} a_i^{\dagger} a_i = \omega \widehat{\Sigma}$, como se tiene $\Sigma$ fijo entonces $\widehat{\Sigma} = \Sigma \widehat{1}$, el hamiltoniano es $\widehat{H} = \omega \Sigma \widehat{1} + \widehat{H}_I$, despejando: $\widehat{H}_I = \widehat{H} - \omega \Sigma \widehat{1}$, esto es posible hacer porque todos los osciladores tienen la misma frecuencia $\omega$; así se puede plantear la ecuación de autovalores para $\widehat{H}_I$ como se muestra:

$$\widehat{H}_I|\phi_{\vec{k}}\rangle = \omega_{\vec{k}}'|\phi_{\vec{k}}\rangle \qquad\qquad \omega_{\vec{k}}' = \omega_{\vec{k}} - \omega\Sigma \qquad (4.9)$$

Para el diagrama a) de la Figura 4.4, se tiene $\Sigma = 1$, en (4.6) se tiene: $|\phi_{\vec{k}}\rangle = \beta_{100,\vec{k}}|100\rangle + \beta_{010,\vec{k}}|010\rangle + \beta_{001,\vec{k}}|001\rangle$, así la acción de $\widehat{H}_I$ en $|\phi_{\vec{k}}\rangle$ es, usando (4.8):

$$\widehat{H}_I|\phi_{\vec{k}}\rangle = \beta_{100,\vec{k}}\widehat{H}_I|100\rangle + \beta_{010,\vec{k}}\widehat{H}_I|010\rangle + \beta_{001,\vec{k}}\widehat{H}_I|001\rangle$$

$$\widehat{H}_I|\phi_{\vec{k}}\rangle = \beta_{100,\vec{k}}(g_1|010\rangle + g_2|001\rangle) + \beta_{010,\vec{k}}g_1^*|100\rangle + \beta_{001,\vec{k}}g_2^*|100\rangle$$

$$\widehat{H}_I|\phi_{\vec{k}}\rangle = (\beta_{010,\vec{k}}g_1^* + \beta_{001,\vec{k}}g_2^*)|100\rangle + g_1\beta_{100,\vec{k}}|010\rangle + g_2\beta_{100,\vec{k}}|001\rangle$$

Expresando $\widehat{H}_I|\phi_{\vec{k}}\rangle$ de forma matricial e igualando a $\omega_{\vec{k}}'|\phi_{\vec{k}}\rangle$ según (4.9), se tiene la ecuación matricial de autovalores para el diagrama a) de la Figura 4.4:

$$\begin{pmatrix} 0 & g_1^* & g_2^* \\ g_1 & 0 & 0 \\ g_2 & 0 & 0 \end{pmatrix} \begin{pmatrix} \beta_{100,\vec{k}} \\ \beta_{010,\vec{k}} \\ \beta_{001,\vec{k}} \end{pmatrix} = \omega'_{\vec{k}} \begin{pmatrix} \beta_{100,\vec{k}} \\ \beta_{010,\vec{k}} \\ \beta_{001,\vec{k}} \end{pmatrix} \qquad (4.10)$$

Donde las componentes del vector $|\phi_{\vec{k}}\rangle$ están ordenados según el orden ascendente-horario de los nodos $\vec{n}$ del diagrama, y en general toma la siguiente forma en $\mathcal{H}(\Sigma)$:

$$\begin{array}{lll} \forall \vec{n}_i \in \mathcal{N}(\Sigma) & A^{\rightarrow}(\vec{n}_i) < A^{\rightarrow}(\vec{n}_{i+1}) & \\ & & |\vec{n}_i\rangle = \begin{pmatrix} 0 \\ \vdots \\ 1 \\ \vdots \\ 0 \end{pmatrix} \\ i = 1 : \# & A^{\rightarrow}: \vec{n} \rightarrow n \quad B^{\rightarrow}: n \rightarrow \vec{n} & \end{array} \qquad (4.11)$$

La columna de la izquierda de (4.11): $\forall \vec{n}_i \in \mathcal{N}(\Sigma)$ significa que $\vec{n}_i$ es el $i$-ésimo vector que pertenece al conjunto $\mathcal{N}(\Sigma)$, donde $i$ va de 1 hasta $\# = C_\Sigma^{N+1}$; en la columna del centro de (4.11) se introduce esta notación para la siguiente operación: $A^{\rightarrow}$ es una función que lee el vector $\vec{n}$ como un número binario de izquierda a derecha y devuelve su valor entero $n$, mientras que $B^{\rightarrow}$ toma un entero $n$, lo transforma en binario y lo devuelve escrito como vector de izquierda a derecha[19]; ambos leen y escriben vectores de longitud $N + 1$, de manera que completan con ceros a la derecha cuando es necesario; $A^{\rightarrow}(\vec{n}_i) < A^{\rightarrow}(\vec{n}_{i+1})$ significa que los vectores $\vec{n}_i$ están ordenados de forma ascendente; esto es, de manera que su entero $n$ también esté ordenado de forma ascendente; y finalmente la columna de la derecha de (4.11) significa que en esta representación matricial de $\mathcal{H}(\Sigma)$, $|\vec{n}_i\rangle$ es la dirección $i$-ésima de $\mathcal{H}(\Sigma)$, representada por un vector con un único uno en la fila $i$.

La matriz cuadrada de (4.10) es $\widehat{H}_I^1$; el desarrollo hecho para plantear (4.10) es generalizable para cualquier otro diagrama, y se puede resumir en trasladar los acoples $g_i$ de cada flecha (que va del nodo $\vec{n}$ al nodo $\vec{n}'$) a la matriz hamiltoniano reducida $\widehat{H}_I^\Sigma \in \mathcal{H}(\Sigma)$ de manera que $\vec{n}$ (nodo origen) corresponde a la columna y $\vec{n}'$ (nodo de llegada) corresponde a la fila, esto se justifica porque se cumple el braket:

$$\forall \vec{n} \neq \vec{n}' : \quad \left(\widehat{H}_I\right)_{\vec{n}',\vec{n}} = \langle \vec{n}' | \widehat{H}_I | \vec{n} \rangle = g_i \qquad g_i : (\vec{n}) \rightarrow (\vec{n}') \qquad (4.12)$$

---

[19] La flecha indica el sentido de la lectura o escritura del número binario: de izquierda a derecha.

(4.12) significa que $\hat{H}_I|\vec{n}\rangle$ va a crear las líneas salientes del nodo $\vec{n}$ en el diagrama, lo que se corresponde en la ecuación como la suma: $\hat{H}_I|\vec{n}\rangle = g_{i_1}|\vec{n}_{i_1}\rangle + g_{i_2}|\vec{n}_{i_2}\rangle + \cdots$, entonces el braket $\langle\vec{n}'|\hat{H}_I|\vec{n}\rangle$ localiza uno de esos acoples $g_{i_k}$ si $|\vec{n}'\rangle = |\vec{n}_{i_k}\rangle$, en efecto, $\langle\vec{n}'|\hat{H}_I|\vec{n}\rangle = g_i$ dice que $g_{i_k}$ es el acople de la flecha que va de $\vec{n}$ a $\vec{n}'$.

Puesto que en los diagramas las flechas solo representan la mitad del hamiltoniano, al construir la matriz a partir de las flechas dibujadas usando (4.12), se le tiene que sumar su propia hermítica conjugada para obtener el Hamiltoniano $\hat{H}_I^\Sigma$ completo.

Así, para el diagrama b) de la Figura 4.4 se tiene que en el subespacio $\mathcal{H}(2)$, la matriz del hamiltoniano $\hat{H}_I^2$ y el autoestado $|\phi_{\vec{k}}\rangle$ vienen a ser, usando (4.12), (4.6) y (4.11):

$$\hat{H}_I^2 = \begin{pmatrix} 0 & 0 & g_2^* \\ 0 & 0 & g_1^* \\ g_2 & g_1 & 0 \end{pmatrix} \qquad |\phi_{\vec{k}}\rangle = \begin{pmatrix} \beta_{110,\vec{k}} \\ \beta_{101,\vec{k}} \\ \beta_{011,\vec{k}} \end{pmatrix} \qquad (4.13)$$

Repitiendo el mismo procedimiento para los diagramas de la Figura 4.5 se tiene:

| Diagramas de la Figura 4.5 | $\hat{H}_I^\Sigma$ | $|\phi_{\vec{k}}\rangle$ | |
|---|---|---|---|
| a) | $\begin{pmatrix} 0 & g_1^* & g_2^* & g_3^* & g_4^* \\ g_1 & 0 & & & \\ g_2 & & 0 & & \\ g_3 & & & 0 & \\ g_4 & & & & 0 \end{pmatrix}$ | $\begin{pmatrix} \beta_{10000,k} \\ \beta_{01000,k} \\ \beta_{00100,k} \\ \beta_{00010,k} \\ \beta_{00001,k} \end{pmatrix}$ | (4.14) |
| b) | $\begin{pmatrix} 0 & & g_2^* & & g_3^* & \\ & 0 & g_1^* & & & g_3^* \\ g_2 & g_1 & 0 & & & \\ & & & 0 & g_1^* & g_2^* \\ g_3 & & & g_1 & 0 & \\ & g_3 & & g_2 & & 0 \end{pmatrix}$ | $\begin{pmatrix} \beta_{1100,k} \\ \beta_{1010,k} \\ \beta_{0110,k} \\ \beta_{1001,k} \\ \beta_{0101,k} \\ \beta_{0011,k} \end{pmatrix}$ | |

c) 
$$\begin{pmatrix} 0 & g_2^* & g_3^* & & g_4^* & & & & \\ & 0 & g_1^* & & g_3^* & & g_4^* & & \\ g_2 & g_1 & 0 & & & & & & \\ & & & 0 & g_1^* & g_2^* & & & g_4^* \\ g_3 & & & g_1 & 0 & & & & \\ & g_3 & & g_2 & & 0 & & & \\ & & & & & & 0 & g_1^* & g_2^* & g_3^* \\ g_4 & & & & & & g_1 & 0 & & \\ & g_4 & & & & & g_2 & & 0 & \\ & & & g_4 & & & g_3 & & & 0 \end{pmatrix} \begin{pmatrix} \beta_{11000,k} \\ \beta_{10100,k} \\ \beta_{01100,k} \\ \beta_{10010,k} \\ \beta_{01010,k} \\ \beta_{00110,k} \\ \beta_{10001,k} \\ \beta_{01001,k} \\ \beta_{00101,k} \\ \beta_{00011,k} \end{pmatrix}$$

d)
$$\begin{pmatrix} 0 & & & g_3^* & & & g_4^* & & & \\ & 0 & & g_2^* & & & & g_4^* & & \\ & & 0 & g_1^* & & & & & g_4^* & \\ g_3 & g_2 & g_1 & 0 & & & & & & \\ & & & & 0 & & g_2^* & g_3^* & & \\ & & & & & 0 & g_1^* & & g_3^* & \\ g_4 & & & & g_2 & g_1 & 0 & & & \\ & & & & & & & 0 & g_1^* & g_2^* \\ & g_4 & & & g_3 & & & g_1 & 0 & \\ & & g_4 & & & g_3 & & g_2 & & 0 \end{pmatrix} \begin{pmatrix} \beta_{11100,k} \\ \beta_{11010,k} \\ \beta_{10110,k} \\ \beta_{01110,k} \\ \beta_{11001,k} \\ \beta_{10101,k} \\ \beta_{01101,k} \\ \beta_{10011,k} \\ \beta_{01011,k} \\ \beta_{00111,k} \end{pmatrix}$$

Si se sigue el orden establecido, al usar la ecuación (4.12) se llenan los elementos de la triangular inferior de cada matriz $\widehat{H}_I^\Sigma$, su adjunta corresponde a la triangular superior (con los acoples conjugados), luego $\widehat{H}_I^\Sigma$ es la suma de ambas matrices triangulares.

### 4.3.2 Hamiltoniano libre de (3.23)

El hamiltoniano de (3.23) viene a ser:

$$\widehat{H} = \widehat{H}_0 + \widehat{H}_I \qquad \widehat{H}_0 = \sum_{i=0}^{N} \omega_i \hat{a}_i^\dagger \hat{a}_i \qquad \widehat{H}_I = \sum_{i=1}^{N} g_i \hat{a} \hat{a}_i^\dagger + hc. \qquad (4.15)$$

En las ecuaciones (4.9) se pudo establecer la ecuación de autovalores para el hamiltoniano de interacción, con sus autovalor restado un valor constante, en este caso es necesario considerar el hamiltoniano libre $\widehat{H}_0$ debido a que cada oscilador tiene una frecuencia propia $\omega_i$, así que se debe construir la matriz de $\widehat{H}_I$ mediante el procedimiento descrito en la sección anterior [4.1.2] y añadirle la matriz correspondiente a $\hat{H}_0$, el cual solo contribuye términos en la diagonal, debido a que $|\vec{n}\rangle$ es autoestado de $\hat{a}_i^\dagger \hat{a}_i$ :

$$\left(\hat{H}_0\right)_{\vec{n}',\vec{n}} = \langle \vec{n}'|\hat{H}_0|\vec{n}\rangle = \omega(\vec{n})\delta_{\vec{n},\vec{n}'} \qquad \omega(\vec{n}) = \sum_{i=0}^{N} \omega_i n_i \qquad (4.16)$$

En la diagonal solo se añaden términos correspondientes a la suma de frecuencias $\omega_i$ de los osciladores excitados $n_i = 1$; la gráfica de $\hat{H}_0$ es una "auto interacción" porque si (4.12) se aplicara también para $\vec{n}' = \vec{n}$, le correspondería varias flechas que salen y entran al mismo nodo; sin embargo, como el sentido de giro es irrelevante, se dibuja un bucle en el nodo, con etiqueta $\omega(\vec{n})$ que es la suma de las frecuencias (4.16), como se muestra:

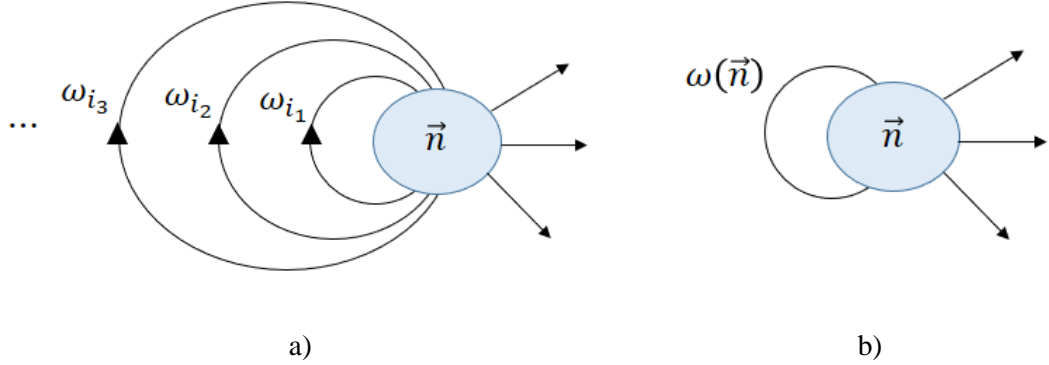

a)          b)

**Figura 4.6** En a) se dibuja cada "auto interacción" con flechas para distintos valores de $i_k$, mientras que en el diagrama b) se dibuja un bucle sin orientación con la suma de frecuencias; ambos diagramas tienen la misma información.

De acuerdo a (4.16), en a) de la Figura 4.6 entran y salen Σ curvas, una por cada estado excitado $|n_i\rangle = |1\rangle$, equivalentemente en b) la etiqueta $\omega(\vec{n})$ suma Σ frecuencias $\omega_i$, los cuales se colocan en la diagonal de $\hat{H}^\Sigma$. Por ejemplo, al considerar el hamiltoniano libre $\hat{H}_0 = \sum_{i=0}^{N} \omega_i \hat{a}_i^\dagger \hat{a}_i$ en a) y b) de la Figura 4.5 se tienen los siguientes diagramas:

| | a) | b) |
|---|---|---|

**Figura 4.7** Diagramas a) y b) de la Figura 4.5 considerando el Hamiltoniano libre

Las etiquetas $\omega(\vec{n})$ en los bucles, según (4.16), se añaden en la diagonal de la matriz $\widehat{H}^\Sigma$ en a) y b) de (4.14); en esta sección solo se añade esta "auto interacción":

$$
\begin{array}{c|cc|c}
\text{Figura 4.7} & \widehat{H}^\Sigma & & |\phi_{\vec{k}}\rangle \\
\hline
\text{a)} & \begin{pmatrix} \omega_0 & g_1^* & g_2^* & g_3^* & g_4^* \\ g_1 & \omega_1 & & & \\ g_2 & & \omega_2 & & \\ g_3 & & & \omega_3 & \\ g_4 & & & & \omega_4 \end{pmatrix} & & \begin{pmatrix} \beta_{10000,k} \\ \beta_{01000,k} \\ \beta_{00100,k} \\ \beta_{00010,k} \\ \beta_{00001,k} \end{pmatrix} \\
\text{b)} & \begin{pmatrix} \omega_0+\omega_1 & & g_2^* & & g_3^* & \\ & \omega_0+\omega_2 & g_1^* & & & g_3^* \\ g_2 & g_1 & \omega_1+\omega_2 & & & \\ & & & \omega_0+\omega_3 & g_1^* & g_2^* \\ g_3 & & & g_1 & \omega_1+\omega_3 & \\ & g_3 & & g_2 & & \omega_2+\omega_3 \end{pmatrix} & & \begin{pmatrix} \beta_{1100,k} \\ \beta_{1010,k} \\ \beta_{0110,k} \\ \beta_{1001,k} \\ \beta_{0101,k} \\ \beta_{0011,k} \end{pmatrix}
\end{array} \quad (4.17)
$$

### 4.3.3 Hamiltoniano total de (3.25)

El hamiltoniano de (3.25) es:

$$\hat{H} = \sum_{i,j=0}^{N} g_{ij} \hat{a}_j^\dagger \hat{a}_i \qquad g_{ji} = g_{ij}^* \qquad g_{ii} = \omega_i \qquad (4.18)$$

$\hat{H}$ tiene interacciones del sistema con cada oscilador del entorno, su auto interacción, y la interacción entre osciladores del entorno dado por $g_{ij} \hat{a}_j^\dagger \hat{a}_i$, $i,j \neq 0$, $i \neq j$, el procedimiento para construir los diagramas y las matrices Hamiltoniano es esencialmente el mismo al descrito en las secciones anteriores pero generalizado al hamiltoniano (4.18), para esto se define un vector $\Delta_{ij}$ según:

$$\Delta_{ij} = (0, \cdots 0, -1, 0, \cdots 0, 1, 0, \cdots 0) \qquad \left(\Delta_{ij}\right)_k = \delta_{j,k} - \delta_{i,k}$$

$$\dim \Delta_{ij} = N + 1 \qquad k = 0, 1, 2, \cdots N \qquad (4.19)$$

$\Delta_{ij}$ es un vector de $N + 1$ componentes, todas son ceros excepto la componentes $i$ y $j$ que tienen los valores $-1$ y $1$ respectivamente; además, $k$ cuenta de $0$ a $N$.

Con (4.19) se generaliza el algoritmo dado en la sección [4.1.2]:

**Algoritmo 2**

**Paso 1**.- Se elige el nodo de partida, se empieza por el primer nodo, se copia su índice vector como $\vec{n}_0$

**Paso 2**.- Se elige el nodo de llegada, se empieza por el nodo que sigue al nodo de partida, se copia su índice vector como $\vec{n}_1$

**Paso 3**.- Se realiza la resta: $\Delta \vec{n} = \vec{n}_1 - \vec{n}_0$

**Paso 4**.- Se evalúa la condición: $\Delta \vec{n} = \Delta_{ij}$ ; Si la condición es verdadera (si existen $i,j$ que cumpla la condición), entonces se dibuja una flecha del nodo de partida al nodo de llegada, y se etiqueta el acople $g_{ij}$ ; si esta condición es falsa no se hace nada.

**Paso 5**.- Se elige el nodo de llegada como el siguiente del nodo de llegada anterior, se copia su índice vector como $\vec{n}_1$ y se regresa al paso 3; si el nodo de llegada anterior es el último nodo, se pasa al paso 6.

**Paso 6.-** Se elige el nodo de partida como el siguiente del nodo de partida anterior, se copia su índice vector como $\vec{n}_0$ y se regresa al paso 2; si el nodo de partida anterior es el penúltimo nodo, se pasa al paso 7.

**Paso 7.-** Se añade la auto interacción en cada nodo $\vec{n}$: un bucle con etiqueta $\omega(\vec{n})$; tras esto se finaliza el algoritmo.

Para el diagrama b) de la Figura 4.7, con $N = 3$ y $\Sigma = 2$, su nuevo diagrama, considerando las interacciones entre osciladores del entorno dado por (4.18), viene a ser:

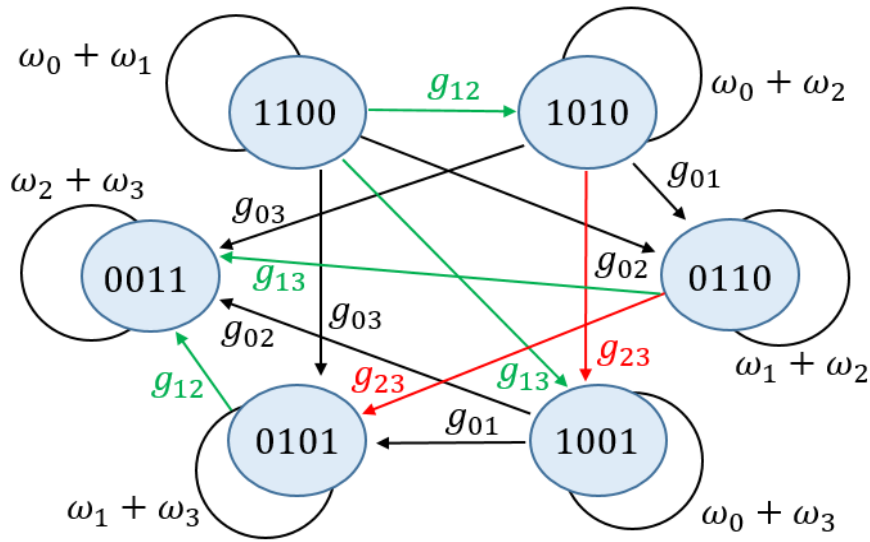

**Figura 4.8** Diagrama de acción del Hamiltoniano (4.18) para $N = 3$ y $\Sigma = 2$; se representa la acción $g_{ij}\hat{a}_j^\dagger \hat{a}_i$ con $j > i$.

En general, usando (4.6) y (4.11) se representan los ket $|\phi_{\vec{k}}\rangle$ en el subespacio $\mathcal{H}(\Sigma)$, en esta misma representación, usando (4.12) y (4.16) se puede construir la matriz del hamiltoniano $\hat{H}^\Sigma$, y su autovector; para la Figura 4.8 viene a ser:

$$\begin{pmatrix} \omega_0 + \omega_1 & g_{12}^* & g_{02}^* & g_{13}^* & g_{03}^{*0} & 0 \\ g_{12} & \omega_0 + \omega_2 & g_{01}^* & g_{23}^* & 0 & g_{03}^* \\ g_{02} & g_{01} & \omega_1 + \omega_2 & 0 & g_{23}^* & g_{13}^* \\ g_{13} & g_{23} & 0 & \omega_0 + \omega_3 & g_{01}^* & g_{02}^* \\ g_{03} & 0 & g_{23} & g_{01} & \omega_1 + \omega_3 & g_{12}^* \\ 0 & g_{03} & g_{13} & g_{02} & g_{12} & \omega_2 + \omega_3 \end{pmatrix} \begin{pmatrix} \beta_{1100,k} \\ \beta_{1010,k} \\ \beta_{0110,k} \\ \beta_{1001,k} \\ \beta_{0101,k} \\ \beta_{0011,k} \end{pmatrix} \quad (4.20)$$

Para $N = 4$ y $\Sigma = 2$ se tiene $\# = 10$ nodos; su diagrama viene a ser:

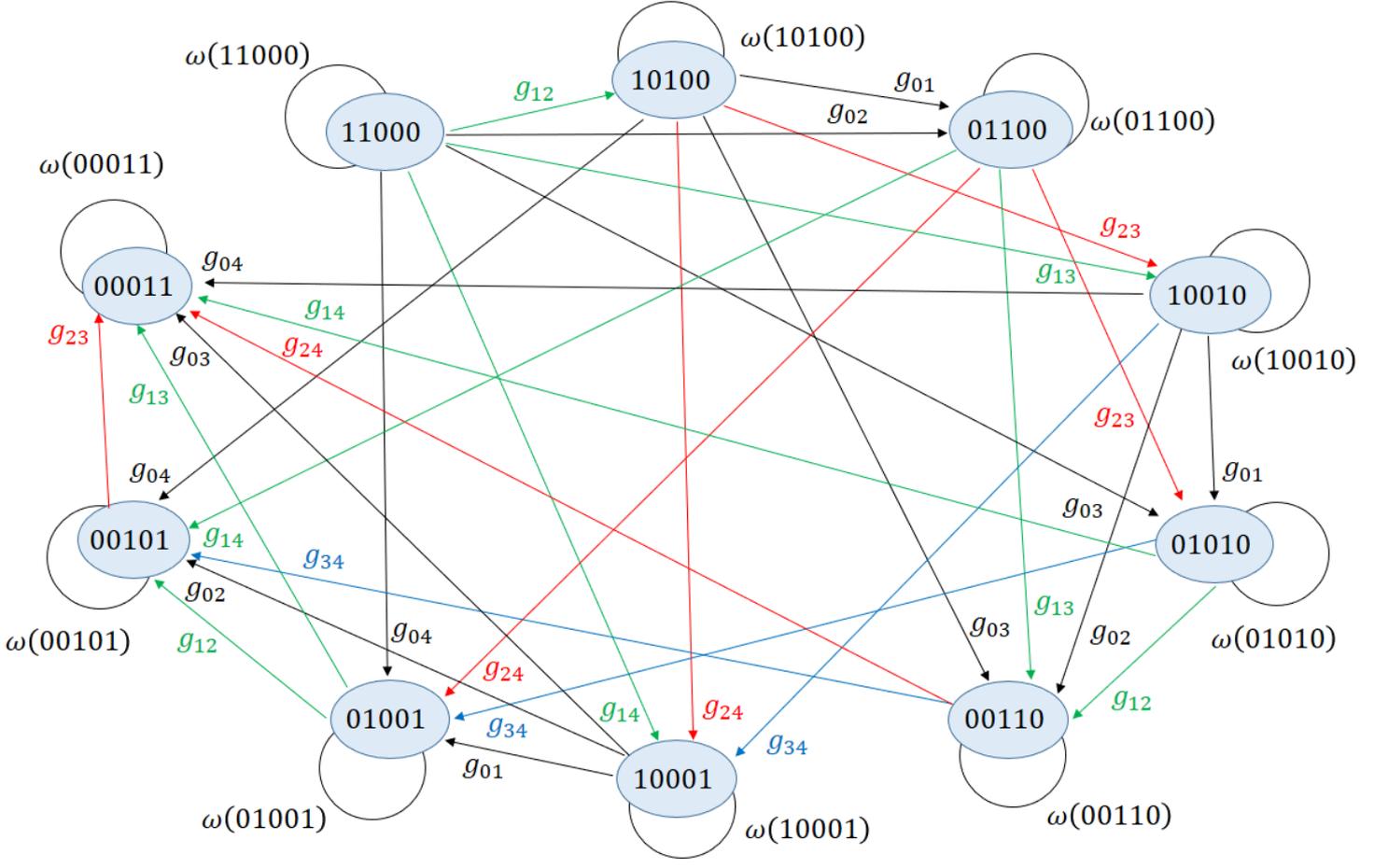

**Figura 4.9** Diagrama para $N = 4$ y $\Sigma = 2$

Su correspondiente hamiltoniano $\widehat{H}^2$ y la transpuesta del ket $|\phi_{\vec{k}}\rangle$ son:

$$\begin{pmatrix}
\omega_0 + \omega_1 & g_{12}^* & g_{02}^* & g_{13}^* & g_{03}^* & 0 & g_{14}^* & g_{04}^* & 0 & 0 \\
g_{12} & \omega_0 + \omega_2 & g_{01}^* & g_{23}^* & 0 & g_{03}^* & g_{24}^* & 0 & g_{04}^* & 0 \\
g_{02} & g_{01} & \omega_1 + \omega_2 & 0 & g_{23}^* & g_{13}^* & 0 & g_{24}^* & g_{14}^* & 0 \\
g_{13} & g_{23} & 0 & \omega_0 + \omega_3 & g_{01}^* & g_{02}^* & g_{34}^* & 0 & 0 & g_{04}^* \\
g_{03} & 0 & g_{23} & g_{01} & \omega_1 + \omega_3 & g_{12}^* & 0 & g_{34}^* & 0 & g_{14}^* \\
0 & g_{03} & g_{13} & g_{02} & g_{12} & \omega_2 + \omega_3 & 0 & 0 & g_{34}^* & g_{24}^* \\
g_{14} & g_{24} & 0 & g_{34} & 0 & 0 & \omega_0 + \omega_4 & g_{01}^* & g_{02}^* & g_{03}^* \\
g_{04} & 0 & g_{24} & 0 & g_{34} & 0 & g_{01} & \omega_1 + \omega_4 & g_{12}^* & g_{13}^* \\
0 & g_{04} & g_{14} & 0 & 0 & g_{34} & g_{02} & g_{12} & \omega_2 + \omega_4 & g_{23}^* \\
0 & 0 & 0 & g_{04} & g_{14} & g_{24} & g_{03} & g_{13} & g_{23} & \omega_3 + \omega_4
\end{pmatrix} \quad (4.21)$$

$(\beta_{11000,k} \quad \beta_{10100,k} \quad \beta_{01100,k} \quad \beta_{10010,k} \quad \beta_{01010,k} \quad \beta_{00110,k} \quad \beta_{10001,k} \quad \beta_{01001,k} \quad \beta_{00101,k} \quad \beta_{00011,k})$

En general, los ceros en $\widehat{H}^\Sigma$ son debidos a que se tiene acople nulo para posibles interacciones de órdenes superior, por ejemplo, $g_{ijkl} a_i a_j a_k^\dagger a_l^\dagger$ o $g_{ijklmn} a_i a_j a_k a_l^\dagger a_m^\dagger a_n^\dagger$ que representan varios procesos de absorción y emisión simultáneamente de energía (sin especificar que se intercambia de dos a dos); si los osciladores fueran de varios niveles, sería posible construir procesos $g_{ijk} a_i a_j a_k^{\dagger 2}$, $g_{ijkl} a_i a_j a_k a_l^{\dagger 3}$, ... , e incluirlos en el Hamiltoniano previa inclusión de kets $|\vec{n}\rangle$ con $n_i = 0, 1, 2, 3, \cdots$, lo cual añaden más nodos a los gráficos aumentando la dimensión de los subespacios; sin embargo, por simplicidad basta con usar $g_{ij} a_i a_j^{\dagger}$ para tener una buena redistribución de la energía.

**4.4 Algoritmos para la construcción del Hamiltoniano reducido**

En la sección [4.3] se emplean unos diagramas esquemáticos que visualizan la acción del hamiltoniano $\widehat{H}$ en un subespacio $\mathcal{H}(\Sigma)$, luego se deriva los hamiltonianos $\widehat{H}^\Sigma$ en su respectiva ecuación de autovalores en $\mathcal{H}(\Sigma)$, reduciendo la complejidad de trabajar en el espacio total $\mathcal{H}$; sin embargo, cuando se tiene valores grandes de $N$ y $\Sigma$ la elaboración de los diagramas también conlleva a un trabajo extenso aunque mecánico, en esta sección se sintetiza la elaboración de los diagramas y se plantea directamente el hamiltoniano $\widehat{H}^\Sigma$.

Para la elaboración de los hamiltonianos $\widehat{H}^\Sigma \in \mathcal{H}(\Sigma)$, se emplean dos algoritmos, el primero partiendo sólo del número $N$, genera las $N+1$ listas $\mathcal{N}(\Sigma)$, que son el conjunto de todos los índices-vectores $\vec{n}$ asociados al subespacio $\mathcal{H}(\Sigma)$, y el segundo actúa para una lista dada $\mathcal{N}(\Sigma)$, creando el hamiltoniano correspondiente; es implícito que el segundo algoritmo se repite para los valores de $\Sigma = 1: N$, creando $N-1$ matrices no triviales $\widehat{H}^\Sigma$. No se necesita conocer todos los $\widehat{H}^\Sigma$, sino solo aquellos para los cuales el estado inicial $|\psi(0)\rangle$ tiene componentes no nulas en sus respectivos subespacios $\mathcal{H}(\Sigma)$.

**Algoritmo 3: Algoritmo para la creación de las listas $\mathcal{N}(\Sigma)$**

Se parte del número $N$

**Paso 1**.- Se inicia con el número $N$.

**Paso 2**.- Se separan las $N+1$ tablas vacías que contendrán todas las listas $\mathcal{N}(\Sigma)$, donde $\Sigma$ va de 0 a $N+1$, cada lista $\mathcal{N}(\Sigma)$ va a contener $\# = C_\Sigma^{N+1}$ vectores $\vec{n}$, de longitud $N+1$.

**Paso 3**.- Se inicia con $\vec{n} = \vec{0} = (\,0\,,0\,,0\,,\cdots 0\,)$.

**Paso 4.-** Se calcula $\Sigma$, que es la suma de todas las componentes del vector $\vec{n}$ dado:
$\Sigma = \sum_{i=0}^{N} n_i$

**Paso 5.-** Se añade el vector $\vec{n}$ a la lista $\mathcal{N}(\Sigma)$

**Paso 6.-** Se hace $\vec{n} \to \vec{n} + 1$, que expresado en las funciones de (4.11) es: $\vec{n} + 1 = B^{\to}(A^{\to}(\vec{n}) + 1)$, que significa: al número binario de $\vec{n}$ (leído de izquierda a derecha) se le aumenta en una unidad.

**Paso 7**.- Se evalúa si $\vec{n} \leq \vec{1} = (\,1\,,1\,,1\,,\cdots 1\,)$; esto es, se lee $\vec{n}$ como un número binario (de izquierda a derecha) y se evalúa si no es mayor que $2^{N+1}$

**Paso 8.-** Si la condicional es afirmativa, se regresa al paso 4, si es negativa se finaliza el algoritmo.

### Algoritmo 4: Algoritmo para la creación del hamiltoniano

Se parte de la lista $\mathcal{N}(\Sigma)$, y de la matriz cuadrada $G$ de orden $N + 1$, cuyos componentes son los acoples $g_{ij}$ (el cual incluye las frecuencias $\omega_i = g_{ii}$ en su diagonal principal).

**Paso 1**.- Se inicia con los parámetros $\Sigma$ y #.

**Paso 2**.- Se inicia un bucle $for$ donde el índice $k$ va de 1 hasta #, cuando se acaba este ciclo **se acaba el algoritmo**.

**Paso 3**.- Mientras $k \leq \#$; de la lista $\mathcal{N}(\Sigma)$ se elige $\vec{n}(k)$, que es el $k-$ésimo vector.

**Paso 4.-** Usando los elementos de la diagonal de la matriz $G$ y del vector $\vec{n}(k)$, se calcula $H_{k,k} = \sum_i g_{ii}\, n_i\,(k)$ y se lo ingresa en la matriz Hamiltoniano $\widehat{H}^{\Sigma}$.

**Paso 5.-** Se inicia otro bucle $for$ para el índice $k'$ que va de $k + 1$ a #, cuando este bucle se acaba se regresa al **paso 2**.

**Paso 6.-** Mientras $k' \leq \#$, de la lista $\mathcal{N}(\Sigma)$ se elige $\vec{n}(k')$, y con $\vec{n}(k)$ se calcula $\Delta\vec{n} = \vec{n}(k') - \vec{n}(k)$.

**Paso 7.-** Se evalúa la condición: $\Delta \vec{n} = \Delta_{ij}$ ; si es verdadero, se define el término $H_{k,k}$ del hamiltoniano como el acople $g_{ij}$ , que es el elemento $i,j$ de la matriz $G$, así mismo se define el término $H_{k,k\prime} = g_{ij}^*$, ambos son ingresados a la matriz Hamiltoniano $\widehat{H}^\Sigma$.

**Paso 8.-** Se regresa al **paso 5**.

Se hace un diagrama de flujo del algoritmo 3:

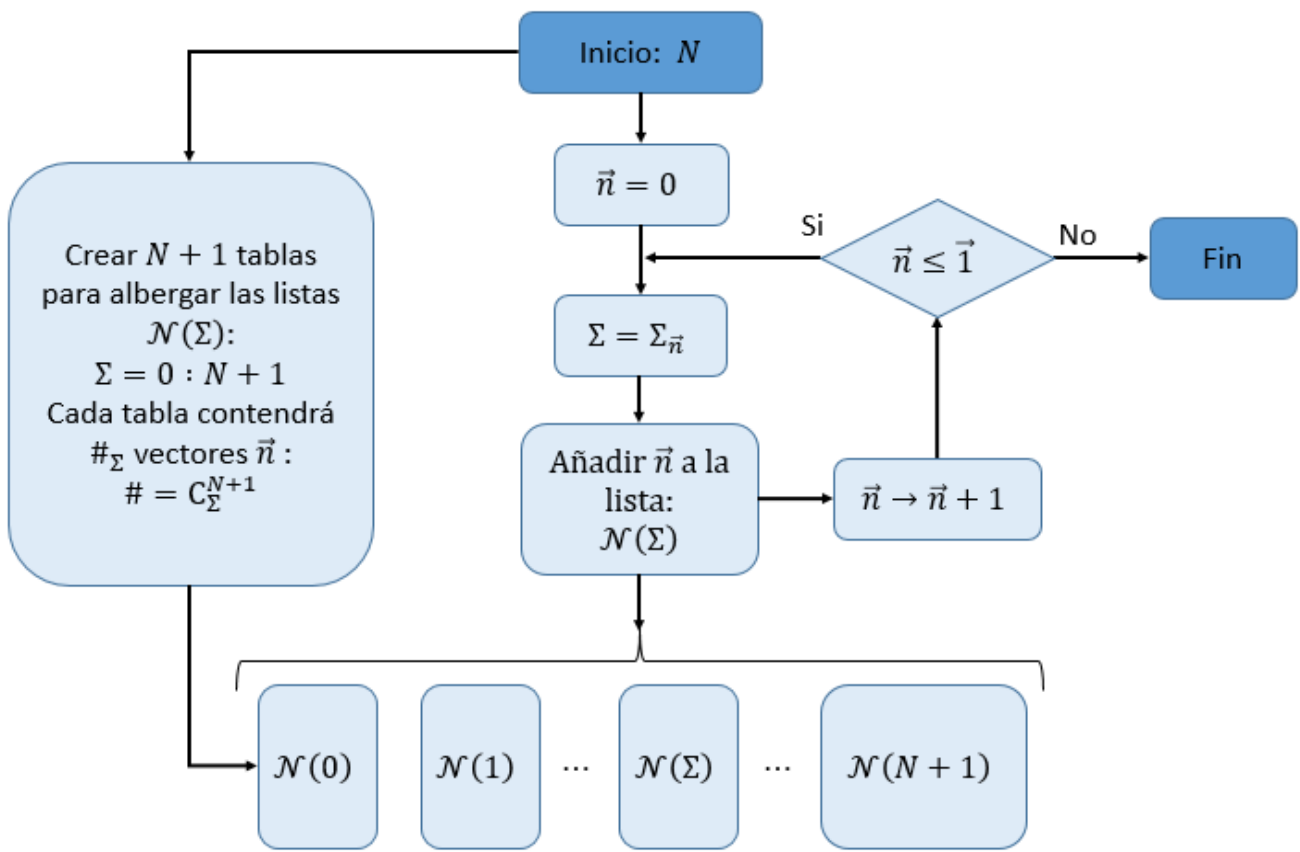

**Figura 4.10**. Diagrama de Flujo del algoritmo 3

Se hace un diagrama de flujo del algoritmo 4:

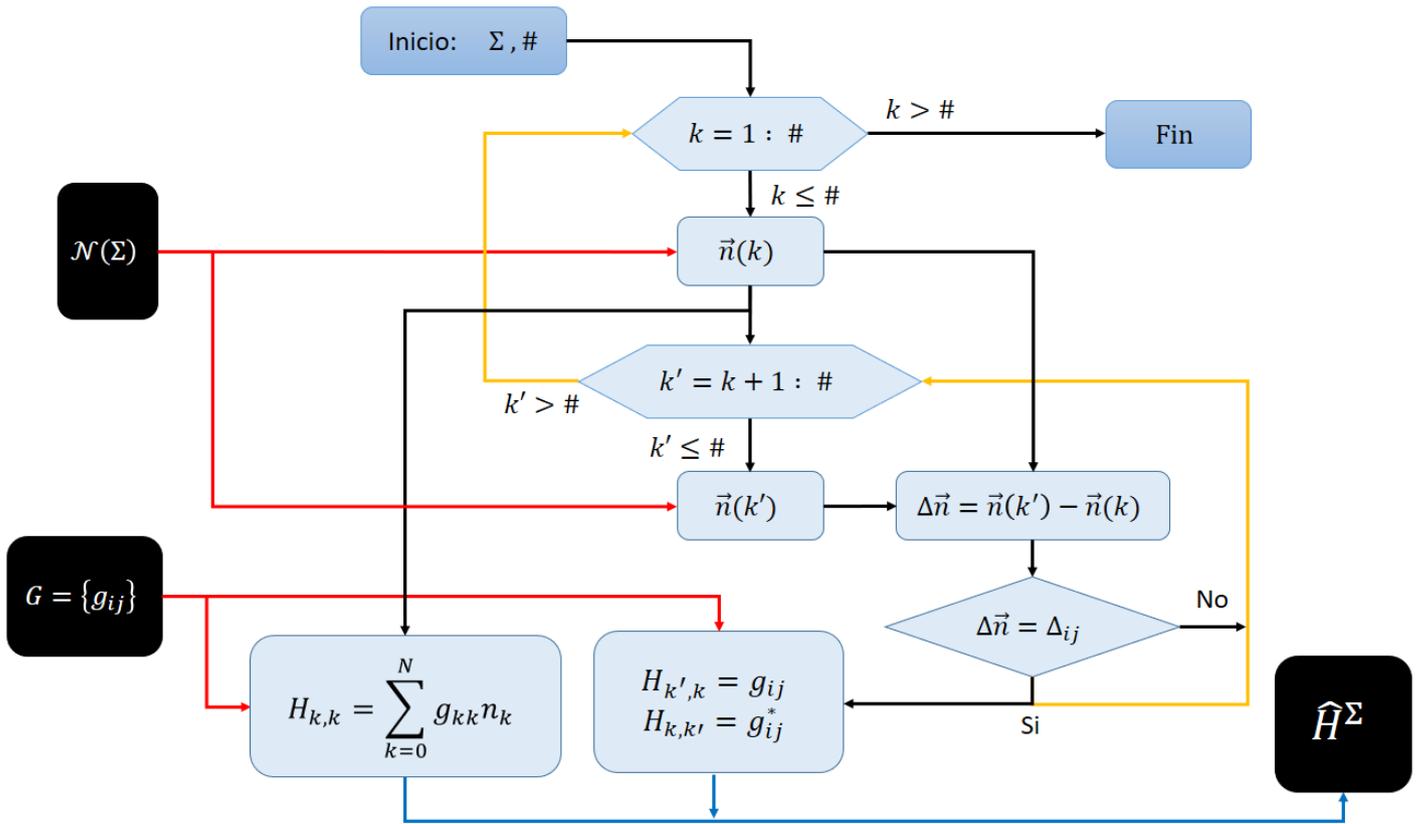

**Figura 4.11** Diagrama de Flujo del algoritmo 4

Entonces, para la ecuación de autovalores se tiene la matriz hamiltoniano $\hat{H}^{\Sigma}$ y la lista $\mathcal{N}(\Sigma)$, esta última tiene la información de los índices $\vec{n}$ de las amplitudes $\beta_{\vec{n},k}$ que conforman el autovector $|\phi_{\vec{k}}\rangle$ a resolver.

# CAPÍTULO V

# HERRAMIENTAS Y NOTACIÓN PARA TRABAJAR EN LOS SUBESPACIOS

## 5.1 Notación de las listas $\mathcal{N}(\Sigma)$ y $\mathcal{K}(\Sigma)$

### 5.1.1 Listas $\mathcal{N}(\Sigma)$ para $\vec{n}$

Para el sistema total de $N+1$ osciladores, cada uno con espacio de Hilbert $\mathcal{H}_i$ y ket $|n_i\rangle \in \mathcal{H}_i$, se tiene el espacio de Hilbert total sistema – entorno $\mathcal{H} = \mathcal{H}_0 \otimes \mathcal{H}_1 \otimes \mathcal{H}_2 \cdots \otimes \mathcal{H}_N$, cuyo ket viene a ser $|\vec{n}\rangle \in \mathcal{H}$, donde $\vec{n} = (n_0, n_1, n_2 \cdots n_N)$ es el índice vector, y pertenece al conjunto $\vec{n} \in \mathcal{N}$ de todas las combinaciones posibles; $\mathcal{H}$ puede ser separado en la suma directa de $N+2$ subespacios de número de excitación definido $\Sigma$:

$$\mathcal{H} = \mathcal{H}(0) \oplus \mathcal{H}(1) \oplus \mathcal{H}(2) \cdots \oplus \mathcal{H}(N+1) \qquad (5.1)$$

Cada subespacio $\mathcal{H}(\Sigma)$ tiene asociado una lista $\mathcal{N}(\Sigma)$ que es un subconjunto de vectores $\vec{n}$, ambos son definidos como se muestra:

$$\mathcal{H}(\Sigma) = \left\{ \sum_{\vec{n}} a_{\vec{n}} |\vec{n}\rangle \in \mathcal{H} \;/\; \hat{\Sigma}|\vec{n}\rangle = \Sigma|\vec{n}\rangle \right\} \qquad \# = C_\Sigma^{N+1}$$

$$\mathcal{N}(\Sigma) = \left\{ \vec{n} \in \mathcal{N} \;/\; \hat{\Sigma}|\vec{n}\rangle = \Sigma|\vec{n}\rangle \right\} \qquad (5.2)$$

Donde # es el número de elementos de $\mathcal{N}(\Sigma)$; $\mathcal{N}$ es una lista, un conjunto ordenado, de manera que los vectores $\vec{n}$ están ordenados de forma ascendente[20], y la lista $\mathcal{N}(\Sigma)$ es un subconjunto también ordenado de $\mathcal{N}$; por ejemplo, para $N = 3$, se tiene $2^{3+1} = 16$ vectores en la lista: $\mathcal{N} = \{0000, 1000, 0100, 1100, 0010, 1010, \cdots 1111\}$, los elementos completos de esta lista son representada en la tabla 5.1 en la que se considera a su derecha su equivalente en número entero $n$, que resulta de aplicar $A^{\rightarrow}(\vec{n}) = n$, de (4.11):

---

[20] Esto es, $\vec{n}$ es leído como un número binario de izquierda a derecha.

| $\vec{n}$ | $n$ |
|---|---|
| $\vec{0} = 0000$ | 0 |
| 1000 | 1 |
| 0100 | 2 |
| 1100 | 3 |
| 0010 | 4 |
| 1010 | 5 |
| 0110 | 6 |
| 1110 | 7 |

| $\vec{n}$ | $n$ |
|---|---|
| 0001 | 8 |
| 1001 | 9 |
| 0101 | 10 |
| 1101 | 11 |
| 0011 | 12 |
| 1011 | 13 |
| 0111 | 14 |
| $\vec{1} = 1111$ | 15 |

**Tabla 5.1**

La lista $\mathcal{N}(\Sigma)$ se forman como un subconjunto de $\mathcal{N}$, cuyos vectores también están ordenados de forma ascendente; por ejemplo, para $N = 3$, se tiene que $\Sigma$ varía de 0 a 4, entonces las listas $\mathcal{N}(\Sigma)$ son representados en la tabla 5.2:

| $\Sigma$ | $\mathcal{N}(\Sigma) = \{\vec{n}\}$ | $\mathcal{N}(\Sigma) = \{n\}$ |
|---|---|---|
| 0 | {0000} | {0} |
| 1 | {1000, 0100, 0010, 0001} | {1, 2, 4, 8} |
| 2 | {1100, 1010, 0110, 1001, 0101, 0011} | {3, 5, 6, 9, 10, 12} |
| 3 | {1110, 1101, 1011, 0111} | {7, 11, 13, 14} |
| 4 | {1111} | {15} |

**Tabla 5.2**

Es importante señalar que las listas $\mathcal{N}$ o $\mathcal{N}(\Sigma)$ se pueden expresar como conjuntos de los vectores $\vec{n}$, o de su equivalente en número entero $n$; también es importante resaltar que en las listas $\mathcal{N}(\Sigma)$ los vectores $\vec{n}$ aunque están ordenados de menor a mayor, pierden la "continuidad" como enteros (por ejemplo, $3, 4, 5, 6, \cdots$) que tenían en $\mathcal{N}$.

### 5.1.2 Listas $\mathcal{K}(\Sigma)$ para $\vec{k}$

Así como el vector $\vec{n}$ varía en todo el espacio de Hilbert $\mathcal{H}$, a través de los kets $|\vec{n}\rangle$, los vectores $\vec{k}$ también varían en todo el espacio $\mathcal{H}$ a través de los autoestados $|\phi_{\vec{k}}\rangle$, la lista $\mathcal{K}$ es el conjunto de todos los $\vec{k}$, también ordenados de forma ascendente, y los subespacios $\mathcal{H}(\Sigma)$ son definidos como:

$$\mathcal{H}(\Sigma) = \left\{ \sum_{\vec{k}} a_{\vec{k}} |\phi_{\vec{k}}\rangle \in \mathcal{H} \ / \ \hat{\Sigma} |\phi_{\vec{k}}\rangle = \Sigma |\phi_{\vec{k}}\rangle \right\} \qquad \# = C_{\Sigma}^{N+1} \qquad (5.3)$$

$$\mathcal{K}(\Sigma) = \left\{ \vec{k} \in \mathcal{K} \ / \ \hat{\Sigma} |\phi_{\vec{k}}\rangle = \Sigma |\phi_{\vec{k}}\rangle \right\}$$

(5.3) establece que $\mathcal{K}(\Sigma)$ es el conjunto de todos los $\vec{k}$ que etiquetan a los autoestados $|\phi_{\vec{k}}\rangle \in \mathcal{H}(\Sigma)$, $\mathcal{K}(\Sigma)$ tiene # elementos; la representación de $\vec{k}$ es en cierto sentido arbitraria, y se elige una forma de dos componentes:

$$\vec{k} = \Sigma, j \qquad \begin{aligned} j &= 1 : \# \\ \Sigma &= 0 : N+1 \end{aligned} \qquad (5.4)$$

Así, para cada $\Sigma$, el índice entero $j$ varía de 1 hasta $\# = C_{\Sigma}^{N+1}$, recorriendo en todos los ket de la base $\{|\phi_{\vec{k}}\rangle\}$ del subespacio $\mathcal{H}(\Sigma)$; también se puede hacer su representación como número entero $k$ que inicia en 0, hasta $2^{N+1} - 1$, como se muestra en la Tabla 5.3 para $N = 4$, la lista $\mathcal{K}$ es:

| $k$ | $\vec{k} = \Sigma, j$ |
|---|---|
| 0 | 0,1 |
| 1 | 1,1 |
| 2 | 1,2 |
| 3 | 1,3 |
| 4 | 1,4 |
| 5 | 2,1 |
| 6 | 2,2 |
| 7 | 2,3 |

| $k$ | $\vec{k} = \Sigma, j$ |
|---|---|
| 8 | 2,4 |
| 9 | 2,5 |
| 10 | 2,6 |
| 11 | 3,1 |
| 12 | 3,2 |
| 13 | 3,3 |
| 14 | 3,4 |
| 15 | 4,1 |

**Tabla 5.3**

De este modo, ordenando por listas $\mathcal{K}(\Sigma)$, vienen a ser:

| $\Sigma$ | $\mathcal{K}(\Sigma) = \{\, k \,\}$ | $\mathcal{K}(\Sigma) = \{\, \vec{k} \,\}$ |
|---|---|---|
| 0 | {0} | {0,1} |
| 1 | {1,2,3,4} | {1,1; 1,2; 1,3; 1,4} |
| 2 | {5,6,7,8,9,10} | {2,1; 2,2; 2,3; 2,4; 2,5; 2,6} |
| 3 | {11,12,13,14} | {3,1; 3,2; 3,3; 3,4} |
| 4 | {15} | {4,1} |

**Tabla 5.4**

A diferencia del vector $\vec{n}$, el vector $\vec{k} = \Sigma, j$ está ordenado de forma ascendente y mantiene una "continuidad" como entero, dentro de cada lista $\mathcal{K}(\Sigma)$.

### 5.1.3 Listas $\mathcal{K}(\Sigma)$: $\vec{n}$ como vector $\Sigma, i$

Los vectores $\vec{k} = \Sigma, j$ y las listas $\mathcal{K}(\Sigma)$ permiten identificar los autoestados $|\phi_{\vec{k}}\rangle$ dentro de cada $\mathcal{H}(\Sigma)$, también es posible contar los kets que conforman la base { $|\vec{n}\rangle$ }, de $\mathcal{H}(\Sigma)$, con esta representación vectorial de dos componentes; así, se plantea que los vectores $\vec{n} \in \mathcal{N}(\Sigma)$ también se pueda representar en dos componentes, análogo a (5.4): $\vec{n} = \Sigma, i$, así en vez de etiquetar los kets $|\phi_{\vec{k}}\rangle$, etiqueta los kets $|\vec{n}\rangle$.

Así por ejemplo, para $N = 3$ se tiene:

| $\Sigma$ | $\mathcal{N}(\Sigma) = \{\vec{n}\}$ | $\mathcal{N}(\Sigma) = \{n\}$ | $\mathcal{K}(\Sigma) = \{\vec{n} = \Sigma, i\}$ |
|---|---|---|---|
| 0 | {0000} | {0} | {0,1} |
| 1 | {1000, 0100, 0010, 0001} | {1,2,4,8} | {1,1; 1,2; 1,3; 1,4} |
| 2 | {1100, 1010, 0110, 1001, 0101, 0011} | {3,5,6,9,10,12} | {2,1; 2,2; 2,3; 2,4; 2,5; 2,6} |
| 3 | {1110, 1101, 1011, 0111} | {7,11,13,14} | {3,1; 3,2; 3,3; 3,4} |
| 4 | {1111} | {15} | {4,1} |

**Tabla 5.5**

En general, para cualquier $N$ dado, en la Tabla 5.6 se muestra los cuatro primeros valores de $\Sigma$:

| $\Sigma$ | $n$ | $\vec{n}$ | $\Sigma, i$ | $\# = C_\Sigma^{N+1}$ |
|---|---|---|---|---|
| 0 | 0 | $00000\cdots 0$ | 0,1 | 1 |
| 1 | 1 | $10000\cdots 0$ | 1,1 | $N+1$ |
|   | 2 | $01000\cdots 0$ | 1,2 |   |
|   | 4 | $00100\cdots 0$ | 1,3 |   |
|   | 8 | $00010\cdots 0$ | 1,4 |   |

|   |   | ⋮ | ⋮ | ⋮ |   |
|---|---|---|---|---|---|
|   | $2^N$ | $00000\cdots1$ | $1,\#$ |   |
| 2 | 3 | $11000\cdots0$ | $2,1$ | $\dfrac{(N+1)N}{2}$ |
|   | 5 | $10100\cdots0$ | $2,2$ |   |
|   | 6 | $01100\cdots0$ | $2,3$ |   |
|   | 9 | $10010\cdots0$ | $2,4$ |   |
|   | ⋮ | ⋮ | ⋮ |   |
|   | $2^N+2^{N-1}$ | $0000\cdots11$ | $2,\#$ |   |
| 3 | 7 | $11100\cdots0$ | $3,1$ | $\dfrac{(N+1)N(N-1)}{6}$ |
|   | 11 | $11010\cdots0$ | $3,2$ |   |
|   | ⋮ | ⋮ | ⋮ |   |
|   | $2^N+2^{N-1}+2^{N-2}$ | $000\cdots111$ | $3,\#$ |   |

**Tabla 5.6** Representación general de $\vec{n}=\Sigma,i$ para los 4 primeros valores de $\Sigma$.

Es importante resaltar que $\vec{n}$ y $\vec{k}$ son índices que denotan dos cualidades distintas, $\vec{n}$ denota la configuración de los ket $|\vec{n}\rangle$ que son autoestados del Hamiltoniano libre $\widehat{H}_0$, y $\vec{k}$ denota los autoestados $|\phi_{\vec{k}}\rangle$ del Hamiltoniano completo $\widehat{H}$, ambos tienen su propia lista $\vec{n}\in\mathcal{N}$ y $\vec{k}\in\mathcal{K}$, y aunque $\mathcal{N}$ y $\mathcal{K}$ tengan el mismo número de elementos, para un $\Sigma$ dado, $\mathcal{N}(\Sigma)$ y $\mathcal{K}(\Sigma)$ no comparten los mismos índices $n$ o $k$ (aunque si el mismo cardinal #); es decir, si bien el índice $n$ en su forma vectorial denota un estado $|\vec{n}\rangle$, la forma vectorial de $k$ no denota algún estado $|\vec{k}\rangle$, sino $|\phi_{\vec{k}}\rangle$.

**5.2 Representación en la nueva notación**

Usando esta notación de las listas y vectores, establecidas en la sección [5.1], se puede hacer una representación en los subespacios, de los elementos de la teoría:

$$\vec{n} \to \Sigma, i \qquad\qquad \vec{k} \to \Sigma, j$$

$$|\vec{n}\rangle \to |\Sigma, i\rangle \qquad\qquad |\phi_{\vec{k}}\rangle \to |\phi_{\Sigma,j}\rangle \qquad (5.6)$$

$$\widehat{H} \to \widehat{H}^{\Sigma} \qquad\qquad \omega_{\vec{k}} \to \omega_{\Sigma,j}$$

Esta notación, la ecuación de autovalores viene a ser:

$$\widehat{H}|\phi_{\vec{k}}\rangle = \omega_{\vec{k}}|\phi_{\vec{k}}\rangle \;\to\; \widehat{H}^{\Sigma}|\phi_{\Sigma,j}\rangle = \omega_{\Sigma,j}|\phi_{\Sigma,j}\rangle \qquad \widehat{H}^{\Sigma} \in \mathcal{H}(\Sigma) \qquad (5.7)$$

Donde $j = 1\!:\!\#$ varía para denotar cada ket $|\phi_{\vec{k}}\rangle \in \mathcal{H}(\Sigma)$, $\widehat{H}^{\Sigma}$ es el hamiltoniano reducido al subespacio $\mathcal{H}(\Sigma)$, y mientras el ket $|\phi_{\vec{k}}\rangle$ tiene dimensión $2^{N+1}$, los ket $|\phi_{\Sigma,j}\rangle$ tienen dimensión $\# = C_{\Sigma}^{N+1}$; esto debido a que cada autoestado $|\phi_{\vec{k}}\rangle$ no ocupa todas las filas $|\vec{n}\rangle$, sino solo $\#$ del total de $2^{N+1}$ filas, siendo ceros el resto de componentes.

Así, de las ecuaciones (3.8) y (5.7), las amplitudes $\beta_{\vec{n},\vec{k}}$ vienen a ser:

$$\langle \vec{n}|\phi_{\vec{k}}\rangle = \beta_{\vec{n},\vec{k}} \;\to\; \langle \Sigma, i|\phi_{\Sigma',j}\rangle = \delta_{\Sigma,\Sigma'}\beta_{i,\Sigma,j} \qquad (5.8)$$

(5.8) es análogo a (4.6), puesto que los kets $|\phi_{\vec{k}}\rangle \in \mathcal{H}(\Sigma)$ son combinaciones lineales de los kets $|\vec{n}\rangle \in \mathcal{H}(\Sigma)$, así, las amplitudes $\beta_{\vec{n},\vec{k}}$ se anulan si $\vec{k}$ y $\vec{n}$ tienen asociados $\Sigma$ distintos; usando (5.8), el ket $|\phi_{\Sigma,j}\rangle$ se expresa como sigue:

$$|\phi_{\Sigma,j}\rangle = \sum_{i=1}^{\#} \beta_{i,\Sigma,j}|i\rangle \in \mathcal{H}(\Sigma) \qquad (5.9)$$

Que es una representación completamente dentro del sub espacio $\mathcal{H}(\Sigma)$; en (5.9) se ha hecho el reemplazo $|\Sigma, i\rangle$ por $|i\rangle$ que es un vector que contiene un uno en la fila $i$, y ceros en el resto de las filas.

Se puede expresar la diagonalización del hamiltoniano $\widehat{H}^{\Sigma}$ mediante una transformación unitaria $\widehat{U}_{\Sigma}$ en el espacio $\mathcal{H}(\Sigma)$:

$$\widehat{H}^{\Sigma} = \sum_{j=1}^{\#} |\phi_{\Sigma,j}\rangle\langle\phi_{\Sigma,j}|\,\omega_{\Sigma,j} \quad \rightarrow \quad \widehat{U}_{\Sigma}^{\dagger}\widehat{H}^{\Sigma}\widehat{U}_{\Sigma} = \sum_{j=1}^{\#} |j\rangle\langle j|\,\omega_{\Sigma,j} \tag{5.10}$$

Donde $\widehat{U}_{\Sigma} : |j\rangle \rightarrow |\phi_{\Sigma,j}\rangle$, esta transformación es representada por una matriz de orden #, la cual es obtenida vía cálculo numérico, como se establece en la sección [4.5.2]; los autovectores $|\phi_{\Sigma,j}\rangle$ vienen a ser las columnas de la matriz $\widehat{U}_{\Sigma}$:

$$\widehat{U}_{\Sigma} = \sum_{j=1}^{\#} |\phi_{\Sigma,j}\rangle\langle j| \qquad \widehat{U}_{\Sigma} = \sum_{j,i=1}^{\#} |j\rangle\langle i|\,\beta_{i,\Sigma,j} \qquad \widehat{U}_{\Sigma} = \begin{pmatrix} |\phi_{\Sigma,1}\rangle & \cdots & |\phi_{\Sigma,\#}\rangle \end{pmatrix} \tag{5.11}$$

Así, $\widehat{U}_{\Sigma}$ viene a ser una matriz de componentes $\beta_{i,\Sigma,j}$ con $i,j = 1:\#$, como se muestra:

$$\widehat{U}_{\Sigma} = \{\beta_{i,\Sigma,j}\} = \begin{pmatrix} \beta_{1\Sigma 1} & \beta_{1\Sigma 2} & \cdots & \beta_{1\Sigma j} & \cdots & \beta_{1\Sigma\#} \\ \beta_{2\Sigma 1} & \beta_{2\Sigma 2} & \cdots & \beta_{2\Sigma j} & \cdots & \beta_{2\Sigma\#} \\ \vdots & \vdots & \ddots & \vdots & & \vdots \\ \beta_{i\Sigma 1} & \beta_{i\Sigma 2} & \cdots & \beta_{i\Sigma j} & \cdots & \beta_{i\Sigma\#} \\ \vdots & \vdots & & \vdots & \ddots & \vdots \\ \beta_{\#\Sigma 1} & \beta_{\#\Sigma 2} & \cdots & \beta_{\#\Sigma j} & \cdots & \beta_{\#\Sigma\#} \end{pmatrix} \tag{5.12}$$

## 5.3 Traza parcial en los grados de libertad del entorno

### 5.3.1 La traza parcial dentro de un solo subespacio

Para tomar la traza parcial, en los grados de libertad del entorno, se separa el vector $\vec{n}$ como: $\vec{n} = \sigma, \vec{\xi}$, con $\sigma$ y $\vec{\xi}$ denotando los números cuánticos del sistema y el entorno respectivamente; así, para un $\sigma$ fijo, el vector $\vec{n} = \sigma, \vec{\xi}$ viene a pertenecer a un sub conjunto $\mathcal{N}_{\sigma}(\Sigma)$ incluido en el total $\mathcal{N}_{\sigma}(\Sigma) \subset \mathcal{N}(\Sigma)$, como se muestra:

$$\mathcal{N}_{\sigma}(\Sigma) = \{\vec{n} \in \mathcal{N}(\Sigma)\ /\ \vec{n} = \sigma, \vec{\xi}\} \qquad\qquad \mathcal{N}(\Sigma) = \bigcup_{\sigma=0}^{1} \mathcal{N}_{\sigma}(\Sigma) \tag{5.13}$$

Las listas $\mathcal{N}_\sigma(\Sigma)$ son disjuntas; $\mathcal{N}_\sigma(\Sigma)$ tiene un sub conjunto asociado $I_\sigma \subset I = \{1, 2, 3, \cdots \#\}$ para el índice $i$ de $\vec{n} = \Sigma, i$, para un $\Sigma$ dado:

$$I_\sigma^\Sigma = \{\, i \in \mathbb{N} \,/\, \sigma, \vec{\xi} = \Sigma, i \,\} \qquad\qquad I^\Sigma = \bigcup_{\sigma=0}^{1} I_\sigma^\Sigma \qquad (5.14)$$

$I_\sigma^\Sigma$ es el conjunto de todos los $i$ de $\vec{n} = \sigma, \vec{\xi} = \Sigma, i$ que pertenecen a un subespacio dado $\mathcal{H}(\Sigma)$, y puede ser creado a partir de un algoritmo descrito en la Figura 5.1: partiendo de la lista $\mathcal{N}(\Sigma)$ se inicia un ciclo $Do\ while$ que genera todos los vectores $\vec{n} \in \mathcal{N}(\Sigma)$ en orden, y a la vez los $i$ que van de 1 a #, y solo si la primera componente del vector $\vec{n}$ tiene el valor deseado $n_0 = \sigma$ se le añade el número $i$ a la lista $I_\sigma$ :

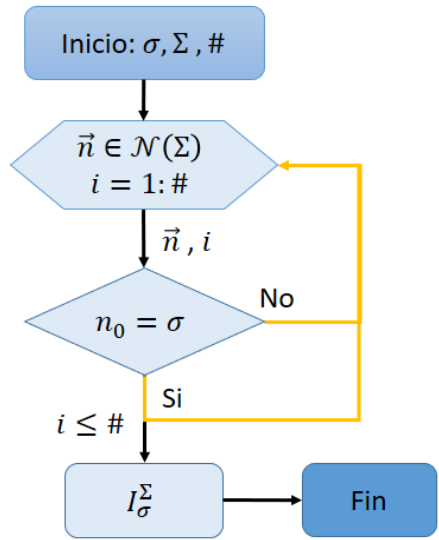

**Figura 5.1**

Así, por ejemplo, de la tabla 5.5, en la columna $\mathcal{N}(\Sigma) = \{\,\vec{n}\,\}$, para la fila $\Sigma = 2$ se tienen los conjuntos $I_0^2 = \{3,5,6\}$ y $I_1^2 = \{1,2,4\}$; con estos subconjuntos $I_\sigma^\Sigma$, se puede expresar la sumatoria en los grados de libertad del entorno para la amplitud $\beta_{\sigma,\vec{\xi},\vec{k}}$ que multiplica otros términos $(\cdots)_{\sigma,\vec{\xi},\vec{k}}$, la sumatoria en $\vec{\xi}$ viene a ser:

$$\Sigma\text{ fijo: }\ \begin{array}{l}\vec{n} = \sigma, \vec{\xi} = \Sigma, i \\ \vec{k} = \Sigma, j\end{array} \qquad \begin{array}{c} \beta_{\sigma,\vec{\xi},\vec{k}} \to \beta_{i,\Sigma,j} \\ \displaystyle\sum_{\vec{\xi}} \beta_{\sigma,\vec{\xi},\vec{k}}(\cdots)_{\sigma,\vec{\xi},\vec{k}} \to \sum_{i \in I_\sigma^\Sigma} \beta_{i,\Sigma,j}(\cdots)_{i,\Sigma,j} \end{array} \qquad (5.15)$$

(5.15) es válida solo dentro de un solo subespacio $\mathcal{H}(\Sigma)$.

En general, una sumatoria en términos que dependen de $\vec{n}, \vec{k}$ y multiplican $\beta_{\vec{n},\vec{k}}$, para cualquier subespacio, se lo puede expresar como:

$$\begin{aligned} \varphi_{\vec{k}} = \sum_{\vec{n}} \beta_{\vec{n},\vec{k}}(\cdots)_{\vec{n},\vec{k}} &\rightarrow \varphi_{\Sigma,j} \\ &= \sum_{\Sigma'} \delta_{\Sigma,\Sigma'} \sum_{i} \beta_{i,\Sigma',j}(\cdots)_{i\Sigma'j} \end{aligned} \qquad \begin{aligned} \Sigma' &= 0:N+1 \\ i &= 1:\# = C_{\Sigma'}^{N+1} \end{aligned} \qquad (5.16)$$

La sumatoria en $\vec{n} = \Sigma', i$ se divide en dos sumatorias, para $\Sigma'$ y para $i$ (asociado a $\Sigma'$); la sumatoria en $\Sigma'$ viene multiplicada por $\delta_{\Sigma,\Sigma'}$ (debido a $\beta_{\vec{n},\vec{k}}$), al ejecutar la suma en $\Sigma'$ en (5.16) se obtiene: $\varphi_{\Sigma,j} = \sum_i \beta_{i,\Sigma,j}(\cdots)_{i\Sigma j}$ donde $i$ está asociado a $\Sigma$, sin embargo es conveniente mantener la forma general (5.16) para cuando los elementos $\beta_{i,\Sigma,j}(\cdots)_{i\Sigma j}$ no se expresan de forma general, sino dependen del subespacio, entonces se desarrolla: $\varphi_{\Sigma,j} = \delta_{\Sigma,0}(\cdots)_{101} + \delta_{\Sigma,1} \sum_i \beta_{i,1,j}(\cdots)_{i1j} + \delta_{\Sigma,2} \sum_i \beta_{i,2,j}(\cdots)_{i2j} + \cdots \delta_{\Sigma,N+1}(\cdots)_{1,N+1,1}$, (5.16) varía en todos los subespacios.

A partir de (5.15) se puede llevar a un caso particular, para $\vec{k}, \vec{n}$ y $\vec{n}'$ que pertenecen a un solo subespacio $\mathcal{H}(\Sigma)$:

$$\begin{aligned} \Sigma \text{ fijo:} \quad \vec{k} = \Sigma, j \quad \vec{k}' = \Sigma, j' \quad \vec{n} = \sigma, \vec{\xi} = \Sigma, i \quad \vec{n}' = \sigma', \vec{\xi} = \Sigma, i' \\ \sum_{\vec{\xi}} \beta_{\sigma,\vec{\xi},\vec{k}} \beta^*_{\sigma',\vec{\xi},\vec{k}'} \rightarrow \delta_{\sigma,\sigma'} \sum_{i \in I_\sigma^\Sigma} \beta_{i,\Sigma,j} \beta^*_{i,\Sigma,j'} \end{aligned} \qquad (5.17)$$

La delta $\delta_{\sigma,\sigma'}$ se debe a que para $\sigma \neq \sigma'$: de (5.13) se tiene que $\vec{n}$ y $\vec{n}'$ pertenecen a $\mathcal{N}_\sigma(\Sigma)$ y $\mathcal{N}_{\sigma'}(\Sigma)$ que no comparten ningún vector en común, también los subconjuntos asociados $I_\sigma$ y $I_{\sigma'}$ son disjuntos, ecuación (5.14), de manera que la sumatoria en $\vec{\xi}$ en (5.17), que corresponde a sumar en $i \in I_\sigma^\Sigma$ y en $i' \in I_{\sigma'}^\Sigma$, se traduce en tener que sumar sólo en los elementos en común $I_\sigma \cap I_{\sigma'} = \emptyset$; es decir, se anula la suma.

### 5.3.2 La traza parcial en subespacios distintos

Desde que se puede pasar de la representación vectorial (binaria) $\vec{n}$ a entero $n$, se puede hacer los mismo con la variable de los grados de libertad del entorno: $\vec{\xi} \to \mathcal{E}$;

$$\vec{n} = \sigma, \vec{\xi}: \begin{array}{l} \sigma\, 0000 \cdots 0 \\ \sigma\, 1000 \cdots 0 \\ \sigma\, 0100 \cdots 0 \\ \vdots \end{array} \qquad \vec{\xi} \to \mathcal{E} \qquad \mathcal{E} = 0, 1, 2, 3, \cdots 2^N - 1 \qquad (5.18)$$

A la izquierda de (5.18) se muestra que, para la primera componente fijada $n_0 = \sigma$, el vector $\vec{\xi}$ puede variar en sus $N$ componentes $\vec{\xi} = n_1, n_2, n_3, \cdots n_N$; al centro de (5.18) se realiza la trasformación del vector $\vec{\xi}$ al número asociado $\mathcal{E}$, el cual puede variar de $0$ hasta $2^{N-1}$; así, la relación entre $n$ (que resulta de transformar $\vec{n}$) y $\mathcal{E}$ viene a ser, para dos valores $\sigma, \sigma'$ de $n_0$:

$$\begin{array}{lll} \vec{n}_1 = \sigma, \vec{\xi} \to n_1 & n_1 = \sigma + 2\mathcal{E} & n_1 = \Sigma_1, i_1 \\ \vec{n}_2 = \sigma', \vec{\xi} \to n_2 & n_2 = \sigma' + 2\mathcal{E} & n_2 = \Sigma_2, i_2 \end{array} \qquad (5.19)$$

(5.19) crea parejas de $n_1 = \Sigma_1, i_1$ y $n_2 = \Sigma_2, i_2$ que tienen el mismo $\mathcal{E}$ o $\vec{\xi}$, que es lo que se necesita para tomar la traza parcial en todos los grados de libertad del entorno, de forma general sin estar restringido a un solo subespacio.

Por ejemplo, si $\vec{n}_1 = 10100 \cdots 0$, se tiene $\sigma = 1$ y $\vec{\xi} = 0100 \cdots 0$, entonces: $\vec{n}_1 \to n_1 = 5$, $\vec{\xi} \to \mathcal{E} = 2$, así cumple: $5 = 1 + 2 \cdot 2$; la obtención de $n_1 = \Sigma_1, i_1 = 2,2$ se puede obtener de la tabla 5.6, que relaciona $\Sigma_1, i_1$ con $n_1$.

También es posible usar los enteros $n$ para expresar la traza parcial a través de cualquier subespacio, usando (5.19):

$$\sigma, \vec{\xi} \to n_1 \qquad \sigma', \vec{\xi} \to n_2 \qquad \vec{k} \to k \qquad \vec{k}' \to k' \qquad (5.20)$$

$$\sum_{\vec{\xi}} \beta_{\sigma,\vec{\xi},\vec{k}} \beta^*_{\sigma',\vec{\xi},\vec{k}'} \to \sum_{\mathcal{E}=0}^{2^N-1} \beta_{n_1,k} \beta^*_{n_2,k'} = \sum_{\mathcal{E}=0}^{2^N-1} \beta_{\sigma+2\mathcal{E},k} \beta^*_{\sigma'+2\mathcal{E},k'}$$

Para pasar (5.20) a la notación de los sub espacios se considera además de $n_r = \Sigma_r, i_r$, $r = 1,2$ de (5.19), y además $\vec{k} = \Sigma, j$ y $\vec{k}' = \Sigma' j'$ :

$$\sum_{\vec{\xi}} \beta_{\sigma,\vec{\xi},\vec{k}} \beta^*_{\sigma',\vec{\xi},\vec{k}'} \to \sum_{(\Sigma_1,i_1;\Sigma_2,i_2)} \beta_{(\Sigma_1,i_1),(\Sigma j)} \beta^*_{(\Sigma_2,i_2),(\Sigma' j')} \qquad (i_1,i_2) \in I_{\sigma,\sigma'}(\Sigma_1,\Sigma_2)$$

Donde $I_{\sigma,\sigma'}(\Sigma_1,\Sigma_2)$ es una generalización de $I_\sigma^\Sigma$, y es el conjunto de todos los pares $(i_1, i_2)$ correlacionados, tales que: $\Sigma_1, i_1 = \sigma, \vec{\xi}$ y $\Sigma_2, i_2 = \sigma', \vec{\xi}$, para todos los valores de $\vec{\xi}$ posibles que permitan $\Sigma_1$ y $\Sigma_2$ dados; así, usando (5.8):

$$\sum_{\vec{\xi}} \beta_{\sigma,\vec{\xi},\vec{k}} \beta^*_{\sigma',\vec{\xi},\vec{k}'} \to \sum_{(\Sigma_1,i_1;\Sigma_2,i_2)} \delta_{\Sigma,\Sigma_1} \delta_{\Sigma',\Sigma_2} \beta_{i_1 \Sigma j} \beta^*_{i_2 \Sigma' j'} \qquad (i_1,i_2) \in I_{\sigma,\sigma'}(\Sigma_1,\Sigma_2)$$

Al ejecutar las deltas eliminan las sumatorias en $\Sigma_1$ y $\Sigma_2$:

$$\sum_{\vec{\xi}} \beta_{\sigma,\vec{\xi},\vec{k}} \beta^*_{\sigma',\vec{\xi},\vec{k}'} \to \sum_{i_1,i_2} \beta_{i_1 \Sigma j} \beta^*_{i_2 \Sigma' j'} \qquad (i_1,i_2) \in I_{\sigma,\sigma'}(\Sigma,\Sigma') \qquad (5.21)$$

los $(\Sigma, i_1; \Sigma', i_2)$ son parejas de $(n_1, n_2)$ que pertenecen al conjunto $I_{\rho,\sigma}(\Sigma,\Sigma')$, que es una generalización de $I_\sigma^\Sigma$ para distintos subespacios:

$$I_{\sigma,\sigma'}(\Sigma,\Sigma') = \left\{ (i_1, i_2) \in I_\sigma^\Sigma \times I_{\sigma'}^{\Sigma'} \,/\, \begin{array}{l} \Sigma, i_1 = \sigma + 2\mathcal{E} \\ \Sigma', i_2 = \sigma' + 2\mathcal{E} \end{array} \right\}$$

$$I_{\sigma,\sigma'} = \bigcup_{\mathcal{E}=0}^{2^{N-1}} \{(\sigma + 2\mathcal{E}, \sigma' + 2\mathcal{E})\} \qquad (5.22)$$

Se puede desarrollar $I_{\sigma,\sigma'}$ como se muestra: $I_{\sigma,\sigma'} = \{(\sigma,\sigma'), (\sigma+2, \sigma'+2), (\sigma+4, \sigma'+4), \cdots, (\sigma+2^{N+1}-2, \sigma'+2^{N+1}-2)\}$, después usando la tabla 5.6 se convierten las parejas $(n_1, n_2)$ a $(\Sigma_1, i_1; \Sigma_2, i_2)$, así a partir de $I_{\sigma,\sigma'}$ se deriva $I_{\sigma,\sigma'}(\Sigma, \Sigma')$ seleccionando solo las parejas $(i_1, i_2)$ tales que $\Sigma_1 = \Sigma$ y $\Sigma_2 = \Sigma'$.

De (5.22) se puede demostrar el rol de la delta $\delta_{\sigma,\sigma'}$ en (5.17) para $\Sigma = \Sigma'$, pues para $\sigma \neq \sigma'$, los índices $\Sigma, i_1 = \sigma + 2\mathcal{E}$ y $\Sigma', i_2 = \sigma' + 2\mathcal{E}$ tienen distintos $\Sigma \neq \Sigma'$ ($2\mathcal{E}$ establece una serie de unos y ceros a la derecha de $\sigma$ o $\sigma'$ en la representación vectorial $\vec{n}$, y si $\sigma \neq \sigma'$, entonces $\sigma + 2\mathcal{E}$ y $\sigma' + 2\mathcal{E}$ tendrán distintos unos, lo que se cuenta con $\Sigma$), pero eso contradice la condición inicial $\Sigma = \Sigma'$, de manera que no hay solución posible, y el conjunto es vacío; en general, la única manera de tener conjuntos no vacíos de $I_{\sigma,\sigma'}(\Sigma, \Sigma')$, para $\sigma \neq \sigma'$, es que se trabaje en distintos subespacios $\Sigma \neq \Sigma'$.

# CAPÍTULO VI

# SOLUCIONES EXACTAS PARA CONDICIONES INICIALES ESPECIALES

**6.1 La matriz de acoples G**

Como se refirió en el algoritmo 4, la matriz $G$ está compuesto de los elementos $g_{ij}$, que es el acople entre el oscilador $i$ y $j$, tales que cumplen (3.25); esto es, $g_{ji} = g_{ij}^*$ y $g_{ii} = \omega_i$, $G$ es una matriz hermitiana.

**6.1.1 Osciladores del entorno sin interacciones entre ellos**

La elección de los elementos de la matriz $G = \{g_{ij}\}$ es compatible con $J(\omega)$ de acuerdo a (2.46); esto es, en esta sección se obtienen los elementos $g_{ij}$ que al reemplazar en (2.46) resulta la densidad espectral $J(\omega)$ de (6.1).

Se considera la siguiente densidad espectral $J(\omega)$, ver referencia [2]:

$$J(\omega) = 2\pi\eta\omega_c \left(\frac{\omega}{\omega_c}\right)^s e^{-\omega/\omega_c} \qquad \omega > 0 \qquad (6.1)$$

Donde $\eta$ es la fuerza de acoplamiento adimensional entre el sistema y el entorno, $s$ da el orden de la densidad espectral (óhmico, sub o súper óhmico), y $\omega_c$ es la frecuencia de corte del espectro, el cual debe ser del orden de la frecuencia del sistema $\omega_c \approx \omega_0$, una elección razonable es hacer $\omega_c = \omega_0$; se tiene que disminuyendo el valor de $\eta$ por debajo de 0.3 se encamina a la aproximación markoviana, de acuerdo a la referencia [2].

El modelo simplificado tiene un espectro discreto de $N$ frecuencias $\omega_i$, $i = 1:N$, pero para el límite $N \gg 1$ se puede aproximar al caso continuo, tomando muchos valores próximos entre sí; de acuerdo a (2.46) se puede elegir una densidad de estados uniforme en un intervalo de frecuencias, sobre el que se define la función acople $k(\omega)$:

$$g(\omega) = \begin{cases} 1 & \omega \in [\omega_{min}, \omega_{max}] \\ 0 & \omega \notin [\omega_{min}, \omega_{max}] \end{cases} \qquad \begin{aligned} \omega_{min} &= \min\{\omega_i\} \\ \omega_{max} &= \max\{\omega_i\} \end{aligned} \qquad (6.2)$$

Así, $\omega_{min}$ es la frecuencia mínima de los osciladores del entorno, y $\omega_{max}$ es la máxima; se considera una frecuencia mínima no nula $\omega_{min}$ que tenga un valor pequeño respecto a la frecuencia del sistema $\omega_0$; mientras que $\omega_{max}$ sea varias veces mayor que $\omega_0$, aunque no necesariamente mucho más grande, pues $J(\omega)$ de (6.1) se desvanece relativamente rápido para frecuencias mucho mayores que $\omega_0$.

Se puede usar (6.2) para crear las frecuencias de los osciladores del entorno mediante una variable aleatoria, se puede crear aleatoriamente los $\omega_i$ con una distribución equiprobable en $[\omega_{min}, \omega_{max}]$, mediante el uso de una variable aleatoria uniforme en el intervalo $r \in [0\,;1]$, no obstante, para el cálculo numérico es deseable que las frecuencias no sean muy próximas entre sí[21] de modo que se las puede generar mediante la siguiente ecuación de recurrencia: $\omega_{i+1} = \omega_i + r'h$, donde $h$ es una constante que fija la escala de separación media entre frecuencias, y $r' = (1-\epsilon)r + \epsilon$, es una variable aleatoria uniforme en $[\epsilon\,;1]$ que nunca se anula ($1 > \epsilon > 0$).

La función $k(\omega)$ de (2.46) es el acople $V_j$ (como una función de la frecuencia $\omega_j$) llevado al límite continuo, $V_j$ es el acople entre el oscilador del sistema y cada oscilador del entorno; es decir, está dado por los acoples $g_j = V_j$ del hamiltoniano de interacción (3.22) o (3.23), de este modo la integral de la densidad espectral $J(\omega)$ de (2.46), multiplicada por una potencia de la frecuencia $\omega^k$ viene a ser:

$$\int J(\omega)\omega^k\,d\omega = 2\pi \sum_j |g_j|^2 \int \omega^k\,\delta(\omega - \omega_j)d\omega$$

Esta integral se puede evaluar en dos intervalos: en todo el dominio de las frecuencias $\Omega$, o en cada celda $\Delta\omega_i$ que son sub conjuntos disjuntos de $\Omega$, cada subconjunto $\Delta\omega_i$ contiene a la frecuencia $\omega_i$, pero no otra frecuencia del espectro discreto $\{\omega_j\}$. Los subconjuntos componen todo el intervalo: $\Omega = \bigcup_i \Delta\omega_i$, la integral de la delta de Dirac en $\Delta\omega_i$ es: $\int_{\Delta\omega_i} f(\omega)\,\delta(\omega - \omega_j)d\omega = f(\omega_j)\delta_{ij}$. Las integrales ya sea en $\Omega$, o en cada $\Delta\omega_i$ con $k = 0$, resultan:

$$\int_\Omega J(\omega)\omega^k\,d\omega = 2\pi \sum_j |g_j|^2 \omega_j^k \qquad \int_{\Delta\omega_i} J(\omega)\,d\omega = 2\pi|g_i|^2$$

---

[21] En el Anexo B, Figura B.1, los autovalores son pequeñas desviaciones de las frecuencias, que no exceden de las frecuencias vecinas, mientras más próximas son dos frecuencias $\omega_i$ los autovalores son más próximos a cada frecuencia, y por lo tanto sus cálculos requieren de mayor precisión numérica.

Se define $|g|_k^2$ como igual a la sumatoria, entonces se despeja $|g|_k^2$ y $|g_i|^2$:

$$|g|_k^2 = \sum_j |g_j|^2 \omega_j^k = \frac{1}{2\pi}\int_\Omega J(\omega)\omega^k\,d\omega \qquad |g_i|^2 = \frac{1}{2\pi}\int_{\Delta\omega_i} J(\omega)\,d\omega \qquad (6.3)$$

Para este caso, de la densidad espectral (6.1), y el límite $N \gg 1$, se tienen los intervalos $\Omega = [\omega_{min}\,;\,\omega_{max}]$ y $\Delta\omega_i = \langle \omega_i - \frac{\epsilon_i^-}{2}\,;\,\omega_i + \frac{\epsilon_i^+}{2}\rangle$, $\epsilon_i^-$ y $\epsilon_i^+$ son definidos en (6.4), su función es darle el ancho a su respectivo intervalo: $|\Delta\omega_i| = \epsilon_i^- + \epsilon_i^+ = \epsilon_i$, entonces cumplen $|\Omega| = \sum_{i=1}^N \epsilon_i$. Para $N$ grande o $\epsilon_i$ pequeño, $J(\omega)$ en el intervalo $\Delta\omega_i$ se puede aproximar a una función constante $J(\omega_i)$ y la integral de la derecha de (6.3) se simplifica a $|g_i|^2 = \frac{1}{2\pi}\int_{\Delta\omega_i} J(\omega)\,d\omega \to \frac{J(\omega)}{2\pi}\epsilon_i$, dando el valor de los acoples $g_j$:

$$g_j = \sqrt{\frac{J(\omega_j)\epsilon_j}{2\pi}}\,e^{i\theta_j} \qquad \theta_j = \arg(g_j) \qquad \epsilon_j = \frac{\omega_{j+1} - \omega_{j-1}}{2} \qquad (6.4)$$

En general los acoples $g_j$ son complejos, entonces se tiene que introducir sus fases $\theta_j$, cuya información no se puede deducir de la función de densidad espectral, de este modo hay una libertad de elegir $\theta_j$ para una densidad espectral dada, y queda a sujeción del modelo físico que se desea describir; en muchos casos, como el de este trabajo, se puede fijar $\theta_j = 0$ dando acoples reales, esta es una elección razonable desde que $J(\omega)$ depende de $|g_j|^2$ y no de $g_j$. Las ecuaciones de (6.4) son válidas para cualquier densidad espectral $J(\omega)$, pero en este modelo se usa la de (6.1); desde que las frecuencias $\omega_j$ son creadas aleatoriamente, los acoples $g_j$ también tienen valores aleatorios. La fórmula de $\epsilon_j$ a la derecha de (6.4) se puede derivar de la Figura 6.1, donde las frecuencias $\omega_i$ no están en general igualmente espaciadas, y se considera que $\epsilon_i^- = \epsilon_{i-1}^+$; es decir, que la celda $\Delta\omega_i$ va del punto medio entre $\omega_i$ y $\omega_{i-1}$ al punto medio entre $\omega_{i+1}$ y $\omega_i$:

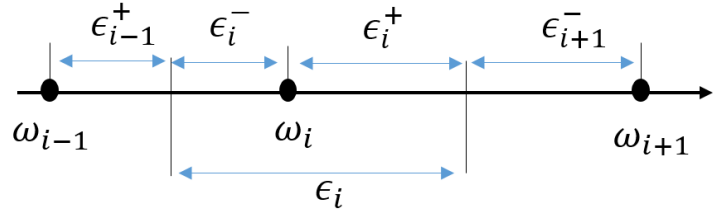

Figura 6.1

Las frecuencias $\omega_j$, $j = 0:N$ se ubican a lo largo de la diagonal principal de $G$, y los acoples entre el sistema y cada oscilador del entorno $g_j = g_{0j}$, $g_{j0} = g_{0j}^*$, $j > 0$ se ubican a lo largo de la primera fila y columna, como se muestra:

$$G = \{g_{ij}\} = \begin{pmatrix} \omega_0 & g_1 & g_2 & g_3 & g_4 & \cdots & g_N \\ g_1^* & \omega_1 & 0 & 0 & 0 & & 0 \\ g_{2^*} & 0 & \omega_2 & 0 & 0 & & 0 \\ g_3^* & 0 & 0 & \omega_3 & 0 & & 0 \\ g_4^* & 0 & 0 & 0 & \omega_4 & & 0 \\ \vdots & & & & & \ddots & \vdots \\ g_N^* & 0 & 0 & 0 & 0 & \cdots & \omega_N \end{pmatrix} \qquad (6.5)$$

Los acoples $g_{ij}$ entre osciladores del entorno, $i, j \neq 0$, $i \neq j$, son nulos $g_{ij} = 0$, entonces $G$ de (6.5) es apropiado para describir un oscilador en interacción con un reservorio de fotones, ya que estos no interactúan entre sí.

**6.1.2 Osciladores del entorno con interacciones entre ellos**

La ecuación (6.4) se puede extender a todos los acoples entre los osciladores del entorno, un primer intento podría ser: $g_{ij} = \sqrt{J_i(\omega_j)\epsilon_j/2\pi}e^{i\alpha_{ij}}$, donde cada oscilador $i$ interactúa con todos los demás osciladores mediante una densidad espectral $J_i(\omega) = 2\pi\eta_i\omega_i\left(\frac{\omega}{\omega_i}\right)^s e^{-\omega/\omega_i}$, el problema con esta expresión es que en general $J_i(\omega_j)$ no es simétrico $J_i(\omega_j) \neq J_j(\omega_i)$ ya que, junto a $\alpha_{ij} = -\alpha_{ij}$, los acoples deben cumplir $g_{ij} = g_{ji}^*$, la solución es simetrizarlo como sigue: $J_i(\omega_j)\epsilon_j \to [J_i(\omega_j)\epsilon_j + J_j(\omega_i)\epsilon_i]/2$, dando:

$$g_{ij} = \sqrt{L_i(\omega_j)\epsilon_j + L_j(\omega_i)\epsilon_i}\, e^{i\alpha_{ij}} \qquad L_i(\omega) = \frac{1}{2}\eta_i\omega_i\left(\frac{\omega}{\omega_i}\right)^s e^{-\omega/\omega_i} \qquad i \neq j > 0 \qquad (6.6)$$

Donde $J_i(\omega_j) = 2\pi \left( L_i(\omega_j) + L_j(\omega_i) \right)$ es la densidad espectral del oscilador $i$ del entorno evaluado en $\omega_j$, $J_i(\omega)$ se puede interpolar como una función continua entre ellas.

Así, con (6.2), (6.4) y (6.6) se puede construir la matriz $G = \{g_{ij}\}$ como se muestra:

$$G = \{g_{ij}\} = \begin{pmatrix} \omega_0 & g_{10}^* & g_{20}^* & g_{30}^* & g_{40}^* & \cdots & g_{i0}^* & \cdots & g_{N0}^* \\ g_{10} & \omega_1 & g_{21}^* & g_{31}^* & g_{41}^* & & g_{i1}^* & & g_{N1}^* \\ g_{20} & g_{21} & \omega_2 & g_{32}^* & g_{42}^* & & g_{i2}^* & & g_{N2}^* \\ g_{30} & g_{31} & g_{32} & \omega_3 & g_{43}^* & & g_{i3}^* & & g_{N3}^* \\ g_{40} & g_{41} & g_{42} & g_{43} & \omega_4 & \cdots & g_{i4}^* & \cdots & g_{N4}^* \\ \vdots & & & & \vdots & \ddots & \vdots & & \vdots \\ g_{i0} & g_{i1} & g_{i2} & g_{i3} & g_{i4} & \cdots & \omega_{ii} & \cdots & g_{iN}^* \\ \vdots & & & & \vdots & & \vdots & \ddots & \vdots \\ g_{N0} & g_{N1} & g_{N2} & g_{N3} & g_{N4} & \cdots & g_{iN} & \cdots & \omega_N \end{pmatrix} \quad (6.7)$$

Hay tres regiones a distinguir en la matriz $G$, la diagonal (correspondiente a las frecuencias de cada oscilador), la primera fila excepto el primer elemento de la diagonal (corresponde a los acoples del sistema con cada oscilador del entorno, y los demás, que corresponde a los acoples entre osciladores del entorno, cada uno de ellos son descritos por las ecuaciones (9.3), (9.2) y (9.4) respectivamente, y señalados en las siguientes imágenes, considerando $g_{ji}^* = g_{ij}$:

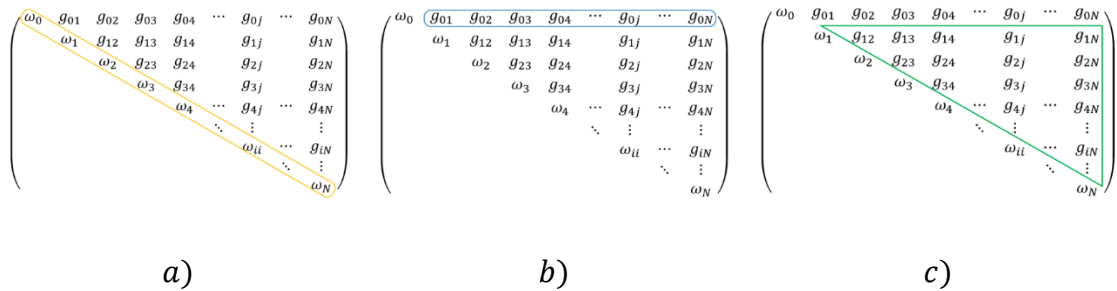

$a)$ $\qquad\qquad b)$ $\qquad\qquad c)$

Figura 6.2

Las fuerzas de acoplamientos $\eta_i$, con $\eta_0 = \eta$, pueden ser elegidas arbitrariamente para $i > 0$; sin embargo, para $i = 0$ se debe elegir cuidadosamente si se desea el acople débil o fuerte entre el sistema y el entorno: si desea una dinámica de acoplamiento débil, entonces se fija $\eta \sim 0.01$, si por el contrario se desea una dinámica de acoplamiento fuerte se puede elegir $\eta \sim 1$. La descripción de esta sección no es desarrollada en este trabajo, pero queda concebida para ser empleada en trabajos posteriores.

**6.2 Las condiciones iniciales**

Algunas características generales son consideradas, como que al inicio, $t = 0$, el estado del sistema y del entorno son factorizables, y que el sistema y el entorno se encuentran en un estado puro.

**6.2.1 Entorno a temperatura nula: $\Sigma = 1$**

La elección de un entorno a temperatura nula es muy conveniente, pues requiere el menor subespacio de Hilbert no trivial $\mathcal{H}(1)$, que tiene dimensión $\# = N + 1$, mientras que para subespacios con $\Sigma$ mayores se tiene un mayor número de dimensiones $\# = C_\Sigma^{N+1}$ que crecen rápidamente. La elección de $\Sigma = 1$ permite evaluar el mayor valor de $N$ para una disposición de recursos computacionales dadas; en $\mathcal{H}(1)$ todos los ket tienen un solo oscilador excitado, y todos los demás $N$ osciladores están en su estado fundamental (de aquí la temperatura nula); así, al inicio se plantea el estado:

$$|\psi(0)\rangle = |1\rangle \otimes |000 \cdots 0\rangle_\xi \qquad \hat{\rho}_0(0) = \begin{pmatrix} 0 & 0 \\ 0 & 1 \end{pmatrix} \qquad (6.8)$$

El sistema se encuentra en $|1\rangle$, un estado puro, y es factorizable del entorno, el cual se encuentra en el estado fundamental $|000 \cdots 0\rangle_\xi$, lo cual corresponde a la temperatura nula $T = 0$; $\hat{\rho}_0(0)$ es la traza parcial de $|\psi(0)\rangle\langle\psi(0)|$ en todos los grados de libertad del entorno; conforme el sistema total evoluciona en el tiempo, el sistema se relaja, ocupando el estado $|0\rangle$, mientras que algún oscilador del entorno va ocupando el estado $|1\rangle$; (6.8) justifica que todos los osciladores del modelo simplificado sean sistemas de dos niveles, pues para la interacción planteada, ningún oscilador se excitará más de dos veces: hay un solo quantum de energía que se disipa entre el sistema y el entorno.

Las ecuaciones de (6.8) representan un estado puro correspondiente a un ket perteneciente al subespacio $\mathcal{H}(1)$; sin embargo, la matriz de densidad inicial de (6.8) no contiene términos de coherencia, fuera de la diagonal, a evolucionar. En el subespacio $\mathcal{H}(1)$ no es posible plantear un estado para todos los osciladores tal que el estado del sistema resulte en una matriz con elementos fuera de la diagonal como en (6.10); por ejemplo, es posible definir el siguiente estado:

$$|\psi\rangle = \alpha|00\cdots 1\cdots 0\rangle + \beta|100\cdots 0\rangle \qquad \hat{\rho}_0(0) = \begin{pmatrix} |\alpha|^2 & 0 \\ 0 & |\beta|^2 \end{pmatrix} \qquad (6.9)$$

Donde $|00\cdots 1\cdots 0\rangle$ indica que algún otro oscilador del entorno está excitado, ahora (6.9) ya no es factorizable como (6.8), y la traza parcial en el entorno da $\hat{\rho}_0(0)$ como un estado mezcla, lo que tampoco tiene coherencia cuántica que explorar.

### 6.2.2 Entorno a temperatura nula: $\Sigma = 0\,, 1$

La necesidad de tener términos de coherencia en el estado inicial del sistema, y que éste sea puro y factorizable del entorno, conduce a plantear que el sistema se encuentre en el estado deseado $|\psi_S\rangle = \alpha|0\rangle + \beta|1\rangle$, y el entorno en su estado fundamental $\big|\vec{0}\big\rangle_\xi$:

$$|\psi\rangle = (\alpha|0\rangle + \beta|1\rangle)\otimes|000\cdots 0\rangle_\xi \qquad \hat{\rho}_0(0) = \begin{pmatrix} |\alpha|^2 & \alpha^*\beta \\ \alpha\beta^* & |\beta|^2 \end{pmatrix} \qquad (6.10)$$

El ket $|\psi\rangle$ pertenece a los dos subespacios $\mathcal{H}(0)$ y $\mathcal{H}(1)$, aunque $\mathcal{H}(0)$ solo es de una dimensión y no complica demasiado la dinámica, pues el estado $|0\rangle|000\cdots 0\rangle_\xi$ no cambia en el tiempo; (6.8) se deriva de (6.10) para $\alpha = 0$.

### 6.2.3 Hamiltoniano reducido a temperatura nula

De las secciones precedentes [6.2.1] y [6.2.2] se ha establecido que el modelo a temperatura nula requiere trabajar en los subespacios $\mathcal{H}(0)$ y $\mathcal{H}(1)$; el subespacio $\mathcal{H}(0)$ es trivial, solo contiene el estado fundamental de todo el sistema $|\vec{0}\rangle$, y su hamiltoniano se anula de acuerdo a (3.1) y (3.3), pues no se considera la energía de punto cero, así se tiene $\widehat{H}^0 = 0$, y en consecuencia $\omega_{01} = 0$.

En el subespacio $\mathcal{H}(1)$ el hamiltoniano reducido $\widehat{H}^1$ puede ser derivado del algoritmo 4 de la sección [4.4], o usando las ecuaciones (4.12) y (4.16), y viene a ser:

$$\widehat{H}^1 \equiv G^T \qquad (6.11)$$

(6.11) también se cumple para el hamiltoniano (4.18) con acoples no nulos $g_{ij}$ entre osciladores del entorno.

### 6.3 Soluciones exactas para las condiciones iniciales

#### 6.3.1 Entorno a temperatura nula: $\Sigma = 1$

Para $\Sigma = 1$ se tiene el estado inicial (6.8), el estado inicial se expresa de acuerdo a (3.4) como sigue:

$$|\psi\rangle = \sum_{\vec{n}} \psi_{\vec{n}} |\vec{n}\rangle \qquad \psi_{\vec{n}} = \delta_{\vec{n},\vec{n}_0} \qquad \left.\begin{array}{l} \vec{n}_0 = 100\cdots 0 \\ \\ \vec{n}_0 \to \Sigma, i = 1,1 \end{array}\right| \qquad (6.12)$$

Como la dinámica está restringida al subespacio $\mathcal{H}(1)$, solo participan los autoestados $|\phi_{\vec{k}}\rangle \in \mathcal{H}(1)$, así los autoestados y sus amplitudes se expresa de acuerdo a (5.8), añadiendo la delta $\delta_{\Sigma,1}$ que indica la restricción de la dinámica:

$$|\phi_{\vec{k}}\rangle \to |\phi_{\Sigma,j}\rangle = \delta_{\Sigma,1}|\phi_{1,j}\rangle \qquad \beta_{\vec{n},\vec{k}} \to \delta_{\Sigma,1}\beta_{i,1,j} \qquad (6.13)$$

Las amplitudes en la base de modos normales están dadas por $\varphi_{\vec{k}} = \langle\phi_{\vec{k}}|\psi(0)\rangle$ de (3.20); así, reemplazando $\psi_{\vec{n}} = \delta_{\vec{n},\vec{n}_0}$ de (6.12) en $\varphi_{\vec{k}}$, y usando (6.13), y $\vec{n}_0 = 1,1$:

$$\varphi_{\vec{k}} = \sum_{\vec{n}} \psi_{\vec{n}} \beta^*_{\vec{n},\vec{k}} = \beta^*_{\vec{n}_0,\vec{k}} \to \varphi_{\Sigma,j} = \delta_{\Sigma,1}\beta^*_{1,1,j} \qquad (6.14)$$

Aplicando (6.13), (6.14) y (5.17) a $\chi_{\sigma,\sigma'}$ de (3.26) (y considerando[22] $D = \emptyset$):

$$\chi_{\sigma,\sigma'} = \sum_{\vec{k}} \left( \sum_{\vec{n}',\vec{n}''} \psi_{\vec{n}'} \psi^*_{\vec{n}''} \beta^*_{\vec{n}',\vec{k}} \beta_{\vec{n}'',\vec{k}} \right) \sum_{\vec{\xi}} \beta_{\sigma,\vec{\xi},\vec{k}} \beta^*_{\sigma',\vec{\xi},\vec{k}} = \sum_{\vec{k}} |\varphi_{\vec{k}}|^2 \sum_{\vec{\xi}} \beta_{\sigma,\vec{\xi},\vec{k}} \beta^*_{\sigma',\vec{\xi},\vec{k}}$$

$$\to \delta_{\sigma,\sigma'} \sum_{\Sigma,j} \delta_{\Sigma,1} |\beta_{1,1,j}|^2 \sum_{i \in I^\Sigma_\sigma} |\beta_{i,\Sigma,j}|^2 = \delta_{\sigma,\sigma'} \sum_j |\beta_{1,1,j}|^2 \sum_{i \in I^1_\sigma} |\beta_{i,1,j}|^2$$

---

[22] En las evaluaciones del modelo no se encuentra degeneración en los autoestados de $\hat{H}$.

$$\chi_{\sigma,\sigma'} \to \delta_{\sigma,\sigma'} \sum_j |\beta_{1,1,j}|^2 \sum_{i \in I_\sigma^1} |\beta_{i,1,j}|^2 \tag{6.15}$$

El conjunto $I_\sigma^1$ se deduce según la ecuación (5.14), o el algoritmo dado en la Figura 5.1: para $\Sigma = 1$, $\mathcal{N}(1)$ y el conjunto de índices $I^1$ son:

$$\mathcal{N}(1) = \{1000\cdots 0\,, 0100\cdots 0\,, 0010\cdots 0\,, \cdots, 000\cdots 01\}$$
$$I^1 = \{\,1\,,2\,,3\,,4\,,\cdots N+1\,\} \tag{6.16}$$

Así, las listas $\mathcal{N}_\sigma(1)$ y los conjuntos $I_\sigma$ vienen a ser:

$$\mathcal{N}_0(1) = \{\,0100\cdots 0\,, 0010\cdots 0\,, \cdots, 000\cdots 01\} \qquad \mathcal{N}_1(1) = \{1000\cdots 0\,\}$$
$$I_0^1 = \{\,2\,,3\,,4\,,\cdots N+1\,\} \qquad\qquad\qquad\qquad I_1^1 = \{\,1\,\} \tag{6.17}$$

(6.17) muestra que $I_1^1$ tiene un solo elemento mientras que $I_0^1$ tiene $N$ elementos; se definen ahora las probabilidades $p_{ij}^\Sigma$:

$$p_{ij}^\Sigma = |\langle i|\phi_{\Sigma,j}\rangle|^2 = |\beta_{i,\Sigma,j}|^2 \qquad \sum_{i=1}^{N+1} p_{ij}^\Sigma = \sum_{j=1}^{N+1} p_{ij}^\Sigma = 1 \qquad p_j = p_{1j}^1 \tag{6.18}$$

$p_{ij}^\Sigma$ es la probabilidad de que al medir un estado $|\Sigma,i\rangle$ colapse al estado $|\phi_{\Sigma,j}\rangle$, o viceversa; Así, (6.17) y (6.18) en (6.15) da, para $\sigma = 0\,, 1$:

$$\chi_{0,\sigma} = \delta_{0,\sigma}(1 - \chi_{1,1}) \qquad\qquad \chi_{1,\sigma} = \delta_{1,\sigma} \sum_j p_j^{\,2} \tag{6.19}$$

Donde $p_j^{\,2} = |\beta_{1,1,j}|^4$, $\chi_{1,\sigma}$ es el promedio de la probabilidad $\langle p_j \rangle$, $\chi_{\sigma,\sigma'}$ forma una matriz de densidad estacionaria, sin términos de coherencia, de forma matricial es:

$$\bar{\bar{\rho}} = \sum_{\sigma,\sigma'} \overline{\rho_{\sigma,\sigma'}} |\sigma\rangle\langle\sigma'| = \begin{pmatrix} 1-p & 0 \\ 0 & p \end{pmatrix} \qquad p = \sum_j |\beta_{1,1,j}|^4 \tag{6.20}$$

En los términos dependientes del tiempo $\Gamma_{\sigma,\sigma'}(t)$ de (3.26), para $D = \emptyset$, se añaden (6.13), (6.14) y (5.17), de forma análoga a (6.15):

$$\Gamma_{\sigma,\sigma'}(t) = \delta_{\sigma,\sigma'} \sum_{j \neq j'} \beta_{1,1,j}^* \beta_{1,1,j'} \, e^{-i(\omega_{1j} - \omega_{1j'})t} \sum_{i \in I_\sigma^1} \beta_{i,1,j} \beta_{i,1,j'}^* \qquad (6.21)$$

(6.17) en (6.21) resuelve la sumatoria en $i \in I_\sigma^1$ para $\sigma = 1$:

$$\Gamma_{1,\sigma}(t) = \delta_{1,\sigma} \sum_{j \neq j'} |\beta_{1,1,j}|^2 |\beta_{1,1,j'}|^2 \, e^{-i(\omega_{1j} - \omega_{1j'})t} \qquad (6.22)$$

Como $\Gamma_{1,1}(t)$ es invariante ante la permutación de $j, j'$ y la compleja conjugada juntas, entonces se puede escribir como la suma de dos sumatorias con $j' > j$ y $j' < j$, en esta última se permuta $j, j'$ quedando igual a la compleja conjugada de la primera sumatoria, así $\Gamma_{1,1}(t)$ viene a ser la primera sumatoria más su conjugada compleja:

$$\Gamma_{1,\sigma}(t) = 2\delta_{1,\sigma} \sum_{j' > j} p_j \, p_{j'} \cos(\omega_{1j} - \omega_{1j'})t$$

$$\Gamma_{0,\sigma}(t) = -\delta_{0,\sigma} \Gamma_{1,1}(t) \qquad (6.23)$$

$\Gamma_{0,\sigma}(t)$ de (6.23) se deduce por consistencia de (6.23) y (6.19), la traza de $\Gamma_{\sigma,\sigma'}(t)$ debe ser nula, en consecuencia $\Gamma_{0,0}(t) = -\Gamma_{1,1}(t)$, y los términos fuera de la diagonal son evidentemente nulos; así, (6.23) y (6.19) dan la solución exacta del modelo simplificado de dos dimensiones, ecuaciones (3.26), para las condiciones iniciales (6.8) o (6.12), incluso con degeneración, porque los términos $\chi_{\sigma,\sigma'}^D$ se cancelan al sumar $\chi_{\sigma,\sigma'}$ y $\Gamma_{\sigma,\sigma'}(t)$; así, considerando $\bar{\bar{\rho}}$ de (6.20), se puede expresar la matriz de densidad del sistema como:

$$\hat{\rho}(t) = \begin{pmatrix} 1 - p - \Gamma_{11}(t) & 0 \\ 0 & p + \Gamma_{11}(t) \end{pmatrix} \qquad (6.24)$$

El cual devuelve el estado inicial para $t = 0$, pues de (6.22) se tiene $\Gamma_{1,1}(0) = \sum_{j \neq j'} |\beta_{1,1,j}|^2 |\beta_{1,1,j'}|^2 = \sum_{j,j'} |\beta_{1,1,j}|^2 |\beta_{1,1,j'}|^2 - \sum_j |\beta_{1,1,j}|^2 |\beta_{1,1,j}|^2$, se observa que la sumatoria $\sum_{j,j'} |\beta_{1,1,j}|^2 |\beta_{1,1,j'}|^2 = \left(\sum_j |\beta_{1,1,j}|^2\right)^2 = 1$ porque las amplitudes $\beta_{1,1,j}$ están normalizadas, y la otra sumatoria es $\sum_j |\beta_{1,1,j}|^2 |\beta_{1,1,j}|^2 = \sum_j |\beta_{1,1,j}|^4 = p$ de acuerdo a (6.20), así $\Gamma_{1,1}(0) = 1 - p$, de este modo se obtiene $\hat{\rho}(0) = \begin{pmatrix} 0 & 0 \\ 0 & 1 \end{pmatrix}$.

### 6.3.2 Entorno a temperatura nula: $\Sigma = 0, 1$

Para el estado inicial (6.10) se tienen las amplitudes del estado inicial total:

$$|\psi\rangle = \alpha|0000\cdots 0\rangle + \beta|1000\cdots 0\rangle \qquad \hat{\rho}(0) = \begin{pmatrix} |\alpha|^2 & \alpha\beta^* \\ \alpha^*\beta & |\beta|^2 \end{pmatrix} \tag{6.25}$$

$$\psi_0(0) = \alpha \qquad \psi_1(0) = \beta \qquad \forall n > 1: \quad \psi_n(0) = 0$$

$n$ es la forma entera de $\vec{n}$, para $n = 0 \to \Sigma, i = 0,1$, y para $n = 1 \to \Sigma, i = 1,1$; la dinámica del sistema total ocurre en los subespacios $\mathcal{H}(0)$ y $\mathcal{H}(1)$, al reemplazar las amplitudes de (6.25) en (3.20), y usando (5.8) se tiene:

$$\varphi_{\vec{k}} = \sum_{\vec{n}} \psi_{\vec{n}} \beta^*_{\vec{n},\vec{k}} = \alpha \beta^*_{0,k} + \beta \beta^*_{1,k} \to \varphi_{\Sigma,j} = \delta_{\Sigma,0}\alpha + \delta_{\Sigma,1}\beta\beta^*_{1,1,j} \tag{6.26}$$

Donde $\beta^*_{0,k} = \delta_{\Sigma,0}$ puesto que $\beta^*_{1,0,j} = 1$, pues el subespacio $\mathcal{H}(0)$ solo tiene un elemento; el módulo al cuadrado de $\varphi_{\vec{k}}$ viene a ser:

$$\left|\varphi_{\vec{k}}\right|^2 \to \left|\varphi_{\Sigma j}\right|^2 = \delta_{\Sigma,0}|\alpha|^2 + \delta_{\Sigma,1}\left|\beta\beta^*_{11j}\right|^2 \tag{6.27}$$

De (5.22) se tiene que calcular $I_{\sigma,\sigma'}(\Sigma, \Sigma')$ para $\Sigma = 0, 1$:

Para $\sigma, \sigma' = 0,0$:

$$I_{0,0} = \{(0,1; 0,1), (1,2; 1,2), (1,3; 1,3), (2,3; 2,3), (1,4; 1,4), \cdots\}$$

$$I_{0,0}(0,0) = \{(0,1; 0,1)\} \qquad I_{0,0}(0,1) = \{\,\}$$

$$I_{0,0}(1,0) = \{\,\} \qquad I_{0,0}(1,1) = \{(1,2; 1,2), (1,3; 1,3), \cdots (1, \#; 1, \#)\} \tag{6.28}$$

Para $\sigma, \sigma' = 0,1$:

$$I_{0,1} = \{(0,1; 1,1), (1,2; 2,1), (1,3; 2,2), (2,3; 3,1), (1,4; 2,4), \cdots\}$$

$$I_{0,1}(0,0) = \{\,\} \qquad I_{0,1}(0,1) = \{(0,1; 1,1)\} \tag{6.29}$$

$$I_{0,1}(1,0) = \{\ \} \qquad\qquad I_{0,1}(1,1) = \{\ \}$$

Para $\sigma, \sigma' = 1,0$ :

$$I_{1,0} = \{(1,1;0,1), (2,1;1,2), (2,2;1,3), (3,1;2,3), (2,4;1,4), \cdots\}$$

$$I_{1,0}(0,0) = \{\ \} \qquad\qquad I_{1,0}(0,1) = \{\ \}$$
$$I_{1,0}(1,0) = \{(1,1;0,1)\} \qquad\qquad I_{1,0}(1,1) = \{\ \} \tag{6.30}$$

Para $\sigma, \sigma' = 1,1$ :

$$I_{1,1} = \{\,(1,1;1,1), (2,1;2,1), (2,2;2,2)\,, (3,1;3,1)\,, (2,4;2,4)\,, \cdots\}$$

$$I_{1,1}(0,0) = \{\ \} \qquad\qquad I_{1,1}(0,1) = \{\ \}$$
$$I_{1,1}(1,0) = \{\ \} \qquad\qquad I_{1,1}(1,1) = \{(1,1;1,1)\} \tag{6.31}$$

Así, de (6.28) a (6.31) en (5.21), se generan los siguientes términos:

Para $\sigma, \sigma' = 0,0$ solo hay términos no nulos para dos casos: $\Sigma, \Sigma' = 0 \wedge \Sigma, \Sigma' = 1$:

$$\sum_{\vec{\xi}} \beta_{0,\vec{\xi},\vec{k}} \beta^*_{0,\vec{\xi},\vec{k}'} \to \delta_{\Sigma,0}\delta_{\Sigma',0} + \delta_{\Sigma,1}\delta_{\Sigma',1} \sum_{i=2}^{N+1} \beta_{i1j}\beta^*_{i1j'}$$

Donde se ha considerado $\vec{k} = \Sigma, j$ y $\vec{k}' = \Sigma', j'$, además que $\beta_{10j} = 1$; se le agrega las deltas de kronecker para cada caso de $I_{0,0}(\Sigma, \Sigma')$.

Para $\sigma, \sigma' = 0,1$ solo hay un término no nulo, cuando $\Sigma = 0 \wedge \Sigma' = 1$:

$$\sum_{\vec{\xi}} \beta_{0,\vec{\xi},\vec{k}} \beta^*_{1,\vec{\xi},\vec{k}'} \to \delta_{\Sigma,0}\delta_{\Sigma',1}\,\beta^*_{11j'}$$

Para $\sigma, \sigma' = 1,0$ solo hay un término no nulo, cuando $\Sigma = 1 \wedge \Sigma' = 0$:

$$\sum_{\vec{\xi}} \beta_{1,\vec{\xi},\vec{k}} \beta^*_{0,\vec{\xi},\vec{k}'} \to \delta_{\Sigma,1}\delta_{\Sigma',0}\,\beta_{11j}$$

Para $\sigma, \sigma' = 1,1$ solo hay un término no nulo, cuando $\Sigma = 1 \wedge \Sigma' = 1$:

$$\sum_{\vec{\xi}} \beta_{1,\vec{\xi},\vec{k}} \beta^*_{1,\vec{\xi},\vec{k}'} \to \delta_{\Sigma,1} \delta_{\Sigma',1}\, \beta_{11j} \beta^*_{11j'}$$

Estas expresiones se pueden resumir en:

$$\sum_{\vec{\xi}} \beta_{\sigma,\vec{\xi},\vec{k}} \beta^*_{\sigma',\vec{\xi},\vec{k}'} \to \begin{cases} \delta_{\Sigma,0}\delta_{\Sigma',0} + \delta_{\Sigma,1}\delta_{\Sigma',1} \sum_{i=2}^{N+1} \beta_{i1j}\beta^*_{i1j'} & \sigma,\sigma' = 0,0 \\ \delta_{\Sigma,0}\delta_{\Sigma',1}\, \beta^*_{11j'} & \sigma,\sigma' = 0,1 \\ \delta_{\Sigma,1}\delta_{\Sigma',0}\, \beta_{11j} & \sigma,\sigma' = 1,0 \\ \delta_{\Sigma,1}\delta_{\Sigma',1}\, \beta_{11j}\beta^*_{11j'} & \sigma,\sigma' = 1,1 \end{cases} \quad (6.32)$$

Las deltas en $\Sigma, \Sigma'$ se mantienen porque se van a integrar a otras ecuaciones que suman sobre $\vec{k}, \vec{k}'$; entonces, cuando $\vec{k} = \vec{k}'$ ambos pertenecen a un mismo subespacio, se tiene $\Sigma = \Sigma'$, y esto anula los términos para $\sigma' \neq \sigma$, y (6.32) se simplifica a:

$$\sum_{\vec{\xi}} \beta_{\sigma,\vec{\xi},\vec{k}} \beta^*_{\sigma',\vec{\xi},\vec{k}'} = \delta_{\sigma,\sigma'} \left( \delta_{\sigma,0} \left( \delta_{\Sigma,0} + \delta_{\Sigma,1} \sum_{i=2}^{N+1} |\beta_{i1j}|^2 \right) + \delta_{\sigma,1}\delta_{\Sigma,1} |\beta_{11j}|^2 \right)$$

Como se vio en (6.18), $p_j = |\beta_{11j}|^2$ y $\sum_{i=2}^{N+1}|\beta_{i1j}|^2 = 1 - p_j$, entonces:

$$\sum_{\vec{\xi}} \beta_{\sigma,\vec{\xi},\vec{k}} \beta^*_{\sigma',\vec{\xi},\vec{k}'} = \delta_{\sigma,\sigma'} \left( \delta_{\sigma,0} \left( \delta_{\Sigma,0} + \delta_{\Sigma,1}(1 - p_j) \right) + \delta_{\sigma,1}\delta_{\Sigma,1}\, p_j \right) \quad \begin{array}{l} \vec{k} = \Sigma, j \\ \\ \Sigma = 0,1 \end{array} \quad (6.33)$$

Así, (6.27) y (6.33) en $\chi_{\sigma,\sigma'}$ de (3.26), para $D = \emptyset$:

$$\chi_{\sigma,\sigma'} = \sum_{\vec{k}} |\varphi_{\vec{k}}|^2 \sum_{\vec{\xi}} \beta_{\sigma,\vec{\xi},\vec{k}} \beta^*_{\sigma',\vec{\xi},\vec{k}'}$$

$$= \delta_{\sigma,\sigma'} \sum_{\Sigma,j} \left( \delta_{\Sigma,0}|\alpha|^2 + \delta_{\Sigma,1}|\beta \beta^*_{11j}|^2 \right) \left( \delta_{\sigma,0} \left( \delta_{\Sigma,0} + \delta_{\Sigma,1}(1 - p_j) \right) + \delta_{\sigma,1}\delta_{\Sigma,1}\, p_j \right)$$

$$\chi_{\sigma,\sigma'} = \delta_{\sigma,\sigma'}\delta_{\sigma,0}\left(\sum_{\Sigma,j}\delta_{\Sigma,0}|\alpha|^2 + \sum_{\Sigma,j}\delta_{\Sigma,1}|\beta|^2 p_j(1-p_j)\right) + \delta_{\sigma,\sigma'}\delta_{\sigma,1}\sum_{\Sigma,j}\delta_{\Sigma,1}|\beta|^2 p_j^2$$

Se ejecuta las sumatorias, para $\Sigma = 0 \to j = 1$, pero para $\Sigma = 1 \to j = 1:N+1$:

$$\chi_{\sigma,\sigma'} = \delta_{\sigma,\sigma'}\delta_{\sigma,0}\left(|\alpha|^2 + |\beta|^2\left(1 - \sum_j p_j^2\right)\right) + \delta_{\sigma,\sigma'}\delta_{\sigma,1}|\beta|^2\sum_j p_j^2$$

De (6.20): $p = \sum_j |\beta_{1,1,j}|^4 = \sum_j p_j^2$, entonces se puede representar matricialmente:

$$\chi_{\sigma,\sigma'} = \delta_{\sigma,\sigma'}\left[\delta_{\sigma,0}\bigl(|\alpha|^2 + |\beta|^2(1-p)\bigr) + \delta_{\sigma,1}|\beta|^2 p\right]$$

$$\bar{\bar{\rho}} = \begin{pmatrix} |\alpha|^2 + |\beta|^2(1-p) & 0 \\ 0 & |\beta|^2 p \end{pmatrix} \qquad p = \sum_j |\beta_{1,1,j}|^4 \tag{6.34}$$

La matriz de $\chi_{\sigma,\sigma'}$ corresponde con la matriz de densidad $\bar{\bar{\rho}}$ promediada en tiempos muy grandes, la cual es un estado mixto, esto significa que al promediar los términos de interferencia, se pierde la coherencia cuántica, pues al inicio se partió del estado puro (6.25); (6.34) devuelve (6.20) para $\alpha = 0$.

$\Gamma_{\sigma,\sigma'}(t)$ es el término dependiente del tiempo de (3.26), para el primer factor:

$$\sum_{\vec{k}\neq\vec{k}'}\varphi_{\vec{k}}\varphi_{\vec{k}'}^* \to \sum_{(\Sigma,j)\neq(\Sigma',j')}\bigl(\alpha\delta_{\Sigma,0} + \beta\beta_{11j}^*\delta_{\Sigma,1}\bigr)\bigl(\alpha^*\delta_{\Sigma',0} + \beta^*\beta_{11j'}\delta_{\Sigma',1}\bigr)$$

$$\sum_{\vec{k}\neq\vec{k}'}\varphi_{\vec{k}}\varphi_{\vec{k}'}^* = \sum_{(\Sigma,j)\neq(\Sigma',j')}|\alpha|^2\delta_{\Sigma,0}\delta_{\Sigma',0} + \sum_{(\Sigma,j)\neq(\Sigma',j')}\alpha\beta^*\beta_{11j'}\delta_{\Sigma,0}\delta_{\Sigma',1}$$
$$+ \sum_{(\Sigma,j)\neq(\Sigma',j')}\alpha^*\beta\beta_{11j}^*\delta_{\Sigma,1}\delta_{\Sigma',0} + \sum_{(\Sigma,j)\neq(\Sigma',j')}|\beta|^2\beta_{11j}^*\beta_{11j'}\delta_{\Sigma,1}\delta_{\Sigma',1}$$

Entonces, en $\Gamma_{\sigma,\sigma'}(t)$:

$$\Gamma_{\sigma,\sigma'}(t) = \sum_{\vec{k}\neq\vec{k}'}\varphi_{\vec{k}}\varphi_{\vec{k}'}^* e^{-i(\omega_{\vec{k}} - \omega_{\vec{k}'})t}\sum_{\vec{\xi}}\beta_{\sigma,\vec{\xi},\vec{k}}\beta_{\sigma',\vec{\xi},\vec{k}'}^*$$

$$\Gamma_{\sigma,\sigma'}(t) \to \left( \sum_{(\Sigma,j)\neq(\Sigma',j')} |\alpha|^2 \delta_{\Sigma,0}\delta_{\Sigma',0}\, e^{-i(\omega_{\Sigma,j}-\omega_{\Sigma',j'})t} \right.$$

$$+ \sum_{(\Sigma,j)\neq(\Sigma',j')} \alpha\beta^* \beta_{11j'} \delta_{\Sigma,0}\delta_{\Sigma',1}\, e^{-i(\omega_{\Sigma,j}-\omega_{\Sigma',j'})t}$$

$$+ \sum_{(\Sigma,j)\neq(\Sigma',j')} \alpha^*\beta \beta_{11j}^* \delta_{\Sigma,1}\delta_{\Sigma',0}\, e^{-i(\omega_{\Sigma,j}-\omega_{\Sigma',j'})t} \quad (6.35)$$

$$\left. + \sum_{(\Sigma,j)\neq(\Sigma',j')} |\beta|^2 \beta_{11j}^* \beta_{11j'} \delta_{\Sigma,1}\delta_{\Sigma',1}\, e^{-i(\omega_{\Sigma,j}-\omega_{\Sigma',j'})t} \right)$$

$$\cdot \sum_{\vec{\xi}} \beta_{\sigma,\vec{\xi},\vec{k}} \beta_{\sigma',\vec{\xi},\vec{k}'}^*$$

El factor $\sum_{\vec{\xi}} \beta_{\sigma,\vec{\xi},\vec{k}} \beta_{\sigma',\vec{\xi},\vec{k}'}^*$ tiene 4 posibles resultados, los cuales contienen deltas en $\Sigma, \Sigma'$, y se eliminarán con las correspondientes deltas del factor $\sum_{\vec{k}\neq\vec{k}'} \varphi_{\vec{k}} \varphi_{\vec{k}'}^*$ dependiendo cómo se evalúe $\sigma, \sigma'$, así usando (6.32) en (6.35) se tiene:

Para $\sigma, \sigma' = 0,0$:

$$\Gamma_{0,0}(t) \to \sum_{(\Sigma,j)\neq(\Sigma',j')} |\alpha|^2 \delta_{\Sigma,0}\delta_{\Sigma',0}\, e^{-i(\omega_{\Sigma,j}-\omega_{\Sigma',j'})t} \left( \delta_{\Sigma,0}\delta_{\Sigma',0} + \delta_{\Sigma,1}\delta_{\Sigma',1} \sum_{i=2}^{N+1} \beta_{i1j}\beta_{i1j'}^* \right)$$

$$+ \sum_{(\Sigma,j)\neq(\Sigma',j')} \alpha\beta^* \beta_{11j'} \delta_{\Sigma,0}\delta_{\Sigma',1}\, e^{-i(\omega_{\Sigma,j}-\omega_{\Sigma',j'})t} \left( \delta_{\Sigma,0}\delta_{\Sigma',0} + \delta_{\Sigma,1}\delta_{\Sigma',1} \sum_{i=2}^{N+1} \beta_{i1j}\beta_{i1j'}^* \right)$$

$$+ \sum_{(\Sigma,j)\neq(\Sigma',j')} \alpha^*\beta \beta_{11j}^* \delta_{\Sigma,1}\delta_{\Sigma',0}\, e^{-i(\omega_{\Sigma,j}-\omega_{\Sigma',j'})t} \left( \delta_{\Sigma,0}\delta_{\Sigma',0} + \delta_{\Sigma,1}\delta_{\Sigma',1} \sum_{i=2}^{N+1} \beta_{i1j}\beta_{i1j'}^* \right)$$

$$+ \sum_{(\Sigma,j)\neq(\Sigma',j')} |\beta|^2 \beta_{11j}^* \beta_{11j'} \delta_{\Sigma,1}\delta_{\Sigma',1}\, e^{-i(\omega_{\Sigma,j}-\omega_{\Sigma',j'})t} \left( \delta_{\Sigma,0}\delta_{\Sigma',0} \right.$$

$$\left. + \delta_{\Sigma,1}\delta_{\Sigma',1} \sum_{i=2}^{N+1} \beta_{i1j}\beta_{i1j'}^* \right)$$

Multiplicando las deltas entre sí, y el que $(\Sigma, j) \neq (\Sigma', j')$, se eliminan todos los términos excepto el último:

$$\Gamma_{0,0}(t) \to \sum_{(\Sigma,j)\neq(\Sigma',j')} |\beta|^2 \beta_{11j}^* \beta_{11j'} \delta_{\Sigma,1}\delta_{\Sigma',1}\, e^{-i(\omega_{\Sigma,j}-\omega_{\Sigma',j'})t} \sum_{i=2}^{N+1} \beta_{i1j}\beta_{i1j'}^*$$

$$\Gamma_{0,0}(t) = \sum_{j\neq j'} |\beta|^2 \beta_{11j}^* \beta_{11j'}\, e^{-i(\omega_{1,j}-\omega_{1,j'})t} \sum_{i=2}^{N+1} \beta_{i1j}\beta_{i1j'}^*$$

Como $\Gamma_{0,0}(t)$ es hermitiano; esto es, que es invariante ante la permutación de $j$ y $j'$ y la conjugación compleja, entonces se puede expresar como:

$$\Gamma_{0,0}(t) = 2|\beta|^2 \mathbb{Re}\left[\sum_{j'>j} \beta_{11j}^* \beta_{11j'}\, e^{-i(\omega_{1,j}-\omega_{1,j'})t} \sum_{i=2}^{N+1} \beta_{i1j}\beta_{i1j'}^*\right] \qquad (6.36)$$

Para $\sigma,\sigma' = 0,1$, de (6.32) se tiene $\sum_{\vec{\xi}} \beta_{\sigma,\vec{\xi},\vec{k}}\beta_{\sigma',\vec{\xi},\vec{k}'}^* = \delta_{\Sigma,0}\delta_{\Sigma',1}\, \beta_{11j'}^*$, solo sobrevive el segundo término en (6.35), debido al producto de sus deltas:

$$\Gamma_{0,1}(t) = \alpha\beta^* \sum_{(\Sigma,j)\neq(\Sigma',j')} \beta_{11j'}\delta_{\Sigma,0}\delta_{\Sigma',1} e^{-i(\omega_{\Sigma,j}-\omega_{\Sigma',j'})t}\, \beta_{11j'}^*$$

Al ejecutar las deltas, $\Sigma = 0 \to j = 1$, y considerando que[23] $\omega_{0,1} = 0$, se reduce a:

$$\Gamma_{0,1}(t) = \alpha\beta^* \sum_{j'} p_{j'}\, e^{i\omega_{1,j'}t} \qquad\qquad p_{j'} = |\beta_{11j'}|^2 \qquad (6.37)$$

Para $\sigma,\sigma' = 1,0$ se tiene $\sum_{\vec{\xi}} \beta_{\sigma,\vec{\xi},\vec{k}}\beta_{\sigma',\vec{\xi},\vec{k}'}^* = \delta_{\Sigma,1}\delta_{\Sigma',0}\, \beta_{11j}$, el producto de las deltas solo permite la supervivencia del tercer término de (6.35):

$$\Gamma_{1,0}(t) = \alpha^*\beta \sum_{(\Sigma,j)\neq(\Sigma',j')} \beta_{11j}^*\delta_{\Sigma,1}\delta_{\Sigma',0}\, e^{-i(\omega_{\Sigma,j}-\omega_{\Sigma',j'})t}\, \beta_{11j}$$

Al ejecutar las deltas, $\Sigma' = 0 \to j' = 1$, y considerando que $\omega_{0,1} = 0$, se reduce a:

$$\Gamma_{1,0}(t) = \alpha^*\beta \sum_{j} p_j\, e^{-i\omega_{1,j}t} \qquad\qquad p_j = |\beta_{11j}|^2 \qquad (6.38)$$

---

[23] Esto debido al Hamiltoniano más general (3.1) y (3.3) que establece autovalor nulo $\omega_{0,1} = 0$ para el estado fundamental del sistema total $|0\rangle$.

Evidentemente, $\Gamma_{1,0}(t) = \Gamma^*_{0,1}(t)$.

Para $\sigma, \sigma' = 1,1$ se tiene $\sum_{\vec{\xi}} \beta_{\sigma,\vec{\xi},\vec{k}} \beta^*_{\sigma',\vec{\xi},\vec{k}'} = \delta_{\Sigma,1} \delta_{\Sigma',1} \beta_{11j} \beta^*_{11j'}$, el producto de las deltas solo permite la supervivencia del cuarto término de (6.35):

$$\Gamma_{1,1}(t) = |\beta|^2 \sum_{j \neq j'} |\beta_{11j'}|^2 |\beta_{11j}|^2 e^{-i(\omega_{1,j}-\omega_{1,j'})t}$$

Como el sumando $|\beta_{11j'}|^2 |\beta_{11j}|^2 e^{-i(\omega_{1,j}-\omega_{1,j'})t}$ es hermitiano (invariante frente a la permutación de $j$ y $j'$, y la conjugada compleja) se lo puede expresar como:

$$\Gamma_{1,1}(t) = 2|\beta|^2 \sum_{j'>j} p_{j'} p_j \cos(\omega_{1,j'} - \omega_{1,j})t \qquad p_j = |\beta_{11j}|^2 \qquad (6.39)$$

Como la traza de $\hat{\rho}(t)$ es invariante en el tiempo, se hace: $\Gamma_{0,0}(t) = -\Gamma_{1,1}(t)$, y considerando $\Gamma_{0,1}(t) = \Gamma^*_{1,0}(t)$, entonces considerando (6.34) la matriz de densidad total dependiente del tiempo $\rho_{\sigma,\sigma'}(t) = \chi_{\sigma,\sigma'} + \Gamma_{\sigma,\sigma'}(t)$ de (3.26), viene a ser:

$$\hat{\rho}(t) = \begin{pmatrix} |\alpha|^2 + |\beta|^2(1-p) - \Gamma_{1,1}(t) & \Gamma^*_{1,0}(t) \\ \Gamma_{1,0}(t) & |\beta|^2 p + \Gamma_{1,1}(t) \end{pmatrix} \qquad p = \sum_j |\beta_{1,1,j}|^4 \quad (6.40)$$

O, se puede expresar de la siguiente forma compacta:

$$\hat{\rho}(t) = \begin{pmatrix} 1 - |\beta|^2 |\Gamma(t)|^2 & \alpha \beta^* \Gamma^*(t) \\ \alpha^* \beta \Gamma(t) & |\beta|^2 |\Gamma(t)|^2 \end{pmatrix} \qquad \Gamma(t) = \sum_{j=1}^{N+1} |\beta_{1,1,j}|^2 e^{-i\omega_{1,j}t} \quad (6.41)$$

La forma de $\rho_{11}(t)$ de (6.41) se obtiene al considerar que $|\beta|^2 p + \Gamma_{1,1}(t)$ es $|\beta|^2 \sum_{j,j'} |\beta_{11j'}|^2 |\beta_{11j}|^2 e^{-i(\omega_{1,j}-\omega_{1,j'})t}$, donde la sumatoria es el producto de $\Gamma(t)$ por su conjugada compleja. La matriz $\hat{\rho}(t)$ queda determinada al calcular $\Gamma(t)$, la cual es la unidad al inicio del tiempo $\Gamma(0) = \sum_{j=1}^{N+1} |\beta_{1,1,j}|^2 = 1$ porque las amplitudes $\beta_{1,1,j}$ están normalizadas, entonces (6.41) retorna el estado inicial (6.25) para $t = 0$. De la ecuación (6.41) se deriva (6.24) para $\alpha = 0$ (lo cual implica $\beta = 1$). Al promediar $\Gamma(t)$ en tiempos largos evidentemente se anula, así los términos de coherencia $\rho_{1,0}(t)$ y $\rho_{0,1}(t)$ son valores complejos oscilantes con promedio nulo. Las probabilidades del estado excitado y

fundamental $\rho_{1,1}(t)$ y $\rho_{0,0}(t)$ respectivamente, son siempre reales pero se promedian en tiempos largos a $p$ y $1-p$ respectivamente. La ecuación (6.40) o (6.41) es la solución exacta y general, incluso ante la presencia de degeneración, pues los términos $\chi_{\sigma,\sigma'}^D$ se cancelan al sumarse $\chi_{\sigma,\sigma'}$ y $\Gamma_{\sigma,\sigma'}(t)$ en (6.40).

### 6.4 Características generales de las soluciones exactas

#### 6.4.1 La probabilidad $p_0$ de encontrarse excitado en el equilibrio

Las soluciones exactas obtenidas en la sección [6.3] dependen de las probabilidades establecidas en (6.18) $p_j = |\langle 1,1|\phi_{1,j}\rangle|^2$, donde $|1,1\rangle$ es el estado $|1000\cdots 0\rangle$ usando la notación de la sección [5.1.3] y la ecuación (5.6), $|\Sigma, i\rangle = |1,1\rangle = |1000\cdots 0\rangle$, en esta notación $\Sigma = 1$ indica que el ket de estado pertenece a $\mathcal{H}(1)$, $i = 1$ indica que el ket es el primer vector de la base de $\mathcal{H}(1)$. El ket $|1,1\rangle$ es el estado inicial (6.12); en consecuencia, $p_j$ es la probabilidad de que el estado inicial $|\psi(0)\rangle$ se encuentre en el $j$-ésimo modo de oscilación colectiva $|\phi_{1,j}\rangle$.

La probabilidad que en el equilibrio el sistema se encuentre nuevamente en el estado excitado $p = \sum_j |\beta_{1,1,j}|^4$ de (6.34), puede ser deducido desde el enfoque de la base de modos normales, sección [3.1.2], partiendo del estado inicial $|1,1\rangle$, la probabilidad de ser encontrado en el estado $|\phi_{1,j}\rangle$ es $p_j = |\beta_{1,1,j}|^2$, este estado permanece inalterado en el tiempo, así en el equilibrio solo se suprimen los términos de coherencia entre los modos normales no degenerados, ecuación (3.20), y asumiendo $D = \emptyset$, la probabilidad de medir al sistema nuevamente en el estado inicial se puede plantear como la suma de todas las probabilidades de que el estado inicial $|\psi(0)\rangle$ se encuentre en el $j$-ésimo modo normal $|\phi_{1,j}\rangle$ y que a su vez este nuevamente se encuentre en el estado inicial[24]:

$$p_0 = \sum_{j=1}^{\#} p_{|\psi(0)\rangle \to |\phi_{1,j}\rangle} \cdot p_{|\phi_{1,j}\rangle \to |\psi(0)\rangle} \qquad p_{|\psi(0)\rangle \to |\phi_{1,j}\rangle} = |\langle\psi(0)|\phi_{1,j}\rangle|^2 \qquad (6.42)$$

---

[24] Este planteamiento "clásico" de las probabilidades es posible porque los términos de coherencia son suprimidos debido al promedio temporal, descrito en (3.20).

Así, para $|\psi(0)\rangle = |1,1\rangle$ se tiene $p_0 = p$, y para $|\psi(0)\rangle = \alpha|0,1\rangle + \beta|1,1\rangle$ se tiene $p_0 = |\beta|^2 p$; (6.42) se puede generalizar a cualquier estado inicial $|\psi_0\rangle$ y final $|\psi'\rangle$:

$$p_0[\psi_0 \to \psi'] = \sum_{j=1}^{\#} p_{|\psi_0\rangle \to |\phi_{1,j}\rangle} \cdot p_{|\phi_{1,j}\rangle \to |\psi'\rangle} \qquad p_{|\psi\rangle \to |\phi_{1,j}\rangle} = |\langle \psi | \phi_{1,j} \rangle|^2 \qquad (6.43)$$

La probabilidad de que, iniciando con sistema excitado, el oscilador $k$ se encuentre excitado en el equilibrio, corresponde a elegir $|\psi_0\rangle = |1,1\rangle$ y $|\psi'\rangle = |1, k+1\rangle$, en (6.43):

$$p_0[1 \to k] = \sum_{j=1}^{N+1} p_j |\beta_{k+1,1,j}|^2 \qquad \sum_{k=0}^{N} p_0[1 \to k] = 1 \qquad (6.44)$$

La probabilidad $p_0[1 \to k]$ corresponde a sumar el producto de las probabilidades correspondientes a la primera y $k$-ésima fila de $\widehat{U}_1$, esta probabilidad corresponde iniciar con un oscilador con energía $\omega_0$ y encontrar otro oscilador con energía $\omega_k$.

Si las probabilidades $p_j$ no tienden a una distribución equiprobable, y por el contrario tiene un pico pronunciado: $\exists a \,/\, 1 > p_a \gg 1/N$, entonces el estado inicial $|\psi(0)\rangle$ tendría una importante componente en un modo de oscilación $|\phi_{1,a}\rangle$, y al realizar el promedio $p = \sum_{j \neq a} p_j^2 + \sum_a p_a^2$, los términos $\sum_a p_a^2$ son predominantes y comparables con la unidad; esto es, un relevante efecto memoria.

### 6.4.2 Evolución en tiempos cortos sin caída exponencial

Los términos de la matriz de densidad (6.41) dependen temporalmente de $\Gamma(t)$, que es una sumatoria y puede aproximarse a una integral en el límite continuo cuando $N \gg 1$; sea la función $\Gamma'(t)$ una generalización de $\Gamma(t)$ donde en vez de las probabilidades $p_j = |\beta_{1,1,j}|^2$ se tiene otra función discreta $q_k$ en general compleja, $k = 1:N+1$, que modula $e^{-i\omega_k t}$; se puede expresar $q_k = q(\omega_{1,k})$, así en el límite continuo $N \to \infty$, $\Gamma'(t)$ se convierte en una transformada inversa de Fourier de $q(\omega)$:

$$\Gamma'(t) = \sum_k q_k e^{-i\omega_k t} \to \tilde{q}(t) = \int q(\omega) e^{-i\omega t} d\omega \qquad \sum_k q_k = 1 \qquad (6.45)$$

Si se desea que conforme $N \to \infty$ la función $\Gamma(t)$ tienda a una caída exponencial en el tiempo, con un parámetro $\lambda$ no negativo, entonces $\tilde{q}(t)$ debe ser definida como sigue:

$$\tilde{q}(t) = e^{-\lambda t} u(t) \qquad \lambda > 0 \qquad u(t) = \begin{cases} 1 & t > 0 \\ 0 & t < 0 \end{cases} \qquad (6.46)$$

Basta con aplicar la transformada de Fourier a $\tilde{q}(t)$ para obtener la función $q(\omega_k) = q_k$ que lo pueda generar mediante la sumatoria (6.45):

$$\mathcal{F}\{\tilde{q}(t)\} = \frac{1}{2\pi} \int_{-\infty}^{\infty} dt\, e^{i\omega t}\, \tilde{q}(t) = \int q(\omega') \left[\frac{1}{2\pi} \int_{-\infty}^{\infty} dt\, e^{i(\omega-\omega')t}\right] d\omega' = q(\omega)$$

Esto es: $\mathcal{F}\{\tilde{q}(t)\} = q(\omega)$, al usar $\tilde{q}(t)$ de (6.45) se obtiene:

$$q(\omega) = \frac{1}{2\pi} \frac{1}{(\lambda - i\omega)} = \frac{1}{2\pi(\lambda^2 + \omega^2)}[\lambda + i\omega] \qquad (6.47)$$

(6.47) establece que $q_k = q(\omega_k)$ debe ser compleja y no solo real[25] para que la sumatoria (6.45) genere una función de caída exponencial en el tiempo, de manera que no existe ninguna distribución de probabilidad $p_j$ que permita a $\Gamma(t)$ ser una función de tipo caída exponencial como la dada en (6.46). Este resultado es muy importante, pues en la solución (2.47) se tiene una caída exponencial para la dinámica markoviana en el tiempo tanto para la probabilidad el estado excitado como para la amplitud de la coherencia cuántica.

Adicionalmente, $\Gamma(-t) = \Gamma^*(t)$, de manera que en general $\Gamma(t)$ no es simétrica en el tiempo; esto es, invariante ante una inversión temporal, sin embargo $|\Gamma(t)|^2$ sí lo es, de manera que la derivada temporal de $|\Gamma(t)|^2$ evaluado en $t = 0$ se anula[26], en efecto de (6.39) se tiene que $\dot{\Gamma}_{1,1}(0) = 0$, lo que conduce a $\dot{\rho}_{1,1}(0) = 0$, mientras que $\dot{\Gamma}(0) = -i\langle \omega_{1,j} \rangle$, y de acuerdo a (6.48) es $\dot{\Gamma}(0) = -i\omega_0$, lo cual corresponde a $\dot{\rho}_{1,0}(0) = -i\alpha^*\beta\omega_0$; es decir, al inicio $t = 0$ los elementos $\rho_{1,1}$ y $\rho_{1,0}$ de la matriz de densidad (6.41) tienen pendientes nula y $-i\alpha^*\beta\omega_0$ respectivamente, mientras que en la solución markoviana (2.47) los respectivos elementos de la matriz de densidad son proporcionales a la densidad espectral evaluada en la frecuencia del sistema $J(\omega_0)$; de este modo, se evidencia que las soluciones de este modelo simplificado no coinciden con la correspondiente solución markoviana.

---

[25] De hecho, la parte real decae más rápido que la parte imaginaria, como función de la frecuencia $\omega$.
[26] Porque si la pendiente en $t = 0$ fuera no nula, entonces $|\Gamma(t)|^2$ se comportaría distinto hacia adelante o atrás en el tiempo, dejando de ser simétrica.

### 6.4.3 Distribución de probabilidad en el espectro de las frecuencias colectivas

Toda la información del operador de densidad a un tiempo $t$ se basa en la función $\Gamma(t)$ de (6.41), que se compone de las probabilidades $p_j = |\beta_{1,1,j}|^2$ y frecuencias colectivas $\omega_{1,j}$, obtenidas de los autovectores y autovalores del hamiltoniano $\widehat{H}^1$; es posible expresar las probabilidades como funciones de las frecuencias $p_j = p(\omega_{1,j})$, entonces $\Gamma(-t)$ viene a ser la transformada discreta de Fourier de $p(\omega_{1,j})$, es posible también extraer suficiente información del hamiltoniano $\widehat{H}^1$ para determinar los principales parámetros de la distribución de probabilidad $p(\omega_{1,j})$; así por ejemplo, en la base de los vectores $|1,i\rangle$ se tienen las componentes del Hamiltoniano como $(\widehat{H}^1)_{ij} = \langle 1,i|\widehat{H}^1|1,j\rangle$, así de acuerdo a (6.5) o (6.7) y (6.11) se tiene:

$$\langle 1,i|\widehat{H}^1|1,j\rangle = \omega_0 \qquad \langle\omega\rangle = \sum_j |\beta_{1,1,j}|^2 \omega_{1,j} = \omega_0 \qquad (6.48)$$

El lado derecho de (6.48) resulta de considerar $\widehat{H}^1 = \sum_{j=1}^{\#} |\phi_{1,j}\rangle\langle\phi_{1,j}| \omega_{1,j}$ de acuerdo a (5.10), entonces $\langle\omega\rangle$ es el promedio de las frecuencias colectivas en esta distribución de probabilidad $p(\omega_{1,j}) = |\beta_{1,1,j}|^2$; en general se tiene:

$$\langle\omega^n\rangle = \sum_j |\beta_{1,1,j}|^2 \omega_{1,j}^n = (\widehat{H}^{1^n})_{1,1} \qquad \langle\omega^n\rangle = \int_{-\infty}^{\infty} p(\omega)\omega^n \, d\omega \qquad (6.49)$$

El lado izquierdo de (6.49) se obtiene al tomar el braket en el estado $|1,i\rangle$ de la potencia $n$ del hamiltoniano $\widehat{H}^{1^n} = \sum_{j=1}^{\#} |\phi_{1,j}\rangle\langle\phi_{1,j}| \omega_{1,j}^n$, el lado derecho es la forma continua del lado izquierdo, y se puede considerar el límite cuando $N \to \infty$; para el hamiltoniano proveniente de (6.11) y (6.5) se tiene calculado en la ecuación (A.9) del Anexo A los elementos $(\widehat{H}^{1^n})_{1,1}$ y por consiguiente los valores medios $\langle\omega^n\rangle$; así por ejemplo considerando (A.10), (A.11) y (A.12) se tiene: $\langle\omega^2\rangle = \omega_0^2 + |g|^2$, entonces considerando (6.48), la variancia de la distribución de probabilidad es:

$$\sigma^2 = \langle \omega^2 \rangle - \langle \omega \rangle^2 = \sum_{i=1}^{N} |g_i|^2 \qquad (6.50)$$

Así, (6.48) y (6.50) muestra que $p(\omega_{1,j}) = |\beta_{1,1,j}|^2$ es una distribución de probabilidad centrada en la frecuencia del sistema $\omega_0$ y con varianza igual a la suma de los cuadrados de los módulos de los acoples; (6.49) y (A.9) permite caracterizar la distribución al definir el promedio de las potencias de las frecuencias colectivas.

Para la densidad espectral (6.1) se tiene en la Tabla A.2 los valores de las componentes 1-1 de las 6 primeras potencias del hamiltoniano, de acuerdo a (6.49) son los promedios $\langle \omega^n \rangle$ que caracterizan la distribución de probabilidad, y que se muestran claramente dependientes del orden óhmico de la distribución espectral, así la varianza (6.50) viene a ser $\sigma^2 = \eta\, \omega_0^2\, s!$, lo cual es proporcional a la fuerza de acoplamiento, la frecuencia del sistema y el factorial del orden óhmico; se evalúa la tabla A.2 para tres valores correspondientes a espectro sub óhmico, óhmico y súper óhmico:

$$\langle \omega^n \rangle = \left(\widehat{H}^{1^n}\right)_{1,1}$$

| $n$ | $s = 0.5$ | $s = 1$ | $s = 3$ |
|---|---|---|---|
| 1 | $\omega_0$ | $\omega_0$ | $\omega_0$ |
| 2 | $\omega_0^2\left(1 + \frac{\sqrt{\pi}}{2}\eta\right)$ | $\omega_0^2(1 + \eta)$ | $\omega_0^2(1 + 6\eta)$ |
| 3 | $\omega_0^3\left(1 + \frac{7\sqrt{\pi}}{4}\eta\right)$ | $\omega_0^3(1 + 4\eta)$ | $\omega_0^3(1 + 36\eta)$ |
| 4 | $\omega_0^4\left(1 + \frac{39\sqrt{\pi}}{8}\eta + \frac{\pi}{4}\eta^2\right)$ | $\omega_0^4(1 + 13\eta + \eta^2)$ | $\omega_0^4(1 + 186\eta + 36\eta^2)$ |
| 5 | $\omega_0^5\left(1 + \frac{233\sqrt{\pi}}{16}\eta + \frac{9\pi}{8}\eta^2\right)$ | $\omega_0^5(1 + 46\eta + 7\eta^2)$ | $\omega_0^5(1 + 1056\eta + 369\eta^2)$ |
| 6 | $\omega_0^6\left(1 + \frac{1511\sqrt{\pi}}{32}\eta + \frac{129\pi}{16}\eta^2 + \frac{\sqrt{\pi}^3}{8}\eta^3\right)$ | $\omega_0^6(1 + 199\eta + 46\eta^2 + \eta^3)$ | $\omega_0^6(1 + 6966\eta + 4536\eta^2 + 216\eta^3)$ |

Tabla 6.1

Es posible emplear los promedios $\langle \omega^n \rangle$ para expresar la función $\Gamma(t)$ de (6.41), en base a la distribución $p(\omega_{1,j}) = |\beta_{1,1,j}|^2$ y (A.13) como una aproximación a primer orden en la fuerza de acoplamiento:

$$\Gamma(t) = \langle e^{-i\omega t} \rangle = \sum_{n=0}^{\infty} \frac{\langle \omega^n \rangle (-it)^n}{n!} = \sum_{n=0}^{\infty} \frac{(-it)^n}{n!} \langle \omega^n \rangle$$

$$\Gamma(t) \sim e^{-i\omega_0 t} + \eta \sum_{n=2}^{\infty} \frac{(-i\omega_0 t)^n}{n!} \left( \sum_{i=1}^{n-1} (n-i) \gamma_{s+i} \right) + \mathcal{O}(\eta^2) \quad (6.53)$$

Es evidente que para $\eta \to 0$ entonces $\Gamma(t) \to e^{-i\omega_0 t}$, lo cual corresponde al sistema sin interacción con el entorno, de manera que la modificación de su evolución libre es perturbativa en la fuerza de acoplamiento débil; (6.53) puede dar una aproximación para tiempos breves, pero se necesitaría considerar a órdenes mayores en $\eta$ para una correcta descripción de su evolución a tiempos más largos, lo cual sumado a la dificultad en el cálculo numérico de sumar potencias con los signos alternados (acumulan el error considerablemente) hace un modo inestable y limitado su cómputo para tiempos mayores.

La opción razonable para obtener $\Gamma(t)$ y $p = \langle p_j \rangle$ es determinar la distribución de probabilidad misma $p_j = p(\omega_{1,j}) = |\beta_{1,1,j}|^2$, en el Anexo B se logra la obtención de los autovalores (frecuencias colectivas $\omega_{1,j}$) y las amplitudes $\beta_{i,1,j} = \beta_{i,j}$ mediante funciones recursivas; no obstante, las probabilidades $p_j = |\beta_{1,j}|^2$ quedan bien definidas en la ecuación (B.9), donde $p_j = p(\omega_{1,j})$ toma valores máximos cuando $\omega_{1,j}$ se aproxima a $\omega_0$ y decae conforme se aleja de este, cumpliendo con los valores medios de potencias superiores $\langle \omega^n \rangle$ según (A.9); la ecuación (B.9) no solo permite el cálculo de $\Gamma(t)$ como una suma de ondas planas, sino también de $p = \langle p_j \rangle$, la probabilidad que el sistema se encuentre excitado en el equilibrio, lo que caracteriza su efecto memoria.

# CAPÍTULO VII

# EVALUACIÓN Y RESULTADOS

## 7.1 Características generales

Como se refirió en el algoritmo 4, la matriz $G$ está compuesta de los elementos $g_{ij}$, tales que cumplen (3.25), así $G$ es una matriz hermitiana. En (6.11) se establece que el hamiltoniano para $\Sigma = 1$ es $\widehat{H}^1 = G^T$. En esta evaluación se trabaja por simplicidad con acoples reales, de manera que los $\beta_{i1j}$ son reales también, y basta con diagonalizar la matriz $G$ para encontrar los autoestados $|\phi_{1j}\rangle$; para la evaluación en acoplamiento débil se elige desde $\eta = 0.001$ hasta $\eta = 0.3$, ver referencia [2]; se escoge una escala de tiempo donde $\omega_0 = 1$; se elige un entorno con $N = 1000$ osciladores, entonces las matrices $G$ y $\widehat{H}^1$ son de 1001 filas y columnas. Sólo para el cálculo de la entropía de Von Neumann se elige $\alpha = \beta = 1/\sqrt{2}$ en (6.41). Estos datos son resumidos en la tabla 7.1:

| Cantidad | Valor o característica | | |
|---|---|---|---|
| Número de osciladores del entorno | $N = 1000$ | | |
| Acoples entre osciladores | $g_{ij} \in \mathbb{R}$ | | |
| Amplitudes de los autoestados | $\beta_{i1j} \in \mathbb{R}$ | | |
| Frecuencia del sistema | $\omega_0 = 1$ | | |
| Fuerza de acoplamiento débil | $\eta = 0.001 \rightarrow 0.3$ | | |
| Orden de la densidad espectral | $s = 0.5$ | $s = 1$ | $s = 2$ |
| Frecuencias máximas y mínimas | $\omega_{min} = 0.05$ | $\omega_{max} = 5\,;7$ | |
| Amplitudes del estado inicial (sólo para el cálculo de la entropía) | $\alpha = \beta = 1/\sqrt{2}$ | | |

**Tabla 7.1**

En la tabla 7.1 el orden de la densidad espectral $s$ se evalúa para los tres casos: sub óhmico, óhmico y súper óhmico; tras considerar las informaciones de la tabla 7.1 se evalúan $p_j = p(\omega_{1,j}) = |\beta_{1,1,j}|^2$ y las funciones $\Gamma(t)$ y $|\Gamma(t)|^2$ dadas en (6.41).

Evaluando $\omega = \omega_0 = 1$ en $J(\omega)$ de (6.1), y considerando que $\omega_c = \omega_0$ [2], se tiene:

$$J(\omega_0) = \frac{2\pi\omega_0}{e}\eta \sim 2.311454\eta \tag{7.1}$$

Que se emplea en la ecuación (2.47) de la dinámica markoviana para comparar con la evolución del modelo simplificado.

### 7.2 Acople débil del sistema y el entorno: $\eta = 0.001$

Se tienen las siguientes figuras sobre la evolución del sistema; se ha encontrado que el cálculo numérico es más estable cuando menos aleatoriedad hay en la separación entre las frecuencias $\omega_i$, correspondiente a los osciladores del entorno, de manera que en (6.4) el parámetro $\epsilon_j$ adquiere un valor constante para esta evaluación; cada figura corresponde a un conjunto de gráficos sobre la distribución de probabilidades $p_j$ y la evolución temporal de los elementos de la matriz de densidad del sistema.

Para $\eta = 0.001$ y $s = 1$ se tiene la siguiente densidad espectral:

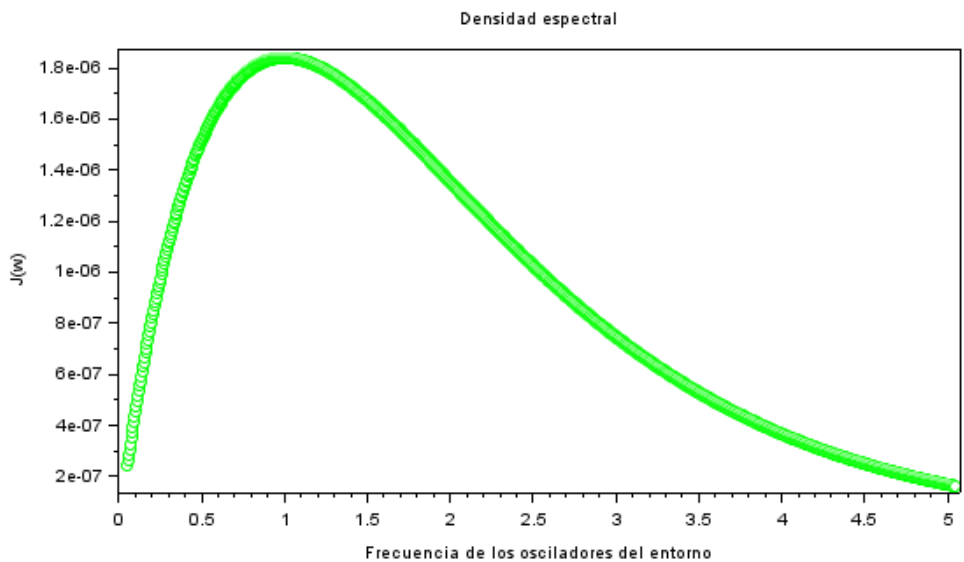

**Figura 7.1** Densidad espectral para $\eta = 0.001$ y $s = 1$

A continuación, los gráficos para esta evaluación:

a1) 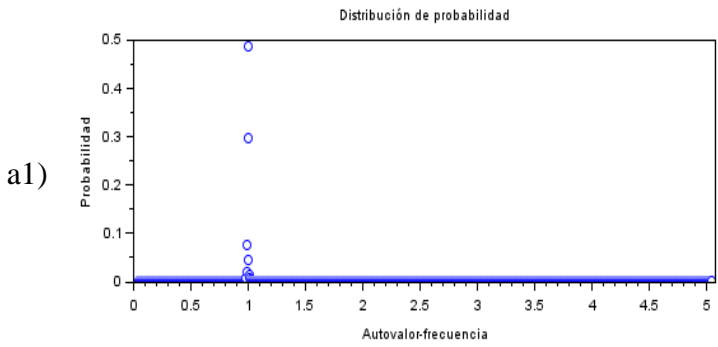

a2) 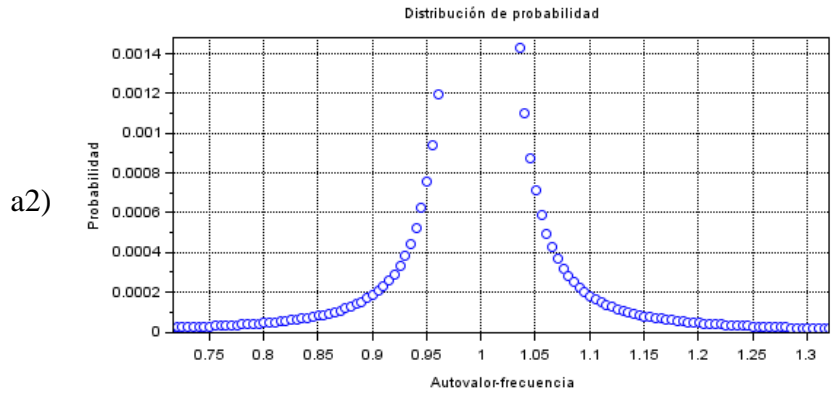

b) 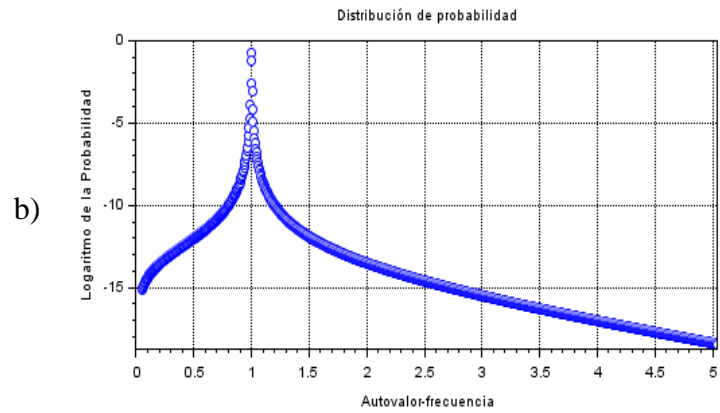

c1) 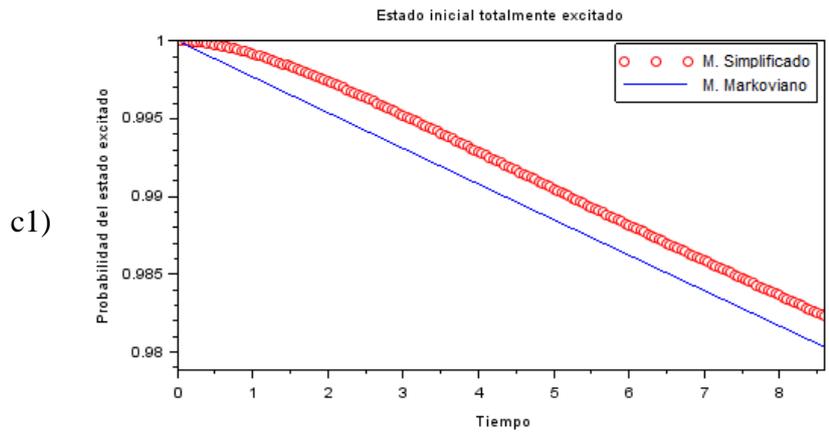

c2) 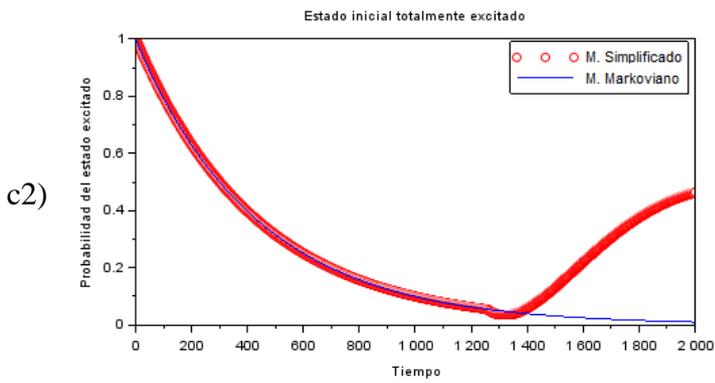

c3) 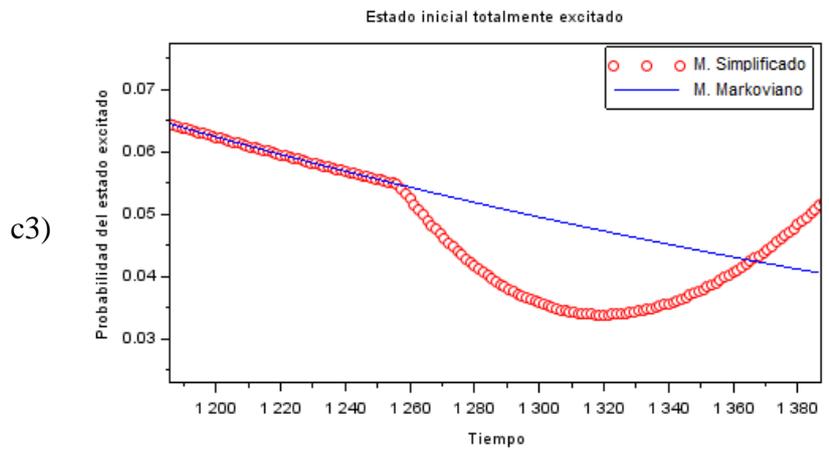

d) 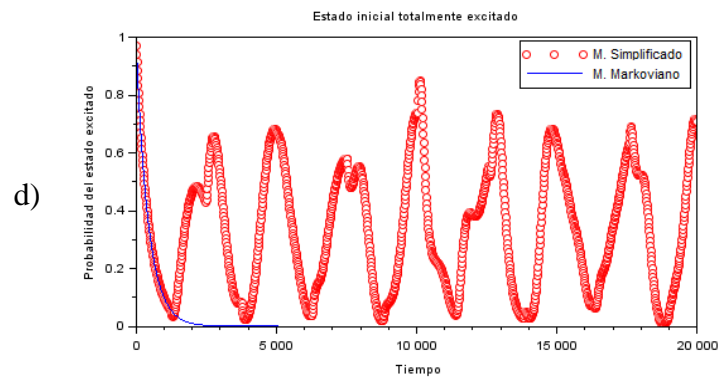

e) 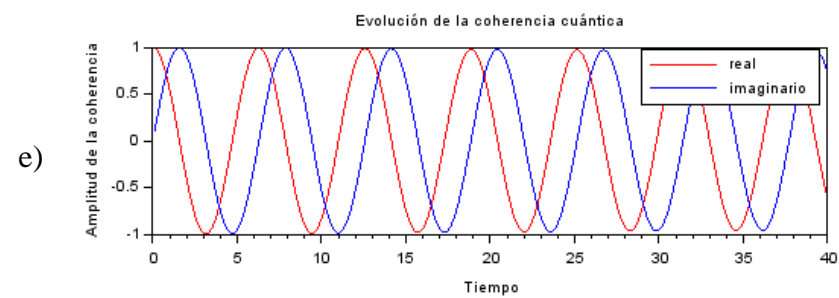

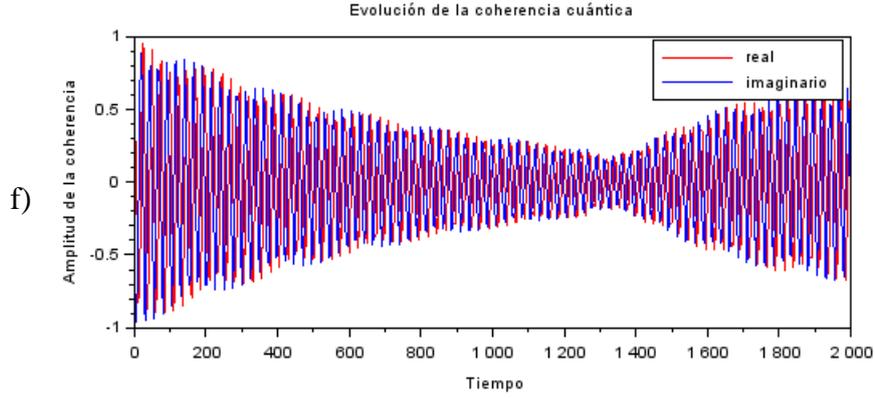

f)

**Figura 7.2** Evaluación para $\eta = 0.001$ y $s = 1$

En a1) de la Figura 7.2 se tiene la distribución de las probabilidades $p_j = p(\omega_{1,j}) = |\beta_{1,1,j}|^2$ como función de los autovalores $\omega_{1,j}$, que se asemeja a la distribución delta de Dirac, en a2) se hace un zoom donde se aprecia mejor su comportamiento alrededor de $\omega_0$, en b) se grafica el logaritmo natural de la probabilidad donde se deduce mejor el comportamiento general de $p_j$; en c1) se grafica $|\Gamma(t)|^2$ para tiempos breves donde se evidencia que tiene pendiente nula al inicio, distinto a la solución markoviana, como se discutió en la sección [6.4.2], en c2) y d) se grafica $|\Gamma(t)|^2$ para escalas de tiempos mayores, en este último se exhiben fuertes "sobrevivencias" [12] de la probabilidad, asociado a un fuerte efecto memoria del sistema general, en c3) se hace un zoom en la región donde $\Gamma(t)$ se aleja notoriamente de la solución markoviana (la cual cambia de color, de azul a gris; en c1), c2), c3) y d) se considera $\alpha = 0$ y $\beta = 1$ en (6.41), de manera que las gráficas corresponden a la evolución de $\rho_{11}(t)$, la probabilidad del estado excitado. En e) y f) se grafican la función $\Gamma(t)$ para distintas escalas de tiempo, en esta última también se evidencia una fuerte sobrevivencia de la coherencia cuántica.

En las gráficas c1), c2) y d) la curva roja representa la función $|\Gamma(t)|^2$ que es la evolución de la probabilidad de un sistema que al inicio se encuentra totalmente excitado, mientras que la curva azul corresponde a la correspondiente solución markoviana, dada en la ecuación (2.47); en las gráficas e) y f) la función $\Gamma(t)$ corresponde a la evolución del término de coherencia $\Gamma_{10}(t)/\alpha^*\beta$ de (6.40) o $\rho_{10}(t)/\alpha^*\beta$ de (6.41).

Se obtiene el valor de $p = \chi_{1,1} = \langle p_j \rangle$, como se discutió en la sección [6.4.1]:

$$p = \langle p_j \rangle = 0.3363172 \qquad (7.2)$$

La evolución de la entropía de Von Neumann de la matriz de densidad (6.41), con $\alpha = \beta = 1/\sqrt{2}$, se realiza mediante el Anexo C, y se presenta a continuación, en comparación con la probabilidad del estado excitado:

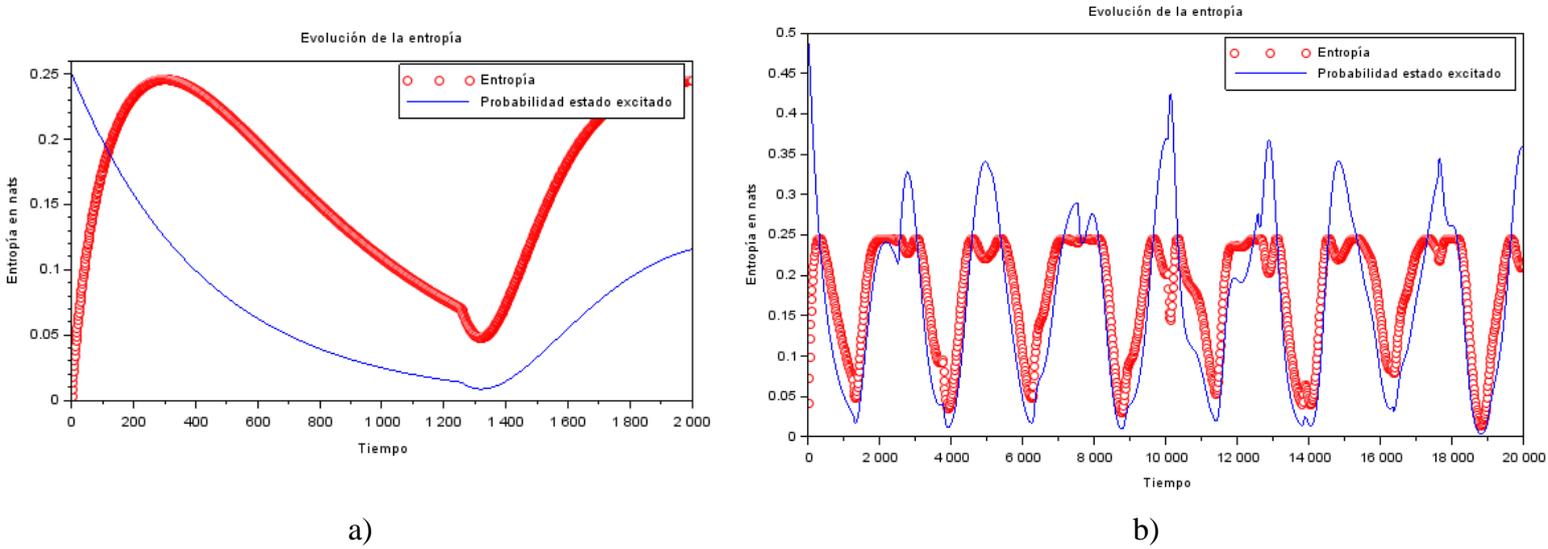

a)　　　　　　　　　　　　　　　　　　　　b)

**Figura 7.3** Evolución de la entropía de Von Neumann del sistema

En la Figura 7.3 se grafica la entropía en nats en dos escalas de tiempo, mientras que la probabilidad del estado excitado está escalada por un factor 1/4 en a) y 1/2 en b).

A continuación, los resultados de las evaluaciones para $s = 0.5\,,1\,,2$ y $\eta = 0.001$:

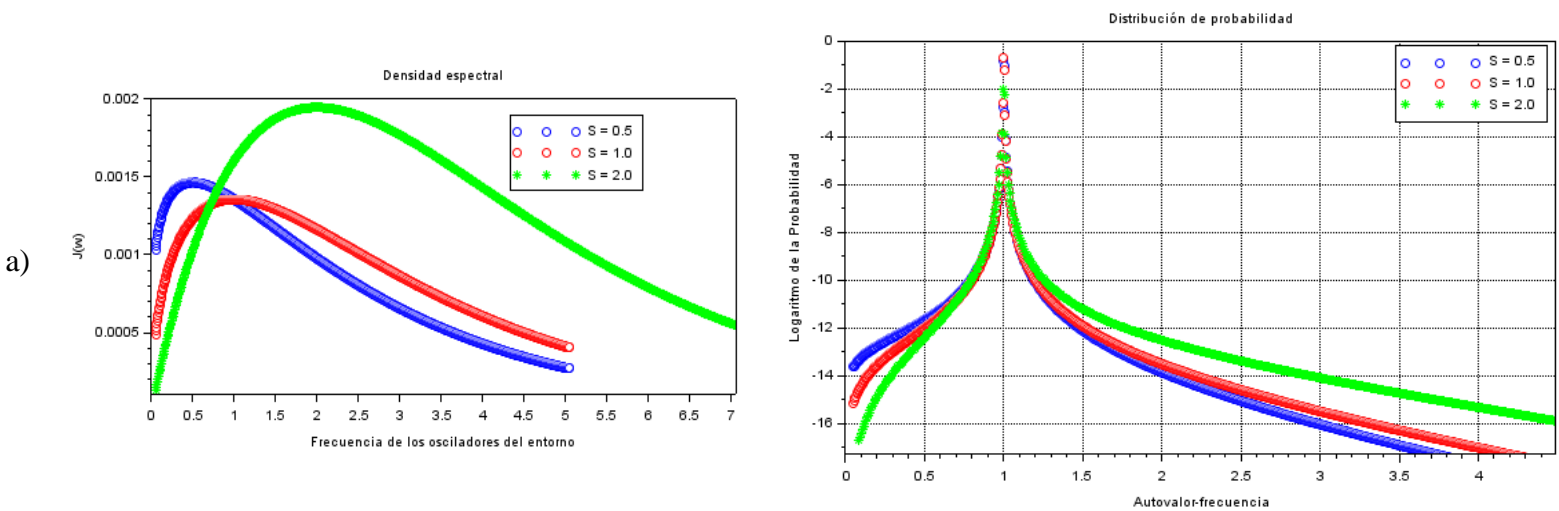

a)

b) 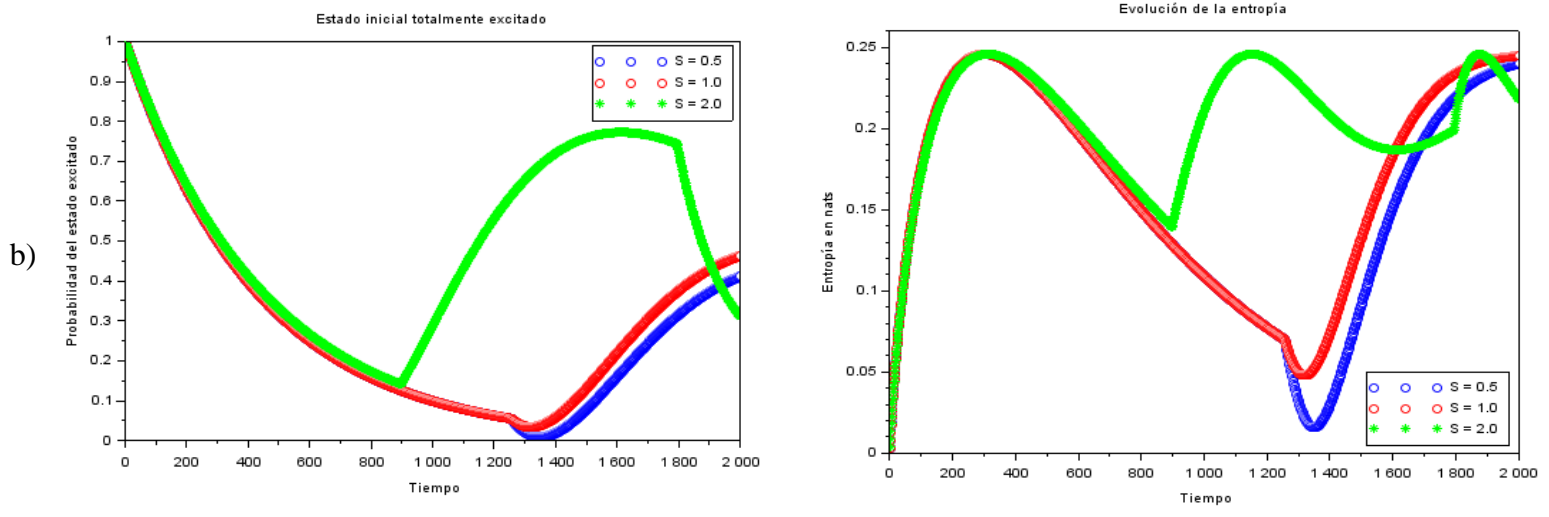

c)

d) 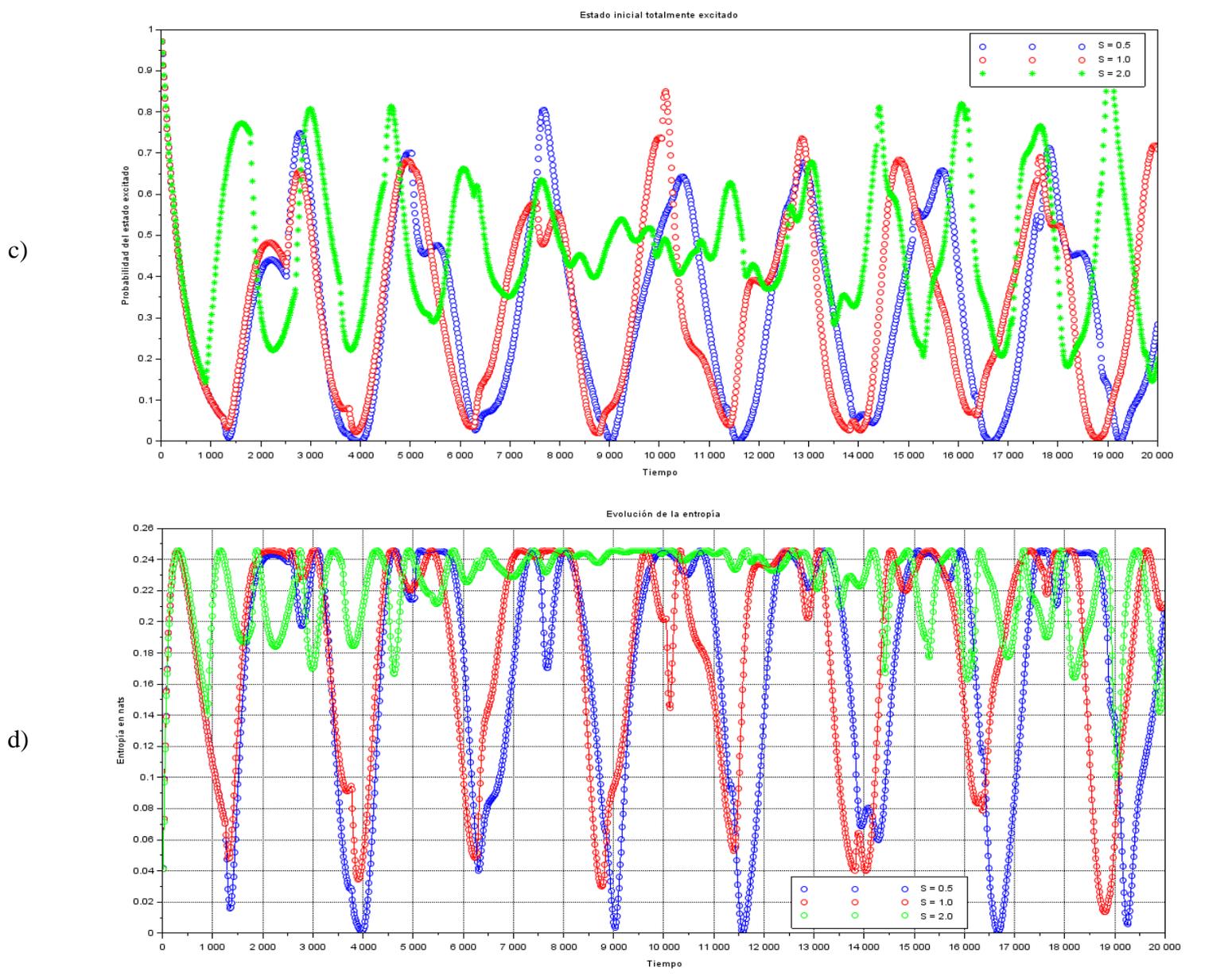

**Figura 7.4** Evaluación para $\eta = 0.001$ y $s = 0.5, 1, 2$

Los promedios de las probabilidades para $s = 0.5$ y $s = 2$ respectivamente son:

$$p_{s=0.5} = \langle p_j \rangle = 0.3250733 \qquad p_{s=2} = \langle p_j \rangle = 0.4646133 \qquad (7.3)$$

A la izquierda de la fila a) de la Figura 7.4 se tiene la densidad espectral para los tres casos: subohmico, óhmico y superohmico; para $s = 2$ se extiende el dominio de las frecuencias para conseguir una mejor representación de la densidad espectral; a la derecha de a) se tiene el logaritmo de la probabilidad $p_j$ con respecto a los autovalores, se observa una diferencia en el comportamiento debido al parámetro $s$. A la izquierda de b) se tiene la evolución de la probabilidad del estado excitado, la cual muestra dependencia del parámetro $s$, a la derecha de b) se tiene la evolución de la entropía del sistema para el mismo intervalo de tiempo, el cual también muestra dependencia del parámetro $s$. En c) se tiene la evolución de la probabilidad del estado excitado para tiempos largos donde se observa la presencia de sobrevivencias (survivals) cuyo comportamiento cambia con el parámetro $s$; en d) se tiene la evolución de la entropía del sistema para tiempos largos, en la misma escala de tiempo que en c), se observa que la entropía se mantiene no nula, con picos de descenso donde se acerca al valor nulo.

## 7.3 Acople débil del sistema y el entorno: $\eta = 0.01 \rightarrow 0.3$

Se consideran los resultados para distintos valores de $\eta$, pero $s = 1$.

a) 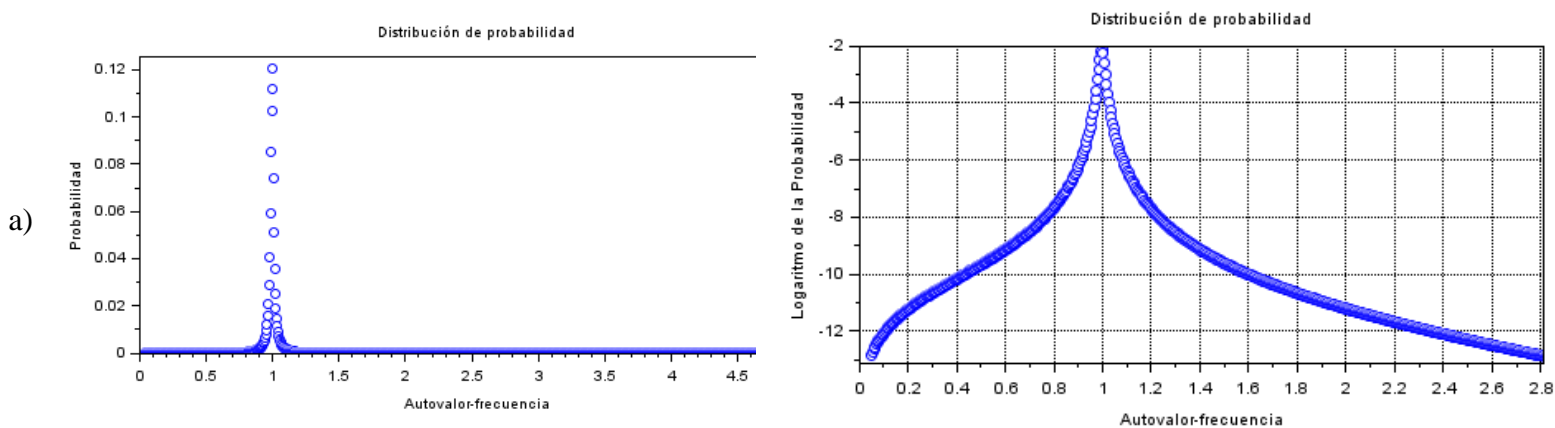

b) 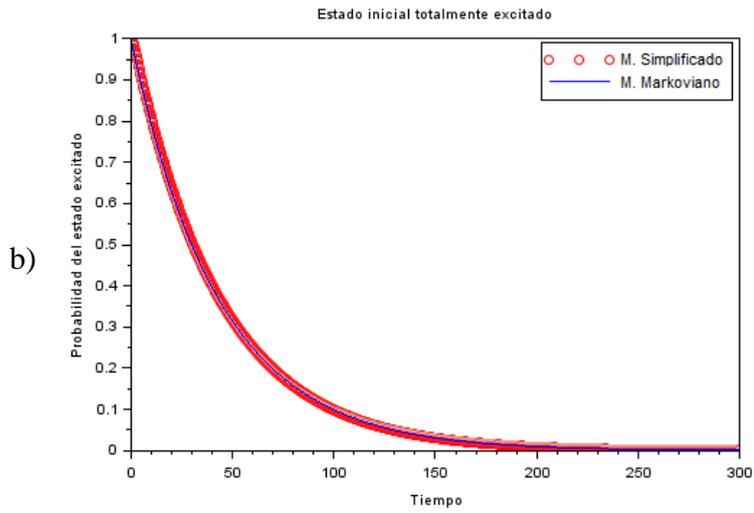 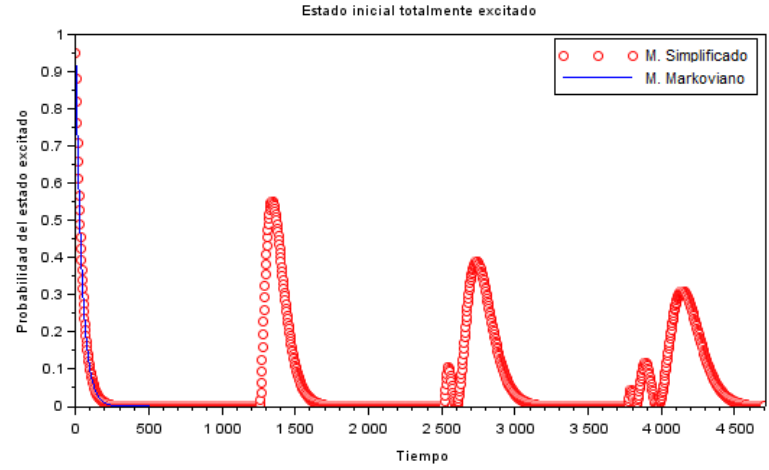

c) 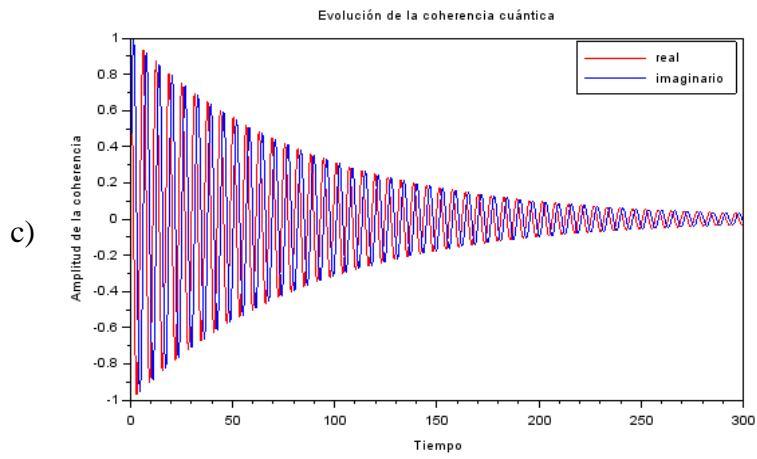 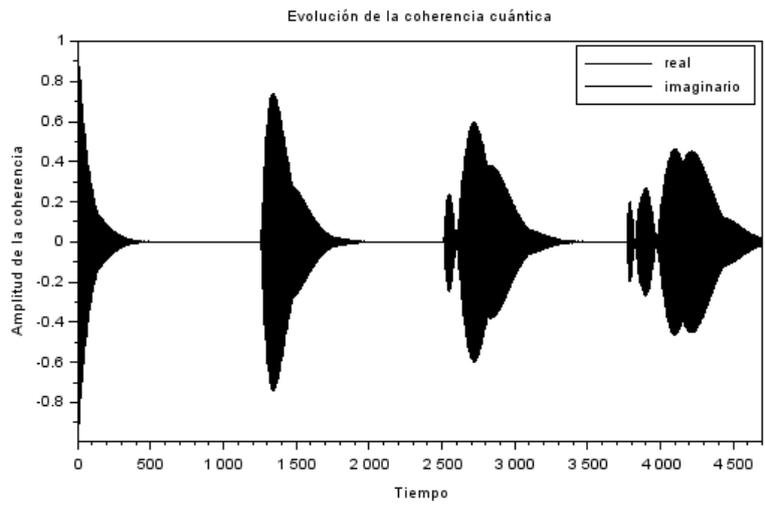

d) 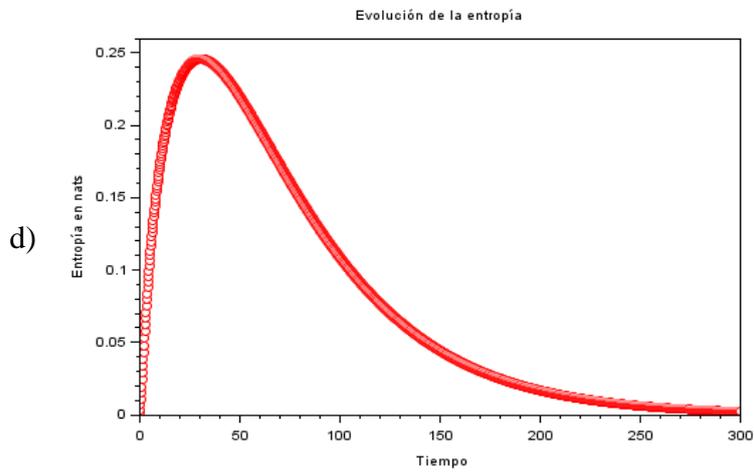 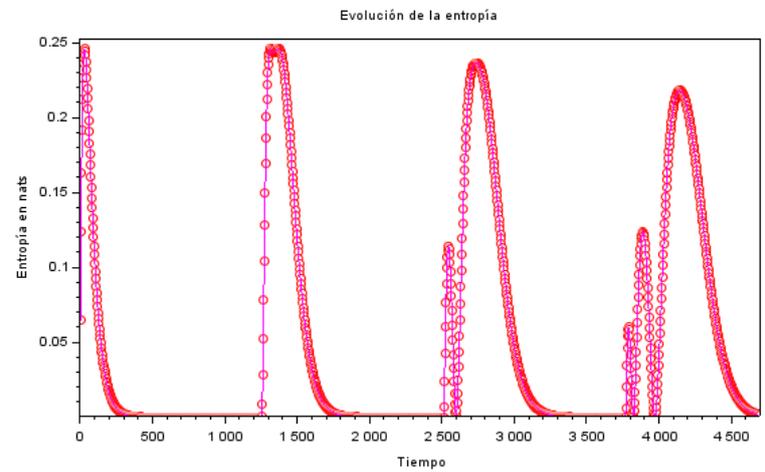

**Figura 7.5** Evaluación para $\eta = 0.01$ y $s = 1$

El promedio de la probabilidad del estado excitado en tiempos largos es:

$$p = \langle p_j \rangle = 0.0631602 \qquad (7.4)$$

Para $\eta = 0.1$ y $s = 1$:

a) 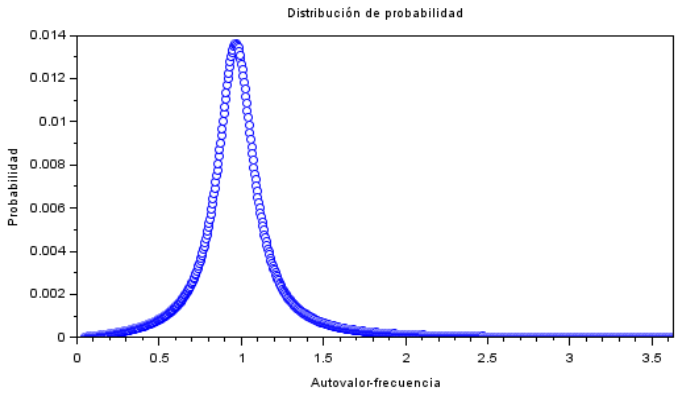 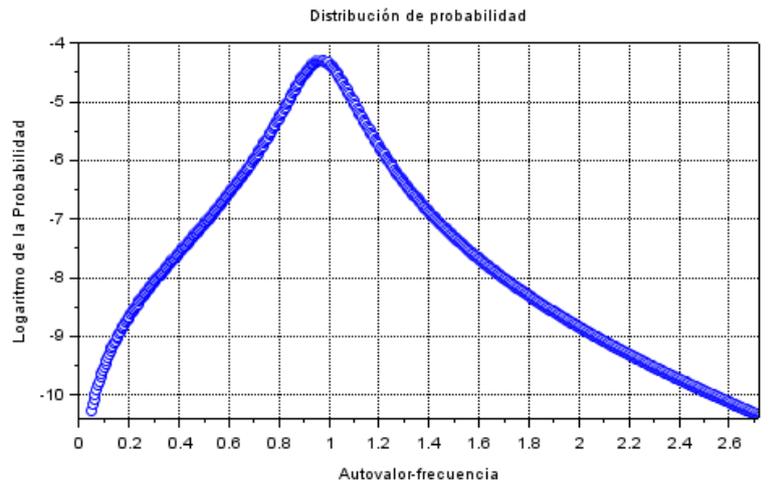

b) 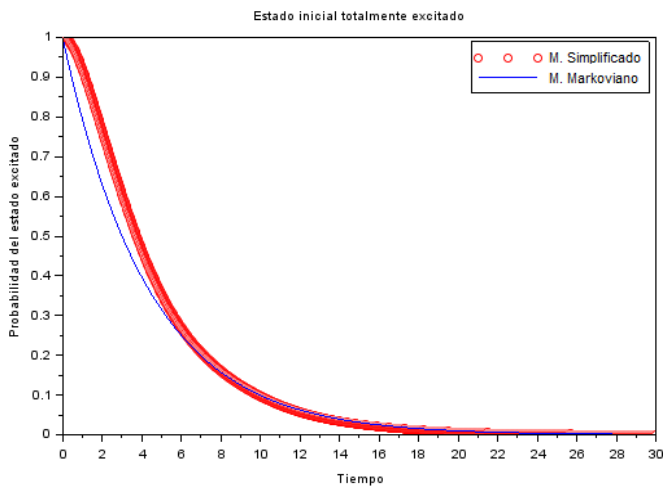 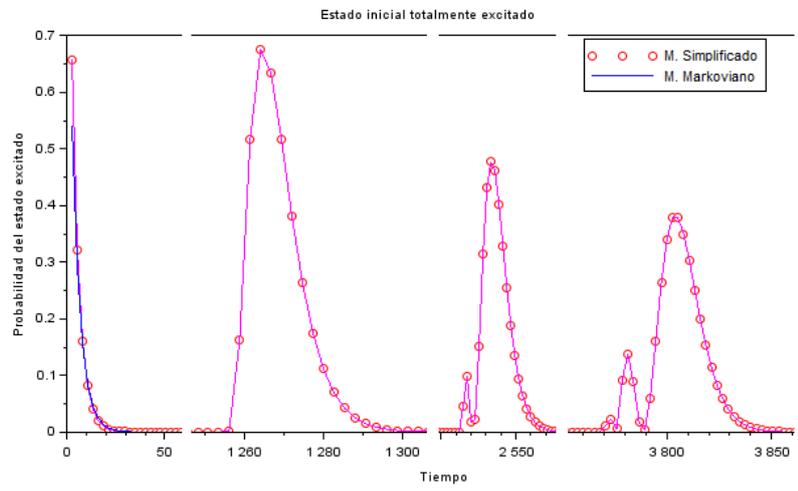

c) 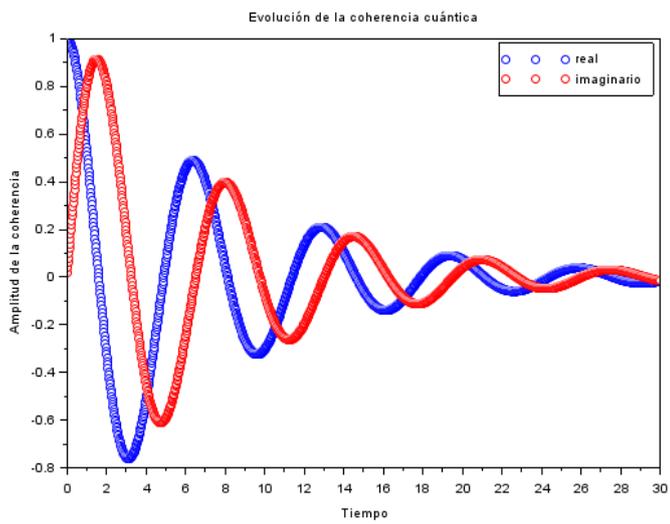 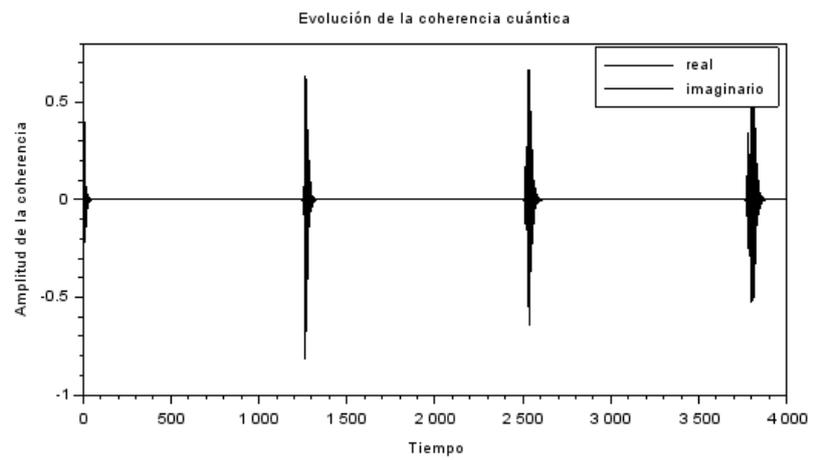

d) 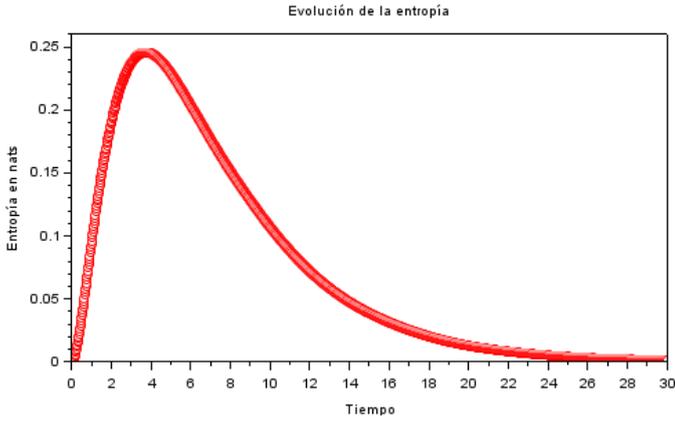 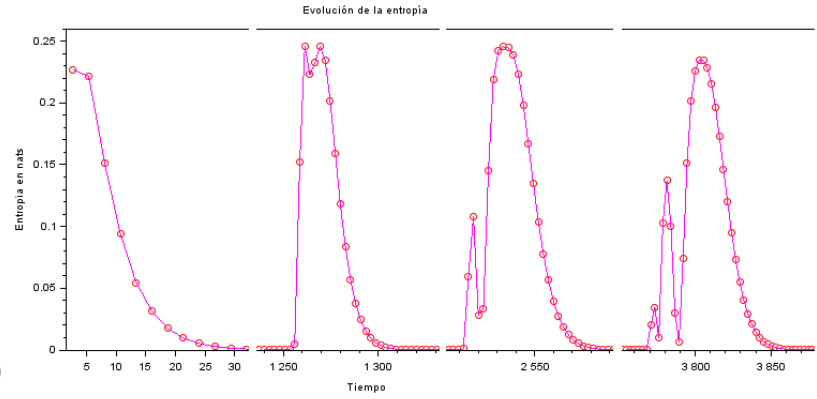

**Figura 7.6** Evaluación para $\eta = 0.1$ y $s = 1$

En los gráficos de la derecha en b) y d) se ha tenido que obviar intervalos de tiempo donde no aparecen sobrevivencias de la probabilidad; además, no hay suficiente resolución para representar correctamente las curvas correspondientes al inicio del tiempo, dichas curvas están bien graficadas en las izquierdas de b) y d), pues corresponden al mismo gráfico, pero en dos escalas de tiempo distintas.

$$p = \langle p_j \rangle = 0.0075569 \tag{7.5}$$

Para $\eta = 0.3$ se tiene:

a) 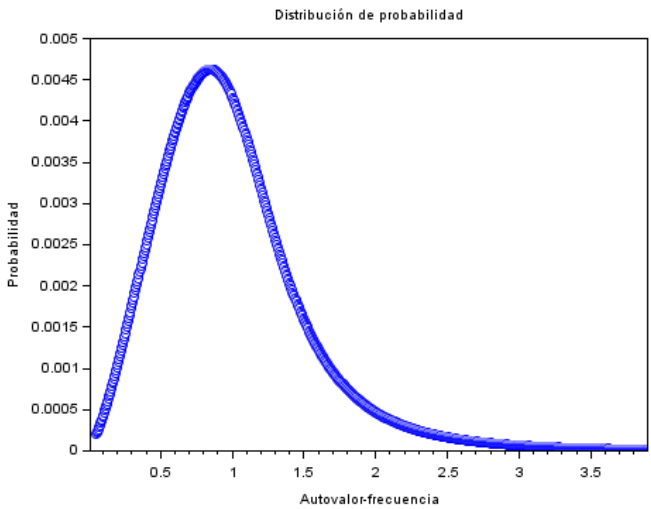 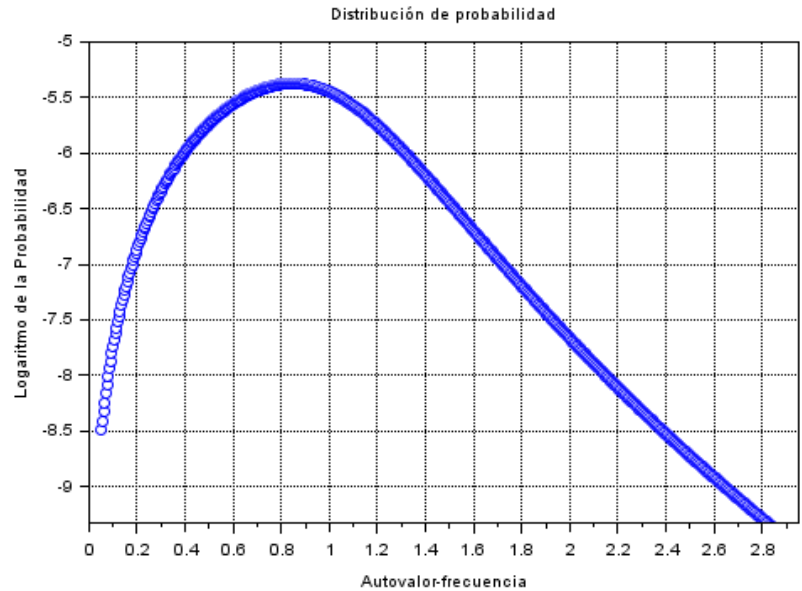

b)

c)

d)

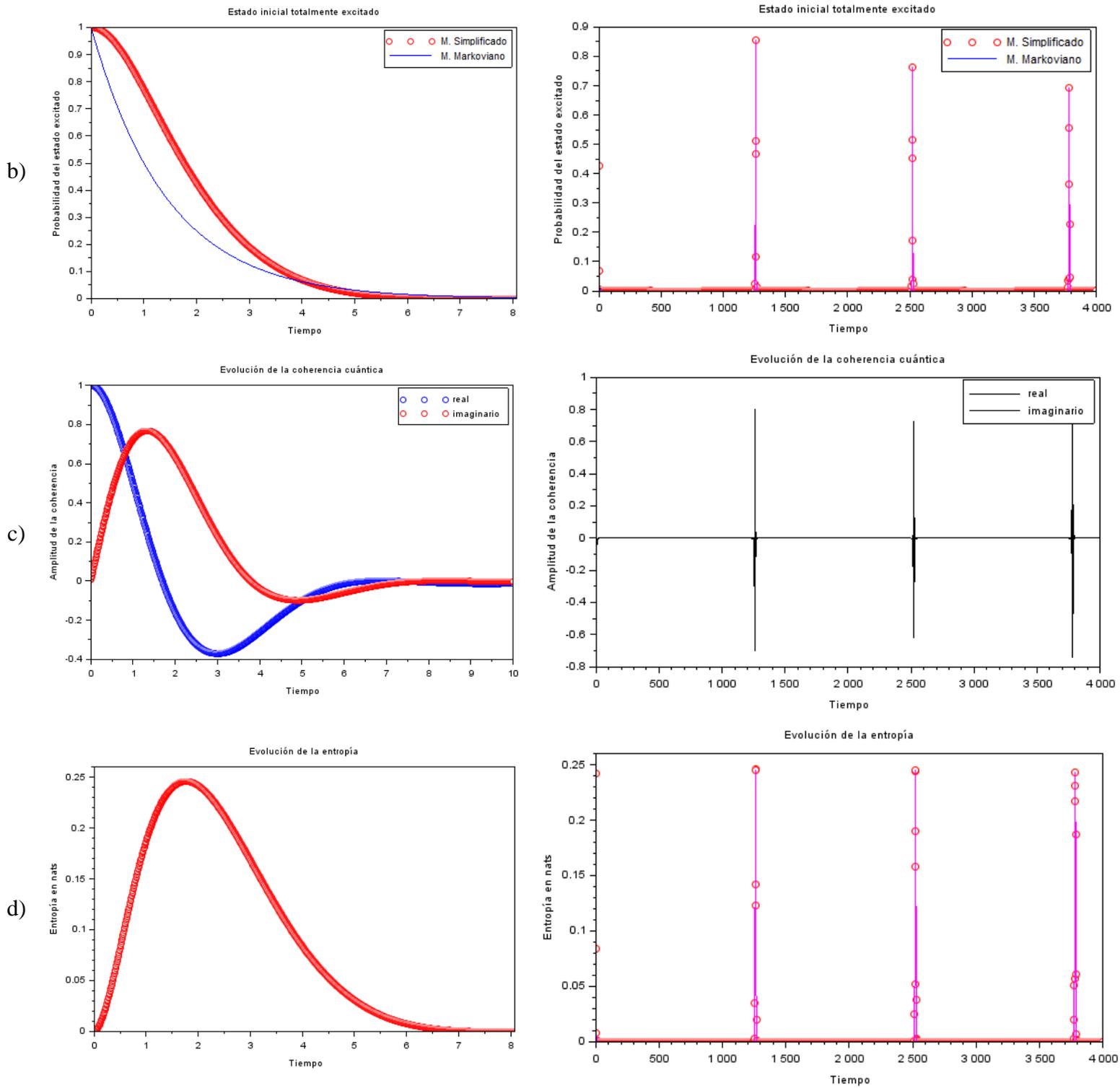

**Figura 7.7** Evaluación para $\eta = 0.3$ y $s = 1$

$$p = \langle p_j \rangle = 0.0031478 \tag{7.6}$$

### 7.4 Síntesis de los resultados

#### 7.4.1 Intervalo de tiempos breves: mayor decaimiento de la probabilidad

El intervalo de tiempos breves es: $t \in [t_0, t_1]$ donde $t_0 = 0$, y $t_1$ corresponde al tiempo donde ambas curvas (la del modelo simplificado y la solución markoviana en los gráficos de la evolución de la probabilidad del estado excitado) se separan, siempre antes de la aparición de las sobrevivencias [12]. El tiempo $t_1$ puede ser considerado aproximadamente como un múltiplo de $\tau_{1/2}$, el tiempo en el que la probabilidad del estado excitado $|\Gamma(t)|^2$ desciende a la mitad:

$$t_1 = n\, \tau_{1/2} \qquad n \lesssim 5 \qquad \forall\, t < t_1:\ |\Gamma(t)|^2 \sim e^{-J(\omega_0)t} \qquad (7.7)$$

$e^{-J(\omega_0)t}$ de (7.7) es la solución markoviana, y se aproxima a $|\Gamma(t)|^2$, $\forall\, t < t_1$.

En las Figuras (7.2) y (7.4), para $\eta = 0.001$ y diversos valores de $s = 0.5, 1, 2$, se evidencia una distribución de probabilidad que tiende a una delta de Dirac (en la escala de autovalores-frecuencias comparable con $\omega_0 = 1$), con una dispersión proporcional a $\eta$, de acuerdo a (6.50) y los valores respectivos de la Tabla 6.1. La pendiente de la curva correspondiente al modelo simplificado inicia horizontal a diferencia de la correspondiente a la solución markoviana (caída exponencial), sin embargo, ambas se aproximan mucho entre sí en el intervalo de tiempos breves $t \in [t_0, t_1]$ donde la mayor parte de la probabilidad del estado excitado decae, en tal trayecto no hay diferencia apreciable frente a la variación del orden óhmico $s$ de la densidad espectral $J(\omega)$, lo cual es consistente con la solución markoviana, cuyo parámetro de decaimiento viene dado por la ecuación (7.1). En la Figura 7.5 se evalúa para un acoplamiento más fuerte $\eta = 0.01$ donde la distribución de probabilidad se hace notoriamente más ancho (y se aleja de aproximarse a una delta de Dirac) también se evidencia la proximidad entre ambas curvas, del modelo simplificado y la solución markoviana; sin embargo, conforme se evalúa para $\eta = 0.1$ y $\eta = 0.3$ en las Figuras 7.6 y 7.7 las curvas presentan una notoria separación entre sí, siendo mayor conforme aumenta la fuerza de acoplamiento. Se considera el intervalo $t \in [t_0, t_1]$ como el dominio donde la aproximación markoviana es válida para acoplamientos débiles, pero deja de serlo para fuerzas de acoplamiento del orden $\eta \sim 0.1$ para este problema físico con densidad espectral dada en la ecuación (6.1).

### 7.4.2 Intervalo de tiempos largos: sobrevivencias de la probabilidad

En los gráficos de la evolución de la probabilidad del estado excitado se evidencia la aparición de importantes sobrevivencias de la probabilidad, con picos que pueden llegar a superar el 80% en tiempos específicos; para fuerzas de acoplamientos muy débiles $\eta = 0.001$ se tiene una alta variabilidad de la sobrevivencia, la cual se promedia para tiempos largos en el orden del 30%, aumentando conforme crece $s$ (orden óhmico de la densidad espectral). Para fuerzas de acoplamientos mayores $\eta \geq 0.01$ siguen habiendo picos en la sobrevivencia de la probabilidad pero alejados entre sí (manteniendo una posición fija o estable en el tiempo), pero sin sobrevivencias entre ellos (la probabilidad tiende a anularse), y cada vez más angostos conforme crece la fuerza de acoplamiento, ensanchándose y decreciendo para tiempos mayores. El promedio de la probabilidad para tiempos largos disminuye conforme aumenta la fuerza de acoplamiento, como se muestra:

| Fuerza de acoplamiento | Promedio de la probabilidad | | |
|---|---|---|---|
| $\eta = 0.001$ ($s = 0.5, 1, 2$) | 0.3250733 | 0.3363172 | 0.4646133 |
| $\eta = 0.01$ | 0.0631602 | | |
| $\eta = 0.1$ | 0.0075569 | | |
| $\eta = 0.3$ | 0.0031478 | | |

**Tabla 7.2**

A diferencia de la solución markoviana que solo presenta un decaimiento exponencial de la probabilidad para este problema, la solución exacta de este modelo simplificado presenta una fuerte sobrevivencia de la probabilidad en la dinámica del acoplamiento débil, que se hace menor conforme más fuerte es el acoplamiento y mayor el orden óhmico de la densidad espectral; aun así, es interesante notar que los picos de las sobrevivencias aumentan al pasar de $\eta = 0.1$ a $\eta = 0.3$, lo que hace suponer que para este problema físico habría un valor de $\eta_0$ según el cual la amplitud de los máximos de la sobrevivencia disminuye conforme $\eta$ crece y se acerca a $\eta_0$ para luego aumentar, conforme crece $\eta$, superando $\eta_0$, esta suposición es consistente con la dinámica no-markoviana donde el acoplamiento fuerte entre el sistema y reservorio permite un mayor efecto memoria, y con ello la presencia de sobrevivencias [2].

### 7.4.3 Evolución de la entropía de Von Neumann del sistema

En todas las gráficas de la evolución de la entropía en tiempos breves $t \in [t_0, t_1]$: a) de la Figura 7.3, la gráfica derecha de b) de la Figura 7.4, y en las izquierdas de d) de la Figura 7.5, Figura 7.6 y Figura 7.7 se tiene un comportamiento característico, siempre se inicia con valor nulo y crece hasta alcanzar un máximo, a partir del cual empieza a decrecer, mientras que la probabilidad del estado excitado solo decrece en ese intervalo de tiempo $t \in [t_0, t_1]$. La entropía de Von Neumann de la matriz de densidad es una medida de la ignorancia del estado mezcla (ver sección [2.1.2]), siendo nulo cuando el estado es puro, tal como corresponde al inicio del tiempo, razón por la cual la entropía inicia con valor nulo, y cuyo crecimiento pone de manifiesto que la matriz de densidad del sistema "se vuelve más mezcla" porque en la interacción con el reservorio se pierde información del estado puro inicial $\hat{\rho}_0 = |\psi_0\rangle\langle\psi_0|$, donde $|\psi_0\rangle = \alpha|0\rangle + \beta|1\rangle$; por otro lado, la disminución de la entropía corresponde a una disminución de la ignorancia de su estado, de su cualidad de mezcla, y la aproximación a un estado puro, el cual no necesariamente corresponde al estado inicial del sistema, sino al estado fundamental $\hat{\rho} = |0\rangle\langle 0|$, pues en su interacción con el reservorio a temperatura nula el sistema "cede" su excitación al entorno, tendiendo a decaer al estado fundamental, lo cual da certeza sobre el estado actual del sistema, pero ya no del estado inicial.

En las gráficas de la evolución de la entropía para tiempos largos $t \gg t_1$, como en b) de la Figura 7.3 (y mejor comportado en las derechas de d) de la Figura 7.5, Figura 7.6 y Figura 7.7) se observa el aumento de la entropía correlacionado a la presencia de las sobrevivencias (aumento de $|\Gamma(t)|$ o su cuadrado), debido a que en dichas sobrevivencias el sistema se aleja del estado fundamental y recupera información del estado inicial, pues de acuerdo a (6.41), si $|\Gamma(t)| \to 0$ entonces $\hat{\rho} \to |0\rangle\langle 0|$, pero si $|\Gamma(t)| \to 1$ entonces $\hat{\rho} \to |\psi_0\rangle\langle\psi_0|$; así, en el intervalo a tiempos largos el sistema evoluciona entre el estado puro fundamental (con entropía nula y sin sobrevivencias) y una mezcla con memoria del estado puro inicial (con sobrevivencias y entropía no nula).

# CAPÍTULO VIII

# CONCLUSIONES

## 8.1 Conclusiones generales

Es factible y útil dividir el espacio de Hilbert total $\mathcal{H} = \bigotimes_{i=0}^{N} \mathcal{H}_i$ en los subespacios de número de excitación definido $\mathcal{H} = \bigoplus_{\Sigma=0}^{N+1} \mathcal{H}(\Sigma)$ para reducir el problema general y obtener la solución exacta de la dinámica irreversible de un sistema cuántico abierto.

El modelo simplificado, aun siendo una reducción de la realidad física, muestra un comportamiento coherente con la mecánica cuántica de sistemas abiertos, al presentar un decaimiento de la probabilidad del estado excitado consistente con la dinámica markoviana, pero también la presencia de sobrevivencias, *revivals* [12], debido al efecto memoria entre el sistema y su entorno con un número finito de osciladores $N$. Del Anexo E se deduce que el tiempo en el que aparecen las primeras sobrevivencias se hace mayor conforme crece $N$, de manera que para un reservorio muy grande $N \to \infty$, las sobrevivencias se alejan hasta el infinito en el tiempo, y la aproximación markoviana para este modelo, como una caída exponencial de la probabilidad del estado excitado, es válida en todo el tiempo finito.

## 8.2 Conclusiones específicas

La aproximación markoviana es válida para acoplamientos débiles en intervalos de tiempo donde ocurre la mayor parte del decaimiento, pero antes de la aparición de sobrevivencias en la solución exacta de este modelo simplificado.

En la solución exacta de este modelo aparecen sobrevivencias de la probabilidad del estado excitado que no aparecen en la solución markoviana, de manera que la aproximación markoviana deja de ser válida en el tiempo donde se presentan tales supervivencias para este modelo. La amplitud del término de coherencia, de la matriz de densidad, también presenta decaimiento y sobrevivencias.

El promedio de la probabilidad del estado excitado disminuye conforme aumenta la fuerza del acoplamiento. Las amplitudes de las supervivencias de la probabilidad del estado excitado y de la coherencia cuántica disminuyen con el paso del tiempo, también tienden a ensancharse.

Se observa que para la fuerza de acoplamiento $\eta = 0.001$ la supervivencia de la probabilidad aumenta con el orden óhmico $s$ de la densidad espectral.

Se observa que la aproximación markoviana deja de ser bien comportada a tiempos breves $t < t_1$ para fuerzas de acoplamientos del orden $\eta = 0.1$ .

Se observa el comportamiento de la entropía de Von Neumann del sistema, descrito en la sección [7.4.3], que inicia con valor nulo porque el estado inicial es puro y evoluciona hasta un máximo local, del cual decrece porque se aproxima al estado puro fundamental (debido a que el reservorio está a temperatura nula). La presencia de sobrevivencias devuelve información sobre el estado inicial al estado actual del sistema, el cual es una mezcla, y aumenta la entropía en correlación con las sobrevivencias. La dinámica de la entropía con las sobrevivencias pone de manifiesto el efecto memoria entre el sistema y el reservorio.

De la Figura E.2 del Anexo E. se evidencia que las apariciones de las sobrevivencias se alejan en el tiempo conforme crece el número de osciladores del reservorio, así para un reservorio con infinitos osciladores la aparición de las sobrevivencias se alejarían hasta un tiempo infinito, eliminando las sobrevivencias de la dinámica del sistema, y con ello el efecto memoria, durante tiempos finitos.

De la Figura E.3 del Anexo E. se evidencia que la evolución de la probabilidad del estado excitado en el tiempo se aproxima mejor a la solución markoviana, de caída exponencial, conforme el número de osciladores del reservorio aumenta, así para un reservorio infinito la aproximación markoviana es bien comportada para acoplamientos débiles con $\eta \sim 0.01$ en este modelo simplificado.

# Anexo A

# Cálculo de $H_{1,1}^n$: el primer elemento del hamiltoniano a la potencia $n$

En la ecuación (6.11) se establece que el hamiltoniano para $\Sigma = 1$ es $\widehat{H}^1 = G^T$, y de acuerdo a (6.5) se puede escribir el hamiltoniano como sigue:

$$\widehat{H}^1 = \sum_{i,j=1}^{N+1} H_{ij}|1,i\rangle\langle 1,j| \qquad \begin{aligned} H_{ij} &= \omega_{i-1}\delta_{ij} + g_{i-1}\delta_{j,1} + g_{j-1}^*\delta_{i,1} \\ g_0 &= 0 \end{aligned} \quad (A.1)$$

Este hamiltoniano es el mismo al que se tiene en (2.44), donde el sistema se acopla a todos los osciladores del reservorio, pero sin acoples entre los osciladores del reservorio. El hamiltoniano está reducido a $\mathcal{H}(1)$, el subespacio con número de excitación uno.

En lo sucesivo, las componentes $i,j$ de la potencia $n$ del hamiltoniano viene a ser:

$$\langle 1,i|\bigl(\widehat{H}^1\bigr)^n|1,j\rangle = H_{ij}^n \qquad \bigl(\widehat{H}^1\bigr)^n = \sum_{i,j=1}^{N+1} H_{ij}^n|1,i\rangle\langle 1,j| \quad (A.2)$$

Así, usando (A.1) y (A.2) se puede expresar las componentes $i,j$ del producto matricial de $\bigl(\widehat{H}^1\bigr)^{n+1} = \widehat{H}^1\bigl(\widehat{H}^1\bigr)^n$ como se muestra:

$$H_{ij}^{n+1} = \omega_{i-1}H_{ij}^n + g_{i-1}H_{1j}^n + \delta_{i,1}\sum_k g_{k-1}^* H_{kj}^n \qquad k = 2\colon N+1 \quad (A.3)$$

El índice $k$ inicia en 2 porque la sumatoria se anula para $k = 1$, con esto se deducen las dos relaciones de recurrencia con las que se puede construir de forma explícita $H_{1,1}^{n+1}$:

$$H_{1,1}^{n+1} = \omega_0 H_{1,1}^n + \sum_{i=2}^{N+1} g_{i-1}^* H_{i,1}^n \qquad \forall i > 1\colon H_{i,1}^{n+1} = \omega_{i-1}H_{i1}^n + g_{i-1}H_{11}^n \quad (A.4)$$

Ambas ecuaciones de (A.4) se refieren mutuamente, y es posible construir con ellas cualquier $H_{1,1}^{n+1}$ partiendo de $H_{ij}^0 = \delta_{i,j}$ (que corresponde a la matriz identidad), por ejemplo, se puede obtener $H_{1,1}$ de (A.1) para $n = 0$ en (A.4); es posible reducir (A.4) a una sola ecuación de recurrencia expresando la segunda ecuación de forma explícita, mediante la siguiente inducción:

| n | |
|---|---|
| | $\forall i > 1: H_{i,1}^{n+1} = \omega_{i-1} H_{i1}^n + g_{i-1} H_{11}^n$ |
| 0 | $H_{i,1}^1 = \omega_{i-1} H_{i1}^0 + g_{i-1} H_{11}^0$ |
| 1 | $H_{i,1}^2 = \omega_{i-1}^2 H_{i1}^0 + g_{i-1}(\omega_{i-1} H_{11}^0 + H_{11}^1)$ |
| 2 | $H_{i,1}^3 = \omega_{i-1}^3 H_{i1}^0 + g_{i-1}(\omega_{i-1}^2 H_{11}^0 + \omega_{i-1} H_{11}^1 + H_{11}^2)$ |
| 3 | $H_{i,1}^4 = \omega_{i-1}^4 H_{i1}^0 + g_{i-1}(\omega_{i-1}^3 H_{11}^0 + \omega_{i-1}^2 H_{11}^1 + \omega_{i-1} H_{11}^2 + H_{11}^3)$ |
| 4 | $H_{i,1}^5 = \omega_{i-1}^5 H_{i1}^0 + g_{i-1}(\omega_{i-1}^4 H_{11}^0 + \omega_{i-1}^3 H_{11}^1 + \omega_{i-1}^2 H_{11}^2 + \omega_{i-1} H_{11}^3 + H_{11}^4)$ |

**Tabla A.1**

Como $i > 1$ entonces $H_{i1}^0 = \delta_{i,1} = 0$, se eliminan los primeros términos en cada desarrollo de la inducción, así se tiene un patrón que se expresa como sigue:

$$\forall i > 1: H_{i,1}^{n+1} = g_{i-1} \sum_{j=0}^{n} \omega_{i-1}^{n-j} H_{11}^j \tag{A.5}$$

Así, (A.5) en la primera ecuación de (A.4) da una única relación de recurrencia:

$$H_{1,1}^{n+1} = \omega_0 H_{1,1}^n + \sum_{i=1}^{N} |g_i|^2 \sum_{j=0}^{n-1} \omega_i^{n-j-1} H_{11}^j \tag{A.6}$$

Donde se hizo el cambio de índice: $(i-1) \to i$ en $g_i$ y $\omega_i^{n-j-1}$; usando la siguiente notación, proveniente de (6.3), se puede re expresar (A.6):

$$|g|_k^2 = \sum_{i=1}^{N} |g_i|^2 \omega_i^k \qquad H_{1,1}^{n+1} = \omega_0 H_{1,1}^n + \sum_{j=0}^{n-1} |g|_{n-j-1}^2 H_{11}^j \tag{A.7}$$

Entonces se desarrolla (A.7) de forma recursiva, como se muestra:

$$H_{1,1}^{n+1} = \omega_0 \left( \omega_0 H_{1,1}^{n-1} + \sum_{j=0}^{n-2} |g|_{n-j-2}^2 H_{11}^j \right) + \sum_{j=0}^{n-1} |g|_{n-j-1}^2 H_{11}^j$$

Se agrupan las dos sumatorias más un término sobrante de una de las sumatorias:

$$H_{1,1}^{n+1} = \omega_0^2 H_{1,1}^{n-1} + \sum_{j=0}^{n-2}(\omega_0|g|_{n-j-2}^2 + |g|_{n-j-1}^2)H_{11}^j + |g|_0^2 H_{11}^{n-1}$$

Este proceso se repite iterativamente, generando términos sobrantes de las sumatorias que se acumulan:

$$H_{1,1}^{n+1} = \omega_0^3 H_{1,1}^{n-2} + \sum_{j=0}^{n-3}(\omega_0^2|g|_{n-j-3}^2 + \omega_0|g|_{n-j-2}^2 + |g|_{n-j-1}^2)H_{11}^j$$
$$+ (\omega_0|g|_0^2 + |g|_1^2)H_{11}^{n-2} + |g|_0^2 H_{11}^{n-1}$$

$$H_{1,1}^{n+1} = \omega_0^4 H_{1,1}^{n-3} + \sum_{j=0}^{n-4}(\omega_0^3|g|_{n-j-4}^2 + \omega_0^2|g|_{n-j-3}^2 + \omega_0|g|_{n-j-2}^2 + |g|_{n-j-1}^2)H_{11}^j$$
$$+ (\omega_0^2|g|_0^2 + \omega_0|g|_1^2 + |g|_2^2)H_{11}^{n-3} + (\omega_0|g|_0^2 + |g|_1^2)H_{11}^{n-2} + |g|_0^2 H_{11}^{n-1}$$

$$\vdots$$

Entonces aparece un patrón que puede ser expresado como sigue, para el $k$-ésimo "descenso" en la potencia del hamiltoniano:

$$H_{1,1}^{n+1} = \omega_0^{k+1} H_{1,1}^{n-k} + \sum_{j=0}^{n-1-k}\left(\sum_{i=1}^{k+1}\omega_0^{i-1}|g|_{n-j-i}^2\right)H_{11}^j + \sum_{j=1}^{k}\left(\sum_{i=0}^{j-1}\omega_0^{j-1-i}|g|_i^2\right)H_{11}^{n-j}$$

Si $k = n$, se tiene $H_{1,1}^0 = \delta_{1,1} = 1$ en el primer término, y la desaparición de la primera sumatoria (no hay un conjunto de números donde sumar el índice $j$), quedando:

$$H_{1,1}^{n+1} = \omega_0^{n+1} + \sum_{j=1}^{n} a_j H_{11}^{n-j} \qquad a_j = \sum_{i=0}^{j-1}\omega_0^{j-1-i}|g|_i^2 \qquad (A.8)$$

De (A.8) es posible encontrar el desarrollo para $H_{1,1}^n$ mediante la iteración para ordenes menores de $H_{11}^{n-1-j}$, proveniente de (A.8) mismo, como se muestra:

$$H_{1,1}^n = \omega_0^n + \sum_{i=1}^{n-1} a_i H_{11}^{n-1-i} = \omega_0^n + \sum_{i=1}^{n-1} a_i \omega_0^{n-1-i} + \sum_{i=1}^{n-1} a_i \sum_{j=1}^{n-2-i} a_j H_{11}^{n-2-i-j}$$

$$H_{1,1}^n = \omega_0^n + \sum_{i=1}^{n-1} a_i\, \omega_0^{n-1-i} + \sum_{i=1}^{n-1} a_i \sum_{j=1}^{n-2-i} a_j\, \omega_0^{n-2-i-j} + \sum_{i=1}^{n-1} a_i \sum_{j=1}^{n-2-i} a_j \sum_{k=1}^{n-3-i-j} a_k\, H_{11}^{n-3-i-j-k}$$

La generalización correspondiente viene a ser:

$$\left. \begin{aligned} H_{1,1}^n &= \omega_0^n + \sum_k b_{k,n} \\ b_{k,n} &= \sum_{i_1=1}^{n-1} a_{i_1} \sum_{i_2=1}^{n-2-i_1} a_{i_2} \sum_{i_3=1}^{n-3-i_1-i_2} a_{i_3} \cdots \sum_{i_k=1}^{n-k-\sum_{j=1}^{k-1} i_j} a_{i_k}\, \omega_0^{n-k-\sum_{j=1}^{k} i_j} \end{aligned} \right| \quad (A.9)$$

(A.9) tiene la forma general de $H_{1,1}^n$ en un desarrollo perturbativo de los acoplamientos $g_i$, de acuerdo a (A.7) y (A.8), $\forall j$: $a_{i_j}$ es proporcional a $|g|_k^2$ el cual es proporcional a $|g_i|^2$, el cual es proporcional a la fuerza de acoplamiento $\eta$ de la densidad espectral $J(\omega)$; es decir, $\forall j$: $a_{i_j}$ es proporcional a $\eta$, de manera que $b_{k,n}$ es proporcional a $\eta^k$, así, se puede aproximar $H_{1,1}^n$ a orden $k$ en $\eta$ tomando los $k$ primeros términos $b_{k,n}$.

Es importante notar que no todos los términos $b_{k,n}$ aparecen para un $n$ dado: la condición de $n$ para que $b_{k,n}$ sea no nulo es cuando la sumatoria en el índice $i_k$ de $b_{k,n}$ en (A.9) no se anula porque existe el conjunto de los índices $i_k$ en los cuales sumar, esto implica que la cota superior de la sumatoria sea mayor o igual a 1:

$$n - k - \sum_{j=1}^{k-1} i_j \geq 1 \qquad\qquad n_{min} = 2k \qquad\qquad (A.10)$$

El lado derecho de (A.10) se deduce del izquierdo para el mínimo valor de $n$, lo cual requiere que todos los índices tomen su mínimo valor: $i_j = 1$, eso significa que recién en $H_{1,1}^2$ aparece $b_{1,2}$ que es proporcional a $\eta$, a partir de $H_{1,1}^4$ se tiene $b_{1,4} + b_{2,4}$ cuyos términos son proporcionales a $\eta$ y $\eta^2$ respectivamente, a partir de $H_{1,1}^6$ se tiene $b_{1,6} + b_{2,6} + b_{3,6}$ cuyos términos son proporcionales a $\eta$, $\eta^2$ y $\eta^3$ respectivamente, y así sucesivamente.

Para una densidad espectral que puede expresarse como $J(\omega) = 2\pi\eta\omega_0\Lambda(\nu)$, donde $\nu = \omega/\omega_0$, $\Lambda$ y $\nu$ son adimensionales, $|g|_k^2$ de (A.7), y usando (6.3), viene a ser:

$$J(\omega) = 2\pi\eta\omega_0\Lambda(\omega/\omega_0) \to |g|_k^2 = \eta\omega_0^{k+2}\int_{\Omega_\nu}\Lambda(\nu)\nu^k\,d\nu = \eta\omega_0^{k+2}\Lambda_k \quad (A.11)$$

Donde $\Omega_\nu$ es el dominio correspondiente a $\nu$, y $\Lambda_k = \int_{\Omega_\nu}\Lambda(\nu)\nu^k\,d\nu$, también adimensional; (A.11) en (A.8) y este a su vez en $b_{1,n}$ de (A.9):

$$a_j = \eta\omega_0^{j+1}\sum_{i=0}^{j-1}\Lambda_i \qquad b_{1,n} = \eta\omega_0^n\sum_{i=1}^{n-1}(n-i)\Lambda_{i-1} \quad (A.12)$$

Considerando (A.10) y (A.12), una aproximación razonable a primer orden en $\eta$ de $H_{1,1}^n$, según (A.9), viene a ser:

$$H_{1,1}^n = \omega_0^n + b_{1,n} + \mathcal{O}(\eta^2) = \omega_0^n\left(1 + \eta\sum_{i=1}^{n-1}(n-i)\Lambda_{i-1}\right) + \mathcal{O}(\eta^2) \quad (A.13)$$

Usando la densidad espectral (6.1) en (A.11), considerando el límite continuo donde las frecuencias toman todos los reales positivos, se tiene:

$$\Lambda_k = \int_0^\infty \nu^{s+k}e^{-\nu}\,d\nu = \gamma_{s+k+1} \quad (A.14)$$

Donde $\gamma_x$ es la función gamma de $x$; al considerar (A.14) en (A.9) se tiene la Tabla A.2 para los primeros 6 valores de $n$, considerando todos los términos del desarrollo:

| $n$ | $H_{1,1}^n = \left(\widehat{H}^{1^n}\right)_{1,1}$ |
| --- | --- |
| 1 | $\omega_0$ |
| 2 | $\omega_0^2(1 + \eta\gamma_{s+1})$ |
| 3 | $\omega_0^3(1 + \eta(2\gamma_{s+1} + \gamma_{s+2}))$ |
| 4 | $\omega_0^4(1 + \eta(3\gamma_{s+1} + 2\gamma_{s+2} + \gamma_{s+3}) + \eta^2\gamma_{s+1}^2)$ |
| 5 | $\omega_0^5(1 + \eta(4\gamma_{s+1} + 3\gamma_{s+2} + 2\gamma_{s+3} + \gamma_{s+4}) + \eta^2(3\gamma_{s+1}^2 + 2\gamma_{s+1}\gamma_{s+2}))$ |
| 6 | $\omega_0^6(1 + \eta(5\gamma_{s+1} + 4\gamma_{s+2} + 3\gamma_{s+3} + 2\gamma_{s+4} + \gamma_{s+5}) + \eta^2(6\gamma_{s+1}^2 + \gamma_{s+2}^2 + 2\gamma_{s+1}(3\gamma_{s+2} + 2\gamma_{s+3})) + \eta^3\gamma_{s+1}^3)$ |

**Tabla A.2**

# Anexo B

# Cálculo numérico de los autovalores y autovectores del hamiltoniano

En esta sección se establece un método recursivo para el cálculo numérico de la ecuación de eigenvalores del hamiltoniano dado en (A.1), aprovechando su estructura particular (fuera de la diagonal solo la primera fila y columna son no nulos) y el límite de acople débil, en el que los elementos $g_i$ fuera de la diagonal sean muy pequeños respecto a los elementos $\omega_i$ de la diagonal, lo que en un sentido vagamente aproximado, el hamiltoniano toma la forma de una matriz diagonal de las frecuencias $\widehat{H}^1 \sim Diag(\omega_0, \omega_1, \cdots \omega_N)$, y en ese mismo sentido sus autovalores se pueden aproximar a las correspondientes frecuencias, y los autovectores a los ket con un uno en la fila correspondiente, y ceros en el resto de las filas.

La ecuación de autovalores, expresado de forma matricial, $\forall j = 1: N+1$, es:

$$\begin{pmatrix} \omega_0 & g_1^* & g_2^* & \cdots & g_i^* & \cdots & g_N^* \\ g_1 & \omega_1 & 0 & 0 & 0 & & 0 \\ g_2 & 0 & \omega_2 & 0 & 0 & & 0 \\ \vdots & & & \ddots & 0 & & 0 \\ g_i & 0 & 0 & 0 & \omega_i & & 0 \\ \vdots & & & & & \ddots & \vdots \\ g_N & 0 & 0 & 0 & 0 & \cdots & \omega_N \end{pmatrix} \begin{pmatrix} \beta_{1,j} \\ \beta_{2,j} \\ \beta_{3,j} \\ \vdots \\ \beta_{i+1,j} \\ \vdots \\ \beta_{N+1,j} \end{pmatrix} = \omega_{1,j} \begin{pmatrix} \beta_{1,j} \\ \beta_{2,j} \\ \beta_{3,j} \\ \vdots \\ \beta_{i+1,j} \\ \vdots \\ \beta_{N+1,j} \end{pmatrix} \quad (B.1)$$

A partir de (B.1) se obtiene para la primera fila: $\omega_0 \beta_{1,j} + \sum_{i=1}^N g_i^* \beta_{i+1,j} = \beta_{1,j} \omega_{1,j}$, y para las demás filas: $g_i \beta_{1,j} + \omega_i \beta_{i+1,j} = \beta_{i+1,j} \omega_{1,j}$; despejando unas amplitudes:

$$\beta_{1,j} = \frac{1}{\omega_{1,j} - \omega_0} \sum_{i=1}^N g_i^* \beta_{i+1,j} \qquad \beta_{i+1,j} = \frac{g_i \beta_{1,j}}{\omega_{1,j} - \omega_i} \quad , i = 1: N \qquad (B.2)$$

La ecuación de la izquierda de (B.2) obtiene solo la primera componente del autovector, mientras que la ecuación de la derecha es válida para todas las demás componentes, ambas ecuaciones se resuelven de manera sucesiva entre sí, de modo que las iteraciones de estas ecuaciones transforman unas amplitudes aproximadas iniciales $\beta_{i,j}^{(0)}$ en otras amplitudes $\beta_{i,j}^{(k)}$ que convergen al valor exacto tanto como se desea; para un eficiente uso iterativo de las ecuaciones (B.2) se requiere precisar el autovalor $\omega_{1,j}$ con el que se está trabajando, para esto se reemplaza la ecuación de la derecha en la izquierda, el resultado es una relación implícita para los autovalores que no depende de las amplitudes, sino de las frecuencias $\omega_i$ y el módulo de los acoples $g_i$, que son conocidos:

$$\omega_{1,j} - \omega_0 = \sum_{i=1}^{N} \frac{|g_i|^2}{\omega_{1,j} - \omega_i} \qquad j = 1:N+1 \qquad (B.3)$$

En la Figura B.1 se representan las intersecciones entre la recta azul, que corresponde a $y = x - \omega_0$, con la curva roja, que corresponde a $f(x) = \sum_{i=1}^{N}|g_i|^2/(x - \omega_i)$, la proyección en el eje $x$ de los $N+1$ puntos donde ambas curvas se cortan corresponden a las soluciones de (B.3): $x = \omega_{1,j}$ que son los autovalores.

En la Figura B.1 las frecuencias de los osciladores del entorno $\omega_i$ están ordenadas de forma ascendente con $i = 1:N$, sin embargo $\omega_0$ se encuentra entre las frecuencias $\omega_{i_0}$ y $\omega_{i_0+1}$, de manera que se generan dos grupos a ambos lados de $\omega_0$, para las frecuencias $\omega_i < \omega_0$ ($i \leq i_0$) los autovalores $\omega_{1,i+1}$ se encuentran muy cerca y por detrás de sus correspondientes frecuencias: $\omega_{1,i+1} < \omega_i$, mientras que para $i > i_0$ se tiene lo contrario: los autovalores se encuentran ligeramente por delante de sus correspondientes frecuencias $\omega_{1,i+1} > \omega_i$, conforme más se alejen de $\omega_0$ los autovalores $\omega_{1,i+1}$ más se aproximarán a sus correspondientes frecuencias $\omega_i$, el autovalor $\omega_{1,1}$ está asociado a la frecuencia del sistema $\omega_0$ y puede encontrarse por delante o por detrás del mismo.

La Figura B.1 demuestra no solo que todos los autovalores $\omega_{1,i+1}$ existen, sino que se encuentran confinados a la proximidad de su frecuencia asociada $\omega_i$ y no pueden extenderse más allá de la frecuencia vecina $\omega_{i-1}$ o $\omega_{1,i+1}$ según sea el caso; el cálculo numérico del autovalor puede hacerse encontrando la raíz de la función $f_0(x) = f(x) + \omega_0 - x$ por el método de Newton Raphson: partiendo de una ligera desviación de la frecuencia $x_j^{(0)} = \omega_{j-1} \pm \varepsilon$ y usando la fórmula de recurrencia $x_j^{(k+1)} = x_j^{(k)} - f_0\left(x_j^{(k)}\right)/f_0'\left(x_j^{(k)}\right)$ hasta que el valor absoluto de la diferencia sea menor a cierta tolerancia $1 \gg \left|x_j^{(k+1)} - x_j^{(k)}\right| \to 0$, la derivada de $f_0(x)$ es siempre menor a $-1$ y por lo tanto nunca provoca una singularidad al estar como denominador; para el caso del autovalor $\omega_{1,1}$ se puede partir de evaluar $f_0(\omega_0)$, si es positivo entonces se inicia desde la izquierda $x_1^{(0)} = \omega_{i_0} + \varepsilon$, y si es negativo se inicia desde la derecha $x_1^{(0)} = \omega_{i_0+1} - \varepsilon$, es muy importante cuidar la elección de los puntos iniciales $x_j^{(0)}$ porque es posible que partiendo de otros puntos del mismo intervalo $\langle \omega_i \, ; \, \omega_{i+1} \rangle$ rápidamente se salte a otros intervalos debido a la presencia de asíntotas que actúan como "catapultas" del cálculo numérico enviándolos lejos del intervalo inicial; así es muy importante asegurar que el cálculo numérico empleado para obtener las raíces de $f_0(x)$ mantienen confinado a $x$ en el intervalo del que inició.

Una aproximación para las raíces de $f_0(x)$ que se encuentran considerablemente alejados del punto $x = \omega_0$, consiste en tomar solo el término que más contribuye a la sumatoria de (B.3), sea $\delta_j = \omega_{1,j+1} - \omega_j$ y $\Delta_j = \omega_j - \omega_0$, entonces:

$$\begin{array}{c} \forall j > 0 \\ |\Delta_j| \gg |\delta_j| \end{array} : \; \delta_j + \Delta_j = \sum_{i=1}^{N} \frac{|g_i|^2}{\omega_{1,j+1} - \omega_i} \;\rightarrow\; \delta_j + \Delta_j \sim \frac{|g_j|^2}{\delta_j} \tag{B.4}$$

Es decir, que toda la sumatoria se reduce al término $i = j$, esta aproximación se puede justificar porque, de acuerdo a la Figura B.1, la recta azul corta a la curva roja en puntos muy próximos a las asíntotas y alejados del eje $x$ (con respecto a distancias de escala $\omega_{i+1} - \omega_i$, que es la separación entre asíntotas) lo cual hace predominante en la sumatoria de (B.3) la contribución del término $\frac{|g_j|^2}{\omega_{1,j+1} - \omega_j}$ (que se aproxima mucho a la asíntota en $x = \omega_j$) sobre los demás términos de la sumatoria (que se alejan de sus respectivas asíntotas, y por lo tanto contribuyen menos a la sumatoria), entonces (B.4) viene a ser: $\delta_j \sim \frac{|g_j|^2}{\Delta_j}(1 + \delta_j/\Delta_j)^{-1}$, y como $|\Delta_j| \gg |\delta_j|$, entonces $\delta_j \sim \frac{|g_j|^2}{\Delta_j}(1 - \delta_j/\Delta_j)$, despejando $\delta_j$ y luego $\omega_{1,j+1}$ se tiene:

$$\begin{array}{c} \forall j > 0 \\ |\omega_j - \omega_0| \gg |\omega_{1,j+1} - \omega_j| \end{array} : \; \omega_{1,j+1} \sim \omega_j + \frac{|g_j|^2}{(\omega_j - \omega_0)\left(1 + \frac{|g_j|^2}{(\omega_j - \omega_0)^2}\right)} \tag{B.5}$$

**Figura B.1**

Para los autovalores que están relativamente próximos a $\omega_0$ la aproximación (B.4) no es válida y se debe calcular numéricamente las raíces de $f_0(x)$, una buena elección es emplear el método de Newton-Raphson, para el que se eligen los valores iniciales:

$$\forall j > 1 : x_j^{(0)} = \begin{cases} \omega_{j-1} - \varepsilon & j-1 < i_0 \\ \omega_{j-1} + \varepsilon & j-1 > i_0 \end{cases} \qquad x_1^{(0)} = \begin{cases} \omega_{i_0} + \varepsilon & f_0(\omega_0) > 0 \\ \omega_{i_0+1} - \varepsilon & f_0(\omega_0) < 0 \end{cases} \tag{B.6}$$

Los $x_j^{(k)}$ son obtenidos mediante la siguiente fórmula de recurrencia:

$$\forall j : x_j^{(k+1)} = x_j^{(k)} - \frac{f_0\left(x_j^{(k)}\right)}{f_0'\left(x_j^{(k)}\right)} \tag{B.7}$$

De esta manera se puede calcular los $N+1$ autovalores $\omega_{1,j}$ a partir de las $N$ frecuencias $\omega_i$ y acoples $g_i$; a partir de cada autovalor $\omega_{1,j}$, $j = 1:N+1$, se puede obtener su respectivo autovector, de componentes $\beta_{i,j}$, mediante el uso recursivo de las ecuaciones (B.2), se inicia creando la primera componente $\beta_{1,j}$ con la ecuación izquierda de (B.2) y luego se crean todas las demás componentes usando la derecha de (B.2), como valores iniciales de las amplitudes se fija la primera componente a la unidad:

$$\begin{aligned} &\beta_{1,j}^{(1)} = 1 \\ &\beta_{i+1,j}^{(k)} = \frac{g_i \beta_{1,j}^{(k)}}{\omega_{1,j} - \omega_i} \quad, i = 1:N \\ &\beta_{i,j}^{(k)} \to \beta_{i,j}^{(k)} \alpha^{-1} \qquad \qquad \alpha = \sqrt{\sum_l \left|\beta_{l,j}^{(k)}\right|^2} \\ &\beta_{1,j}^{(k+1)} = \frac{1}{\omega_{1,j} - \omega_0} \sum_{i=1}^{N} g_i^* \beta_{i+1,j}^{(k)} \end{aligned} \tag{B.8}$$

(B.8) establece el proceso iterativo en el orden mostrado: los valores iniciales se definen solo para las primeras componentes según $\beta_{1,j}^{(1)} = 1$, a partir de esto se generan todas las demás amplitudes $\beta_{i+1,j}^{(1)}$ y se las normaliza, luego se evalúa $\beta_{1,j}^{(2)}$ y nuevamente se evalúan las amplitudes $\beta_{i+1,j}^{(2)}$ y se normalizan, entonces se evalúa el producto escalar entre las amplitudes $\beta_{i,j}^{(1)}$ y $\beta_{i,j}^{(2)}$ (para un $j$ dado) y se evalúa su convergencia a la unidad, el proceso se repite iterativamente, empleando las fórmulas de recurrencia, normalizando y evaluando el producto escalar, hasta que el producto escalar se aproxima a la unidad tanto como se desea.

De (B.8), las amplitudes $\beta_{i,j}^{(1)}$ ya son una buena aproximación (no normalizada), conforme estén mejor calculados los autovalores[27] $\omega_{1,j}$ ; de acuerdo a (B.2) si se elige $\beta_{1,j} = 1$ como un factor de escala, las demás amplitudes son $\beta_{i+1,j} = \frac{g_i}{\omega_{1,j}-\omega_i}$ y se emplean para normalizar $\beta_{1,j}$, cuyos módulos al cuadrado vienen a ser:

$$|\beta_{1,j}|^2 = \left(1 + \sum_{i=1}^{N} \left|\frac{g_i}{\omega_{1,j} - \omega_i}\right|^2\right)^{-1} \tag{B.9}$$

(B.9) da la distribución de probabilidad como el módulo al cuadrado de las primeras componentes $\beta_{1,j}$ de cada autovector (o los cuadrados de los módulos de la primera fila de la matriz unitaria que realiza la diagonalización). Es evidente que el término entre paréntesis es: $1 + \sum_{i=1}^{N} \left|\frac{g_i}{\omega_{1,j}-\omega_i}\right|^2 = -f_0'(x = \omega_{1,j})$, de manera que $|\beta_{1,j}|^2 = -1/f_0'(x_j)$, donde $x_j = \omega_{1,j}$ son las raíces de $f_0(x)$; el término $-1/f_0'(x_j)$ es la pendiente de la recta normal a la curva $f_0(x)$ en $x_j$ .

---

[27] En principio, para los valores exactos de los autovalores solo bastaría una iteración y la correspondiente normalización, pero se realizan algunas iteraciones más para precisar el valor de las amplitudes.

# Anexo C

# Cálculo de la entropía de Von Neumann del sistema

En esta sección se establece el cálculo de la entropía de Von Neumann $S(\hat{\rho})$ establecido en la ecuación (2.5), la matriz de densidad del sistema $\hat{\rho}(t)$ viene dado por la ecuación (6.41). La ecuación (2.5) establece la invariancia de la entropía frente a transformaciones unitarias, de manera que $S(\hat{\rho})$ puede ser evaluado en otra matriz de densidad $\hat{\rho}_D(t)$ que resulta de diagonalizar $\hat{\rho}(t)$, mediante $\hat{U}$, como se muestra:

$$S(t) = S(\hat{\rho}(t)) = S(\hat{\rho}_D(t)) \qquad \hat{\rho}_D(t) = \hat{U}\hat{\rho}(t)\hat{U}^\dagger \qquad (C.1)$$

En general, para una matriz de densidad $\hat{\rho}$ de orden 2x2, se tienen los autovalores:

$$\hat{\rho} = \begin{pmatrix} 1-p & q^* \\ q & p \end{pmatrix} \qquad \Delta = (1-2p)^2 + 4|q|^2$$

$$p_1 = \frac{(1-\sqrt{\Delta})}{2} \qquad p_2 = \frac{(1+\sqrt{\Delta})}{2} \qquad (C.2)$$

$$|\psi_1\rangle = \begin{pmatrix} 1-2p-\sqrt{\Delta} \\ 2q \end{pmatrix} \qquad |\psi_2\rangle = \begin{pmatrix} 1-2p+\sqrt{\Delta} \\ 2q \end{pmatrix}$$

En (C.2) los kets $|\psi_i\rangle$ son los autovectores ortogonales no normalizados de $\hat{\rho}$, con autovalores $p_i$ que son las probabilidades de que $\hat{\rho}$ se encuentre en el estado puro $|\psi_i\rangle\langle\psi_i|$; Así, se tiene la mezcla: $\hat{\rho}_D = p_1|\psi_1\rangle\langle\psi_1| + p_2|\psi_2\rangle\langle\psi_2|$, y de acuerdo a (2.5) y (C.1) la entropía $S(\hat{\rho})$ viene a ser:

$$S(\hat{\rho}) = -p_1 \ln p_1 - p_2 \ln p_2 \qquad (C.3)$$

Para la matriz de densidad de interés (6.41), al comparar con (C.2) se tiene $p \to p(t) = |\beta|^2|\Gamma(t)|^2$ y $q \to q(t) = \alpha^*\beta\,\Gamma(t)$, así $\Delta(t) = (1-2|\beta|^2|\Gamma(t)|^2)^2 + 4|\alpha^*\beta\,\Gamma(t)|^2$, entonces $\Delta(t)$ y los autovalores $p_1$ y $p_2$ de (C.2) son:

$$\Delta(t) = 1 + 4|\beta|^4(|\Gamma(t)|^4 - |\Gamma(t)|^2) \qquad p_{1,2}(t) = \frac{(1 \mp \sqrt{\Delta(t)})}{2} \qquad (C.4)$$

Reemplazando (C.4) en (C.3) se obtiene $S(t) = S(\hat{\rho}(t))$ la entropía de Von Neumann del sistema, para la matriz de densidad (6.41), de forma analítica y exacta.

# Anexo D

# Caso particular de $N = 4$

En esta sección se evalúa la solución exacta para un caso particular de un oscilador (sistema) de frecuencia $\omega_0$ acoplándose con 4 osciladores de frecuencias $\omega_i$ (entorno), mediante un mismo valor de acople $g_s$ a la vez que los osciladores del entorno se acoplan entre sí mediante un mismo valor de acople $g_e$, evaluando la razón $g_e/g_s$ desde valores pequeños hasta próximos a la unidad. Así, el hamiltoniano (4.18) viene a ser:

$$\widehat{H} = \sum_{i,j=0}^{4} g_{ij} \hat{a}_j^\dagger \hat{a}_i \qquad g_{ij} = \begin{cases} g_s & i \vee j = 0 \quad i \neq j \\ g_e & i \wedge j > 0 \quad i \neq j \\ \omega_i & i = j \end{cases} \qquad (D.1)$$

De acuerdo a la sección [4.3.3], el diagrama de acción del Hamiltoniano en $\mathcal{H}(1)$ es:

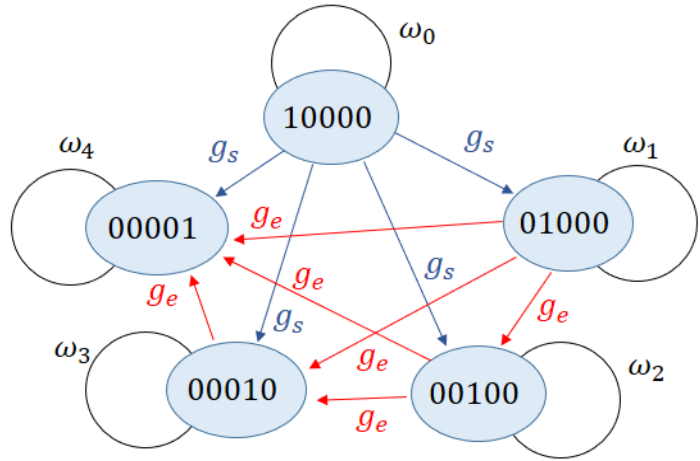

**Figura D.1**

La ecuación de autovalores en este subespacio es:

$$\begin{pmatrix} \omega_0 & g_s & g_s & g_s & g_s \\ g_s & \omega_1 & g_e & g_e & g_e \\ g_s & g_e & \omega_2 & g_e & g_e \\ g_s & g_e & g_e & \omega_3 & g_e \\ g_s & g_e & g_e & g_e & \omega_4 \end{pmatrix} \begin{pmatrix} \beta_{1,1,j} \\ \beta_{2,1,j} \\ \beta_{3,1,j} \\ \beta_{4,1,j} \\ \beta_{5,1,j} \end{pmatrix} = \omega_{1,j} \begin{pmatrix} \beta_{1,1,j} \\ \beta_{2,1,j} \\ \beta_{3,1,j} \\ \beta_{4,1,j} \\ \beta_{5,1,j} \end{pmatrix} \qquad (D.2)$$

La ecuación (D.2) se resuelve para 5 autovalores y autovectores mediante cálculo numérico, no analítico, lo que demanda definir los elementos de la matriz hamiltoniano:

$$\omega_1 = 0.5 \qquad \omega_2 = 0.75 \qquad \omega_0 = 1 \qquad \omega_3 = 1.25 \qquad \omega_4 = 1.5$$
$$0 \leq g_e \leq g_s \tag{D.3}$$

Con el valor numérico de las amplitudes $\beta_{1,1,j}$ y autovalores $\omega_{1,j}$ se puede evaluar la función $\Gamma(t)$ con $N = 4$ de (6.41) que caracteriza la matriz de densidad del sistema $\hat{\rho}(t)$; en esta sección se grafica $|\Gamma(t)|^2$ en contraste con la probabilidad del estado excitado $p(t) = e^{-J(\omega_0)t}$ de la solución markoviana asociada para este caso ($\beta = 1$), dada por la ecuación (2.47), la densidad espectral $J(\omega)$ es definida según (2.46) para acoples $V_j = g_s$, así se puede asociar $J(\omega_0) = 2\pi|g_s|^2$; entonces las funciones que participan son:

$$\Gamma(t) = \sum_{j=1}^{5} |\beta_{1,1,j}|^2 e^{-i\omega_{1,j}t} \qquad\qquad p(t) = e^{-2\pi|g_s|^2 t} \tag{D.4}$$

A continuación, los gráficos resultantes de evaluar $|\Gamma(t)|^2$ (curva azul) y $p(t)$ (curva naranja), ambas describen la evolución temporal de la probabilidad del estado excitado del sistema, para un conjunto de valores de los acoples $0 \leq g_e \leq g_s = 0.1$ donde la razón $g_e/g_s$ toma los valores cero y uno, no hay autovalores degenerados:

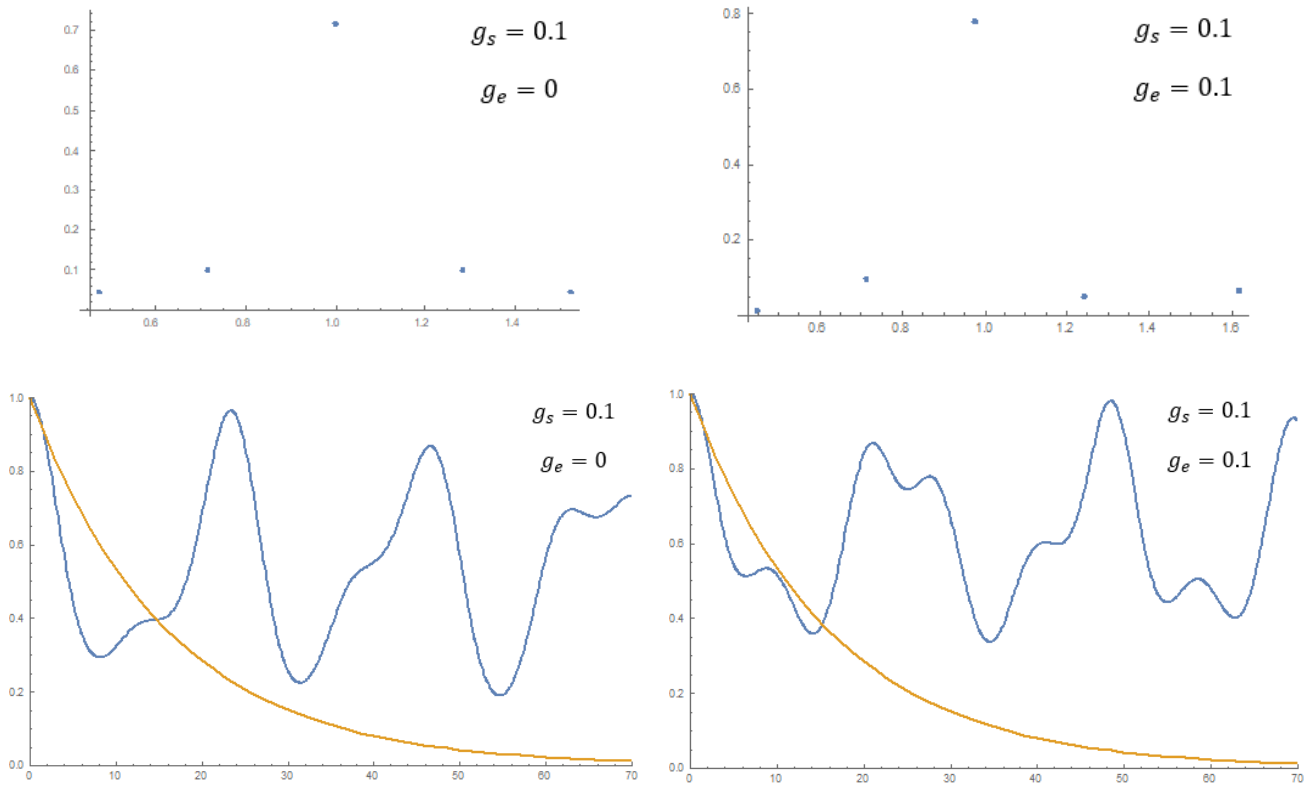

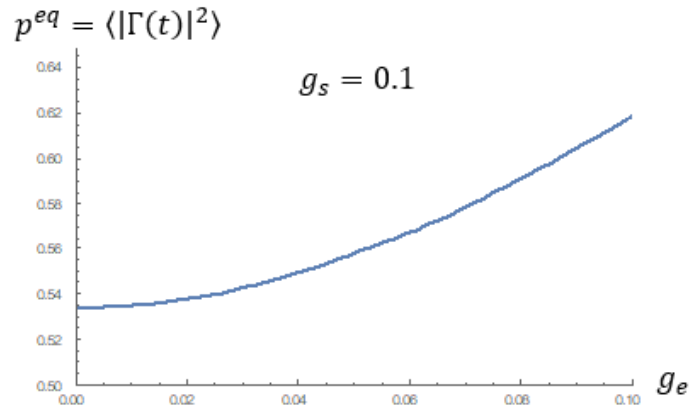

**Figura D.2**

Los gráficos de la Figura D.2 están ordenados como sigue: la columna de la izquierda y derecha (para las dos primeras filas) corresponden a las razones $g_e/g_s = 0$ y $g_e/g_s = 1$ respectivamente, las gráficas de la fila superior corresponden a las probabilidades $|\beta_{1,1,j}|^2$ en el eje vertical, y a los autovalores $\omega_{1,j}$ en el eje horizontal; los gráficos en la segunda fila comparan las funciones $|\Gamma(t)|^2$ y $p(t)$ de (D.4), y el gráfico de la tercera fila es el promedio de la probabilidad en tiempos largos $p^{eq} = \langle|\Gamma(t)|^2\rangle$ en el eje vertical, y el acople $g_e$ en el eje horizontal, observándose un crecimiento continuo.

Es importante resaltar que los acoples $g_s$ son todos iguales, a diferencia de los acoples $g_i$ del hamiltoniano (B.1) los cuales tienden a anularse conforme $i$ se acerca a 1 o $N$, lo cual conduce a que las probabilidades $|\beta_{1,1,j}|^2$ disminuyan conforme $j \to 2$ por la derecha[28] y $j \to N$, que es lo que se aborda en esta tesis, y considerando además el reducido número de partículas, donde $N = 4$, no se justifica el uso de la aproximación markoviana $p(t) = e^{-J(\omega_0)t}$ (curva naranja) para este conjunto de osciladores porque el efecto memoria es muy fuerte debido a la constancia de los acoples $g_s$ y el reducido número de osciladores, de modo que la presencia de la curva naranja $p(t) = e^{-J(\omega_0)t}$ es meramente referencial para ambos casos $g_e = 0$ y $g_e = g_s$; no obstante, un mejor ajuste de ambas funciones $|\Gamma(t)|^2$ y $p(t)$ para $g_e = 0$, yendo desde valores pequeños de $N \sim 1$ hasta valores próximos a $N = 1000$, con los acoples variados $g_i$, en los que sí se compara con la correspondiente solución markoviana, se realiza en el Anexo E.

---

[28] Para $j = 1$ se tiene uno de los valores más altos para $|\beta_{1,1,j}|^2$.

Se evidencia un cambio sutil en la distribución de probabilidad $\left|\beta_{1,1,j}\right|^2$ al aumentar el acople $g_e$ desde 0 hasta $g_s = 0.1$ (primera fila de la Figura D.2), lo cual se expresa en una modificación (no dramática) de la evolución de $|\Gamma(t)|^2$ (curva azul) al aumentar $g_e$ de 0 hasta $g_s$ ; sin embargo, se observa un incremento significativo del promedio de la probabilidad en tiempos largos $p^{eq} = \langle|\Gamma(t)|^2\rangle$, tercera fila de la Figura D.2, conforme se aumenta el acople $g_e$ , cuyos valores extremos en la correspondiente gráfica son:

$$p^{eq}[g_e = 0] = 0.5335 \qquad\qquad p^{eq}[g_e = g_s] = 0.6185 \qquad (D.5)$$

Se deduce que al "activar" los acoplamientos entre osciladores del entorno, se ayuda a redistribuir mejor la energía e información cuántica entre todos los osciladores a la vez que se favorecen las sobrevivencias (*survivals*) de la probabilidad $|\Gamma(t)|^2$, esto es asociado al efecto memoria en el entorno, precisamente contrario al sentido de la aproximación markoviana; se puede decir que al aumentar el valor de los acoples entre los osciladores del entorno aumenta el efecto memoria y mengua cualquier carácter markoviano que pueda ser asociado.

# Anexo E

# Evaluación de $|\Gamma(t)|^2$ para $N: 4 \to 499$

En esta sección se evalúa la función $|\Gamma(t)|^2$ de (6.41), que describe la probabilidad del estado excitado del sistema, para distintos valores de $N$, iniciando en 4 y finalizando en 499, para una densidad espectral (6.1) con fuerza de acoplamiento $\eta = 0.1$ y de $N = 5 \to 99$ para $\eta = 0.01$, en todos los casos son de orden óhmico $s = 1$ y se contrasta con la correspondiente solución markoviana $p(t) = e^{-J(\omega_0)t}$, donde $J(\omega_0) = \frac{2\pi\omega_0}{e}\eta$ de acuerdo a (7.1), aunque sea referencial para valores $N \sim 4$; se elige un intervalo de frecuencias para los osciladores del entorno entre 0 y 2 (en unidades $\omega_0 = 1$). En la Figura E.1 se muestra la densidad espectral para $\eta = 0.1$ y $N = 199$ (izquierda), y la distribución de probabilidad de los autovalores para $\eta = 0.1$ y $N = 99$ (derecha):

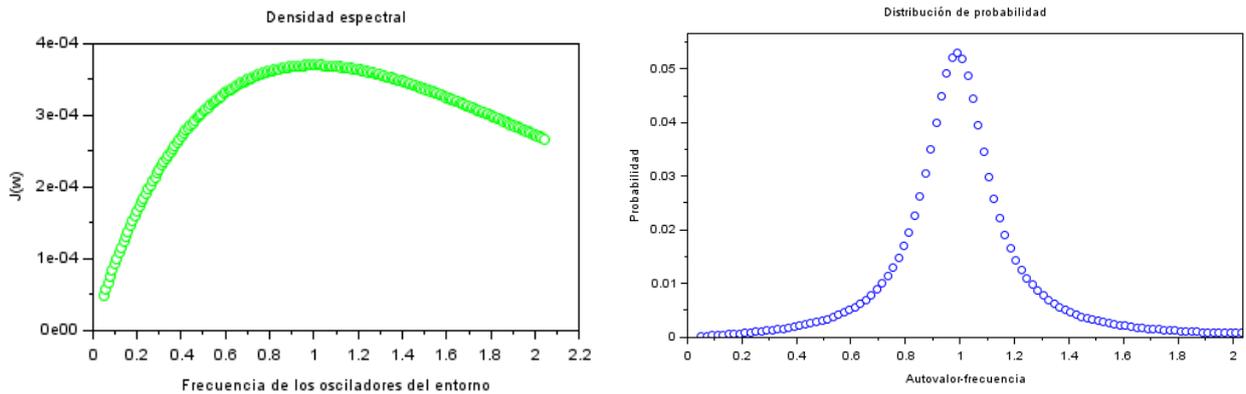

**Figura E.1**

A continuación, las gráficas para $\eta = 0.1$:

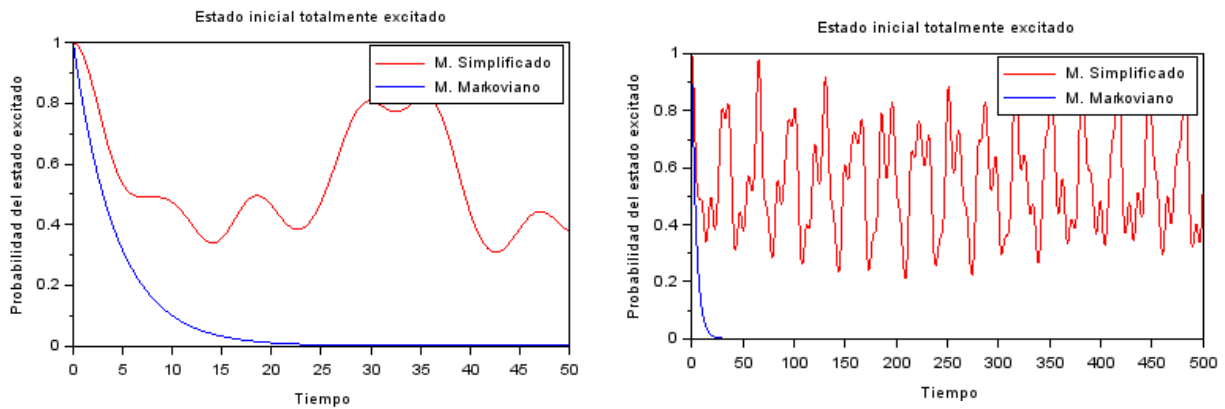

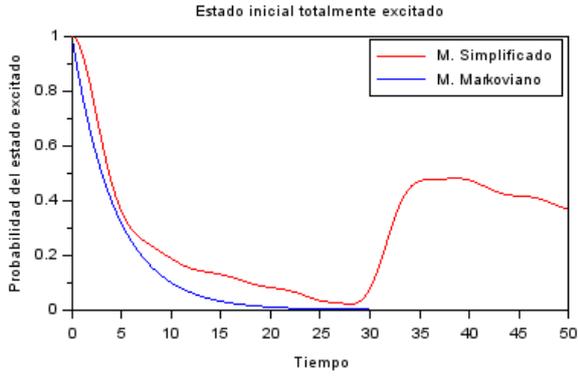
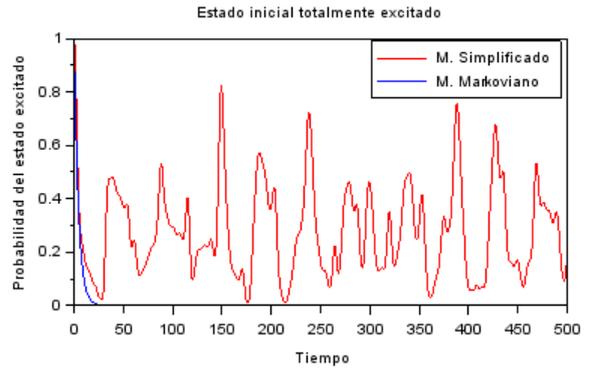
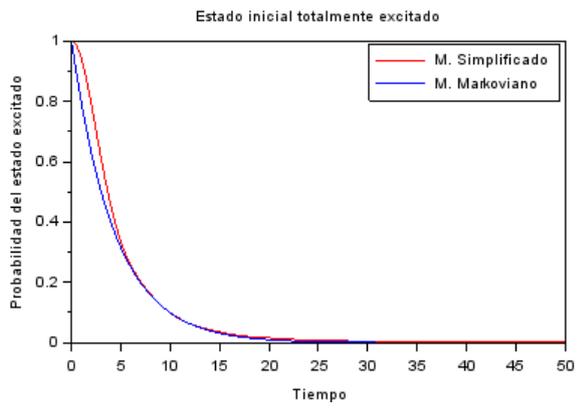
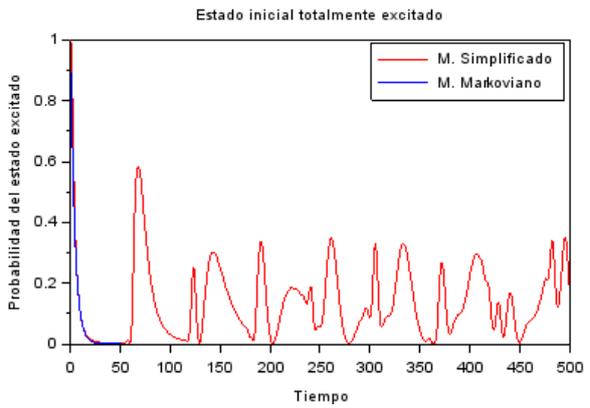
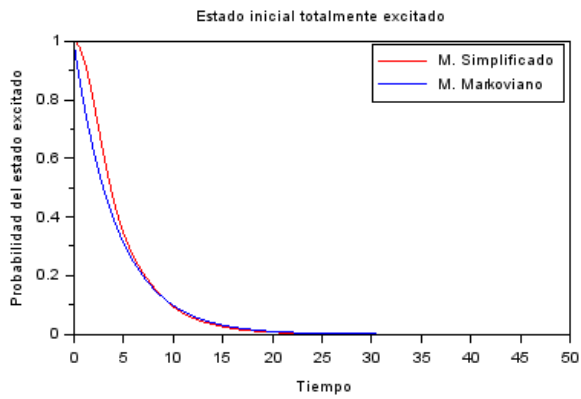
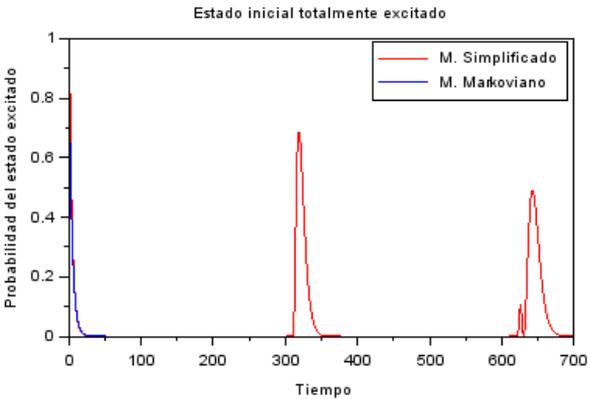
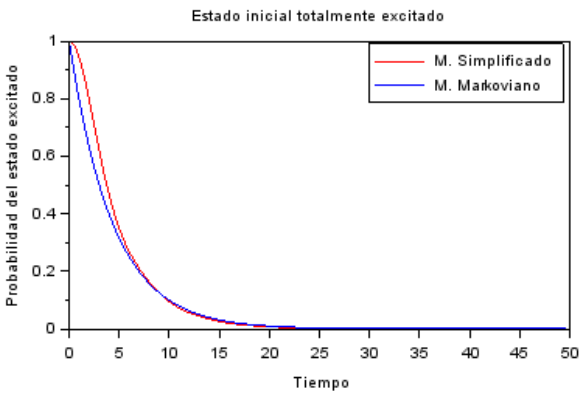
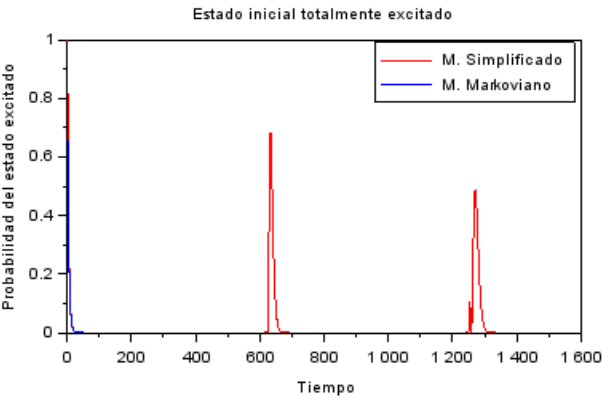

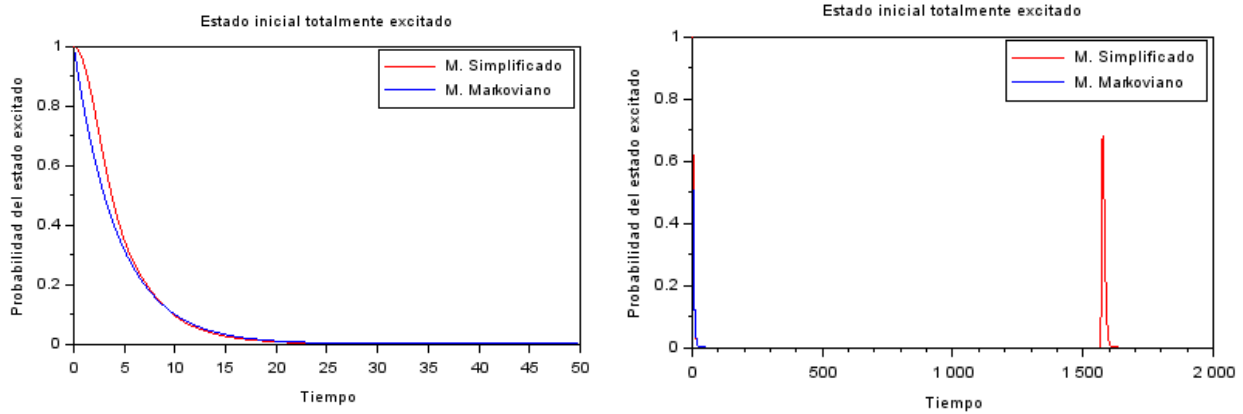

**Figura E.2**

En la Figura E.2 cada fila corresponde a un mismo número $N$, en el siguiente orden: $N = 4, 9, 19, 99, 199, 499$; todas las gráficas de la columna de la izquierda corresponde a una evaluación en tiempos breves, y las de la columna de la derecha corresponde a tiempos largos; Es muy importante notar que conforme aumenta el número $N$ las sobrevivencias disminuyen y la aparición de las mismas se alejan en el tiempo, lo que significa que la aproximación markoviana, de caída exponencial, se mantiene válido en un intervalo mayor de tiempo; esto hace suponer que en el límite $N \to \infty$ las supervivencias de la probabilidad se alejarán hasta el infinito en el tiempo, dejando a la aproximación markoviana válida en todo el dominio del tiempo. Los valores del promedio de la probabilidad son presentados en la siguiente Tabla E.1:

| $N$ | $p^{eq} = \langle |\Gamma(t)|^2 \rangle$ |
|---|---|
| 4 | 0.5532913 |
| 9 | 0.2776405 |
| 19 | 0.1535004 |
| 99 | 0.0300308 |
| 199 | 0.0152131 |
| 499 | 0.0061340 |

**Tabla E.1**

De la tabla E.1 se puede intuir que en el límite $N \to \infty$, el promedio temporal de la probabilidad se anula: $p^{eq} = \langle |\Gamma(t)|^2 \rangle \to 0$.

Para este acoplamiento la solución exacta no coincide del todo con la curva markoviana, esto se consigue para una fuerza de acoplamiento más débil $\eta = 0.01$:

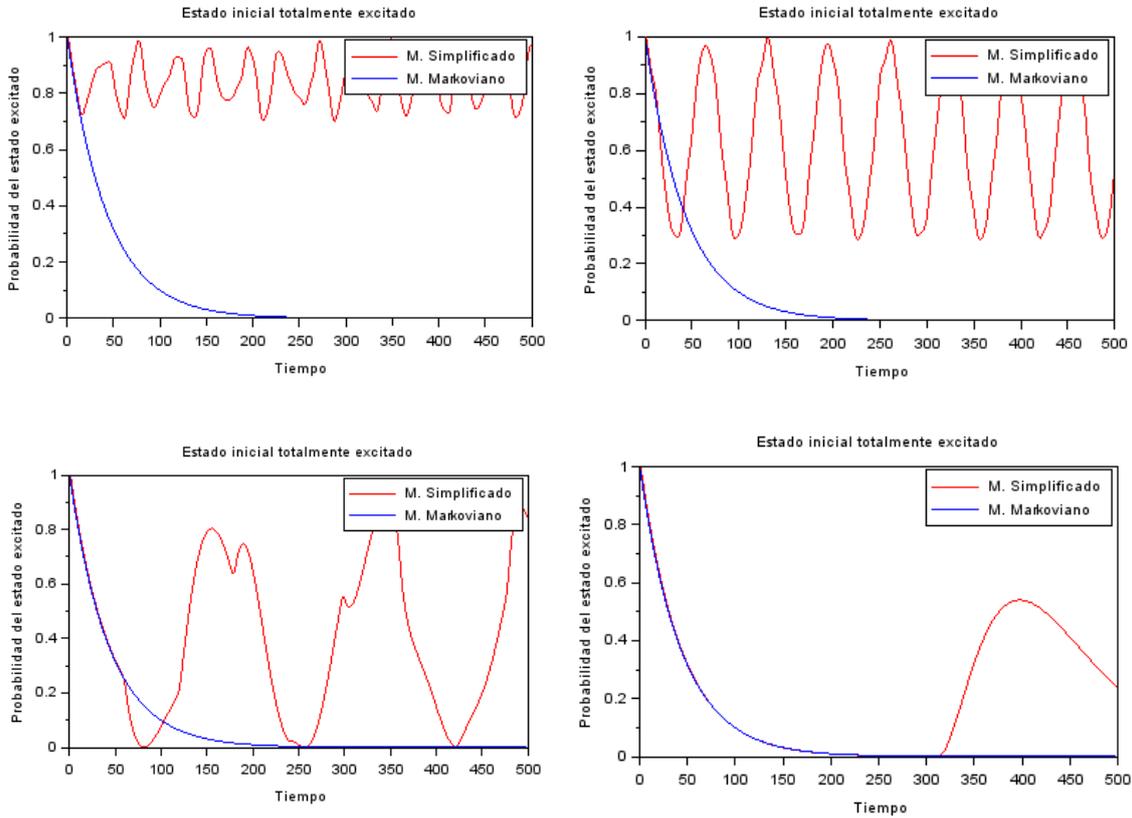

**Figura E.3**

En la Figura E.3 se grafica $|\Gamma(t)|^2$ para los siguientes valores de $N$: superior izquierda $N = 5$, superior derecha $N = 9$, inferior izquierda $N = 19$ e inferior derecha $N = 99$; los valores respectivos de $p^{eq} = \langle |\Gamma(t)|^2 \rangle$ son dados en la Tabla E.2:

| $N$ | $p^{eq} = \langle |\Gamma(t)|^2 \rangle$ | $N$ | $p^{eq} = \langle |\Gamma(t)|^2 \rangle$ |
|---|---|---|---|
| 5 | 0.8422344 | 19 | 0.4090875 |
| 9 | 0.6808924 | 99 | 0.1994525 |

**Tabla E.2**

# BIBLIOGRAFÍA

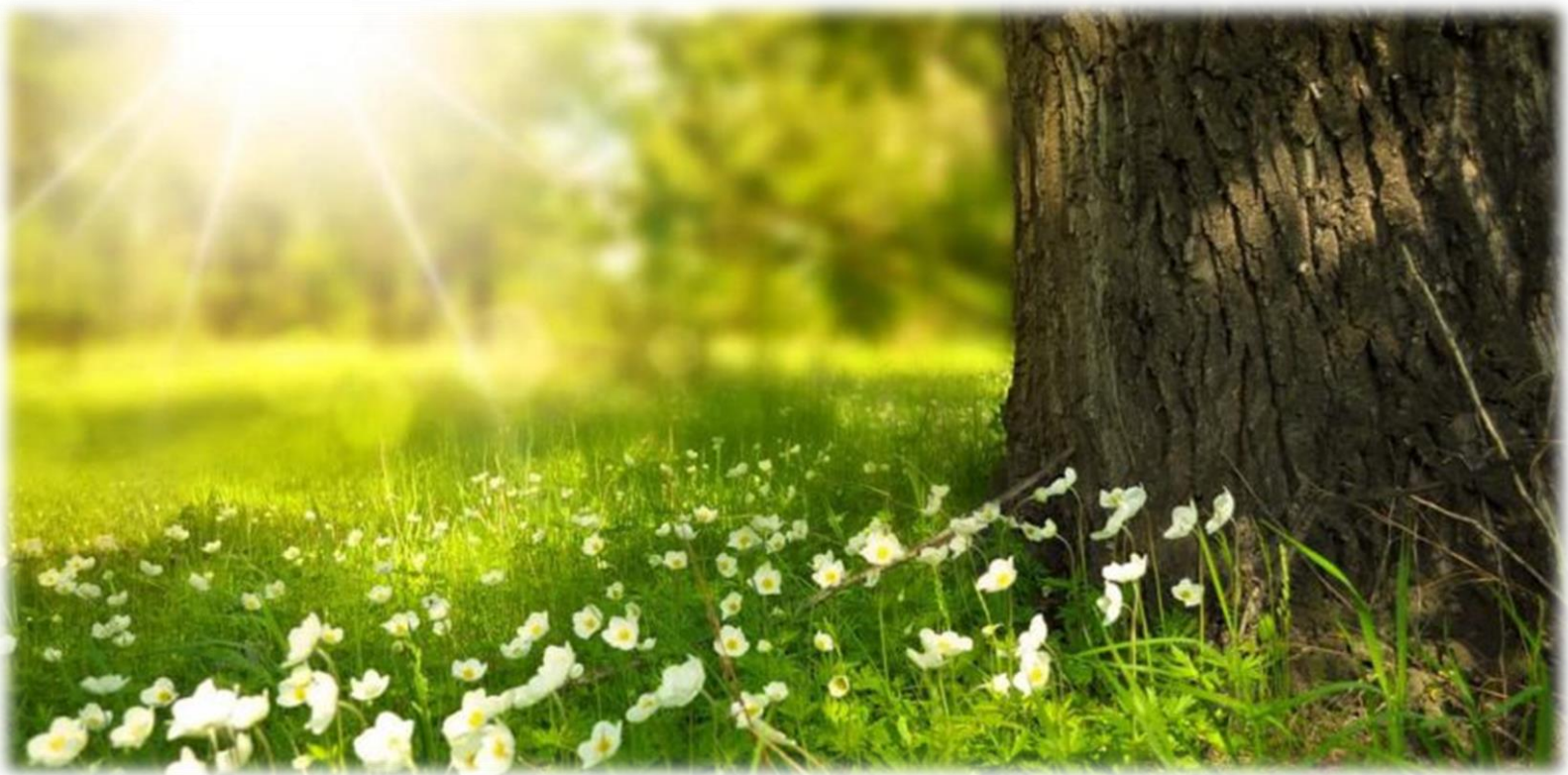

La Física es, probablemente, la disciplina académica más hermosa y elegante de la ciencia, un auténtico triunfo del intelecto humano desde que aparecimos en las cavernas y luchamos por abrirnos paso en el mundo natural mediante el conocimiento y la técnica.

La física es un conjunto de teorías, que superan la cosmovisión aristotélica de la que nacieron debido a que son sometidas a la depuración experimental: si la teoría no coincide con los experimentos es rechazada, sin importar cuán atractiva sea, la autoridad o la inteligente de quien la propuso, y por eso está dotada de poderosas herramientas matemáticas "el lenguaje de la naturaleza" para darle formalidad a las mediciones y observaciones. Las mediciones y el lenguaje formal otorgan la robustez necesaria para soportar el conocimiento más fundamental del que dispone nuestra especie.

Eduardo Sotelo